\documentclass[journal,article,submit,moreauthors,pdftex]{Definitions/mdpi} 

\firstpage{1} 
\pubvolume{xx}
\issuenum{1}
\articlenumber{5}
\pubyear{2020}
\copyrightyear{2020}
\history{Received: date; Accepted: date; Published: date}

\Title{Tetra- and penta-quark structures in the constituent quark model}

\Author{Gang Yang $^{1}$\orcidA, Jialun Ping $^{2}$ and Jorge Segovia $^{3}$}

\AuthorNames{Firstname Lastname, Firstname Lastname and Firstname Lastname}

\address{%
$^{1}$ \quad Department of Physics, Zhejiang Normal University, Jinhua 321004, China; yanggang@zjnu.edu.cn\\
$^{2}$ \quad Department of Physics, Nanjing Normal University, Nanjing 210023, China; jlping@njnu.edu.cn\\
$^{3}$ \quad Dpto. Sistemas F\'isicos, Qu\'imicos y Naturales, U. Pablo de Olavide, E-41013 Sevilla, Spain; jsegovia@upo.es}

\abstract{With the development of high energy physics experiments, a large amount of exotic states in the hadronic sector have been observed. In order to shed some insights on the nature of the tetraquark and pentaquark candidates, a constituent quark model, along with the gaussian expansion method, has been employed systematically in the real- and complex-range investigations. We review herein the double- and full-heavy tetraquarks, but also the hidden-charm, -bottom and doubly charmed pentaquarks. Several experimentally observed exotic hadrons are well reproduced within our approach; moreover, their possible compositeness are suggested and other properties analyzed accordingly such as their decay widths and general patterns in the spectrum. Besides, we report also some predictions not yet seen by experiment within the studied tetraquark and pentaquark sectors.}


\begin{document}


\section{Introduction}

A Large amount of conventional hadrons, 3-quarks baryons and quark-antiquark mesons, could be described well in a constituent quark model~\cite{Vijande:2004he, YYC2008BS, Yang:2017xpp} which was firstly proposed by M. Gell-Mann in 1964~\cite{MGMPL1964}. Especially, this 'toy model' had successfully predicted the existence of the $\Omega$ baryon. However, with decades of experimental efforts in high energy physics, many exotic states have been observed by each collaboration, e.g., BABAR, Belle, BES, CDF, CLEO-c, LHCb, $etc$. The remarkable example should be the $X(3872)$ which was firstly announced through the $B^\pm \rightarrow K^\pm\pi^+ \pi^- J/\psi $ decay by the Belle Collaboration in 2003~\cite{skc:2003prl} and later on, CDF~\cite{CDFX3872}, D0~\cite{D0X3872} and BABAR~\cite{BABARX3872} collaborations confirmed this charmoniumlike state. Meanwhile, dozens of so-called XYZ particles have also been observed worldwide in B factories, $\tau$-charm facilities and hadron-hadron colliders. In particular, $Y(3940)$ was discovered by the Belle collaboration through analyzing the $\omega J/\psi$ invariant mass spectrum in 2004~\cite{BelleY3940}. Besides, a charged charmoniumlike state $Z^+(4430)$ was observed in the $B\rightarrow K\pi^{\pm}\psi(3686)$ decay by the same collaboration in 2008~\cite{BellZ4430}. $Y(4260)$ was firstly observed in the $e^+ e^-\rightarrow \gamma \pi^+ \pi^- J/\psi$ channel by the BABAR collaboration in 2005~\cite{BABARY4260} and this state was confirmed by the CLEO~\cite{CLEOY4260} and Belle~\cite{BelleY4260} collaborations within the same decay process. Additionally, the other exotic states such as $Y(4140)$, $Y(4274)$, $Z^+(4051)$ and $Z^+(4200)$, $etc$., are all collected in the Particle Data Group~\cite{PDG2018}. Among the tetraquark sector, a fully heavy 4-body system $QQ\bar{Q}\bar{Q}$ is quite appealing. In 2017, a benchmark measurement of the $\Upsilon(1S)$ pair production at $\sqrt{s}$=8 TeV in $pp$ collision was implemented by the CMS collaboration~\cite{vkams:2017jhep}, then an excess at 18.4 GeV in the $\Upsilon \ell^+ \ell^-$ decay channel was proposed in a subsequent preliminary investigation using the CMS data~\cite{SD2018PHD, SD2018II, KY2018}. Moreover, a significant peak at $\sim$$18.2\,\text{GeV}$ was also seen in Cu+Au collisions at RHIC~\cite{LCBland2019}. However, no evidence has been found in the $\Upsilon(1S)\mu^+ \mu^-$ invariant mass spectrum by the investigation of LHCb collaboration~\cite{raba:2018jhep}. Hence, a fully-bottom tetraquark state will be possible if this structure could be confirmed by more experiments in the future. Nevertheless, a new possible fully-charm structure at 6.9 GeV and a broad one around $6.2$$\sim$$6.8\,\text{GeV}$ were reported in the di-$J/\psi$ invariant mass spectrum by the LHCb collaboration very recently~\cite{LA2020CERNINDICO}. 

The achievements in baryon sector are also plentiful. In $2015$, the LHCb Collaboration reported the observation of two hidden-charm pentaquark states in the $J/\psi p$ invariant mass spectrum through the $\Lambda^{0}_{b} \rightarrow J/\psi K^{-}p$ decay~\cite{Aaij:2015tga}. One state is labeled as $P_{c}(4380)^+$ whose mass is $(4380\pm8\pm29)\,\text{MeV}$ and width is $(205\pm18\pm86)\,\text{MeV}$, another one is $P_{c}(4450)^+$ with mass and width $(4449.8\pm1.7\pm2.5)\,\text{MeV}$ and $(39\pm5\pm19)\,\text{MeV}$, respectively. These resonances were supported by the same collaboration in a subsequent model-independent study~\cite{A2016MI}. Furthermore, with much more statistical significance in the same decay channel, $\Lambda^{0}_{b} \rightarrow J/\psi K^{-}p$, made by the LHCb collaboration in 2019~\cite{lhcb:2019pc}, the original observed structure at 4450 MeV is resolved into two narrow peaks at 4440 and 4457 MeV. Accordingly, these two pentaquark states are marked as $P^+_c(4440)$ and $P^+_c(4457)$. Their widths are $20.6\pm4.9^{+8.7}_{-10.1}\,\text{MeV}$ and $6.4\pm2.0^{+5.7}_{-1.9}\,\text{MeV}$, respectively. Besides, a new narrow pentaquark $P^+_c(4312)$ ($\Gamma=9.8\pm2.7^{+3.7}_{-4.5}\,\text{MeV}$) was also reported in the $\Lambda^0_b$ decay, $\Lambda^0_b \rightarrow J/\psi K^- p$. Meanwhile, according to the $b\to c\overline{c}s$ weak decay process, $b$-flavored pentaquark states were searched in the final states $J/\psi K^+\pi^- p$, $J/\psi K^- \pi^- p$, $J/\psi K^- \pi^+ p$, and $J/\psi \phi p$ by the LHCb collaboration~\cite{Aaij:2017jgf}.

These prominently experimental findings triggered extensive theoretical investigations on the multiquark systems. Firstly, in the double-heavy tetraquarks sector, the dimeson with $bb\bar{u}\bar{d}$ constituents had already been proposed~\cite{jclh:1988prd} in 1988. After then, a narrow $bb\bar{u}\bar{d}$ tetraquark state with the $J^P=1^+$ is obtained in heavy quark limit~\cite{ejecq:2017prl} and the theoretical mass is $10389\pm 12\,\text{MeV}$~\cite{mkjlr:2017prl}. Additionally, Ref.~\cite{EHJVAVJMR2019} also calculated their mass, lifetime and decay modes  in a quark model formalism. Furthermore, this double-bottom tetraquark state with $I(J^P)=0(1^+)$ is also predicted by a relativistic quark model~\cite{dernfvogwl2007} and to be seen in heavy-ion collisions at the LHC~\cite{cefgk:2019prd}. For the antiparticle case, a $\bar{b}\bar{b}ud$ bound state which is stable against the  strong decay is proposed by Lattice QCD~\cite{llsm:2019prd}, the predicted mass is $10476\pm 24\pm 10\,\text{MeV}$ and also has the same spin-parity of $J^P=1^+$. Besides, this deeply bound tetraquark state is also supported by the investigations with same formalism in Refs.~\cite{afrjhrlkm2017, pjnmmp2019}. For the other kinds of tetraquarks, $e. g.$, $bc\bar{u}\bar{d}$ axial-vector tetraquark state is predicted at $7105\pm 155\,\text{MeV}$ in the QCD sum rules~\cite{ssaka:2019arx}, $ud\bar{c}\bar{d}$ tetraquark in $I(J^P)=0(1^+)$ state with $15$ to $61\,\text{MeV}$ binding energy respecting to $\bar{D}B^*$ threshold is claimed by Ref.~\cite{afrjhrlkm2019}, decay properties of open-bottom and doubly heavy tetraquarks are investigated in Refs.~\cite{ssakahs:2019prd, yyjp:2019prd, zgw:2019arx, YXRZ2018, YXFSYRZ2019}, the production of double-heavy tetraquarks at a Tera-Z factory and the LHC are estimated by Monte Carlo simulation~\cite{aaaypqqww2018, aaqqww2018}. Moreover, a new $QQ\bar{s}\bar{s}$ tetraquark states are investigated in the chiral quark model in Ref.~\cite{GYangQQSS, YTWLJP2020} recently.

The debates on fully-heavy tetraquarks are even more intensely. Numerous theoretical investigations through various approaches expect the confirmation of $QQ\bar{Q}\bar{Q}$ tetraquark states. In particular, the fully-charm and -bottom tetraquarks are the most concerned. For instance, the ground state of $bb\bar{b}\bar{b}$ tetraquarks mass is $18.72\pm0.02\,\text{GeV}$ by a non-relativistic effective field theory investigation~\cite{mna:2018epjc}. Similarly, theoretical mass of fully-bottom tetraquarks $\sim$$18.8\,\text{GeV}$ within $0^+$ and $2^+$ states are suggested in QCD sum rules~\cite{zgwqqqq:2017epjc}. Besides, this exotic meson in $0^+$ state also with a predicted mass at $\sim$$18.8\,\text{GeV}$ is confirmed by a diffusion Monte Carlo approach~\cite{yb:2019plb} and symmetry analysis~\cite{mksnjl:2017prd}. Moreover, the existence of $bb\bar{b}\bar{b}$ tetraquark is supported by relativized~\cite{mabjfcdres2019}, non-relativistic quark model~\cite{avb:2012prd} and QCD sum rules~\cite{wchxc:2017plb} too. This tetraquark state is estimated around $0.1\,\text{GeV}$ below threshold in diquark-antidiquark configuration studied by Ref.~\cite{aeap:2018epjc}. Very recently, we also performed a systematic study on the $QQ\bar{Q}\bar{Q}$ tetraquark states in an effective potential model which is based on the results of Lattice QCD investigation on heavy quark pairs~\cite{TKSSPRD2012} and there is a hint for possible deeply bound $bb\bar{b}\bar{b}$ tetraquark states around $18.0\,\text{GeV}$~\cite{GYangFHT2020}. However, we still do not robustly conclude the existence of this exotic state for the potential is obtained only in two-body sector and no color-dependent interaction is considered therein.

The fully-charm tetraquarks even draws more attention due to the recent report on an observation of new structure at $6.9\,\text{GeV}$ and another broad one around $6.2$$\sim$$6.8\,\text{GeV}$ in the di-$J/\psi$ invariant mass spectrum by the LHCb collaboration~\cite{LA2020CERNINDICO}, besides, there is also a hint for structure $\sim$7.2 GeV. Actually, the possibility of finding $cc\bar{c}\bar{c}$ tetraquark at LHC had already been investigated on the cross section of $pp\rightarrow 2J/\psi+X$ at $\sqrt{s}=7\,\text{TeV}$ by Ref.~\cite{avbakl:2011prd} in 2011. In additional, a narrow $cc\bar{c}\bar{c}$ tetraquark resonance in the mass region $5-\,6\,\text{GeV}$ was predicted by means of Bethe-Salpeter~\cite{whge:2012plb} in 2012, and this fully-charm tetraquarks within $0^+$ and $2^+$ states at $\sim$$6.0\,\text{GeV}$ were also predicted in QCD sum rules~\cite{zgwqqqq:2017epjc} and Ref.~\cite{mksnjl:2017prd}. The conclusion is supported by a non-relativistic model studies~\cite{avb:2012prd, vrdfsn:2019cpc} too. 

In the recently theoretical investigations, mass of the S-wave $cc\bar{c}\bar{c}$ tetraquark within a energy region from 5.96 to 6.32 GeV is concluded by a non-relativistic diquark-antidiquark model~\cite{PLTO2006}. Besides, the two structures at 6.5 GeV and 6.9 GeV are identified as the S- and P-wave fully-charm tetraquarks, respectively, in a potential model~\cite{MSLFXLXHZQZ2020}. However, the opposite conclusions are obtained in a dynamical study by means of diquark model~\cite{JFGRFL2008}. Furthermore, the radial excitation states favor the two structures in both QCD sum rules~\cite{ZGW2020FHT}, relativized quark model~\cite{QDYFHT2020} and string junction picture~\cite{MKJLR2020FC}. In constituent quark model~\cite{GYangFHT2020, XJTX2020}, spin-parity of the structure at 6.9 GeV is suggested to be $2^+$. Molecular configurations for these two structures are proposed by perturbation QCD~\cite{RMASNARDRGR2008}, however, compact tetraquark structure is obtained in Holography inspired stringy hadron model~\cite{JSDW2008}. Meanwhile, the experimental data and theoretical importance on the $cc\bar{c}\bar{c}$ tetraquarks are reviewed in the articles~\cite{JR2008FHT, KCSZ2008}. Apart from the spectrum of the fully-charm tetraquarks, their strong decay properties~\cite{ HWXSFHT2020}, productions via $\bar{p}p$ annihilation reaction~\cite{XWQLHXYXYHXQ2007} and dynamical simulation~\cite{JZDCXLTM2008} are also extensively investigated. However, there are still a large amount of investigations against the observation of a bound state in fully-charm and -bottom tetraquarks, $e. g.$, within effective model investigations~\cite{jmrav:2017prd, jwyrl:2018prd, xc:2019epja, mslqfl:2019prd, gjw:2019arx, jmravjv2018} and Lattice QCD study~\cite{cheec:2018prd}. Nevertheless, possible stable or narrow states in the $bb\bar{b}\bar{c}$ and $bc\bar{b}\bar{c}$ sectors are available by quark models studies~\cite{jmrav:2017prd, jwyrl:2018prd}. The features of color-magnetic and Coulomb interactions, along with different color configurations within the fully-heavy tetraquark states are studied in Ref.~\cite{CDHCJP2019} by various models. In addition, the charmoniumlike and bottomoniumlike states $Z_c(3900)$, $Z_c(4020)$, $Z_b(10610)$ and $Z_b(10650)$ are well identified as the $D^{(*)}\bar{D}^*$ and $B^{(*)}\bar{B}^*$ molecular resonances, respectively, in a chiral effective field theory study~\cite{BWLMSZ2020}.

As for the single heavy tetraquark state, the LHCb collaboration recently reported two structures, $X_0(2900)$ and $X_1(2900)$, by the $B^{\pm} \rightarrow D^+D^-K^{\pm}$ decay channel with a model-depend~\cite{SALHCbM20201} and a model-independent~\cite{SALHCbM20202} investigations, respectively.
  This fact may indicate the first evidence of a open-charm tetraquark state which quark content is $cs\bar{u}\bar{d}$. Meanwhile, this open charm state has been predicted at 2850 MeV in color-magnetism model~\cite{JCSLYLZSTY2020} and a coupled channel unitary approach~\cite{QMTBEO2010}, which is in advance of the experimental results. Recently, extensive theoretical investigations also devote to this subject, and the spin-parity for the lower state, $X_0(2900)$, is generally favored to be $0^+$. In particular, $X_0(2900)$ can be identified as a radial excitation of $ud\bar{s}\bar{c}$ tetraquark state in a two-body potential model~\cite{XHWWRZ2020}. Besides, QCD sum rules~\cite{HCWCRDNS2020, SSAKAHS2020}, one-boson-exchange model~\cite{MLJXLG2020, JHDC2020}, quark delocalization color screening model~\cite{YXXJHHJP2020}, effective field theory~\cite{RMEO2020} and strong decay investigation~\cite{YHJLJXLG2020} all suggest the explanation of S-wave $\bar{D}^*K^*$ ($D^*\bar{K}^*$) molecule state. Refs.~\cite{JZ2020OC, ZW2020OC}, which are also in a QCD sum rules approach, confirm the $0^+$ quantum state, but obtain a compact tetraquark state. Besides, this $cs\bar{u}\bar{d}$ tetraquark in $00^+$ state is also supported by ref.~\cite{MKJLR2020OC}. Furthermore, there are still several different views on these structures reported by the LHCb collaboration, $e. g.$, $X_0(2900)$ is identified as a $\bar{D}^*K^*$ molecule state with $I(J^P)=0(1^+)$~\cite{MHXLPLQW2020OC}, triangle singularity induces the $X$ exotic states~\cite{XLMYHKGLJX2020OC, TJBESS2020}, and the compact $ud\bar{s}\bar{c}$ tetraquark in $0^+$ state disfavors $X_0(2900)$ in an extended relativized quark model~\cite{LDCYD2020OC}. Additionally, several investigations~\cite{XHWWRZ2020, HCWCRDNS2020, JHDC2020} prefer the $J^P=1^-$ for higher state, $X_1(2900)$.
  
 The observations of the hidden-charm pentaquarks $P^+_c(4380)$, $P^+_c(4312)$, $P^+_c(4440)$ and $P^+_c(4457)$ bring great interest in theoretical investigations. Especially, during 2010 to 2013 which is before the announcement of $P^+_c(4380)$ by the LHCb collaboration~\cite{Aaij:2015tga}, several narrow hidden-charm resonances $\sim$4.3 GeV were predicted by means of coupled-channel unitary studies~\cite{PRL1051, PRL1052, Oset1, Oset2, Oset3} and possible loosely bound hidden-charm molecular states were discussed in the one-boson-exchange model~\cite{CPC36}. Then great deals of subsequent theoretical works devoted to the interpretation of the nature this exotic state, particularly, $\Sigma^{(*)}_c\bar{D}^{(*)}$ molecular state with quantum numbers $I(J^P)=\frac{1}{2}({\frac{3}{2}}^-)$ is preferred by the boson exchange model~\cite{RChen}, the constituent quark model~\cite{Yang:2015bmv, HHXPC2016}, the Bethe-Salpeter equation~\cite{JHe}, QCD sum rules~\cite{HXChen, ZGWang, KAYSHS2017}, $etc$. The spin-parity of $\frac{3}{2}^-$ is also suggested in a diquark-triquark model investigation~\cite{RZCQ2016}. Furthermore, some other non-resonance explanations were also proposed such as kinematic effects and triangle singularities~\cite{MBPCTA2016, XHLPCTS20162, XHLPCTS2016}. Strong decay property of $P^+_c(4380)$ is studied in a molecular configuration~\cite{KAYSHS2018SD}.
 
 Furthermore, in 2019 the three newly announced pentaquarks $P^+_c(4312)$, $P^+_c(4440)$ and $P^+_c(4457)$ by the LHCb collaborattion~\cite{lhcb:2019pc} triggered many theoretical investigations again. The main interpretation with $\Sigma^{(*)}_c\bar{D}^{(*)}$ molecular configurations are provided by effective field theory~\cite{MZL190311560, JH190311872}, QCD sum rules~\cite{ZGW190502892}, potential models~\cite{ZHG190400851, HH190400221, HM1904.09756, RZ190410285, MIE190411616, XZW190409891}, heavy quark spin multiplet structures~\cite{YS190400587, CWX190401296} and heavy hadron chiral perturbation theory~\cite{LM190504113}, $etc.$ Moreover, the production~\cite{XC190406015, XYW190411706} and decay properties~\cite{CJX190400872} of these pentaquarks were also investigated. Therein, through Tables III and IV presented in our systematically study on the hidden-charm pentaquark states in 2017~\cite{Yang:2015bmv}, one could notice that the new reported states are described well if with the following assignments that $P^+_c(4312):$ $\frac{1}{2}^- \Sigma_c\bar{D}$, $P^+_c(4440):$ $\frac{1}{2}^- \Sigma_c\bar{D}^*$ and $P^+_c(4457):$ $\frac{3}{2}^- \Sigma_c\bar{D}^*$, their isospins are all of $I=\frac{1}{2}$. These conclusions are supported by the other subsequent theoretical investigations~\cite{MZL190311560, JH190311872, CWX190401296, CJX190400872}.
  Meanwhile, there are also many works devoted to the investigations on other kinds of pentaquark states. $E. g.$, $\bar{Q}qqqq$ bound state is unavailable in a quark model formalism~\cite{JMR190103578}. However, narrow resonances in doubly heavy pentaquarks are possible in potential models~\cite{QSZ180104557, FG190304430, gy:2020dcp}. Besides, triply charmed pentaquarks like $\Xi_{cc}D^{(*)}$ molecular state is suggested within one-boson-exchange model~\cite{FLW190101542} and QCD sum rules~\cite{KAYSHS2018STP}.
  Additionally, some general reviews on the exotic states of tetraquark and pentaquark can be referenced in Refs.~\cite{JV190209799, YRL190311976}.
  
  Apparently, much more investigations on the exotic hadronic states in the future experiments $e. g.$, ATLAS, CMS and LHCb collaborations are necessary. In this review, we mainly focus on a summary of the doubly-, fully-heavy tetraquarks and hidden-, doubly-heavy pentaquarks which were all systematically studied in the framework of constituent quark model. By comparing various calculated bound and resonance states of tetraquark and pentaquark in one theoretical framework, some general or universal features on multi-quark systems are expected, and with a purpose in shedding some insights on the future investigations in hadron physics both experimentally and theoretically.
  
The structure of this article is organized as follows: Sec.~\ref{sec:model} devotes to the theoretical framework where our constituent quark model and wave-functions of tetra- and penta-quark states are illustrated. Then theoretical results along with discussions on each kinds of tetraquarks and pentaquarks are presented in Sec.~\ref{sec:results}, respectively. The last section is about a summary.
 

\section{Theoretical Framework}\label{sec:model}


Among all the methods in dealing with the issues on hadron physics, which is located in the QCD's non-perturbative energy region, the QCD-inspired phenomenology approach, constituent quark model is still a powerful and major way applied to the baryon and meson spectra~\cite{Valcarce:1995dm, Vijande:2004he, Segovia:2008zza, Segovia:2008zz, Ortega:2016hde, Yang:2017xpp}, hadron-hadron interaction~\cite{Fernandez:1993hx, Valcarce:1994nr, Ortega:2009hj, Ortega:2016mms, Ortega:2016pgg} and exotic states~\cite{Vijande:2006jf, Yang:2015bmv, Yang:2017rpg, gy:2020dht, Yang:2018oqd}, $etc$. Therefore, the S-wave systems of doubly-heavy tetraquark, hidden-, doubly-heavy pentaquark in each allowed quantum numbers $I(J^P)$ states are analyzed in this formalism. Meanwhile, the widely accepted Lattice QCD which is based on the first principle has also made prominent achievements in studing the multi-quark systems~\cite{Alexandrou:2001ip, Okiharu:2004wy} and the hadronic interactions~\cite{Prelovsek:2014swa, Lang:2014yfa, Briceno:2017max}. Hence a potential model which is according to the Lattice QCD investigations on the interaction of $Q\bar{Q}$ pair is also employed in the investigation on fully-heavy tetraquark states. Particularly, the doubly-, fully-heavy tetrauqarks with spin-parity $J^P=0^+$, $1^+$ and $2^+$, and in $I=0$ or $1$ isospin sectors; the hidden-charm, bottom and doubly charmed pentaquarks  with quantum numbers $J^P=\frac12^-$, $\frac32^-$ and $\frac52^-$, and in the $I=\frac12$ or $\frac32$ isospin sectors are investigated.

\begin{figure}[H]
\centering
\includegraphics[width=9 cm]{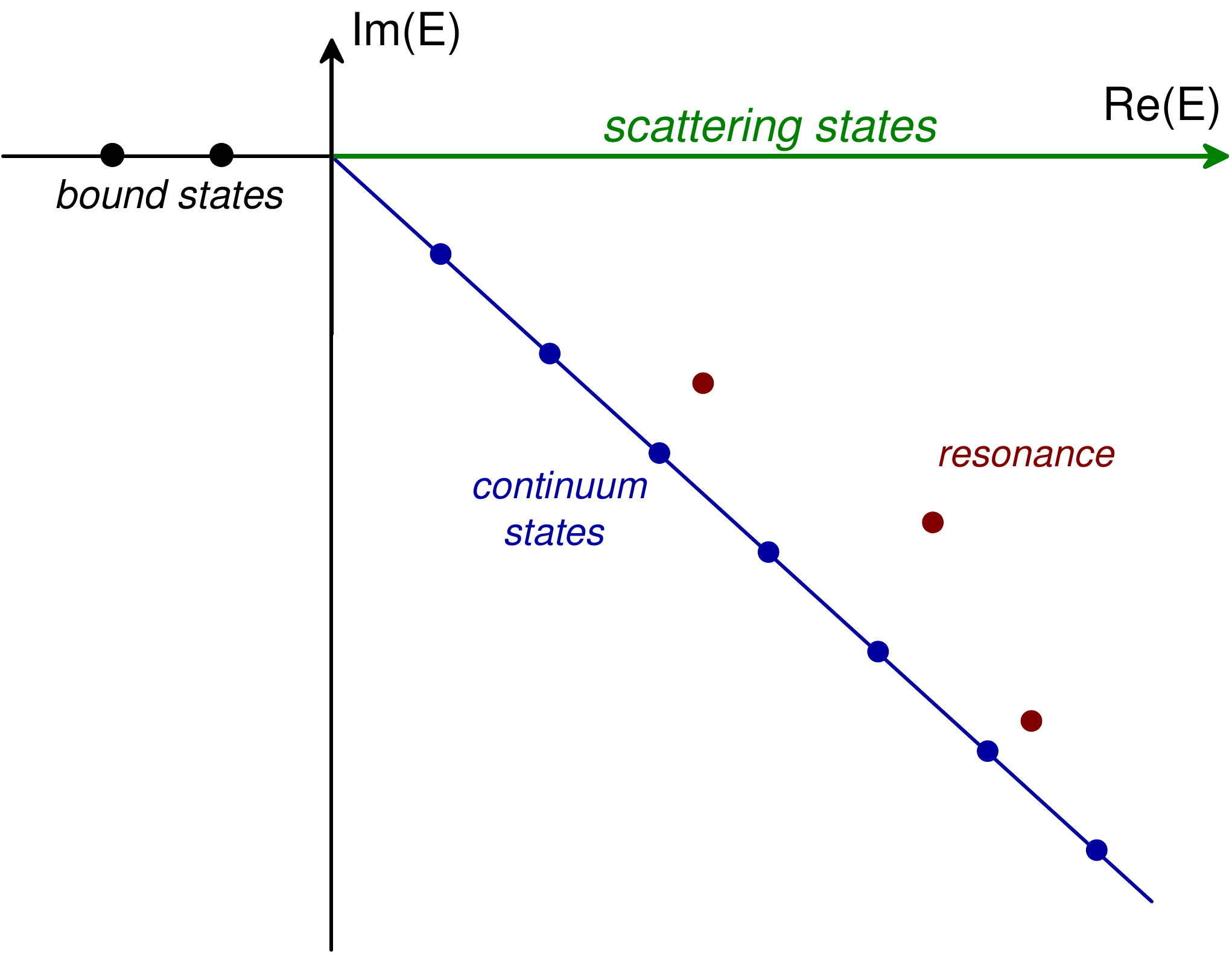}
\caption{Schematic complex energy distribution in the single-channel two-body system.} \label{CSMPP}
\end{figure}

The wave-function of multi-quark systems are exactly constructed in the non-relativistic quantum mechanics range. Specifically, wave-functions of the color, spin, flavor and spatial degrees of freedom are considered in all possible meson-meson color-singlet, hidden-color, diquark-antidiquark and K-types channels for 4-quark systems, baryon-meson color-singlet and hidden-color channels for 5-quark systems. Besides, the couplings of these different configurations in one system are also considered. When solving the eigenvalue problem on 4- and 5-quark systems, both the real- and complex-range calculations are implemented. In particular, the bound, resonance and scattering states can be classified simultaneously in the later framework according to the so-called ABC theorem~\cite{JA22269, EB22280}. 
The crucial manipulation in the complex scaling method (CSM) is to transform the coordinates of relative motions between quarks with a  complex rotation $\vec{r} \rightarrow \vec{r} e^{i\theta}$, and then a complex scaled Schr\"{o}dinger equation $\left[ H(\theta)-E(\theta) \right] \Psi(\theta)=0$ will be solved. In particular, Fig.~\ref{CSMPP} presents a schematic distribution of the complex energy of 2-body system by the CSM according to Ref.~\cite{TMPPNP7912014}.

\subsection{Chiral Quark Model}
The general form of a N-body Hamiltonian can be written as
\begin{equation}
H = \sum_{i=1}^{N}\left( m_i+\frac{\vec{p\,}^2_i}{2m_i}\right) - T_{\text{CM}} + \sum_{j>i=1}^{N} V(\vec{r}_{ij}) \,,
\label{eq:Hamiltonian}
\end{equation}
where the kinetic energy of central mass $T_{\text{CM}}$ is subtracted during calculation and this is owing to the internal relative motions of systems are crucial. Besides the two-body potential in a chiral quark model
\begin{equation}
\label{CHQMVV}
V(\vec{r}_{ij}) = V_{\text{CON}}(\vec{r}_{ij}) + V_{\text{OGE}}(\vec{r}_{ij}) + V_{\chi}(\vec{r}_{ij}) \,,
\end{equation}
contains the color confinement, one gluon exchange and Goldstone-Boson exchange interactions. Furthermore, only the central parts of potential listed in Eq.~(\ref{CHQMVV}) are considered, the spin-orbit and tensor contributions are ignored at present.

Firstly, color-confining should be encoded in the non-Abelian gauge feature of QCD. On one hand, multi-gluon exchanges induce a linearly rising attractive potential which is proportional to the distance between two infinite-heavy quarks has been demonstrated by the investigation of Lattice QCD~\cite{Bali:2005fu}. On the other hand, light-quark pairs spontaneously created in the QCD vacuum may also lead to a breakup of the created color flux-tube at the same scale~\cite{Bali:2005fu}. Accordingly, these two phenomenological features are mimicked in the expression:
\begin{equation}
V_{\text{CON}}(\vec{r}_{ij}\,)=\left[-a_{c}(1-e^{-\mu_{c}r_{ij}})+\Delta \right] 
(\vec{\lambda}_{i}^{c}\cdot\vec{\lambda}_{j}^{c}) \,,
\label{eq:conf}
\end{equation}
particularly, $a_{c}$, $\mu_{c}$ and $\Delta$ are the chiral quark model parameters, $\lambda^c$ represents the Gell-Mann matrices in SU(3) color. One can see that we have a linear potential with an effective confinement strength $\sigma = -a_{c} \, \mu_{c} \, (\vec{\lambda}^{c}_{i}\cdot \vec{\lambda}^{c}_{j})$ if two quarks are extremely close; however, it will turns to be a constant at large distance.

Secondly, the one gluon exchange potential which includes a coulomb interaction and a color-magnetism one is given by
\begin{align}
&
V_{\text{OGE}}(\vec{r}_{ij}) = \frac{1}{4} \alpha_{s} (\vec{\lambda}_{i}^{c}\cdot
\vec{\lambda}_{j}^{c}) \Bigg[\frac{1}{r_{ij}} 
-\frac{1}{6m_{i}m_{j}} (\vec{\sigma}_{i}\cdot\vec{\sigma}_{j}) 
\frac{e^{-r_{ij}/r_{0}(\mu)}}{r_{ij}r_{0}^{2}(\mu)} \Bigg] \,,
\end{align}
where $m_{i}$ is the constituent quark mass and the Pauli matrices in spin degree of freedom are denoted by $\vec{\sigma}$. The contact term of spatial part in color-magnetism interaction has been regularized as
\begin{equation}
\delta(\vec{r}_{ij})\sim\frac{1}{4\pi r_{0}^{2}}\frac{e^{-r_{ij}/r_{0}}}{r_{ij}} \,,
\end{equation}
with $r_{0}(\mu_{ij})=\hat{r}_{0}/\mu_{ij}$ a regulator which depends on the reduced quark mass $\mu_{ij}$.

According to Ref.~\cite{Segovia:2013wma}, a parameterized scheme for the QCD strong coupling constant $\alpha_s$ is used herein and the detail is 
\begin{equation}
\alpha_{s}(\mu_{ij})=\frac{\alpha_{0}}{\ln\left(\frac{\mu_{ij}^{2}+\mu_{0}^{2}}{\Lambda_{0}^{2}} \right)} \,,
\end{equation}
in which $\alpha_{0}$, $\mu_{0}$ and $\Lambda_{0}$ are all of the model parameters.

Finally, the central parts of chiral potentials which include the pion, kaon, $\eta$ and $\sigma$ exchange interactions are written as below
\begin{align}
&
V_{\pi}\left( \vec{r}_{ij} \right) = \frac{g_{ch}^{2}}{4\pi}
\frac{m_{\pi}^2}{12m_{i}m_{j}} \frac{\Lambda_{\pi}^{2}}{\Lambda_{\pi}^{2}-m_{\pi}
^{2}}m_{\pi} \Bigg[ Y(m_{\pi}r_{ij})
-\frac{\Lambda_{\pi}^{3}}{m_{\pi}^{3}}
Y(\Lambda_{\pi}r_{ij}) \bigg] (\vec{\sigma}_{i}\cdot\vec{\sigma}_{j})\sum_{a=1}^{3}(\lambda_{i}^{a}
\cdot\lambda_{j}^{a}) \,, \\
& 
V_{\sigma}\left( \vec{r}_{ij} \right) = - \frac{g_{ch}^{2}}{4\pi}
\frac{\Lambda_{\sigma}^{2}}{\Lambda_{\sigma}^{2}-m_{\sigma}^{2}}m_{\sigma} \Bigg[
Y(m_{\sigma}r_{ij})
-\frac{\Lambda_{\sigma}}{m_{\sigma}}Y(\Lambda_{\sigma}r_{ij})
\Bigg] \,, \\
& 
V_{K}\left( \vec{r}_{ij} \right)= \frac{g_{ch}^{2}}{4\pi}
\frac{m_{K}^2}{12m_{i}m_{j}} \frac{\Lambda_{K}^{2}}{\Lambda_{K}^{2}-m_{K}^{2}}m_{
K} \Bigg[ Y(m_{K}r_{ij})
-\frac{\Lambda_{K}^{3}}{m_{K}^{3}}Y(\Lambda_{K}r_{ij})
\Bigg] (\vec{\sigma}_{i}\cdot\vec{\sigma}_{j})\sum_{a=4}^{7}(\lambda_{i}^{a}
\cdot\lambda_{j}^{a}) \,, \\
& 
V_{\eta}\left( \vec{r}_{ij} \right) = \frac{g_{ch}^{2}}{4\pi}
\frac{m_{\eta}^2}{12m_{i}m_{j}} \frac{\Lambda_{\eta}^{2}}{\Lambda_{\eta}^{2}-m_{
\eta}^{2}}m_{\eta} \Bigg[ Y(m_{\eta}r_{ij})
-\frac{\Lambda_{\eta}^{3}}{m_{\eta}^{3}
}Y(\Lambda_{\eta}r_{ij}) \Bigg] (\vec{\sigma}_{i}\cdot\vec{\sigma}_{j})
\Big[\cos\theta_{p} \left(\lambda_{i}^{8}\cdot\lambda_{j}^{8}
\right)-\sin\theta_{p} \Big] \,,
\end{align}
where $Y(x)=e^{-x}/x$ is the Yukawa function and $\lambda^{a}$ is the SU(3) flavor matrix of Gell-Mann.
By introducing an angle $\theta_p$, the physical $\eta$ meson is considered, meanwhile $m_{\pi}$, $m_{K}$ and $m_{\eta}$ are the experimental masses of the SU(3) Goldstone-bosons. As for the $\sigma$ term which is simulated according to the $\pi \pi$ resonance, its value is determined by the PCAC relation $m_{\sigma}^{2}\simeq m_{\pi}^{2}+4m_{u,d}^{2}$~\cite{Scadron:1982eg}.
Finally, the chiral coupling constant, $g_{ch}$ is determined from the $\pi NN$ coupling constant as following
\begin{equation}
\frac{g_{ch}^{2}}{4\pi}=\frac{9}{25}\frac{g_{\pi NN}^{2}}{4\pi} \frac{m_{u,d}^{2}}{m_{N}^2} \,,
\end{equation}
which assumes that SU(3) flavor is an exact symmetry only broken by the different mass of the strange quark.

The chiral quark model parameters are listed in Table~\ref{modelChQM} and they have been fixed in advance reproducing hadron~\cite{Valcarce:1995dm, Vijande:2004he, Segovia:2008zza, Segovia:2008zz, Ortega:2016hde, Yang:2017xpp}, hadron-hadron ~\cite{Fernandez:1993hx, Valcarce:1994nr, Ortega:2009hj, Ortega:2016mms, Ortega:2016pgg} and multiquark~\cite{Vijande:2006jf, Yang:2015bmv, Yang:2017rpg} phenomenology.
In particular, with an application in the study of hidden-charm pentaquark states in Ref.~\cite{Yang:2015bmv}, not only the $P^+_c(4380)$ but also the later three newly observed $P^+_c$ particles are all successfully interpreted.

\begin{table}[H]
\caption{\label{modelChQM} Chiral quark model parameters.}
\centering
\begin{tabular}{cccc}
\toprule
~~~~Quark masses~~~~     & ~~~~$m_u=m_d$ (MeV)~~~~ &  ~~~~313~~~~ \\
                 & $m_s$ (MeV)     & 555 \\
                 & $m_c$ (MeV)     & 1752 \\
                 & $m_b$ (MeV)     & 5100 \\[2ex]
Goldstone-bosons & $\Lambda_\pi=\Lambda_\sigma~$ (fm$^{-1}$) &   4.20 \\
                 & $\Lambda_\eta=\Lambda_K$ (fm$^{-1}$)     &   5.20 \\
                 & $g^2_{ch}/(4\pi)$                         &   0.54 \\
                 & $\theta_P(^\circ)$                        & -15 \\[2ex]
Confinement      & $a_c$ (MeV)         & 430\\
                 & $\mu_c$ (fm$^{-1})$ &   0.70\\
                 & $\Delta$ (MeV)      & 181.10 \\[2ex]
                 & $\alpha_0$              & 2.118 \\
                 & $\Lambda_0~$(fm$^{-1}$) & 0.113 \\
OGE              & $\mu_0~$(MeV)        & 36.976\\
                 & $\hat{r}_0~$(MeV~fm) & 28.170\\
\bottomrule
\end{tabular}
\end{table}

\subsection{Cornell Potential}
In the fully-heavy tetraquark systems $QQ\bar{Q}\bar{Q}$ ($Q=c, b$), the interplay between a pair of heavy quarkonium can be well approximated by the Cornell potential (linear confinement and Coulomb interactions) along with a spin-spin dependent interaction according to the Lattice QCD investigation~\cite{TKSSPRD2012}. Generally, this concise character can be incorporated into the following form for four-body systems,
\begin{equation}
H = \sum_{i=1}^{4}\left( m_i+\frac{\vec{p\,}^2_i}{2m_i}\right) - T_{\text{CM}} + \sum_{j>i=1}^{4} V(\vec{r}_{ij}) \,,
\label{eq:Hamiltonian}
\end{equation}
where the center-of-mass kinetic energy $T_{\text{CM}}$ is also subtracted without losing a generality
as the case in the chiral quark model. Besides, the two-body interactions read as 
\begin{equation}\label{CQMV}
V_{Q\bar{Q}}(\vec{r}_{ij}) = -\frac{\alpha}{\vec{r}_{ij}}+\sigma \vec{r}_{ij}+\beta e^{-\vec{r}_{ij}}(\vec{s}_{i}\cdot\vec{s}_{j})  \,.
\end{equation}
The three parameters $\alpha$, $\sigma$ and $\beta$ in Eq.~(\ref{CQMV}) which relate to the coulomb, confinement and spin-spin interactions are determined by Ref.~\cite{HFUECHENC2020} and their values are listed in Table~\ref{modelLQCDP}. Additionally, Table~\ref{Mmeson} presents the theoretical and experimental masses of the S-wave $Q\bar{Q}$ mesons, apparently, the deviations for each states are acceptable.
 Meanwhile, based on the investigations by quark model and lattice QCD~\cite{CAPFOJ2002}, the quark-quark interaction $V_{QQ}$ is just half of $V_{Q\bar{Q}}$ . This conclusion will be employed in our study in the fully-heavy tetraquark states.

\begin{table}[H]
\caption{\label{modelLQCDP} Potential model parameters.}
\centering
\begin{tabular}{cccc}
\toprule
~~~~Quark masses~~~~    & ~~~~$m_c$ $(MeV)$~~~~     & ~~~~1290~~~~ \\
                 & $m_b$ $(MeV)$     & 4700 \\[2ex]
Coulomb  & $\alpha$    & 0.4105 \\[2ex]
Confinement   & $\sigma$($GeV^2$)    & 0.2 \\[2ex]
Spin-Spin    & $\gamma$ $(GeV)$         & 1.982\\
                 & $\beta_{c}$ $(GeV)$      & 2.06\\
                 & $\beta_{b}$ $(GeV)$          & 0.318\\
\bottomrule
\end{tabular}
\end{table}

\begin{table}[H]
\caption{\label{Mmeson} Theoretical and experimental masses of the S-wave $Q\bar{Q}$ mesons, unit in MeV.}
\centering
\begin{tabular}{ccc}
\toprule
  ~~~~\textbf{State}~~~~ & ~~~~$\textit{\textbf{M}}_{th}$~~~~    &  ~~~~$\textit{\textbf{M}}_{exp}$~~~~ \\
\midrule
  $\eta_c(1S)$  & 2968  & 2981 \\
  $\eta_c(2S)$  & 3655  & 3639 \\
  $J/\psi(1S)$  & 3102  & 3097 \\
  $\psi(2S)$  & 3720  & 3686 \\
  $\eta_b(1S)$  & 9401  & 9398 \\
  $\eta_b(2S)$  & 9961  & 9999 \\
  $\Upsilon(1S)$  & 9463  & 9460 \\
  $\Upsilon(2S)$  & 9981  & 10023 \\
\bottomrule
\end{tabular}
\end{table}

\subsection{Wave-function of Multi-quark System}
There are four degrees of freedom in the quark level: color, spin, flavor and spatial. A complete antisymmetry N-quarks wave-function which fulfills the Pauli principle is written as
\begin{equation}
\label{TWs}
\Psi_{JM_J,I,i,j,k}={\cal A} \left[ \left[ \psi_{L} \chi^{\sigma_i}_{S} \right]_{JM_J} \chi^{f_j}_I \chi^{c}_k \right] \,.
\end{equation}
In Eq.~(\ref{TWs}) $\psi_{L}$, $\chi^{\sigma_i}_{S}$, $\chi^{f_j}_I$ and $\chi^{c}_k$ stand for the spatial, spin, flavor and color wave-functions, respectively. Besides, $\cal{A}$ is the antisymmetry operator of system by considering the nature of two identical particles interchange $i. e.$, $qq$ in SU(3) flavor and $QQ$ in charm or bottom sectors, $etc$.

\begin{figure}[H]
\centering
\includegraphics[width=10 cm]{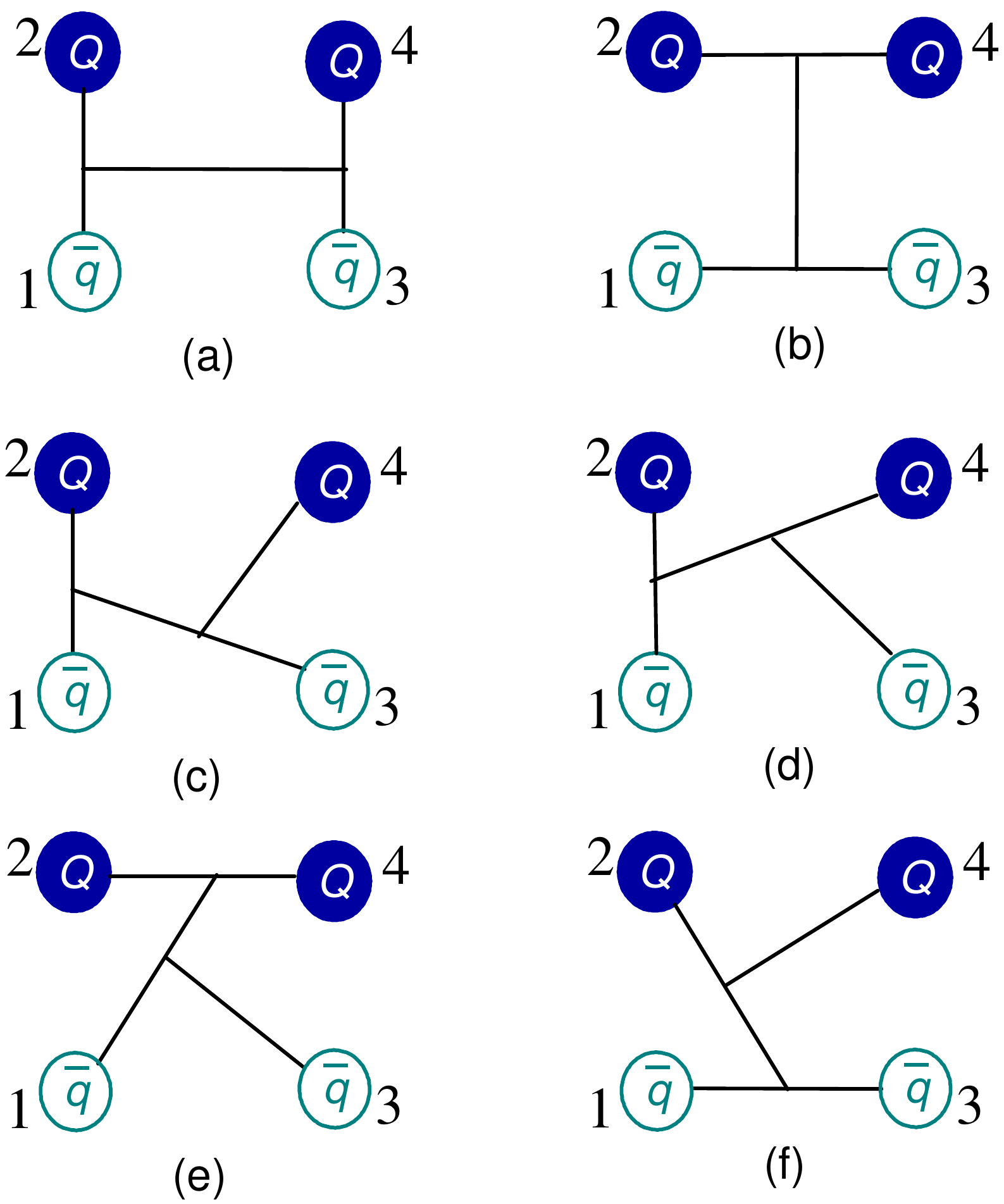}
\caption{Six types of configurations in $QQ\bar{q}\bar{q}$ tetraquarks. Panel $(a)$ is meson-meson structure, panel $(b)$ is diquark-antidiquark one and the other K-type structures are from panel $(c)$ to $(f)$. $(Q=c,b; q=u, d, s)$.} \label{4QCOF}
\end{figure}

Fig.~\ref{4QCOF} shows six configurations in the double-heavy tetraquark states. In particular, Panel (a) is the dimeson structure, panel (b) is the diquark-antidiquark configuration and the other panels are of the K-types. Furthermore, the light flavor antiquarks $\bar{q}$ can be naturally switched with $\bar{Q}$ in the fully-heavy sectors. Accordingly, there are four exchange terms included in the antisymmetry operator for both the double-heavy and fully-heavy tetraquark states which the two quarks and antiquarks should be the same flavor and read as
\begin{equation}
{\cal{A}} = 1-(13)-(24)+(13)(24) \,.
\end{equation}
However, due to the asymmetry between $c$- and $b$-quark, there are only two exchanges for the $\bar{q}c\bar{q}b$ system, namely
\begin{equation}
{\cal{A}} = 1-(13) \,.
\end{equation}

\begin{figure}[H]
\centering
\includegraphics[width=10 cm]{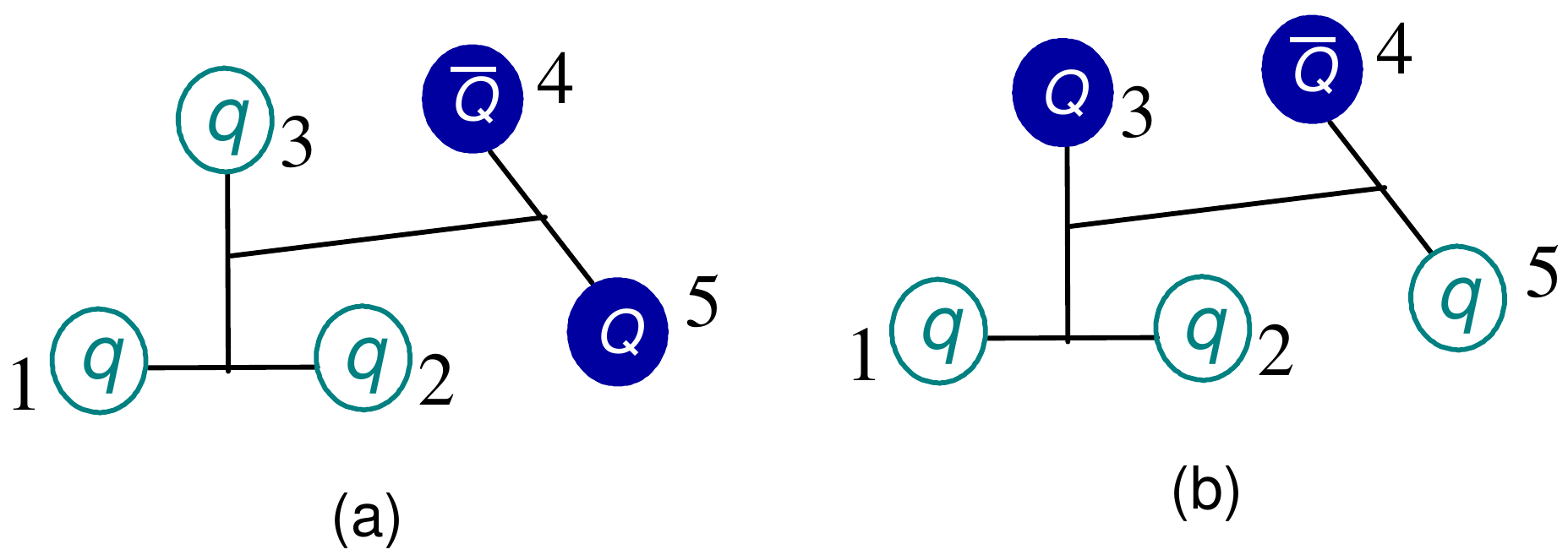}
\caption{Two types of configurations in $qqq\bar{Q}Q$ hidden-flavor pentaquarks. $(Q=c,b; q=u, d, s)$.} \label{5QCOF1}
\end{figure}

\begin{figure}[H]
\centering
\includegraphics[width=10 cm]{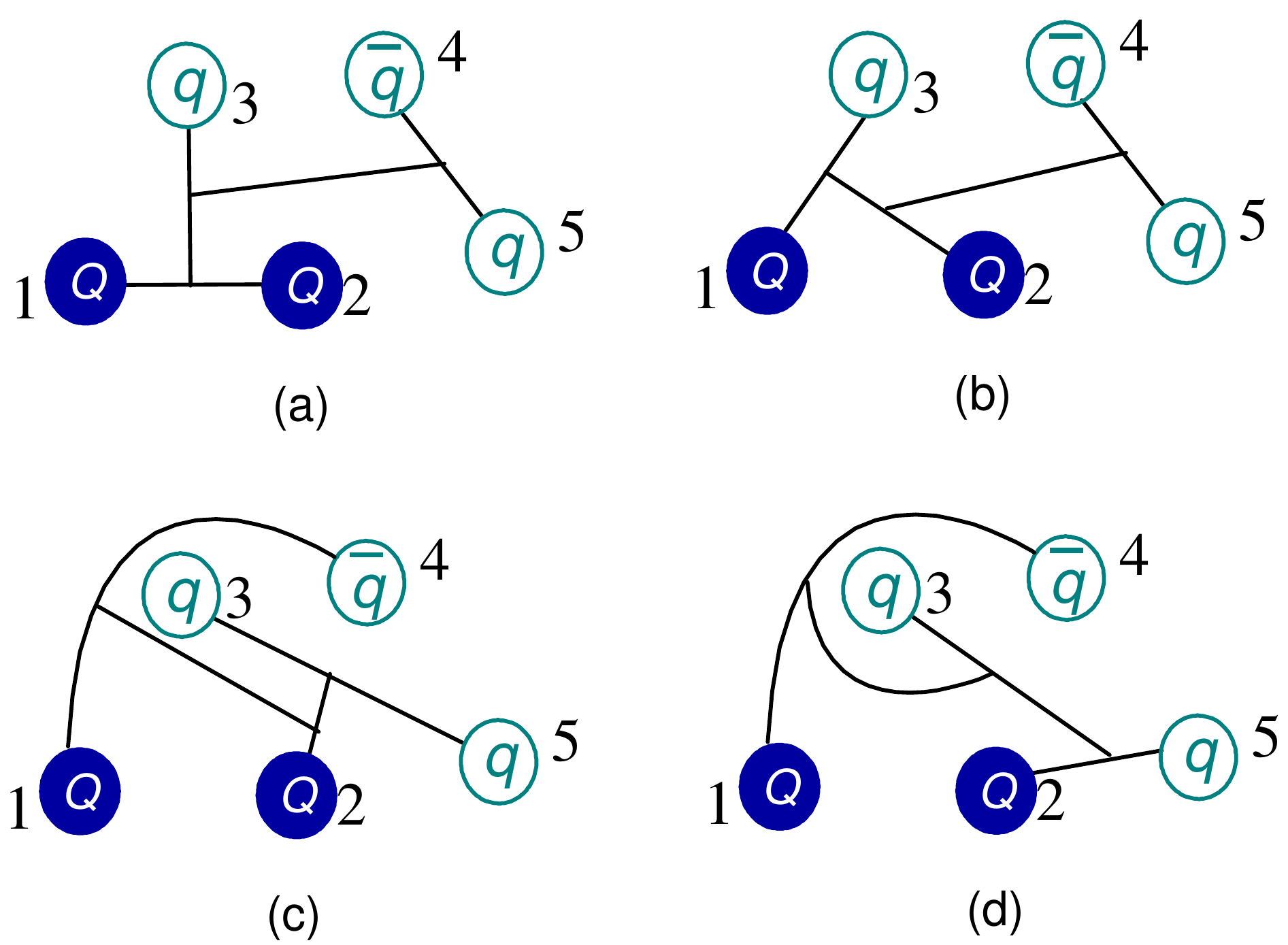}
\caption{Four types of configurations in $QQq\bar{q}q$ doubly-flavor pentaquarks. $(Q=c,b; q=u, d, s)$.} \label{5QCOF2}
\end{figure}

Fig.~\ref{5QCOF1} and~\ref{5QCOF2} present the configurations of hidden-flavor and doubly-heavy flavor pentaquarks, respectively. All of them along with the couplings in calculations are considered. 
For the hidden-flavor pentaquarks presented in Fig.~\ref{5QCOF1}, the quark arrangements $(qqQ)(\bar{Q}q)$ and $(qqq)(\bar{Q}Q)$, we have 
\begin{equation}
{\cal{A}}_1 = 1-(15)-(25) \,,
\end{equation}
for the $(udQ)(\bar{Q}u)+(uuQ)(\bar{Q}d)$ structure, and
\begin{equation}
{\cal{A}}_2 = 1-(13)-(23) \,,
\end{equation}
for the $(uud)(\bar{Q}Q)$ configuration, respectively. However, for the doubly-heavy case in Fig.~\ref{5QCOF2}, the antisymmetry operators are 
\begin{equation}
{\cal{A}}_1 = 1-(35) \,, \label{EE1}
\end{equation}
\begin{equation}
{\cal{A}}_2 = 1-(12)-(35)+(12)(35) \,, \label{EE2}
\end{equation}
\begin{equation}
{\cal{A}}_3 = 1-(12) \,, \label{EE3}
\end{equation}
\begin{equation}
{\cal{A}}_4 = {\cal{A}}_2 \,, \label{EE4}
\end{equation}
and the above equations from (\ref{EE1}) to (\ref{EE4}) represent the results in configurations (a) to (d) of Fig.~\ref{5QCOF2}, respectively. In the following parts, we will introduce the wave-functions of tetraquarks and pentaquarks in each four degrees of freedom.

\subsubsection{Color wave-function}
Much richer color structures in multi-quark systems we will have than those in conventional hadrons ($q\bar{q}$ meson and $qqq$ baryon). The wave-functions in color degree of freedom for each configurations are discussed according to the classification of tetraquark and pentaquark states, respectively. 
\begin{itemize}[leftmargin=*,labelsep=5.8mm, listparindent=2em]
\item	Tetraquark
\end{itemize}

Theoretically, the colorless wave-function of tetraquark in meson-meson configuration presented in Fig.~\ref{4QCOF}(a) can be obtained through two channels, a color-singlet and a hidden-color. However, just the former channel is enough if all spatial excitation states are considered for the multi-quark systems~\cite{Harvey:1980rva, Vijande:2009kj}. Herein, a more economical approach by employing all possible color configurations along with their couplings is favored. Therefore, in the group of $SU(3)$ color, the wave-functions of color-singlet (two color-singlet clusters coupling, $1\times 1$) and hidden-color (two color-octet clusters coupling, $8\times 8$) channels in meson-meson configuration of Fig.~\ref{4QCOF}(a) are marked with $\chi^c_1$ and $\chi^c_2$, respectively,
\begin{align}
\label{Color1}
\chi^c_1 &= \frac{1}{3}(\bar{r}r+\bar{g}g+\bar{b}b)\times (\bar{r}r+\bar{g}g+\bar{b}b) \,,
\end{align}
\begin{align}
\label{Color2}
\chi^c_2 &= \frac{\sqrt{2}}{12}(3\bar{b}r\bar{r}b+3\bar{g}r\bar{r}g+3\bar{b}g\bar{g}b+3\bar{g}b\bar{b}g+3\bar{r}g\bar{g}r
\nonumber\\
&+3\bar{r}b\bar{b}r+2\bar{r}r\bar{r}r+2\bar{g}g\bar{g}g+2\bar{b}b\bar{b}b-\bar{r}r\bar{g}g
\nonumber\\
&-\bar{g}g\bar{r}r-\bar{b}b\bar{g}g-\bar{b}b\bar{r}r-\bar{g}g\bar{b}b-\bar{r}r\bar{b}b) \,.
\end{align}
Meanwhile, as for the diquark-antidiquark channel shown in Fig.~\ref{4QCOF}(b), the color wave-functions  are $\chi^c_3$ (color triplet-antitriplet clusters coupling, $3\times \bar{3}$) and $\chi^c_4$ (color sextet-antisextet clusters coupling, $6\times \bar{6}$), respectively. In particular, it is symmetry for the two quarks (antiquarks) interchange in eq.~(\ref{Color3}) and antisymmetry in eq.~(\ref{Color4}).
\begin{align}
\label{Color3}
\chi^c_3 &= \frac{\sqrt{3}}{6}(\bar{r}r\bar{g}g-\bar{g}r\bar{r}g+\bar{g}g\bar{r}r-\bar{r}g\bar{g}r+\bar{r}r\bar{b}b-\bar{b}r\bar{r}b
\nonumber\\
&+\bar{b}b\bar{r}r-\bar{r}b\bar{b}r+\bar{g}g\bar{b}b-\bar{b}g\bar{g}b+\bar{b}b\bar{g}g-\bar{g}b\bar{b}g) \,,
\end{align}
\begin{align}
\label{Color4}
\chi^c_4 &= \frac{\sqrt{6}}{12}(2\bar{r}r\bar{r}r+2\bar{g}g\bar{g}g+2\bar{b}b\bar{b}b+\bar{r}r\bar{g}g+\bar{g}r\bar{r}g
\nonumber\\
&+\bar{g}g\bar{r}r+\bar{r}g\bar{g}r+\bar{r}r\bar{b}b+\bar{b}r\bar{r}b+\bar{b}b\bar{r}r
\nonumber\\
&+\bar{r}b\bar{b}r+\bar{g}g\bar{b}b+\bar{b}g\bar{g}b+\bar{b}b\bar{g}g+\bar{g}b\bar{b}g) \,.
\end{align}
The rest four structures from Fig.~\ref{4QCOF}(c) to \ref{4QCOF}(f) are about K-types which the 4-quark wave-functions are constructed through the coupled-quarks in turn. Especially, their color bases are obtained by the following coupling coefficients according to the $SU(3)$ color group\footnote{The group chain of K-type is obtained in sequence of quark number (1234), each quark and antiquark is represented with [1] and [11], respectively in the group theory.}.
\begin{itemize}
\setlength{\baselineskip}{25pt}
\item	$K_1$-type of Fig.~\ref{4QCOF}(c): \big[$C^{[21]}_{[11],[1]}C^{[221]}_{[21],[11]}C^{[222]}_{[221],[1]}$\big]$_5$; 
~~\big[$C^{[111]}_{[11],[1]}C^{[221]}_{[111],[11]}C^{[222]}_{[221],[1]}$\big]$_6$;
\item	$K_2$-type of Fig.~\ref{4QCOF}(d): \big[$C^{[111]}_{[11],[1]}C^{[211]}_{[111],[1]}C^{[222]}_{[211],[11]}$\big]$_7$;
~~\big[$C^{[21]}_{[11],[1]}C^{[211]}_{[21],[1]}C^{[222]}_{[211],[11]}$\big]$_8$;
\item	$K_3$-type of Fig.~\ref{4QCOF}(e): \big[$C^{[2]}_{[1],[1]}C^{[211]}_{[2],[11]}C^{[222]}_{[211],[11]}$\big]$_9$;
~~\big[$C^{[11]}_{[1],[1]}C^{[211]}_{[11],[11]}C^{[222]}_{[211],[11]}$\big]$_{10}$;
\item	$K_4$-type of Fig.~\ref{4QCOF}(f): \big[$C^{[22]}_{[11],[11]}C^{[221]}_{[22],[1]}C^{[222]}_{[221],[1]}$\big]$_{11}$;
~~\big[$C^{[211]}_{[11],[11]}C^{[221]}_{[211],[1]}C^{[222]}_{[221],[1]}$\big]$_{12}$.
\end{itemize}
These eight group chains will generate the following wave-functions for K-types which subscripts correspond to the numbers labeled in the brackets above,
\begin{align}
\label{Color5}
\chi^c_5 &= \chi^c_2\,.
\end{align}
\begin{align}
\label{Color6}
\chi^c_6 &= \chi^c_1\,.
\end{align}
\begin{align}
\label{Color7}
\chi^c_7 &= \chi^c_1\,.
\end{align}
\begin{align}
\label{Color8}
\chi^c_8 &= \chi^c_2\,.
\end{align}
\begin{align}
\label{Color9}
\chi^c_9 &= \frac{1}{2\sqrt{6}}(\bar{r}b\bar{b}r+\bar{r}r\bar{b}b+\bar{g}b\bar{b}g+\bar{g}g\bar{b}b+\bar{r}g\bar{g}r+\bar{r}r\bar{g}g+
\nonumber\\
&\bar{b}b\bar{g}g+\bar{b}g\bar{g}b+\bar{g}g\bar{r}r+\bar{g}r\bar{r}g+\bar{b}b\bar{r}r+\bar{b}r\bar{r}b)+
\nonumber\\
&\frac{1}{\sqrt{6}}(\bar{r}r\bar{r}r+\bar{g}g\bar{g}g+\bar{b}b\bar{b}b) \,.
\end{align}
\begin{align}
\label{Color10}
\chi^c_{10} &= \frac{1}{2\sqrt{3}}(\bar{r}b\bar{b}r-\bar{r}r\bar{b}b+\bar{g}b\bar{b}g-\bar{g}g\bar{b}b+\bar{r}g\bar{g}r-\bar{r}r\bar{g}g-
\nonumber\\
&\bar{b}b\bar{g}g+\bar{b}g\bar{g}b-\bar{g}g\bar{r}r+\bar{g}r\bar{r}g-\bar{b}b\bar{r}r+\bar{b}r\bar{r}b) \,.
\end{align}
\begin{align}
\label{Color11}
\chi^c_{11} &= \chi^c_9\,.
\end{align}
\begin{align}
\label{Color12}
\chi^c_{12} &= -\chi^c_{10}\,.
\end{align}

\begin{itemize}[leftmargin=*,labelsep=5.8mm, listparindent=2em]
\item	Pentaquark
\end{itemize}

It is the same as 4-quarks system we discussed before, hadron-hadron structures, diquark-diquark-antiquark ones and even much more color configurations are involved in the 5-quarks sector. However, due to enormous computation in exactly solving the 5-body Schr\"{o}dinger equation, only baryon-meson configuration which the color singlet channels ($k=1$) and the hidden color ones ($k=2,3$) along with their couplings are considered in this work. The details of color wave-functions $\chi^c_k$ are as below,
\begin{eqnarray}
\chi^c_1 & = & \frac{1}{\sqrt{18}}(rgb-rbg+gbr-grb+brg-bgr) \nonumber \\
  & & ~~~~~~(\bar r r+\bar gg+\bar bb), \\
\chi^{c}_k & = & \frac{1}{\sqrt{8}}(\chi^k_{3,1}\chi_{2,8}-\chi^k_{3,2}\chi_{2,7}-\chi^k_{3,3}\chi_{2,6}+\chi^k_{3,4}\chi_{2,5} \nonumber \\
  & &    +\chi^k_{3,5}\chi_{2,4}-\chi^k_{3,6}\chi_{2,3}-\chi^k_{3,7}\chi_{2,2}+\chi^k_{3,8}\chi_{2,1}),
\end{eqnarray}
while $k=2$ and $3$ is of a symmetry and antisymmetry wave-function, respectively. The sub-clusters bases are
\begin{eqnarray}
&& \chi^2_{3,1}=\frac{1}{\sqrt{6}}(2rrg-rgr-grr), ~~\chi^3_{3,1}=\frac{1}{\sqrt{2}}(rgr-grr), \\
&& \chi^2_{3,2}=\frac{1}{\sqrt{6}}(rgg+grg-2ggr),~~ \chi^3_{3,2}=\frac{1}{\sqrt{2}}(rgg-grg), \\
&& \chi^2_{3,3}=\frac{1}{\sqrt{6}}(2rrb-rbr-brr), ~~\chi^3_{3,3}=\frac{1}{\sqrt{2}}(rbr-brr), \\
&& \chi^2_{3,4}=\frac{1}{\sqrt{12}}(2rgb-rbg+2grb-gbr-brg-bgr), \\
& & \chi^3_{3,4}=\frac{1}{\sqrt{4}}(rbg+gbr-brg-bgr), \\
&& \chi^2_{3,5}=\frac{1}{\sqrt{4}}(rbg-gbr+brg-bgr), \\
& & \chi^3_{3,5}=\frac{1}{\sqrt{12}}(2rgb+rbg-2grb-gbr-brg+bgr),  \\
&& \chi^2_{3,6}=\frac{1}{\sqrt{6}}(2ggb-gbg-bgg), ~~\chi^3_{3,6}=\frac{1}{\sqrt{2}}(gbg-bgg),  \\
&& \chi^2_{3,7}=\frac{1}{\sqrt{6}}(rbb+brb-2bbr), ~~\chi^3_{3,7}=\frac{1}{\sqrt{2}}(rbb-brb),  \\
&& \chi^2_{3,8}=\frac{1}{\sqrt{6}}(gbb+bgb-2bbg),~~ \chi^3_{3,8}=\frac{1}{\sqrt{2}}(gbb-bgb),  \\
&& \chi_{2,1}=\bar{b}r,  ~~ \chi_{2,2}=\bar{b}g,    \\
&& \chi_{2,3}=-\bar{g}r, ~~\chi_{2,4}=\frac{1}{\sqrt{2}}(\bar{r}r-\bar{g}g),   \\
&& \chi_{2,5}=\frac{1}{\sqrt{6}}(2\bar{b}b-\bar{r}r-\bar{g}g), ~~\chi_{2,6}=\bar{r}g,   \\
&& \chi_{2,7}=-\bar{g}b, ~~\chi_{2,8}=\bar{r}b.
\end{eqnarray}

\subsubsection{Spin wave-function}
In the 4-quark and 5-quark systems, the total spin $S$ can take values from $0$ to $2$ for the former case and $\frac{1}{2}$ to $\frac{5}{2}$ for the later one, respectively. Their spin wave-functions for one certain configuration listed from Fig.~\ref{4QCOF} to \ref{5QCOF2} are obtained by the couplings of Clebsh-Gordon coefficients in the spin SU(2) group. We now proceed to describe them in the tetraquark and pentaquark states.
\begin{itemize}[leftmargin=*,labelsep=5.8mm, listparindent=2em]
\item	Tetraquark
\end{itemize}

The spin wave-function $\chi^{\sigma_i}_{S, M_S}$ of 4-quark system is organized by two sub-clusters for the dimeson and the diquark-antidiquark structures, and couplings in an increased sequence of quark numbers for K-types. Furthermore, because no spin-orbital dependent potential is included in the model, the third component $(M_S)$ of total spin can be taken the same value as $S$ without losing generality. The details are written as

\begin{align}
\label{SWF1}
\chi_{0,0}^{\sigma_{l1}}(4) &= \chi^\sigma_{00}\chi^\sigma_{00} \\
\chi_{0,0}^{\sigma_{l2}}(4) &= \frac{1}{\sqrt{3}}(\chi^\sigma_{11}\chi^\sigma_{1,-1}-\chi^\sigma_{10}\chi^\sigma_{10}+\chi^\sigma_{1,-1}\chi^\sigma_{11}) \\
\chi_{0,0}^{\sigma_{l3}}(4) &= \frac{1}{\sqrt{2}}\big((\sqrt{\frac{2}{3}}\chi^\sigma_{11}\chi^\sigma_{\frac{1}{2}, -\frac{1}{2}}-\sqrt{\frac{1}{3}}\chi^\sigma_{10}\chi^\sigma_{\frac{1}{2}, \frac{1}{2}})\chi^\sigma_{\frac{1}{2}, -\frac{1}{2}} \\ \nonumber
&-(\sqrt{\frac{1}{3}}\chi^\sigma_{10}\chi^\sigma_{\frac{1}{2}, -\frac{1}{2}}-\sqrt{\frac{2}{3}}\chi^\sigma_{1, -1}\chi^\sigma_{\frac{1}{2}, \frac{1}{2}})\chi^\sigma_{\frac{1}{2}, \frac{1}{2}}\big) \\
\chi_{0,0}^{\sigma_{l4}}(4) &= \frac{1}{\sqrt{2}}(\chi^\sigma_{00}\chi^\sigma_{\frac{1}{2}, \frac{1}{2}}\chi^\sigma_{\frac{1}{2}, -\frac{1}{2}}-\chi^\sigma_{00}\chi^\sigma_{\frac{1}{2}, -\frac{1}{2}}\chi^\sigma_{\frac{1}{2}, \frac{1}{2}}) \\
\chi_{1,1}^{\sigma_{m1}}(4) &= \chi^\sigma_{00}\chi^\sigma_{11} \\ 
\chi_{1,1}^{\sigma_{m2}}(4) &= \chi^\sigma_{11}\chi^\sigma_{00} \\
\chi_{1,1}^{\sigma_{m3}}(4) &= \frac{1}{\sqrt{2}} (\chi^\sigma_{11} \chi^\sigma_{10}-\chi^\sigma_{10} \chi^\sigma_{11}) \\
\chi_{1,1}^{\sigma_{m4}}(4) &= \sqrt{\frac{3}{4}}\chi^\sigma_{11}\chi^\sigma_{\frac{1}{2}, \frac{1}{2}}\chi^\sigma_{\frac{1}{2}, -\frac{1}{2}}-\sqrt{\frac{1}{12}}\chi^\sigma_{11}\chi^\sigma_{\frac{1}{2}, -\frac{1}{2}}\chi^\sigma_{\frac{1}{2}, \frac{1}{2}}-\sqrt{\frac{1}{6}}\chi^\sigma_{10}\chi^\sigma_{\frac{1}{2}, \frac{1}{2}}\chi^\sigma_{\frac{1}{2}, \frac{1}{2}} \\
\chi_{1,1}^{\sigma_{m5}}(4) &= (\sqrt{\frac{2}{3}}\chi^\sigma_{11}\chi^\sigma_{\frac{1}{2}, -\frac{1}{2}}-\sqrt{\frac{1}{3}}\chi^\sigma_{10}\chi^\sigma_{\frac{1}{2}, \frac{1}{2}})\chi^\sigma_{\frac{1}{2}, \frac{1}{2}} \\
\chi_{1,1}^{\sigma_{m6}}(4) &= \chi^\sigma_{00}\chi^\sigma_{\frac{1}{2}, \frac{1}{2}}\chi^\sigma_{\frac{1}{2}, \frac{1}{2}} \\
\label{SWF2}
\chi_{2,2}^{\sigma_{1}}(4) &= \chi^\sigma_{11}\chi^\sigma_{11} 
\end{align}

In the above equations, the superscripts $l_1$...$l_4$ and $m_1$...$m_6$ are signs for each structures presented in Fig.~\ref{4QCOF}, their specific assignments are summarized in Table~\ref{SpinIndex}. Meanwhile, the necessary sub-clusters bases are read as
\begin{align}
\label{SCSpin}
\chi^\sigma_{11} &= \chi^\sigma_{\frac{1}{2}, \frac{1}{2}} \chi^\sigma_{\frac{1}{2}, \frac{1}{2}} \,,  ~\chi^\sigma_{1,-1} = \chi^\sigma_{\frac{1}{2}, -\frac{1}{2}} \chi^\sigma_{\frac{1}{2}, -\frac{1}{2}} \\
\chi^\sigma_{10} &= \frac{1}{\sqrt{2}}(\chi^\sigma_{\frac{1}{2}, \frac{1}{2}} \chi^\sigma_{\frac{1}{2}, -\frac{1}{2}}+\chi^\sigma_{\frac{1}{2}, -\frac{1}{2}} \chi^\sigma_{\frac{1}{2}, \frac{1}{2}}) \\
\chi^\sigma_{00} &= \frac{1}{\sqrt{2}}(\chi^\sigma_{\frac{1}{2}, \frac{1}{2}} \chi^\sigma_{\frac{1}{2}, -\frac{1}{2}}-\chi^\sigma_{\frac{1}{2}, -\frac{1}{2}} \chi^\sigma_{\frac{1}{2}, \frac{1}{2}})  
\end{align}
with  $\chi^\sigma_{\frac{1}{2}, \frac{1}{2}}$ and $\chi^\sigma_{\frac{1}{2}, -\frac{1}{2}}$ could be defined as $\alpha$ and $\beta$, respectively.

\begin{table}[H]
\caption{\label{SpinIndex} Index of spin-wave functions from Eq.~(\ref{SWF1}) to (\ref{SWF2}), their numbers are listed in the column according to each configuration, respectively.}
\centering
\begin{tabular}{ccccccc}
\toprule
    & ~~\textbf{Dimeson}~~ & ~~\textbf{Diquark-antidiquark}~~ & ~~$\textbf{K}_1$~~ & ~~$\textbf{K}_2$~~ & ~~$\textbf{K}_3$~~ & ~~$\textbf{K}_4$~~\\
\midrule
$l_1$ & 1   & 3 & & & & \\
$l_2$ & 2  & 4 & & & & \\
$l_3$ &   &  & 5 & 7 & 9 & 11 \\
$l_4$ &   &  & 6 & 8 & 10 & 12 \\[2ex]
$m_1$  & 1   & 4 & & & & \\
$m_2$ & 2  & 5 & & & & \\
$m_3$ & 3  & 6 & & & & \\
$m_4$ &   &  & 7 & 10 & 13 & 16 \\
$m_5$ &   &  & 8 & 11 & 14 & 17 \\
$m_6$ &   &  & 9 & 12 & 15 & 18 \\
\bottomrule
\end{tabular}
\end{table}

\begin{itemize}[leftmargin=*,labelsep=5.8mm, listparindent=2em]
\item	Pentaquark
\end{itemize}

Total spin for 5-quark systems is considered within a region from $1/2$ to $5/2$. Also based on a baryon and meson sub-clusters couplings formalism, the wave-functions can be
\begin{align}
\label{Spin}
\chi_{\frac12,\frac12}^{n \sigma 1}(5) &= \sqrt{\frac{1}{6}} \chi_{\frac32,-\frac12}^{n \sigma}(3) \chi_{11}^{\sigma}
-\sqrt{\frac{1}{3}} \chi_{\frac32,\frac12}^{n \sigma}(3) \chi_{10}^{\sigma}
+\sqrt{\frac{1}{2}} \chi_{\frac32,\frac32}^{n \sigma}(3) \chi_{1-1}^{\sigma} \,, \\
\chi_{\frac12,\frac12}^{n \sigma 2}(5) &= \sqrt{\frac{1}{3}} \chi_{\frac12,\frac12}^{n \sigma 1}(3) \chi_{10}^{\sigma} -\sqrt{\frac{2}{3}} \chi_{\frac12,-\frac12}^{n \sigma 1}(3) \chi_{11}^{\sigma} \,, \\
\chi_{\frac12,\frac12}^{n \sigma 3}(5) &= \sqrt{\frac{1}{3}} \chi_{\frac12,\frac12}^{n \sigma 2}(3) \chi_{10}^{\sigma} - \sqrt{\frac{2}{3}} \chi_{\frac12,-\frac12}^{n \sigma 2}(3) \chi_{11}^{\sigma} \,, \\
\chi_{\frac12,\frac12}^{n \sigma 4}(5) &= \chi_{\frac12,\frac12}^{n \sigma 1}(3) \chi_{00}^{\sigma} \,, \\
\chi_{\frac12,\frac12}^{n \sigma 5}(5) &= \chi_{\frac12,\frac12}^{n \sigma 2}(3) \chi_{00}^{\sigma} \,,
\end{align}
for $S=1/2$, and
\begin{align}
\chi_{\frac32,\frac32}^{n \sigma 1}(5) &= \sqrt{\frac{3}{5}}
\chi_{\frac32,\frac32}^{n \sigma}(3) \chi_{10}^{\sigma} -\sqrt{\frac{2}{5}} \chi_{\frac32,\frac12}^{n \sigma}(3) \chi_{11}^{\sigma} \,, \\
\chi_{\frac32,\frac32}^{n \sigma 2}(5) &= \chi_{\frac32,\frac32}^{n \sigma}(3) \chi_{00}^{\sigma} \,, \\
\chi_{\frac32,\frac32}^{n \sigma 3}(5) &= \chi_{\frac12,\frac12}^{n \sigma 1}(3) \chi_{11}^{\sigma} \,, \\
\chi_{\frac32,\frac32}^{n \sigma 4}(5) &= \chi_{\frac12,\frac12}^{n \sigma 2}(3) \chi_{11}^{\sigma} \,,
\end{align}
for $S=3/2$, and
\begin{align}
\chi_{\frac52,\frac52}^{n \sigma 1}(5) &= \chi_{\frac32,\frac32}^{n \sigma}(3) \chi_{11}^{\sigma} \,,
\end{align}
for $S=5/2$. These expressions can be obtained easily using SU(2) algebra and considering the 3-quark and quark-antiquark sub-clusters individually. The details read as
\begin{eqnarray}
&& \chi_{\frac32,\frac32}^{\sigma}(3) =\alpha\alpha\alpha,~~  \\
&&   \chi_{\frac32,\frac12}^{\sigma}(3) = \frac{1}{\sqrt{3}}
   (\alpha\alpha\beta+\alpha\beta\alpha+\beta\alpha\alpha),  \\
&&   \chi_{\frac32,-\frac12}^{\sigma}(3) = \frac{1}{\sqrt{3}}
   (\alpha\beta\beta+\beta\alpha\beta+\beta\beta\alpha),   \\
&& \chi_{\frac12,\frac12}^{\sigma 1}(3) =\frac{1}{\sqrt{6}}
   (2\alpha\alpha\beta-\alpha\beta\alpha-\beta\alpha\alpha),  \\
&& \chi_{\frac12,\frac12}^{\sigma 2}(3) =\frac{1}{\sqrt{2}}
   (\alpha\beta\alpha-\beta\alpha\alpha),  \\
&& \chi_{\frac12,-\frac12}^{\sigma 1}(3) =\frac{1}{\sqrt{6}}
   (\alpha\beta\beta-\alpha\beta\beta-2\beta\beta\alpha),  \\
&& \chi_{\frac12,-\frac12}^{\sigma 2}(3) =\frac{1}{\sqrt{2}}
   (\alpha\beta\beta-\beta\alpha\beta),  \\
&& \chi_{11}^{\sigma} =\alpha\alpha,~~
   \chi_{10}^{\sigma} =\frac{1}{\sqrt{2}}
   (\alpha\beta+\beta\alpha),~~\chi_{1-1}^{\sigma} =\beta\beta,  \\
&& \chi_{00}^{\sigma} =\frac{1}{\sqrt{2}} (\alpha\beta-\beta\alpha). 
\end{eqnarray}

\subsubsection{Flavor wave-function}
A similar procedure can be implemented in the iso-spin space and the total flavor wave-function of multi-quark system is introduced according to each configuration of state.
\begin{itemize}[leftmargin=*,labelsep=5.8mm, listparindent=2em]
\item	Tetraquark
\end{itemize}

Generally, there are two kinds of 4-quark systems that we are dealing with, the doubly- and fully-heavy tetraquark states. Hence, the well defined isospin quantum number $I$ can be taken either 0 or 1 for $QQ\bar{q}\bar{q}$ systems but only the iso-scalar state $I=0$ will be considered for $QQ\bar{Q}\bar{Q}$ sectors ($Q=c, b$ and $q=u, d, s$). Herein, we use $\chi^{fi}_{I, M_I}$ to represent the flavor wave-functions and the superscript $i=1,~2$ and $3$ stand for $cc\bar{q}\bar{q}$, $bb\bar{q}\bar{q}$ and $cb\bar{q}\bar{q}$ systems, respectively.
 The specific expressions are as following,
\begin{align}
&
\chi_{0,0}^{f1} = \sqrt{\frac{1}{2}} (\bar{u}c\bar{d}c-\bar{d}c\bar{u}c) \,, \\
&
\chi_{1,-1}^{f1} = \bar{u}c\bar{u}c \,, \\
&
\chi_{0,0}^{f2} = \sqrt{\frac{1}{2}} (\bar{u}b\bar{d}b-\bar{d}b\bar{u}b) \,, \\
&
\chi_{1,-1}^{f2} = \bar{u}b\bar{u}b \,, \\
&
\chi_{0,0}^{f3} = \sqrt{\frac{1}{2}} (\bar{u}c\bar{d}b-\bar{d}c\bar{u}b) \,,  \\
&
\chi_{1,-1}^{f3} = \bar{u}c\bar{u}b \,, \\
&
\chi_{0,0}^{f1} = \bar{s}c\bar{s}c ,\,\,\,
\chi_{0,0}^{f2} = \bar{s}b\bar{s}b ,\,\,\,
\chi_{0,0}^{f3} = \bar{s}c\bar{s}b \,. 
\end{align}

The third component of the isospin $M_I$ is also set to be equal to the absolute value of total isospin $I$ as the case in spin space. This is reasonable since no symmetry broken interaction on isospin included in our model. Meanwhile, it is a trivial results for the fully-heavy tetraquark states $i. e.$, $cc\bar{c}\bar{c}$ and $bb\bar{b}\bar{b}$.

\begin{itemize}[leftmargin=*,labelsep=5.8mm, listparindent=2em]
\item	Pentaquark
\end{itemize}

Three 5-quark systems are studied in this work, namely, hidden-charm, -bottom and doubly charmed pentaquarks. Accordingly, isospin equals to $\frac{1}{2}$ and $\frac{3}{2}$ are both allowed. However, only the hidden-charm pentaquark state in $I=\frac{1}{2}$ sector is discussed in our earliest work. The total 5-quark flavor wave-function is obtained by a coupling with the bases of two sub-clusters which are baryon and meson, respectively.

 In particular, in $uudQ\bar{Q}$ ($Q=c, b$) systems, we have two kinds of separation, one is $(udQ)(\bar{Q}u)+(uuQ)(\bar{Q}d)$ and the other is $(uud)(\bar{Q}Q)$ as illustrated in Fig.~\ref{5QCOF1}. The wave-functions are as below
\begin{eqnarray}
\chi^f_1  & = & \sqrt{\frac{2}{3}} B_{11} M_{\frac12,-\frac12}
 -\sqrt{\frac{1}{3}} B_{10} M_{\frac12,\frac12},  \\
\chi^f_2  & = & B_{00} M_{\frac12,\frac12}, \\
\chi^f_3  & = & B_{\frac12,\frac12}^1 M_{00},  \\
\chi^f_4  & = & B_{\frac12,\frac12}^2 M_{00}, \\
\chi^f_5  & = & B_{\frac32,\frac32} M_{00}, \\
\chi^f_6  & = & B_{\frac12,\frac12} M_{11}, 
\end{eqnarray}
where the necessary bases on sub-clusters are 
\begin{eqnarray}
&& B_{11} =uuQ,~~  B_{10} = \frac{1}{\sqrt{2}}(ud+du)Q,
~~B_{1-1} = ddQ,  \\
&& B_{00} = \frac{1}{\sqrt{2}}(ud-du)Q,  \\
&&B_{\frac12,\frac12}^1 = \frac{1}{\sqrt{6}}(2uud-udu-duu),  \\
&&B_{\frac12,\frac12}^2 = \frac{1}{\sqrt{2}}(ud-du)u,  \\
&&B_{\frac32,\frac32} = uuu,  \\
&&M_{\frac12,\frac12} = \bar{Q}u, ~~~~
M_{\frac12,-\frac12} = \bar{Q}d, ~~~~
M_{00} = \bar{Q}Q. 
\end{eqnarray}

As for the $QQqq\bar{q}$ ($Q=c, b$) tetraquarks shown in Fig.~\ref{5QCOF2} where the complete configurations in baryon-meson sector are considered, their flavor wave-functions with $I=1/2$ and $3/2$ read as 
\begin{align}
&
\chi_{\frac12,\frac12}^{n f1}(5) = \sqrt{\frac{2}{3}} B^n_{11} M_{\frac12,-\frac12} -\sqrt{\frac{1}{3}} B^n_{10} M_{\frac12,\frac12} \,, \\
&
\chi_{\frac12,\frac12}^{n f2}(5) = B^n_{00} M_{\frac12,\frac12} \,,\\
&
\chi_{\frac12,\frac12}^{n f3}(5) = B^n_{\frac12,\frac12} M_{00} \,, \\
&
\chi_{\frac12,\frac12}^{n f4}(5) = -\sqrt{\frac{2}{3}} B^n_{\frac12,-\frac12} M_{11} +\sqrt{\frac{1}{3}} B^n_{\frac12,\frac12} M_{10}  \,, \\
& 
\chi_{\frac32,\frac32}^{n f1}(5) = B^n_{\frac12,\frac12} M_ {1,1}\,, \\
& 
\chi_{\frac32,\frac32}^{n f2}(5) = B^n_{1,1} M_{\frac12,\frac12} \,, 
\end{align}
where the third component of isospin is still chosen to be the same as total one, and the superscript $n$ which value is from 1 to 4 marks each four configurations in Fig.~\ref{5QCOF2}. The flavor wave functions for the baryon and meson clusters are
\begin{align}
B^3_{11}  &= uuc \,,  \,\,\,\,\, B^3_{1-1} = ddc \,, \\ 
B^4_{11}  &= ucu \,,  \,\,\,\,\, B^4_{1-1} = dcd \,, \\ 
B^3_{10}  &= \frac{1}{\sqrt{2}}(ud+du)c \,, \\ 
B^4_{10}  &= \frac{1}{\sqrt{2}}(ucd+dcu) \,, \\ 
B^3_{00} &= \frac{1}{\sqrt{2}}(ud-du)c \,, \\
B^4_{00} &= \frac{1}{\sqrt{2}}(ucd-dcu) \,, \\ 
B^1_{\frac12,\frac12} &= ccu \,,   \,\,\,\,\, B^1_{\frac12,-\frac12} = ccd \,, \\
B^2_{\frac12,\frac12} &= cuc \,,   \,\,\,\,\, B^2_{\frac12,-\frac12} = cdc \,, \\
M_{\frac12,\frac12} &= \bar{d}c \,,  \, M_{\frac12,-\frac12} = -\bar{u}c \,, \\ 
M_{11} &= \bar{d}u \,,   \,\,\,\,\, M_{1-1} = -\bar{u}d \,, \\ 
M_{10} &= -\frac{1}{\sqrt{2}}(\bar{u}u-\bar{d}d) \,, \\ 
M_{00} &= -\frac{1}{\sqrt{2}}(\bar{u}u+\bar{d}d) \,,
\end{align}

\subsubsection{Spatial wave-function}
The few-body bound state problem is solved in an exact and efficient variational method, Gaussian expansion method (GEM)~\cite{Hiyama:2003cu}. In this theoretical framework, the intrinsic spatial wave-function of state is fitted by various widths ($\nu_n$) of Gaussian bases which are taken as the geometric progression form. Eq.~(\ref{phiOWF}) presents a general expression of the orbital wave-function,
\begin{align}
\label{phiOWF}
\phi_{nlm}(\vec{r}) = N_{nl} (r)^{l} e^{-\nu_{n} r^2} Y_{lm}(\hat{r}) \,,
\end{align}
where $N_{nl}$ is the normalization constants
\begin{equation}
 N_{nl}=\left[\frac{2^{l+2}(2\nu_n)^{l+\frac{3}{2}}}{\sqrt \pi(2l+1)}\right]^{\frac{1}{2}}.
\end{equation}
 The angle part of space is trivial in the S-wave multi-quark state, therein the angular matrix element is just a constant due to $Y_{00}=\sqrt{1/4\pi}$. However, as to avoid laborious Racah algebra in solving the angular excitation state, a powerful technique named infinitesimally shifted Gaussian (ISG)~\cite{Hiyama:2003cu} is employed. With the spherical harmonic function absorbed into a shifted vector $\vec{D}$, the new function is
 \begin{equation}
  \phi_{nlm}(\vec{r})=N_{nl}\lim_{\varepsilon\to 0}\frac{1}{(\nu_n \varepsilon)^l}
  \sum_{k=1}^{k_{max}}C_{lm,k}e^{{-\nu_n (\vec{r}-\varepsilon \vec{D}_{lm,k})}^2}\,.
\end{equation}
Their applications in the tetraquark and pentaquark states will be discussed individually.

\begin{itemize}[leftmargin=*,labelsep=5.8mm, listparindent=2em]
\item	Tetraquark
\end{itemize}

The 4-body system is investigated in a set of relative motion coordinates, the spatial wave function is,
\begin{equation}
\label{eq:WFexp4}
\psi_{LM_L}= \left[ \left[ \phi_{n_1l_1}(\vec{\rho}\,) \phi_{n_2l_2}(\vec{\lambda}\,)\right]_{l} \phi_{n_3l_3}(\vec{R}\,) \right]_{L M_L} \,.
\end{equation}
Particularly, the three internal Jacobi coordinates for Fig.~\ref{4QCOF}(a) of meson-meson configuration are read as
\begin{align}
\vec{\rho} &= \vec{x}_1-\vec{x}_2 \,, \\
\vec{\lambda} &= \vec{x}_3 - \vec{x}_4 \,, \\
\vec{R} &= \frac{m_1 \vec{x}_1 + m_2 \vec{x}_2}{m_1+m_2}- \frac{m_3 \vec{x}_3 + m_4 \vec{x}_4}{m_3+m_4} \,,
\end{align}
and the diquark-antdiquark structure of Fig.~\ref{4QCOF}(b) are defined as,
\begin{align}
\vec{\rho} &= \vec{x}_1-\vec{x}_3 \,, \\
\vec{\lambda} &= \vec{x}_2 - \vec{x}_4 \,, \\
\vec{R} &= \frac{m_1 \vec{x}_1 + m_3 \vec{x}_3}{m_1+m_3}- \frac{m_2 \vec{x}_2 + m_4 \vec{x}_4}{m_2+m_4} \,.
\end{align}
Moreover, the other K-type configurations from Fig.~\ref{4QCOF}(c) to \ref{4QCOF}(f) are,
\begin{align}
\vec{\rho} &= \vec{x}_i-\vec{x}_j \,, \\
\vec{\lambda} &= \vec{x}_k- \frac{m_i \vec{x}_i + m_j \vec{x}_j}{m_i+m_j} \,, \\
\vec{R} &= \vec{x}_l- \frac{m_i \vec{x}_i + m_j \vec{x}_j+m_k \vec{x}_k}{m_i+m_j+m_k} \,,
\end{align}
values of the above subscripts $i, j, k, l$ are according to the definitions of each configuration in Fig.~\ref{4QCOF}.
Obviously, the center-of-mass kinetic term $T_{CM}$ can be completely eliminated for a nonrelativistic system in these sets of relative motion coordinates.

\begin{itemize}[leftmargin=*,labelsep=5.8mm, listparindent=2em]
\item	Pentaquark
\end{itemize}

The spatial wave-function of 5-body system is also constructed in the relative motion coordinates, eq.~(\ref{eq:WFexp5}) presents a general form.
\begin{equation}
\label{eq:WFexp5}
\psi_{LM_L}=\left[ \left[ \left[ \phi_{n_1l_1}(\vec{\rho}\,) \phi_{n_2l_2}(\vec{\lambda}\,)\right]_{l} \phi_{n_3l_3}(\vec{r}\,) \right]_{l^{\prime}} \phi_{n_4l_4}(\vec{R}\,) \right]_{LM_L} \,,
\end{equation}
where in a baryon-meson configuration, the four Jacobi coordinates are defined as
\begin{align}
\vec{\rho} &= \vec{x}_1-\vec{x}_2 \,, \\
\vec{\lambda} &= \vec{x}_3 - \left( \frac{m_1\vec{x}_1+m_2\vec{x}_2}{m_1+m_2} \right) \,, \\
\vec{r} &= \vec{x}_4 - \vec{x}_5 \,, \\
\vec{R} &= \left( \frac{m_1 \vec{x}_1 + m_2 \vec{x}_2 + m_3 \vec{x}_3}{m_1+m_2+m_3} \right)-\left( \frac{m_4 \vec{x}_4 + m_5 \vec{x}_5}{m_4+m_5} \right) \,.
\end{align}
Besides, the $T_{CM}$ part can also be entirely deducted in a 5-quark non-relativistic system by this set of Jacobi coordinates.


\section{Results and Discussions}\label{sec:results}

In the constituent quark model formalism, the possible low-lying bound and resonance states of doubly-, fully-heavy tetraquarks, hidden-charm, -bottom and doubly charmed pentaquarks are systematically investigated by means of the computational approach, Gaussian expansion method. The obtained results along with their corresponding discussions are organized as follows.


\subsection{Doubly and Fully Heavy Tetraquarks}

In this part, the S-wave $QQ\bar{q}\bar{q}$ and $QQ\bar{Q}\bar{Q}$ ($Q=c, b$, $q=u, d, s$) tetraquark states with $J^P=0^+$, $1^+$ and $2^+$, the isospin $I=0$ or $1$ are studied in the chiral quark model and Cornell potential model, respectively. We will discuss them one by one.
\unskip
\subsubsection{$QQ\bar{q}\bar{q}$ Tetraquarks}

\begin{itemize}[leftmargin=*,labelsep=5.8mm, listparindent=2em]
\item	Double-charm tetraquarks

According to the Pauli principle, all possible couplings in spin, flavor and color degrees of freedom for the S-wave tetraquark states are considered. Table~\ref{GDDCC} presents the allowed meson-meson and diqurak-antidiquark channels for doubly charmed tetraquarks in $J^P=0^+$, $1^+$ and $2^+$, $I=0$ and $1$ states. However, bound and resonance states are only obtained in the $I(J^P)=0(1^+)$ quantum state. Their calculated masses are listed in Table~\ref{GresultCC1} where two dimeson channels, $D^+ D^{*0}$ and $D^{*+} D^{*0}$, two diquark-antidiquark channels, $(cc)^*(\bar{u}\bar{d})$ and $(cc)(\bar{u}\bar{d})^*$ along with their couplings are all considered. Particularly, the first column lists the allowed channels, their related experimental threshold values ($E^{ex}_{th}$)  are also marked in the parentheses. The color-singlet (S), hidden-color (H) channels and their couplings for dimeson configurations are listed in the second column. The computed masses (M) for each channels along with their binding energy ($E_B$), which is obtained by calculating the difference between the theoretical threshold ($E_{th}$) and the tetraquark mass (M), $E_B=M-E_{th}$, are presented in the 3rd and 4th columns, respectively.  Then, the re-scaled masses ($M'$), whose theoretical uncertainties coming from the model calculation of meson spectra are avoided, for meson-meson structures are listed in the last column, and  they are obtained by comparing the experimental threshold values and binding energies, $M'=E^{ex}_{th}+E_B$.

 Firstly, in the single channel computation for color-singlet (S) and hidden-color cases (H) of the $D^{(*)+} D^{*0}$ structures, the lowest masses are all above threshold values. However, loosely bound states of $D^+ D^{*0}$ and $D^{*+} D^{*0}$ are available in a coupled-channels calculation (S+H). Then after a mass shift correction which is according to the difference between the theoretical and experimental thresholds, the re-scaled masses of these two bound states are $3876\,\text{MeV}$ and $4017\,\text{MeV}$, respectively. Furthermore, these two bound state can be identified as the molecule states of $D^{(*)+} D^{*0}$ due to more than 95\% contributions come from the color-singlet channels.

Deeply bound diquark-antidiquark channel $(cc)^*(\bar{u}\bar{d})$ with a binding energy $\sim$$-140\,\text{MeV}$ is found and the theoretical mass is $3778\,\text{MeV}$. However, another diquark-antidiquark $(cc)(\bar{u}\bar{d})^*$ state is unbound and its mass is above the $D^+ D^{*0}$ and $D^{*+} D^{*0}$ theoretical thresholds with $E_B=+305\,\text{MeV}$ and $+186\,\text{MeV}$, respectively. In a further step, we performed a complete coupled-channels calculation for the channels listed in Table~\ref{GDDCC}, and the lowest-lying bound state mass is $3726\,\text{MeV}$. By analyzing the distance between any two quarks of $cc\bar{q}\bar{q}$ system in Table~\ref{tab:disDD}, the nature of compact double-charm tetraquark state is clearly presented. The general size of this tetraquark state is around $0.67\,\text{fm}$. Meanwhile, tightly bound and compact structure of the obtained tetraquark state is also confirmed in Table~\ref{GresultCompDD} where each component in the coupled-channels calculation is presented and the two dominant channels are the color-singlet channel $D^+ D^{*0}$ (25.8\%) and diquark-antidiquark $(cc)^*(\bar{q}\bar{q})$ one (36.7\%).

As to find possible double-charm tetraquark resonance in excitation state, the complex scaling method is employed in the complete coupled channels calculation too. Fig.~\ref{PP1} shows the distributions of calculated complex energies in the $I(J^P)$= $0(1 ^+)$ channel. Apparently, the bound stat is independent of the rotated angle which is varied from $0^\circ$ to $6^\circ$ and still locates at $3726$ MeV of real-axis. The other energy points are generally aligned along the $D^{(*)+}D^{*0}$ threshold lines which are scattering states. However, one possible resonance state whose mass and width are $\sim$4312 MeV and $\sim$16 MeV, respectively, is obtained in the complex plane and it is marked in a big orange circle with three calculated pole almost overlapping. This unchanged pole is far from the $D^+D^{*0}$ threshold lines, therefore, it can be identified as a $D^{*+}D^{*0}$ resonance.

\begin{table}[H]
\caption{\label{GDDCC} All possible channels for $cc\bar{q}\bar{q}$ $(q=u~or~d)$ tetraquark systems.}
\centering
\begin{tabular}{cccccc}
\toprule
& & \multicolumn{2}{c}{$\textbf{I=0}$} & \multicolumn{2}{c}{$\textbf{I=1}$} \\
~~$\textit{\textbf{J}}^\textbf{P}$~~&~~\textbf{Index}~~ & ~~$\chi_J^{\sigma_i}$;~$\chi_I^{f_j}$;~$\chi_k^c$~~ & ~~\textbf{Channel}~~ & ~~$\chi_J^{\sigma_i}$;~$\chi_I^{f_j}$;~$\chi_k^c$~~ & ~~\textbf{Channel}~~ \\
&&$[i; ~j; ~k]$& &$[i; ~j; ~k]$&  \\
\midrule
$0^+$ & 1  & $[1; ~1; ~1]$   & $(D^+ D^0)^1$ & $[1; ~1; ~1]$   & $(D^0 D^0)^1$ \\
&  2 & $[2; ~1; ~1]$ & $(D^{*+} D^{*0})^1$ & $[2; ~1; ~1]$  & $(D^{*0} D^{*0})^1$ \\
&  3 & $[1; ~1; ~2]$   & $(D^+ D^0)^8$  & $[1; ~1; ~2]$   & $(D^0 D^0)^8$ \\
&  4 & $[2; ~1; ~2]$ & $(D^{*+} D^{*0})^8$ & $[2; ~1; ~2]$   & $(D^{*0} D^{*0})^8$   \\
&  5  &                     &                              & $[3; ~1; ~4]$   & $(cc)(\bar{u}\bar{u})$ \\
&  6 &                      &                              & $[4; ~1; ~3]$   & $(cc)^*(\bar{u}\bar{u})^*$ \\[2ex]
$1^+$ & 1  & $[1; ~1; ~1]$   & $(D^+ D^{*0})^1$ & $[1; ~1; ~1]$   & $(D^0 D^{*0})^1$\\
& 2  & $[3; ~1; ~1]$     & $(D^{*+} D^{*0})^1$  & $[3; ~1; ~1]$   & $(D^{*0} D^{*0})^1$ \\
& 3  & $[1; ~1; ~2]$  & $(D^+ D^{*0})^8$ & $[1; ~1; ~2]$ & $(D^0 D^{*0})^8$ \\
& 4  & $[3; ~1; ~2]$  & $(D^{*+} D^{*0})^8$  & $[3; ~1; ~2]$   & $(D^{*0} D^{*0})^8$  \\
& 5  & $[4; ~1; ~3]$     & $(cc)^*(\bar{u}\bar{d})$  & $[6; ~1; ~3]$   & $(cc)^*(\bar{u}\bar{u})^*$ \\
& 6  & $[5; ~1; ~4]$  & $(cc)(\bar{u}\bar{d})^*$  &                          &   \\[2ex]
$2^+$ & 1  & $[1; ~1; ~1]$   & $(D^{*+} D^{*0})^1$ & $[1; ~1; ~1]$   & $(D^{*0} D^{*0})^1$ \\
& 2  & $[1; ~1; ~2]$  & $(D^{*+} D^{*0})^8$ & $[1; ~1; ~2]$  & $(D^{*0} D^{*0})^8$ \\
& 3  &   &   & $[1; ~1; ~3]$   & $(cc)^*(\bar{u}\bar{u})^*$ \\
\bottomrule
\end{tabular}
\end{table}

\begin{table}[H]
\caption{\label{GresultCC1} Lowest-lying states of double-charm tetraquarks with quantum numbers $I(J^P)=0(1^+)$, unit in MeV.}
\centering
\begin{tabular}{ccccc}
\toprule
~~\textbf{Channel}~~   & ~~\textbf{Color}~~ & ~~$\textbf{M}$~~ & ~~$\textbf{E}_\textbf{B}$~~ & ~~$\textbf{M'}$~~ \\
\midrule
$D^+ D^{*0}$ & S   & $3915$ & $0$  & $3877$ \\
$(3877)$           & H   & $4421$ & $+506$ & $4383$ \\
                    & S+H & $3914$ & $-1$  & $3876$ \\
                    & \multicolumn{4}{c}{Percentage (S;H): 97.3\%; 2.7\%} \\[2ex]
$D^{*+} D^{*0}$ & S   & $4034$ & $0$  & $4018$ \\
$(4018)$             & H   & $4390$ & $+356$ & $4374$ \\
                      & S+H & $4033$ & $-1$ & $4017$ \\
                      & \multicolumn{4}{c}{Percentage (S;H): 95.5\%; 4.5\%} \\[2ex] 
$(cc)^*(\bar{u}\bar{d})$ &    & $3778$ &  & \\[2ex] 
$(cc)(\bar{u}\bar{d})^*$ &    & $4220$ &  & \\[2ex]            
Mixed  & & $3726$ & & \\
\bottomrule
\end{tabular}
\end{table}

\begin{table}[H]
\caption{{Component of each channel in coupled-channels calculation with $I(J^P)=0(1^+)$, the numbers $1$ and $8$ of superscript are for singlet-color and hidden-color channel respectively.}  \label{GresultCompDD}}
\centering
\begin{tabular}{cccc}
\toprule
  ~~$(D^+ D^{*0})^1$~~  & ~~$(D^{*+} D^{*0})^1$~~   & ~~$(D^+ D^{*0})^8$~~ &
   ~~$(D^{*+} D^{*0})^8$~~ \\
 ~~25.8\%~~  & ~~15.4\%~~  & 10.7\%  & 11.2\% ~~\\[2ex]
  ~~$(cc)^*(\bar{u}\bar{d})$~~ & ~~$(cc)(\bar{u}\bar{d})^*$~~  \\ 
 ~~36.7\%  & 0.2\% \\
\bottomrule
\end{tabular}
\end{table}

\begin{table}[H]
\caption{\label{tab:disDD} The distance, in fm, between any two quarks of the found tetraquark bound-states in coupled-channels calculation $(q=u, d)$.}
\centering
\begin{tabular}{ccc}
\toprule
  ~~~~$\textit{\textbf{r}}_{\bar{u}\bar{d}}$~~~~ & ~~~~$\textit{\textbf{r}}_{\bar{q}c}$~~~~ & ~~~~$\textit{\textbf{r}}_{cc}$~~~~  \\
\midrule
  0.658 & 0.666 & 0.522 \\
\bottomrule
\end{tabular}
\end{table}

\begin{figure}[H]
\centering
\includegraphics[width=12 cm]{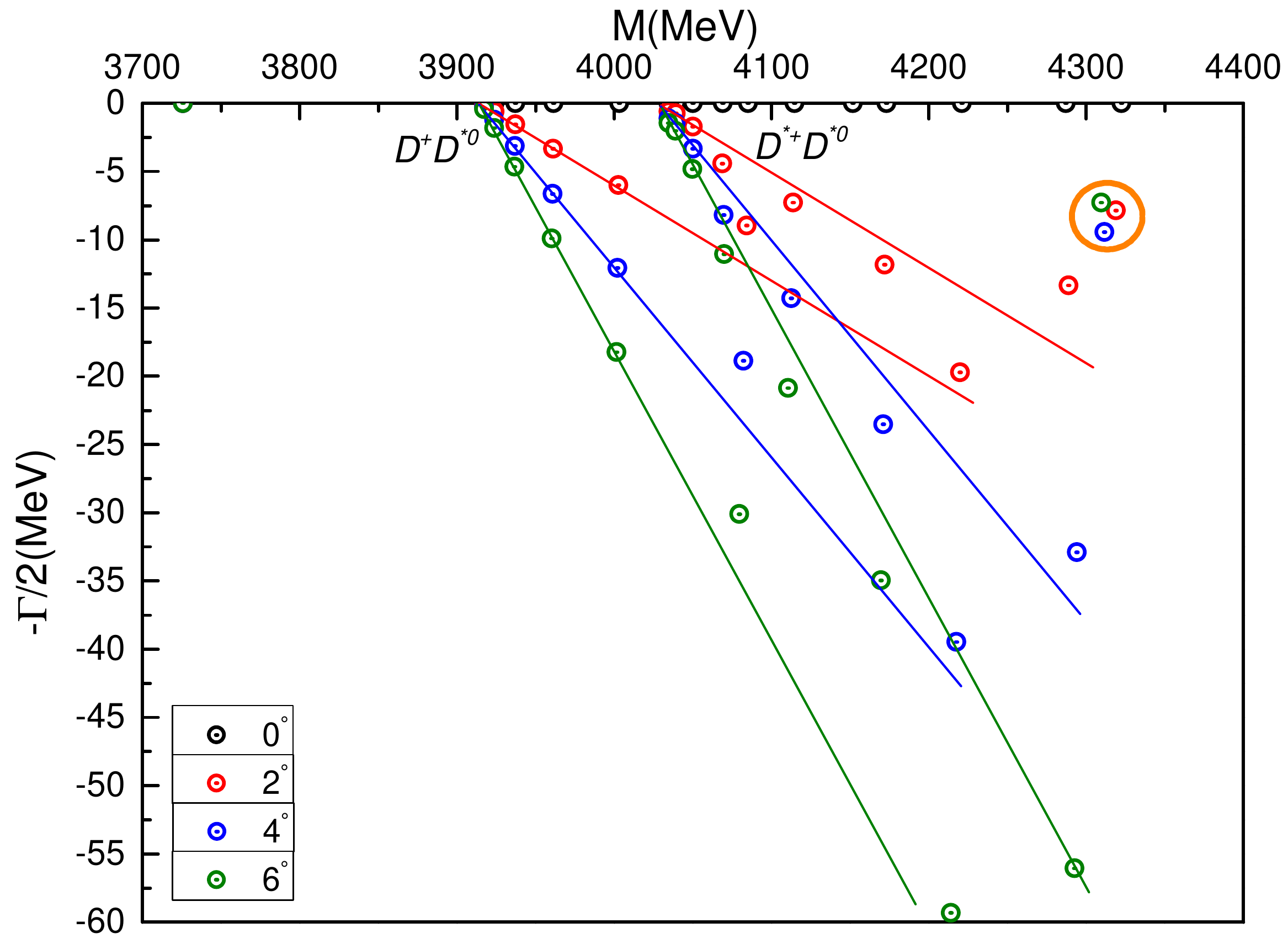}
\caption{Complex energies of double-charm tetraquarks with $I(J^P)=0(1^+)$ in the coupled channels calculation, $\theta$ varying from $0^\circ$ to $6^\circ$ .} \label{PP1}
\end{figure}   

\item	Double-bottom tetraquarks

In the $bb\bar{q}\bar{q}$ ($q=u, d$) sector, possible $B^{(*)-} \bar{B}^{*0}$ meson-meson channels and $(bb)^{(*)}(\bar{q}\bar{q})^{(*)}$ diquark-antidiquark structures in each quantum states are listed in Table~\ref{GBB}. However, it is similar to the doubly charmed case, bound and resonance states are only found in the $I(J^P)=0(1^+)$ state. Table~\ref{GresultBB1} shows the calculated results, and the arrangements of each columns are similar to Table~\ref{GresultCC1}. Firstly, one can notice that bound states of $B^- \bar{B}^{*0}$ and $B^{*-} \bar{B}^{*0}$ in color-singlet channels are obtained, and the $\sim$$-10$ MeV binding energy is owing to much heavier $b$-flavored quarks included. Additionally, in a coupled-channels calculation with hidden-color channels included, a deeper binding energies ($\sim$$-35$ MeV) are obtained for these two dimeson channels. The percentages of color-singlet channel and hidden-color on are around 80\% and 20\%, respectively. By considering the systematic uncertainty during calculation, the modified masses for $B^- \bar{B}^{*0}$ and $B^{*-} \bar{B}^{*0}$ bound state are $10569\,\text{MeV}$ and $10613\,\text{MeV}$ respectively.

There are two diquark-antidiquark channels under investigated, $(bb)^*(\bar{u}\bar{d})$ and $(bb)(\bar{u}\bar{d})^*$, the calculated masses are 10261 MeV and 10787 MeV, respectively. Clearly, the former structure is a tightly bound tetraquark state with binding energy $E_B=-336\,\text{MeV}$. However, the other one is 190 MeV above the $B^- \bar{B}^{*0}$ theoretical threshold. Our result on this diquark-antidiquark bound state is supported by Refs.~\cite{ejecq:2017prl, mkjlr:2017prl, cefgk:2019prd, jclh:1988prd}, and only $\sim$$130\,\text{MeV}$ lower than the calculated value in Ref.~\cite{mkjlr:2017prl}.

In the third step, a complete coupled-channels calculation is performed. Particularly, two bound states which masses are 10238 MeV and 10524 MeV are obtained. The first state is close to the $(bb)^*(\bar{u}\bar{d})$ channel and 23 MeV lower by the coupling effect. The second bound state is below the $B^- \bar{B}^{*0}$ theoretical threshold with $73\,\text{MeV}$. Furthermore, Table~\ref{GresultCompBB} presents the components of these two bound states in coupled-channels computation. There are both around 42\% $(bb)^*(\bar{u}\bar{d})$ channel and $\sim$20\% $B^{(*)-} \bar{B}^{*0}$ in the color-singlet channels of these tetraquark states. Accordingly, they can be identified as compact bound states in an analysis of the internal structure which the distance between any two quarks are calculated in Table~\ref{tab:disBB}. Therein, the general size is less than $0.83\,\text{fm}$ and the values ($0.328\,\text{fm}$ and $0.711\,\text{fm}$) on two bottom quarks are even small for the two obtained bound states.

In a complex range investigation on the $bb\bar{q}\bar{q}$ tetraquark in $I(J^P)=0(1^+)$ state, apart from the original  two bound states, one narrow resonance state is also found. Fig.~\ref{PP2} shows the distributions of the calculated energy points in the complete coupled case and the rotated angle $\theta$ is also taken from $0^\circ$ to $6^\circ$. In this range, the threshold lines of two meson-meson channels $B^- \bar{B}^{*0}$ and $B^{*-} \bar{B}^{*0}$ are well established and the two bound states is stable in the real-axis at 10238 MeV and 10524 MeV, respectively. Meanwhile, a fixed resonance pole at $\sim$10.8 GeV is obtained with the variation of $\theta$. We marked it with a big orange circle in Fig.~\ref{PP2}, besides the theoretical mass and width of this narrow resonance is 10814$\,\text{MeV}$ and $2\,\text{MeV}$, respectively. Because it is closer to the $B^{*-} \bar{B}^{*0}$ threshold lines, this meson-meson resonance is expected to be confirmed in the future experiment.

\begin{table}[H]
\caption{\label{GBB} All possible channels for $bb\bar{q}\bar{q}$ $(q=u~or~d)$ tetraquark systems.}
\centering
\begin{tabular}{cccccc}
\toprule
& & \multicolumn{2}{c}{$I=0$} & \multicolumn{2}{c}{$I=1$} \\
~~$\textit{\textbf{J}}^\textbf{P}$~~&~~\textbf{Index}~~ & ~~$\chi_J^{\sigma_i}$;~$\chi_I^{f_j}$;~$\chi_k^c$~~ & ~~\textbf{Channel}~~ & ~~$\chi_J^{\sigma_i}$;~$\chi_I^{f_j}$;~$\chi_k^c$~~ & ~~\textbf{Channel}~~ \\
&&$[i; ~j; ~k]$& &$[i; ~j; ~k]$&  \\
\midrule
$0^+$ & 1  & $[1; ~2; ~1]$   & $(B^- \bar{B}^0)^1$ & $[1; ~2; ~1]$   & $(B^- B^-)^1$ \\
&  2 & $[2; ~2; ~1]$ & $(B^{*-} \bar{B}^{*0})^1$ & $[2; ~2; ~1]$  & $(B^{*-} B^{*-})^1$ \\
&  3 & $[1; ~2; ~2]$   & $(B^- \bar{B}^0)^8$  & $[1; ~2; ~2]$   & $(B^- B^-)^8$ \\
&  4 & $[2; ~2; ~2]$ & $(B^{*-} \bar{B}^{*0})^8$ & $[2; ~2; ~2]$   & $(B^{*-} B^{*-})^8$   \\
&  5  &                     &                              & $[3; ~2; ~4]$   & $(bb)(\bar{u}\bar{u})$ \\
&  6 &                      &                              & $[4; ~2; ~3]$   & $(bb)^*(\bar{u}\bar{u})^*$ \\[2ex]
$1^+$ & 1  & $[1; ~2; ~1]$   & $(B^- \bar{B}^{*0})^1$ & $[1; ~2; ~1]$   & $(B^- B^{*-})^1$\\
& 2  & $[3; ~2; ~1]$     & $(B^{*-} \bar{B}^{*0})^1$  & $[3; ~2; ~1]$   & $(B^{*-} B^{*-})^1$ \\
& 3  & $[1; ~2; ~2]$  & $(B^- \bar{B}^{*0})^8$ & $[1; ~2; ~2]$ & $(B^- B^{*-})^8$ \\
& 4  & $[3; ~2; ~2]$  & $(B^{*-} \bar{B}^{*0})^8$  & $[3; ~2; ~2]$   & $(B^{*-} B^{*-})^8$  \\
& 5  & $[4; ~2; ~3]$     & $(bb)^*(\bar{u}\bar{d})$  & $[6; ~2; ~3]$   & $(bb)^*(\bar{u}\bar{u})^*$ \\
& 6  & $[5; ~2; ~4]$  & $(bb)(\bar{u}\bar{d})^*$  &                          &   \\[2ex]
$2^+$ & 1  & $[1; ~2; ~1]$   & $(B^{*-} \bar{B}^{*0})^1$ & $[1; ~2; ~1]$   & $(B^{*-} B^{*-})^1$ \\
& 2  & $[1; ~2; ~2]$  & $(B^{*-} \bar{B}^{*0})^8$ & $[1; ~2; ~2]$  & $(B^{*-} B^{*-})^8$ \\
& 3  &   &   & $[1; ~2; ~3]$   & $(bb)^*(\bar{u}\bar{u})^*$ \\
\bottomrule
\end{tabular}
\end{table}

\begin{table}[H]
\caption{\label{GresultBB1} Lowest-lying states of double-bottom tetraquarks with quantum numbers $I(J^P)=0(1^+)$, unit in MeV.}
\centering
\begin{tabular}{ccccc}
\toprule
~~\textbf{Channel}~~   & ~~\textbf{Color}~~ & ~~$\textbf{M}$~~ & ~~$\textbf{E}_\textbf{B}$~~ & ~~$\textbf{M'}$~~ \\
\midrule
$B^- \bar{B}^{*0}$ & S   & $10585$ & $-12$  & $10592$ \\
$(10604)$           & H   & $10987$ & $+390$ & $10994$ \\
                    & S+H & $10562$ & $-35$  & $10569$ \\
                    & \multicolumn{4}{c}{Percentage (S;H): 83.0\%; 17.0\%} \\[2ex]
$B^{*-} \bar{B}^{*0}$ & S   & $10627$ & $-11$  & $10639$ \\
$(10650)$             & H   & $10974$ & $+336$ & $10986$ \\
                      & S+H & $10601$ & $-37$ & $10613$ \\
                      & \multicolumn{4}{c}{Percentage (S;H): 79.6\%; 20.4\%} \\[2ex] 
$(bb)^*(\bar{u}\bar{d})$ &    & $10261$ &  & \\[2ex] 
$(bb)(\bar{u}\bar{d})^*$ &    & $10787$ &  & \\[2ex]                   
Mixed  & & $10238^{1st}$ & & \\
                    & & $10524^{2nd}$ & & \\
\bottomrule
\end{tabular}
\end{table}

\begin{table}[H]
\caption{{Component of each channel in coupled-channels calculation with $I(J^P)=0(1^+)$, the numbers $1$ and $8$ of superscript are for singlet-color and hidden-color channel respectively.}  \label{GresultCompBB}}
\centering
\begin{tabular}{cccc}
\toprule
  &  ~~$(B^- \bar{B}^{*0})^1$~~  & ~~$(B^{*-} \bar{B}^{*0})^1$~~   & ~~$(B^- \bar{B}^{*0})^8$~~ \\
 ~~$1st$~~  & ~~20.7\%~~  & ~~17.9\%~~  & 9.3\% \\
 ~~$2nd$~~  & ~~25.6\%~~  & ~~14.8\%~~  & 9.5\% \\[2ex]
  & ~~$(B^{*-} \bar{B}^{*0})^8$~~ & ~~$(bb)^*(\bar{u}\bar{d})$~~ & ~~$(bb)(\bar{u}\bar{d})^*$~~  \\ 
  ~~$1st$~~  & ~~9.4\%  & 42.6\%  & 0.1\% \\
  ~~$2nd$~~  & ~~9.1\% & 40.2\%  & 0.8\%  \\
\bottomrule
\end{tabular}
\end{table}

\begin{table}[H]
\caption{\label{tab:disBB} The distance, in fm, between any two quarks of the found tetraquark bound-states in coupled-channels calculation, $(q=u,d)$.}
\centering
\begin{tabular}{cccc}
\toprule
 & ~~~~$\textit{\textbf{r}}_{\bar{u}\bar{d}}$~~~~ & ~~~~$\textit{\textbf{r}}_{\bar{q}b}$~~~~ & ~~~~$\textit{\textbf{r}}_{bb}$~~~~  \\
\midrule
 $1st$ & 0.604 & 0.608 & 0.328 \\
 $2nd$ & 0.830 & 0.734 & 0.711 \\
\bottomrule
\end{tabular}
\end{table}
  
\begin{figure}[H]
\centering
\includegraphics[width=12 cm]{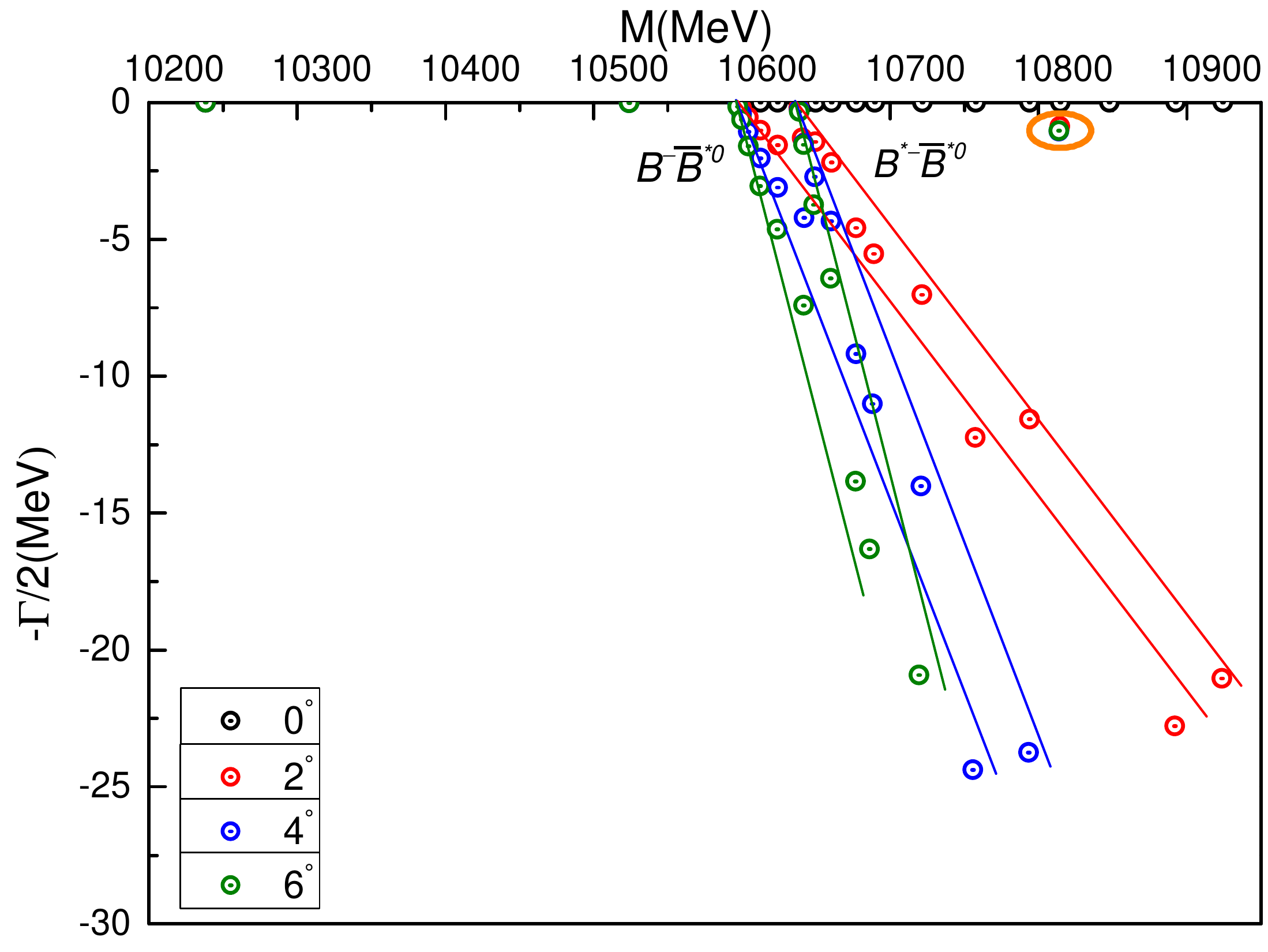}
\caption{Complex energies of double-bottom tetraquarks with $I(J^P)=0(1^+)$ in the coupled channels calculation, $\theta$ varying from $0^\circ$ to $6^\circ$ .} \label{PP2}
\end{figure}

\item	Charm-bottom tetraquarks

Table~\ref{GDB} lists the allowed channels of $cb\bar{q}\bar{q}$ tetraquark with $J^P=0^+$, $1^+$ and $2^+$, $I=0$ and $1$, respectively. However, some bound and resonance states are only found in the iso-scalar sector, besides the calculated results on meson-meson configurations are supported by the investigation of  Ref.~\cite{tfcjvav2019}. We will discuss these tetraquark states according to $I(J^P)$ quantum numbers respectively. Meanwhile, the arrangements of each columns in Tables~\ref{GresultCB1},~\ref{GresultCB2} and \ref{GresultCB3} are still the same as those in Table~\ref{GresultCC1}.

\begin{table}[H]
\caption{\label{GDB} All possible channels for $cb\bar{q}\bar{q}$ $(q=u~or~d)$ tetraquark systems. For a brief purpose, only the $D^{(*)0} B^{(*)0}$ structures are listed and the corresponding $D^{(*)+} \bar{B}^{(*)-}$ ones are absent in $I=0$. However, all these configurations are still employed in constructing the wavefunctions of 4-quark systems.}
\centering
\begin{tabular}{cccccc}
\toprule
& & \multicolumn{2}{c}{$\textbf{I=0}$} & \multicolumn{2}{c}{$\textbf{I=1}$} \\
~~$\textit{\textbf{J}}^\textbf{P}$~~&~~\textbf{Index}~~ & ~~$\chi_J^{\sigma_i}$;~$\chi_I^{f_j}$;~$\chi_k^c$~~ & ~~\textbf{Channel}~~ & ~~$\chi_J^{\sigma_i}$;~$\chi_I^{f_j}$;~$\chi_k^c$~~ & ~~\textbf{Channel}~~ \\
&&$[i; ~j; ~k]$& &$[i; ~j; ~k]$&  \\
\midrule
$0^+$ & 1  & $[1; ~3; ~1]$   & $(D^0 \bar{B}^0)^1$ & $[1; ~3; ~1]$   & $(D^0 B^-)^1$ \\
&  2 & $[2; ~3; ~1]$ & $(D^{*0} \bar{B}^{*0})^1$ & $[2; ~3; ~1]$  & $(D^{*0} B^{*-})^1$ \\
&  3 & $[1; ~3; ~2]$   & $(D^0 \bar{B}^0)^8$  & $[1; ~3; ~2]$   & $(D^0 B^-)^8$ \\
&  4 & $[2; ~3; ~2]$ & $(D^{*0} \bar{B}^{*0})^8$ & $[2; ~3; ~2]$   & $(D^{*0} B^{*-})^8$   \\
&  5  & $[3; ~3; ~3]$ & $(cb)(\bar{u}\bar{d})$  & $[3; ~3; ~4]$   & $(cb)(\bar{u}\bar{u})$ \\
&  6 & $[4; ~3; ~4]$  & $(cb)^*(\bar{u}\bar{d})^*$ & $[4; ~3; ~3]$   & $(cb)^*(\bar{u}\bar{u})^*$ \\[2ex]
$1^+$ & 1  & $[1; ~3; ~1]$   & $(D^0 \bar{B}^{*0})^1$ & $[1; ~3; ~1]$   & $(D^0 B^{*-})^1$\\
& 2  & $[2; ~3; ~1]$  & $(D^{*0} \bar{B}^0)^1$ & $[2; ~3; ~1]$  & $(D^{*0} B^-)^1$ \\
& 3  & $[3; ~3; ~1]$     & $(D^{*0} \bar{B}^{*0})^1$  & $[3; ~3; ~1]$   & $(D^{*0} B^{*-})^1$ \\
& 4  & $[1; ~3; ~2]$  & $(D^0 \bar{B}^{*0})^8$ & $[1; ~3; ~2]$ & $(D^0 B^{*-})^8$ \\
& 5  & $[2; ~3; ~2]$     & $(D^{*0} \bar{B}^0)^8$  & $[2; ~3; ~2]$   & $(D^{*0} B^-)^8$ \\
& 6  & $[3; ~3; ~2]$  & $(D^{*0} \bar{B}^{*0})^8$  & $[3; ~3; ~2]$   & $(D^{*0} B^{*-})^8$  \\
& 7  & $[4; ~3; ~3]$     & $(cb)^*(\bar{u}\bar{d})$  & $[4; ~3; ~4]$   & $(cb)^*(\bar{u}\bar{u})$ \\
& 8  & $[5; ~3; ~4]$  & $(cb)(\bar{u}\bar{d})^*$  &  $[5; ~3; ~3]$    & $(cb)(\bar{u}\bar{u})^*$   \\
& 9  & $[6; ~3; ~4]$  & $(cb)^*(\bar{u}\bar{d})^*$  &  $[6; ~3; ~3]$    & $(cb)^*(\bar{u}\bar{u})^*$   \\[2ex]
$2^+$ & 1  & $[1; ~3; ~1]$   & $(D^{*0} \bar{B}^{*0})^1$ & $[1; ~3; ~1]$   & $(D^{*0} B^{*-})^1$ \\
& 2  & $[1; ~3; ~2]$  & $(D^{*0} \bar{B}^{*0})^8$ & $[1; ~3; ~2]$  & $(D^{*0} B^{*-})^8$ \\
& 3  & $[1; ~3; ~4]$   & $(cb)^*(\bar{u}\bar{d})^*$ & $[1; ~3; ~3]$   & $(cb)^*(\bar{u}\bar{u})^*$ \\
\bottomrule
\end{tabular}
\end{table}

\begin{enumerate}[leftmargin=*,labelsep=4.9mm, listparindent=2em]
\item	$I(J^P)=0(0^+)$ state

Table~\ref{GresultCB1} summarizes the calculated results of each meson-meson, diquark-antiquark channels along with their couplings. Weakly bound states of $D^0 \bar{B}^0$ and $D^{*0} \bar{B}^{*0}$ in color-singlet channels are obtained firstly, the binding energies are $-4\,\text{MeV}$ and $-9\,\text{MeV}$, respectively. Then in a coupled-channels computation which the hidden-color channels are included, bound state in the $D^{*0} \bar{B}^{*0}$ channel is further pushed with $E_B=-39\,\text{MeV}$, however this coupling effect is quite weak in the $D^0 \bar{B}^0$ channel. These features on binding energies are confirmed by the investigation of each components shown in Table~\ref{GresultCB1}, where the proportions of color-singlet channels in the $D^0 \bar{B}^0$ and $D^{*0} \bar{B}^{*0}$ are 96.4\% and 87.8\%, respectively.

In the diquark-antidiquark sector, by comparing with the $D^0 \bar{B}^0$ theoretical threshold value, one tightly bound state $(cb)(\bar{u}\bar{d})$ with $E_B=-148\,\text{MeV}$ and one excited state $(cb)^*(\bar{u}\bar{d})^*$ with $E_B=+306\,\text{MeV}$ are obtained. Furthermore, the lowest bound state is found at 6980 MeV in the fully coupled-channels investigation. Obviously, this should be a compact charm-bottom tetraquark state, since its nature can be confirmed by analyzing the component and inner structure presented in Table~\ref{GresultCompDB1} and \ref{tab:disDB1}, respectively. In particular, size of the tetraquark in $0(0^+)$ state is less than $0.66\,\text{fm}$ and almost 50\% is the $(cb)(\bar{u}\bar{d})$ channel, the sub-dominant components are 26.4\% $(D^0 \bar{B}^0)^1$ and 21.5\% $(D^{*0} \bar{B}^{*0})^1$ channels.

The fully coupled-channels calculation is performed in a complex-range where the rotated angle $\theta$ is taken from $0^\circ$ to $6^\circ$, and then, the nature of bound tetraquark state is clearly shown in Fig.~\ref{PP3}, where the calculated dots are always fixed in the real-axis and at 6980 MeV. The other energy points which are generally aligned along the corresponding $D^0 \bar{B}^0$ and $D^{*0} \bar{B}^{*0}$ threshold lines are of scattering states. However, we also find a narrow resonance state which mass is around $7.7\,\text{GeV}$ and width is $\sim$$12\,\text{MeV}$. Although this resonance pole, marked with orange circle, is both above $D^0 \bar{B}^0$ and $D^{*0} \bar{B}^{*0}$ thresholds, we can still identify it as a $D^{*0} \bar{B}^{*0}$ molecule resonance which is farther away from $D^0 \bar{B}^0$ lines.

\begin{table}[H]
\caption{\label{GresultCB1} Lowest-lying states of charm-bottom tetraquarks with quantum numbers $I(J^P)=0(0^+)$, unit in MeV.}
\centering
\begin{tabular}{ccccc}
\toprule
~~\textbf{Channel}~~   & ~~\textbf{Color}~~ & ~~$\textbf{M}$~~ & ~~$\textbf{E}_\textbf{B}$~~ & ~~$\textbf{M'}$~~ \\
\midrule
$D^0 \bar{B}^0$ & S   & $7172$ & $-4$  & $7143$ \\
$(7147)$         & H   & $7685$ & $+509$ & $7656$ \\
                  & S+H & $7171$ & $-5$  & $7142$ \\
                  & \multicolumn{4}{c}{Percentage (S;H): 96.4\%; 3.6\%}  \\[2ex]
$D^{*0} \bar{B}^{*0}$ & S   & $7327$ & $-9$  & $7325$ \\
$(7334)$           & H   & $7586$ & $+250$ & $7584$ \\
                    & S+H & $7297$ & $-39$  & $7295$ \\
                    & \multicolumn{4}{c}{Percentage (S;H): 87.8\%; 12.2\%} \\[2ex]
$(cb)(\bar{u}\bar{d})$ &    & $7028$ &  & \\[2ex]
$(cb)^*(\bar{u}\bar{d})^*$ &    & $7482$ &  & \\[2ex]
Mixed  & & $6980$ & & \\
\bottomrule
\end{tabular}
\end{table}

\begin{figure}[H]
\centering
\includegraphics[width=12 cm]{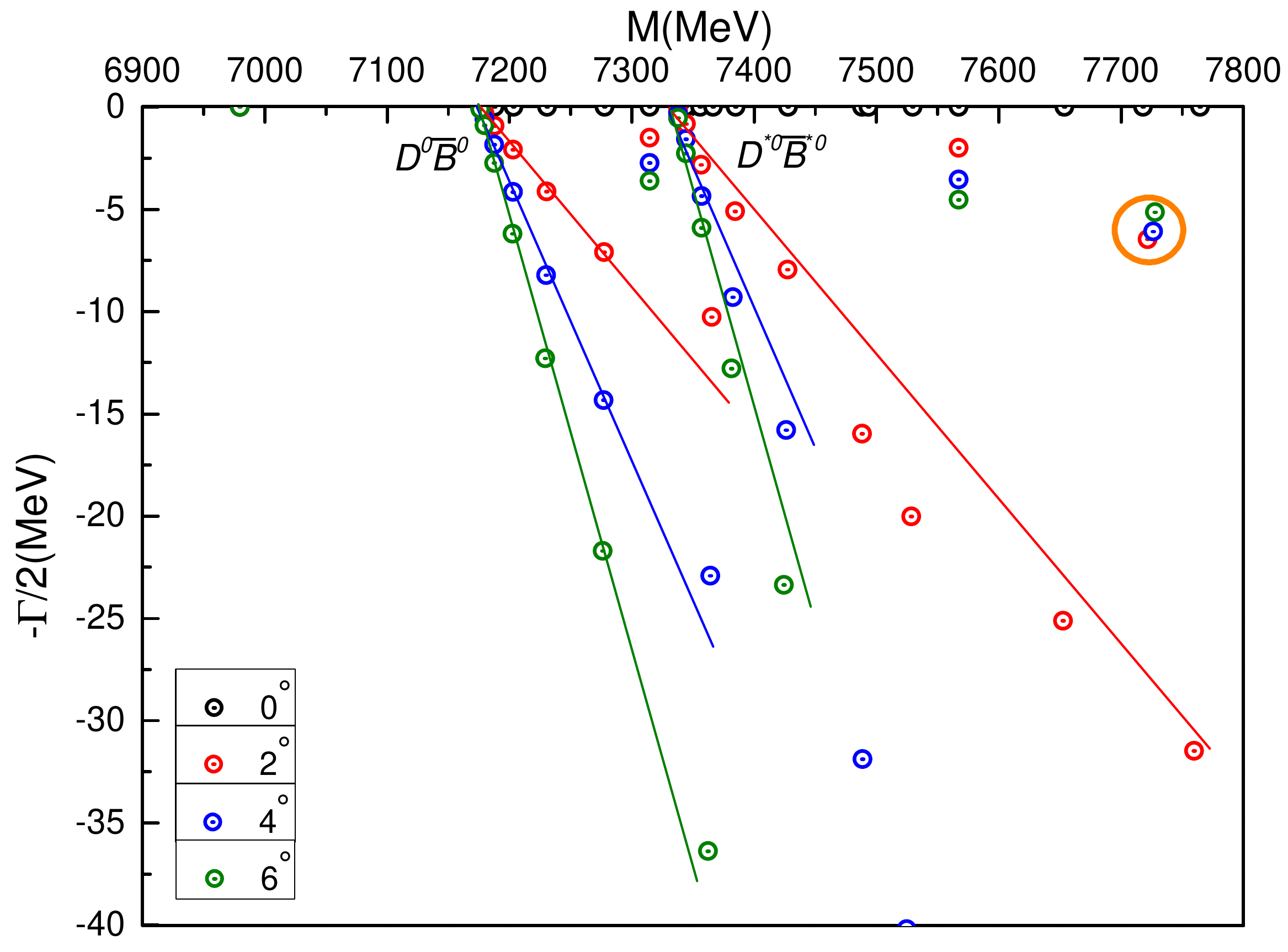}
\caption{Complex energies of charm-bottom tetraquarks with $I(J^P)=0(0^+)$ in the coupled channels calculation, $\theta$ varying from $0^\circ$ to $6^\circ$ .} \label{PP3}
\end{figure}

\item	$I(J^P)=0(1^+)$ state

There are three channels in the dimeson $D^{(*)0} \bar{B}^{(*)0}$ and diquark-antidiquark $(cb)^{(*)}(\bar{u}\bar{d})^{(*)}$ configurations, respectively. Besides, the single channel and the coupled results are all listed in Table~\ref{GresultCB2}. Similar to the above discussed doubly heavy tetraquarks, four conclusions can be drawn in a real-range calculation. (I) Loosely bound states are obtained in the color-singlet channels of $D^0 \bar{B}^{*0}$, $D^{*0} \bar{B}^0$ and $D^{*0} \bar{B}^{*0}$, their weak binding energies are $E_B=-3\,\text{MeV}$, $-2\,\text{MeV}$ and $-2\,\text{MeV}$, respectively. (II) The coupling between singlet- and hidden-color channels in meson-meson configuration are weak with the majority component ( more than 90\%) is the former channel. (III) One tightly bound diquark-antidiquark channel $(cb)^*(\bar{u}\bar{d})$ is found and the theoretical mass is 7039 MeV, the other two diquark-antidiquark channels' masses are above 7.5 GeV. (IV) In a fully coupled-channels calculation, the lowest mass of bound state reduced to 6997 MeV.

In order to have a better insight into the nature of the obtained bound state in the complete coupled-channels calculation, we may also focus on the results on the components and structures of the tetraquark bound state. As shown in Table~\ref{GresultCompDB1} and \ref{tab:disDB1}, the dominant contribution 46.4\% is from $(cb)^*(\bar{u}\bar{d})$ channel and other three sub-dominant channels are the color-singlet channels of $D^0 \bar{B}^{*0}$, $D^{*0} \bar{B}^0$ and $D^{*0} \bar{B}^{*0}$, their contributions are 20.2\%, 11.6\% and 16.8\%, respectively. This strong coupling effect leads to a compact structure which size is less than 0.67 fm again.

With the complete coupled-channels computation extended to a complex-range which $\theta$ is chosen still from $0^\circ$ to $6^\circ$, the bound state is confirmed again. Moreover, one more resonance state is found. In Fig.~\ref{PP4} one can notice that apart from the most scattering points which are the $D^0 \bar{B}^{*0}$, $D^{*0} \bar{B}^0$ and $D^{*0} \bar{B}^{*0}$ channels, one bound state at $6997\,\text{MeV}$ of real-axis and one narrow resonance state with mass and width is $7327\,\text{MeV}$ and $2.4\,\text{MeV}$ are obtained. Due to the resonance pole is located in the region between $D^{*0} \bar{B}^0$ and $D^{*0} \bar{B}^{*0}$ thresholds, it can be identified as the $D^{*0} \bar{B}^0$ resonance state according to the definition in CSM.

\begin{table}[H]
\caption{\label{GresultCB2} Lowest-lying states of charm-bottom tetraquarks with quantum numbers $I(J^P)=0(1^+)$, unit in MeV.}
\centering
\begin{tabular}{ccccc}
\toprule
~~\textbf{Channel}~~   & ~~\textbf{Color}~~ & ~~$\textbf{M}$~~ & ~~$\textbf{E}_\textbf{B}$~~ & ~~$\textbf{M'}$~~ \\
\midrule
$D^0 \bar{B}^{*0}$ & S   & $7214$ & $-3$  & $7190$ \\
$(7193)$           & H   & $7694$ & $+477$ & $7670$ \\
                    & S+H & $7213$ & $-4$  & $7189$ \\
                    & \multicolumn{4}{c}{Percentage (S;H): 96.8\%; 3.2\%} \\[2ex]
$D^{*0} \bar{B}^0$ & S   & $7293$ & $-2$ & $7286$ \\
$(7288)$           & H   & $7707$ & $+412$ & $7700$ \\
                    & S+H & $7292$ & $-3$ & $7285$ \\
                    & \multicolumn{4}{c}{Percentage (S;H): 96.8\%; 3.2\%} \\[2ex]
$D^{*0} \bar{B}^{*0}$ & S   & $7334$ & $-2$  & $7332$ \\
$(7334)$             & H   & $7691$ & $+354$ & $7688$ \\
                      & S+H & $7326$ & $-10$ & $7324$ \\
                      & \multicolumn{4}{c}{Percentage (S;H): 89.3\%; 10.7\%} \\[2ex] 
$(cb)^*(\bar{u}\bar{d})$ &    & $7039$ &  & \\[2ex] 
$(cb)(\bar{u}\bar{d})^*$ &    & $7531$ &  & \\[2ex] 
$(cb)^*(\bar{u}\bar{d})^*$ &    & $7507$ &  & \\[2ex]                   
Mixed  & & $6997$ & & \\
\bottomrule
\end{tabular}
\end{table}

\begin{figure}[H]
\centering
\includegraphics[width=12 cm]{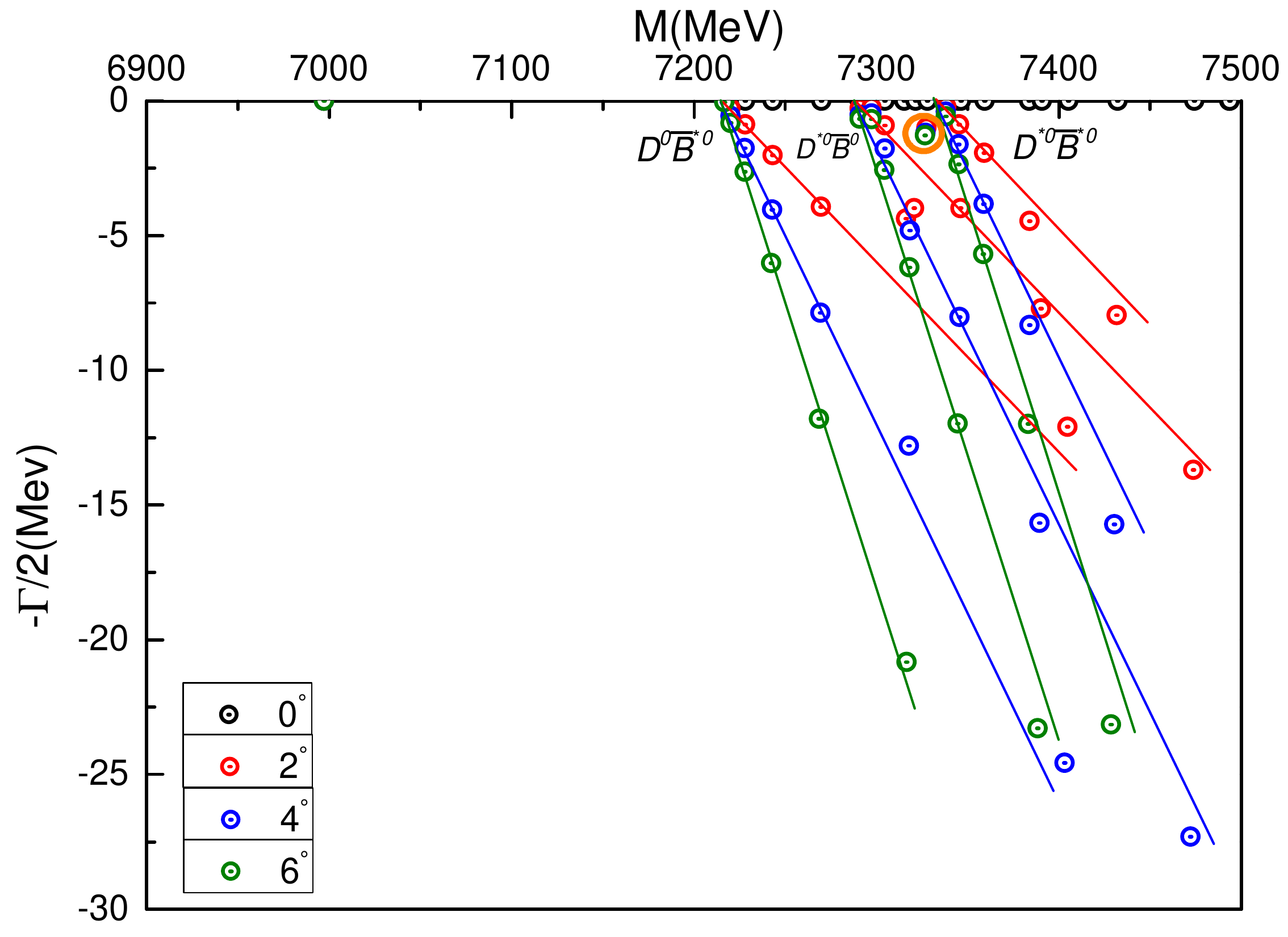}
\caption{Complex energies of charm-bottom tetraquarks with $I(J^P)=0(1^+)$ in the coupled channels calculation, $\theta$ varying from $0^\circ$ to $6^\circ$.} \label{PP4}
\end{figure}

\item	$I(J^P)=0(2^+)$ state 

For the highest spin state, only one meson-meson $D^{*0} \bar{B}^{*0}$ channel and one diquark-antidiquark $(cb)^*(\bar{u}\bar{d})^*$. Firstly, in the single channel calculations, only the $D^{*0} \bar{B}^{*0}$ color-singlet channel is loosely bound with $E_B=-2\,\text{MeV}$. Furthermore, this fact is not changed by the coupling with a hidden-color channel and only 1 MeV decreased in the complete coupled-channels case, the lowest mass is 7333 MeV. Hence, a molecular-type structure is possible for the obtained bound state and it has been supported by calculating the quark distances which is already beyond $1.6\,\text{fm}$ from Table~\ref{tab:disDB1} and $\sim$99\% contributions come from color-singlet channel of $D^{*0} \bar{B}^{*0}$.

In contrast to the previous tetraquark states, no resonance is found in $02^+$ state. It is clearly shown in Fig.~\ref{PP5} that all of the calculated poles are aligned along the $D^{*0} \bar{B}^{*0}$ threshold lines except the weakly bound state at 7333 MeV.

\begin{table}[H]
\caption{\label{GresultCB3} Lowest-lying states of charm-bottom tetraquarks with quantum numbers $I(J^P)=0(2^+)$, unit in MeV.}
\centering
\begin{tabular}{ccccc}
\toprule
~~\textbf{Channel}~~   & ~~\textbf{Color}~~ & ~~$\textbf{M}$~~ & ~~$\textbf{E}_\textbf{B}$~~ & ~~$\textbf{M'}$~~ \\
\midrule
$D^{*0} \bar{B}^{*0}$ & S   & $7334$ & $-2$  & $7332$ \\
$(7334)$             & H   & $7720$ & $+384$ & $7718$ \\
                      & S+H & $7334$ & $-2$  & $7332$ \\
                      & \multicolumn{4}{c}{Percentage (S;H): 99.8\%; 0.2\%} \\[2ex]
$(cb)^*(\bar{u}\bar{d})^*$ &    & $7552$ &  & \\[2ex]                   
Mixed  & & $7333$ & & \\
\bottomrule
\end{tabular}
\end{table}

\begin{figure}[H]
\centering
\includegraphics[width=12 cm]{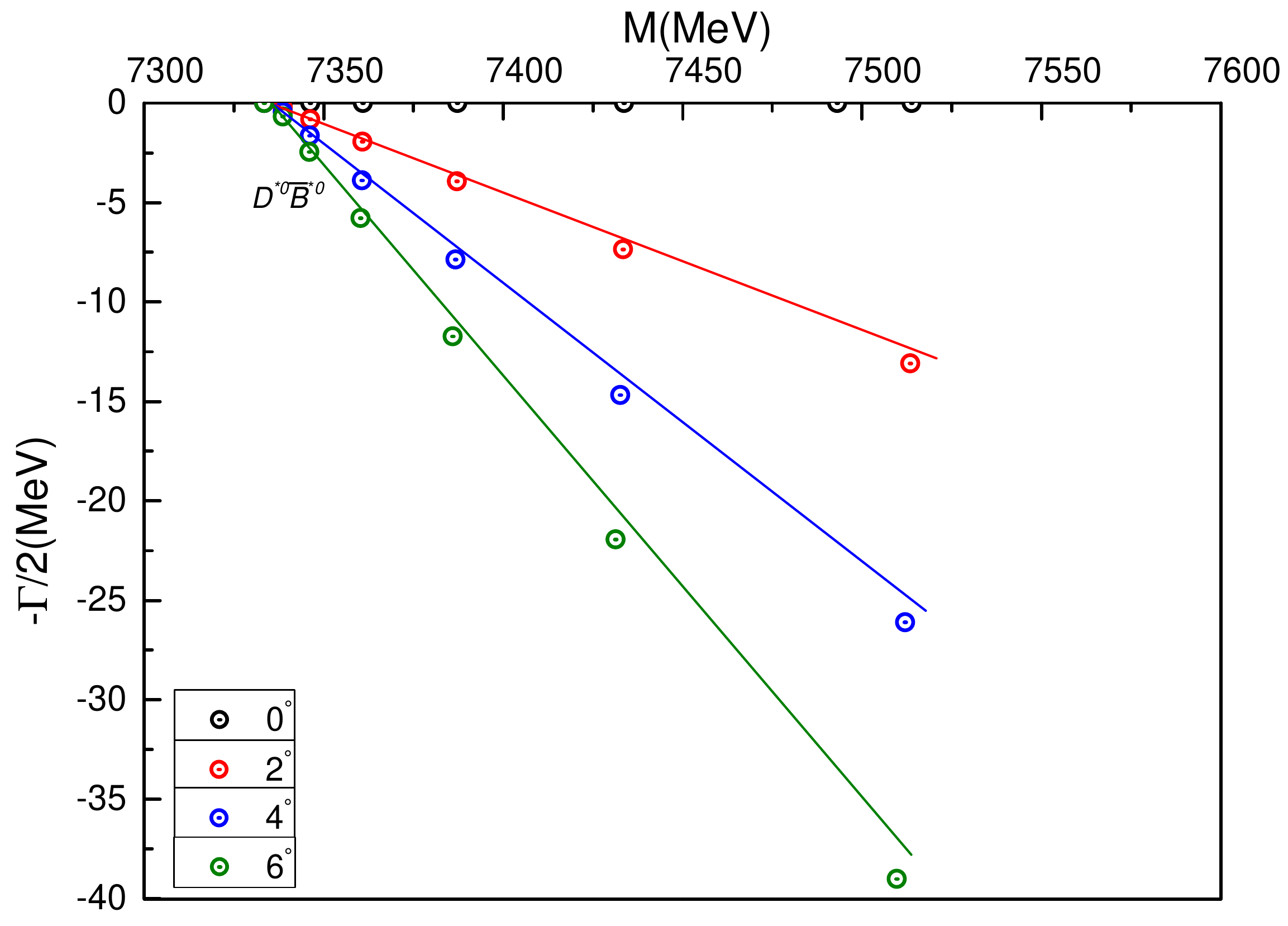}
\caption{Complex energies of charm-bottom tetraquarks with $I(J^P)=0(2^+)$ in the coupled channels calculation, $\theta$ varying from $0^\circ$ to $6^\circ$ .} \label{PP5}
\end{figure}

\end{enumerate}

\end{itemize}

\begin{table}[H]
\caption{{Component of each channel in coupled-channels calculation, the numbers $1$ and $8$ of superscript are for singlet-color and hidden-color channel respectively, $(q=u,d)$.}  \label{GresultCompDB1}}
\centering
\begin{tabular}{ccccc}
\toprule
 $I(J^P)$ & ~~$(D^0 \bar{B}^0)^1$~~  & ~~$(D^{*0} \bar{B}^{*0})^1$~~   & ~~$(D^0 \bar{B}^0)^8$~~ &
   ~~$(D^{*0} \bar{B}^{*0})^8$~~ \\
$0(0^+)$ & ~~26.4\%~~  & ~~21.5\%~~  & 1.6\%~~  & 1.9\% ~~\\[2ex]
 & ~~$(cb)(\bar{u}\bar{d})$~~ & ~~$(cb)^*(\bar{u}\bar{d})^*$~~  \\ 
 & ~~48.5\%~~  & ~~0.1\%~~ \\[2ex]
 $0(1^+)$ & ~~$(D^0 \bar{B}^{*0})^1$~~  & ~~$(D^{*0} \bar{B}^0)^1$~~   & ~~$(D^{*0} \bar{B}^{*0})^1$~~ &
   ~~$(D^0 \bar{B}^{*0})^8$~~ \\
 & ~~20.2\%~~  & ~~11.6\%~~  & ~~16.8\%~~  & ~~1.4\% ~~\\[2ex]
 & ~~$(D^{*0} \bar{B}^0)^8$~~ & ~~$(D^{*0} \bar{B}^{*0})^8$~~ & ~~$(cb)^*(\bar{u}\bar{d})$~~ &
  ~~$(cb)(\bar{u}\bar{d})^*$~~  \\ 
 & ~~1.3\%~~  & 1.8\%~~ & ~~46.4\%~~  & ~~0.1\%~~ \\[2ex]
 & ~~$(cb)^*(\bar{u}\bar{d})^*$~~ \\
 & ~~0.4\%~~ \\[2ex]
 $0(2^+)$ & ~~$(D^{*0} \bar{B}^{*0})^1$~~  & ~~$(D^{*0} \bar{B}^{*0})^8$~~   & ~~$(cb)^*(\bar{u}\bar{d})^*$~~\\
 & ~~98.6\%~~  & ~~0.3\%~~  & ~~1.1\%~~\\
\bottomrule
\end{tabular}
\end{table}

\begin{table}[H]
\caption{\label{tab:disDB1} The distance, in fm, between any two quarks of the found tetraquark bound-states in coupled-channels calculation, $(q=u,d)$.}
\centering
\begin{tabular}{ccccc}
\toprule
  ~~$\textbf{IJ}^\textbf{P}$~~& ~~$\textit{\textbf{r}}_{\bar{u}\bar{d}}$~~ & ~~$\textit{\textbf{r}}_{\bar{q}c}$~~ & ~~$\textit{\textbf{r}}_{\bar{q}b}$~~  & ~~$\textit{\textbf{r}}_{cb}$~~  \\
\midrule
  $00^+$ & 0.635 & 0.653 & 0.610  & 0.428 \\[2ex]
  $01^+$ & 0.632 & 0.661 & 0.616  & 0.434 \\[2ex]
  $02^+$ & 2.248 & 1.612 & 1.597  & 2.102 \\
\bottomrule
\end{tabular}
\end{table}

\subsubsection{$QQ\bar{s}\bar{s}$ Tetraquarks}

As for a natural extension of the work on $QQ\bar{q}\bar{q}$ ($Q=c, b$ and $q=u, d$) systems, the double-heavy tetraquark state in strange quark sector is investigated herein.
 In addition, for this 4-quark system, a complete set of configurations including meson-meson, diquark-antidiquark and K-type structures (Fig.~\ref{4QCOF}) is included.
  As for a clarify purpose, masses and mean square radii of the $Q\bar{s}$ mesons are listed in Table~\ref{MesonMass}. These results will be useful in identifying possible $QQ\bar{s}\bar{s}$ bound or resonance states. Furthermore, Tables ranging from~\ref{GresultCCSS1} to~\ref{GresultCCSS9} summarized our theoretical findings. Particularly, in those tables, the first column shows the allowed channels and, in the parenthesis, the noninteracting meson-meson threshold value of experiment. Color-singlet (S), hidden-color (H) along with other configurations are indexed in the second column, respectively, the third column lists the necessary bases in spin, flavor and color degrees of freedom, the fourth and fifth columns refer to the theoretical mass of each channels and their couplings.

\begin{table}[H]
\caption{\label{MesonMass} Theoretical and experimental masses of $D^{(*)+}_s$ and $B^{(*)}_s$ mesons, their theoretical sizes are also calculated.}
\centering
\begin{tabular}{ccccc}
\toprule
~~\textbf{Meson}~~   & ~~\textbf{\textit{nL}}~~ & ~~\textbf{The.}~~ & ~~\textbf{Exp.}~~\\
\midrule
$D^+_s$ & $1S$   & $1989\,MeV$; $0.47\,fm$ & $1969\,MeV$ \\
               & $2S$   & $2703\,MeV$; $1.06\,fm$ & - \\[2ex]
$D^{*+}_s$ & $1S$   & $2116\,MeV$; $0.55\,fm$ & $2112\,MeV$ \\
                   & $2S$   & $2767\,MeV$; $1.14\,fm$ & - \\[2ex]
$\bar{B}^0_s$ & $1S$   & $5355\,MeV$; $0.47\,fm$ & $5367\,MeV$ \\
               & $2S$   & $6017\,MeV$; $1.01\,fm$ & - \\[2ex]
$\bar{B}^*_s$ & $1S$   & $5400\,MeV$; $0.50\,fm$ & $5415\,MeV$ \\
                   & $2S$   & $6042\,MeV$; $1.04\,fm$ & -  \\
\bottomrule
\end{tabular}
\end{table}

\begin{itemize}[leftmargin=*,labelsep=5.8mm, listparindent=2em]
\item	$cc\bar{s}\bar{s}$ tetraquarks

There is no bound state in the doubly charmed $cc\bar{s}\bar{s}$ system, but resonances are found in the $I(J^P)=0(0^+)$ and $0(2^+)$ quantum states. Obviously, these results are different from the $cc\bar{q}\bar{q}$ tetraquark states. We will discuss them in the following parts.

\begin{enumerate}[leftmargin=*,labelsep=4.9mm, listparindent=2em]
\item	$I(J^P)=0(0^+)$ state

Table~\ref{GresultCCSS1} presents all of the possible channels in $cc\bar{s}\bar{s}$ system with $I(J^P)=0(0^+)$ state. It is clearly to notice that the meson-meson channels of $D^+_s D^+_s$ and $D^{*+}_s D^{*+}_s$ both in color-singlet and hidden-color states are unbound. Moreover, the coupled results in these two color configurations are still unchanged with the obtained masses are 3978 MeV and 4377 MeV, respectively. As for the two diquark-antidiquark channels, masses of $(cc)(\bar{s}\bar{s})$ and $(cc)^{*}(\bar{s}\bar{s})^{*}$ are both around 4.4 GeV which is above the $D^{(*)+}_s D^{(*)+}_s$ threshold values. Their coupled-mass 4379 MeV is quite close to the value of hidden-color channels. However,  although there is a strong coupling between them, it is still not enough to have a bound state. Additionally, bound state is still unavailable in the K-type configurations, mass of the four K-type channels are in the region from 4.2 GeV to 4.8 GeV and there is a degeneration at 4.4 GeV for $(cc)^{(*)}(\bar{s}\bar{s})^{(*)}$, $K_3$ and $K_4$ channels. Finally, the lowest mass (3978 MeV) in the complete coupled-channels calculation is the same as that in the color-singlet channels. Therefore, no bound state is found in $cc\bar{s}\bar{s}$ tetraquark with $I(J^P)=0(0^+)$ state.

Apart from the above real-range study, Fig.~\ref{PPCC1} shows the fully coupled-channels computation in the complex scaling method. Particularly, in the mass region from 3.9 GeV to 5.0 GeV, three meson-meson scattering states, $(1S)D^+_s (1S)D^+_s$, $(1S)D^{*+}_s (1S)D^{*+}_s$ and $(1S)D^+_s (2S)D^+_s$ are well presented. The majority energy points are basically aligned along the cut lines with rotated angle $\theta$ varied less than $6^\circ$. However, there is a stable resonance pole circled with orange in the complex plane. The calculated mass and width for this state is 4902 MeV and 3.54 MeV, respectively. Since it is about 0.6 GeV above the ground state of two non-interacting $D^{*+}_s$ mesons threshold and 0.2GeV above the first radial excitation $(1S)D^+_s (2S)D^+_s$ state, this narrow resonance can be identified as the $D^+_s D^+_s$ molecular state.

\begin{table}[H]
\caption{\label{GresultCCSS1} The lowest-lying eigen-energies of $cc\bar{s}\bar{s}$ tetraquarks with $I(J^P)=0(0^+)$ in the real range calculation. (unit: MeV)}
\centering
\begin{tabular}{ccccc}
\toprule
~~\textbf{Channel}~~   & ~~\textbf{Index}~~ & ~~$\chi_J^{\sigma_i}$;~$\chi_I^{f_j}$;~$\chi_k^c$~~ & ~~$\textbf{Mass}$~~ & ~~\textbf{Mixed}~~ \\
             &  &$[i; ~j; ~k]$&  \\
\midrule
$(D^+_s D^+_s)^1 (3938)$          & 1(S)  & $[1; ~1; ~1]$  & $3978$ & \\
$(D^{*+}_s D^{*+}_s)^1 (4224)$  & 2(S)  & $[2; ~1; ~1]$  & $4232$ & $3978$ \\[2ex]
$(D^+_s D^+_s)^8$          & 3(H) & $[1; ~1; ~2]$   & $4619$ & \\
$(D^{*+}_s D^{*+}_s)^8$  & 4(H) & $[2; ~1; ~2]$   & $4636$ & $4377$ \\[2ex]
$(cc)(\bar{s}\bar{s})$      & 5 & $[3; ~1; ~4]$   & $4433$ & \\
$(cc)^*(\bar{s}\bar{s})^*$  & 6 & $[4; ~1; ~3]$   & $4413$ & $4379$ \\[2ex]
$K_1$  & 7 & $[5; ~1; ~5]$   & $4802$ & \\
$K_1$  & 8 & $[5; ~1; ~6]$   & $4369$ & \\
$K_1$  & 9 & $[6; ~1; ~5]$   & $4698$ & \\
$K_1$  & 10 & $[6; ~1; ~6]$   & $4211$ & $4201$ \\[2ex]
$K_2$  & 11 & $[7; ~1; ~7]$   & $4343$ & \\
$K_2$  & 12 & $[7; ~1; ~8]$   & $4753$ & \\
$K_2$  & 13 & $[8; ~1; ~7]$   & $4166$ & \\
$K_2$  & 14 & $[8; ~1; ~8]$   & $4838$ & $4158$ \\[2ex]
$K_3$  & 15 & $[9; ~1; ~10]$   & $4414$ & \\
$K_3$  & 16 & $[10; ~1; ~9]$   & $4427$ & $4373$ \\[2ex]
$K_4$  & 17 & $[11; ~1; ~12]$   & $4413$ & \\
$K_4$  & 18 & $[12; ~1; ~11]$   & $4439$ & $4379$ \\[2ex]
\multicolumn{3}{c}{All of the above channels:} & $3978$ \\
\bottomrule
\end{tabular}
\end{table}

\begin{figure}[H]
\centering
\includegraphics[width=14 cm]{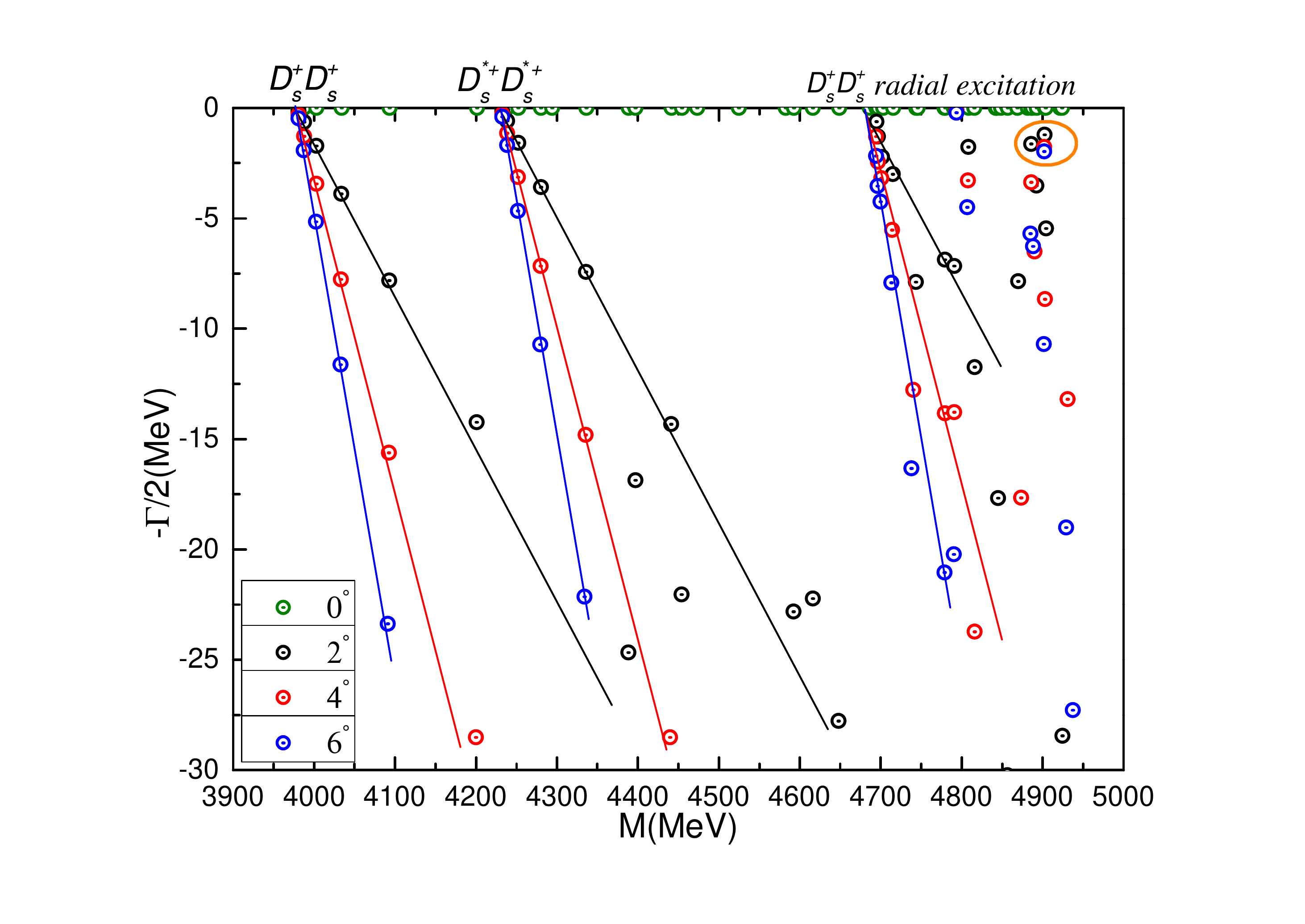}
\caption{Complex energies of $cc\bar{s}\bar{s}$ tetraquarks with $I(J^P)=0(0^+)$ in the complete coupled channels calculation, $\theta$ varying from $0^\circ$ to $6^\circ$ .} \label{PPCC1}
\end{figure}

\item	$I(J^P)=0(1^+)$ state

Two dimeson channels, $D^+_s D^{*+}_s$ and $D^{*+}_s D^{*+}_s$, one diquark-antidiquark channels, $(cc)^*(\bar{s}\bar{s})^*$ and four-K-types configurations are studied in Table~\ref{GresultCCSS2}.  Firstly, each of the channels along with the couplings in one certain configuration are all unbound. Specifically, the couplings are quite weak in both the color-singlet and hidden-color channels of meson-meson configurations. However, several to hundreds of MeV decreased are obtained in the coupling computations of K-types and their coupled-masses are $\sim$4.4 GeV. Then in a complete coupled-channels investigation, the lowest state is still unbound with mass equals to the theoretical threshold of $D^+_s D^{*+}_s$, 4105 MeV. Meanwhile, in comparison with the results of $cc\bar{q}\bar{q}$ tetraquarks in Table~\ref{GresultCC1}, one can find that character of the $D^+_s D^{*+}_s$ state is opposite to $D^+D^{*0}$ case which has a binding energy $\sim$200 MeV.

Fig.~\ref{PPCC2} shows the distributions of complex energies of the $D^+_s D^{*+}_s$ and $D^{*+}_s D^{*+}_s$ scattering states in the complete coupled-channels calculation. In 4.1 GeV to 5.0 GeV energy gap, the ground states and first radial excitation ones are clearly shown. In spite of three slowly descended poles between 4.55 GeV and 4.70 GeV, the nature of resonance state is in contrast to them and neither bound state nor resonance one can be obtained in this quantum state.

\begin{table}[H]
\caption{\label{GresultCCSS2} The lowest-lying eigen-energies of $cc\bar{s}\bar{s}$ tetraquarks with $I(J^P)=0(1^+)$ in the real range calculation. (unit: MeV)}
\centering
\begin{tabular}{ccccc}
\toprule
~~\textbf{Channel}~~   & ~~\textbf{Index}~~ & ~~$\chi_J^{\sigma_i}$;~$\chi_I^{f_j}$;~$\chi_k^c$~~ & ~~$\textbf{Mass}$~~ & ~~\textbf{Mixed}~~ \\
        &      &$[i; ~j; ~k]$ &  \\
\midrule
$(D^+_s D^{*+}_s)^1 (4081)$      & 1(S) & $[1; ~1; ~1]$   & $4105$ & \\
$(D^{*+}_s D^{*+}_s)^1 (4224)$  & 2(S) & $[3; ~1; ~1]$    & $4232$ & $4105$ \\[2ex]
$(D^+_s D^{*+}_s)^8$          & 3(H) & $[1; ~1; ~2]$   & $4401$ & \\
$(D^{*+}_s D^{*+}_s)^8$     & 4(H) & $[3; ~1; ~2]$   & $4607$ & $4400$ \\[2ex]
$(cc)^*(\bar{s}\bar{s})^*$   & 5  & $[6; ~1; ~3]$  & $4424$ & $4424$ \\[2ex]
$K_1$  & 6  & $[7; ~1; ~5]$  & $4537$ & \\
$K_1$  & 7 & $[8; ~1; ~5]$    & $4536$ & \\
$K_1$  & 8 & $[9; ~1; ~5]$   & $4528$ & \\
$K_1$  & 9 & $[7; ~1; ~6]$   & $4440$ & \\
$K_1$  & 10 & $[8; ~1; ~6]$   & $4445$ & \\
$K_1$  & 11 & $[9; ~1; ~6]$   & $4371$ & $4305$ \\[2ex]
$K_2$  & 12 & $[10; ~1; ~7]$   & $4417$ & \\
$K_2$  & 13 & $[11; ~1; ~7]$   & $4419$ & \\
$K_2$  & 14 & $[12; ~1; ~7]$   & $4326$ & \\
$K_2$  & 15 & $[10; ~1; ~8]$   & $4699$ & \\
$K_2$  & 16 & $[11; ~1; ~8]$   & $4787$ & \\
$K_2$  & 17 & $[12; ~1; ~8]$   & $4802$ & $4266$ \\[2ex]
$K_3$  & 18  & $[13; ~1; ~10]$  & $4442$ & \\
$K_3$  & 19 & $[14; ~1; ~10]$   & $4443$ & \\
$K_3$  & 20 & $[15; ~1; ~9]$   & $5013$ & $4424$ \\[2ex]
$K_4$  & 21 & $[16; ~1; ~12]$  & $4427$ & \\
$K_4$  & 22 & $[17; ~1; ~12]$   & $4426$ & \\
$K_4$  & 23 & $[18; ~1; ~11]$   & $4953$ & $4423$ \\[2ex]
\multicolumn{3}{c}{All of the above channels:} & $4105$ \\
\bottomrule
\end{tabular}
\end{table}

\begin{figure}[H]
\centering
\includegraphics[width=14 cm]{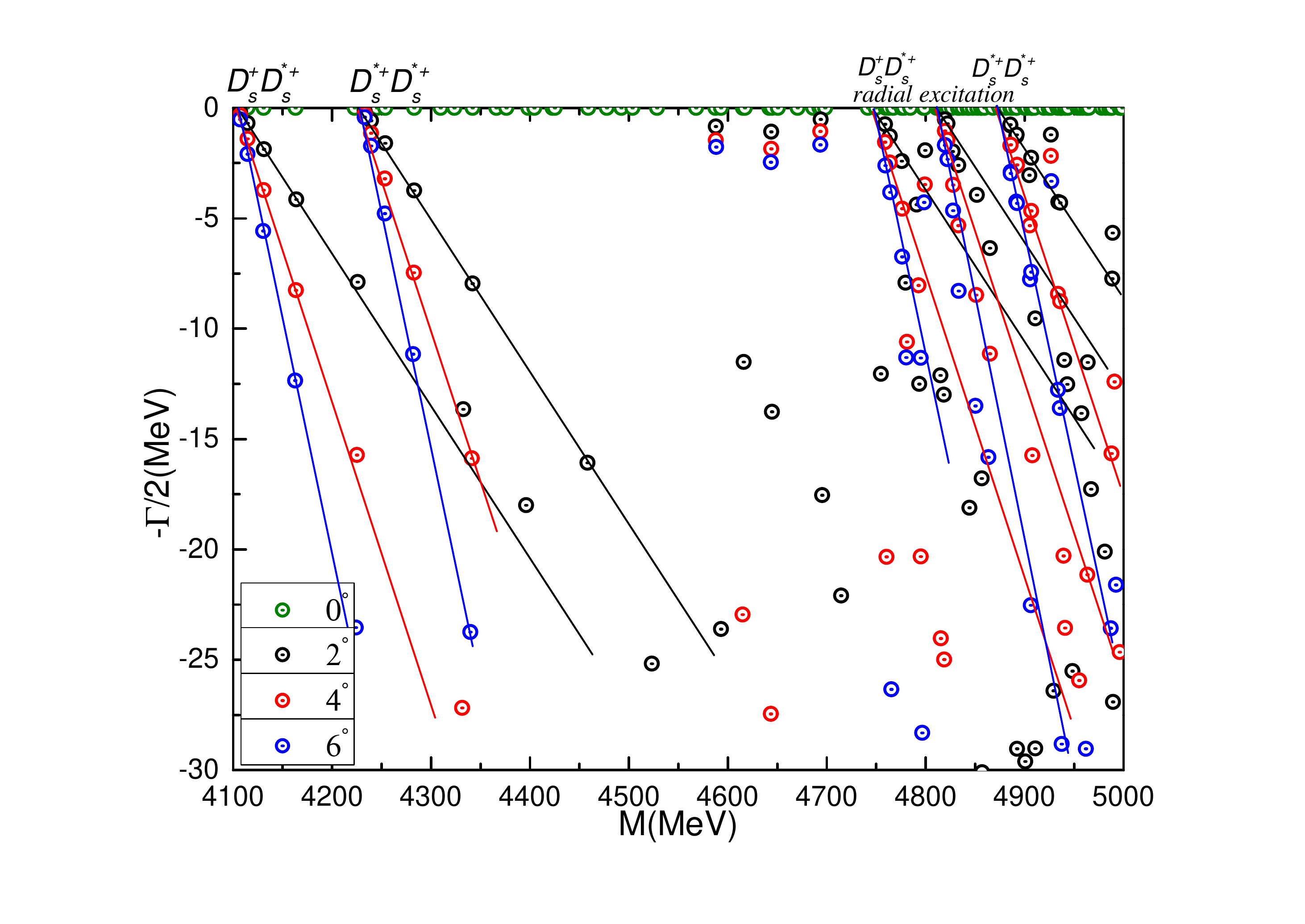}
\caption{Complex energies of $cc\bar{s}\bar{s}$ tetraquarks with $I(J^P)=0(1^+)$ in the coupled channels calculation, $\theta$ varying from $0^\circ$ to $6^\circ$ .} \label{PPCC2}
\end{figure}

\item	$I(J^P)=0(2^+)$ state

In the highest spin state, only the $D^{*+}_s D^{*+}_s$ channel in meson-meson configuration need to be considered. Besides, there is one diquark-antidiquark channel $(cc)^*(\bar{s}\bar{s})^*$. The calculated masses in color-singlet and hidden-color channels are 4232 MeV and 4432 MeV, respectively. The $(cc)^*(\bar{s}\bar{s})^*$ channel mass is very close to the hidden-color one with $M=4446\,\text{MeV}$. As for the other K-types configurations, their theoretical masses are also around 4.38 GeV for the $K_1$, $K_2$ structures and 4.45 GeV for the rest two ones. Obviously, all of them are above the $D^{*+}_s D^{*+}_s$ threshold value and this fact do not change in a fully coupled-channels calculation.

Nevertheless, three resonance states are found in the complex-range calculation which all of the channels listed in Table~\ref{GresultCCSS3} are considered. In Fig.~\ref{PPCC3} we can see it clearly that apart from the continuum states of $D^{*+}_s D^{*+}_s$ in the ground and first radial excitation states, three almost fixed poles are obtained at $\sim$4.8 GeV. In particular, the three orange circles marked the obtained resonance states of $D^{*+}_s D^{*+}_s$ molecule state, their masses and widths are $(4821\,MeV,\,5.58\,MeV)$, $(4846\,MeV,\,10.68\,MeV)$ and $(4775\,MeV,\,23.26\,MeV)$, respectively. These poles are also around 0.6 GeV above two non-interacting $D^{*+}_s$ mesons threshold, and $\sim$0.1 GeV below its first radial excitation state.

\begin{table}[H]
\caption{\label{GresultCCSS3} The lowest-lying eigen-energies of $cc\bar{s}\bar{s}$ tetraquarks with $I(J^P)=0(2^+)$ in the real range calculation. (unit: MeV)}
\centering
\begin{tabular}{ccccc}
\toprule
~~\textbf{Channel}~~   & ~~\textbf{Index}~~ & ~~$\chi_J^{\sigma_i}$;~$\chi_I^{f_j}$;~$\chi_k^c$~~ & ~~$\textbf{Mass}$~~ & ~~\textbf{Mixed}~~ \\
            &  &$[i; ~j; ~k]$&  \\
\midrule
$(D^{*+}_s D^{*+}_s)^1 (4224)$  & 1(S)  & $[1; ~1; ~1]$  & $4232$ & $4232$ \\[2ex]
$(D^{*+}_s D^{*+}_s)^8$  & 2(H)  & $[1; ~1; ~2]$  & $4432$ & $4432$ \\[2ex]
$(cc)^*(\bar{s}\bar{s})^*$  & 3  & $[1; ~1; ~3]$  & $4446$ & $4446$ \\[2ex]
$K_1$  & 4 & $[1; ~1; ~5]$   & $4522$ & \\
$K_1$  & 5  & $[1; ~1; ~6]$  & $4385$ & $4381$ \\[2ex]
$K_2$  & 6  & $[1; ~1; ~7]$  & $4355$ & \\
$K_2$  & 7 & $[1; ~1; ~8]$   & $4666$ & $4354$ \\[2ex]
$K_3$  & 8  & $[1; ~1; ~10]$ & $4448$ & $4448$ \\[2ex]
$K_4$  & 9 & $[1; ~1; ~12]$  & $4446$ & $4446$ \\[2ex]
\multicolumn{3}{c}{All of the above channels:} & $4232$ \\
\bottomrule
\end{tabular}
\end{table}

\begin{figure}[H]
\centering
\includegraphics[width=14 cm]{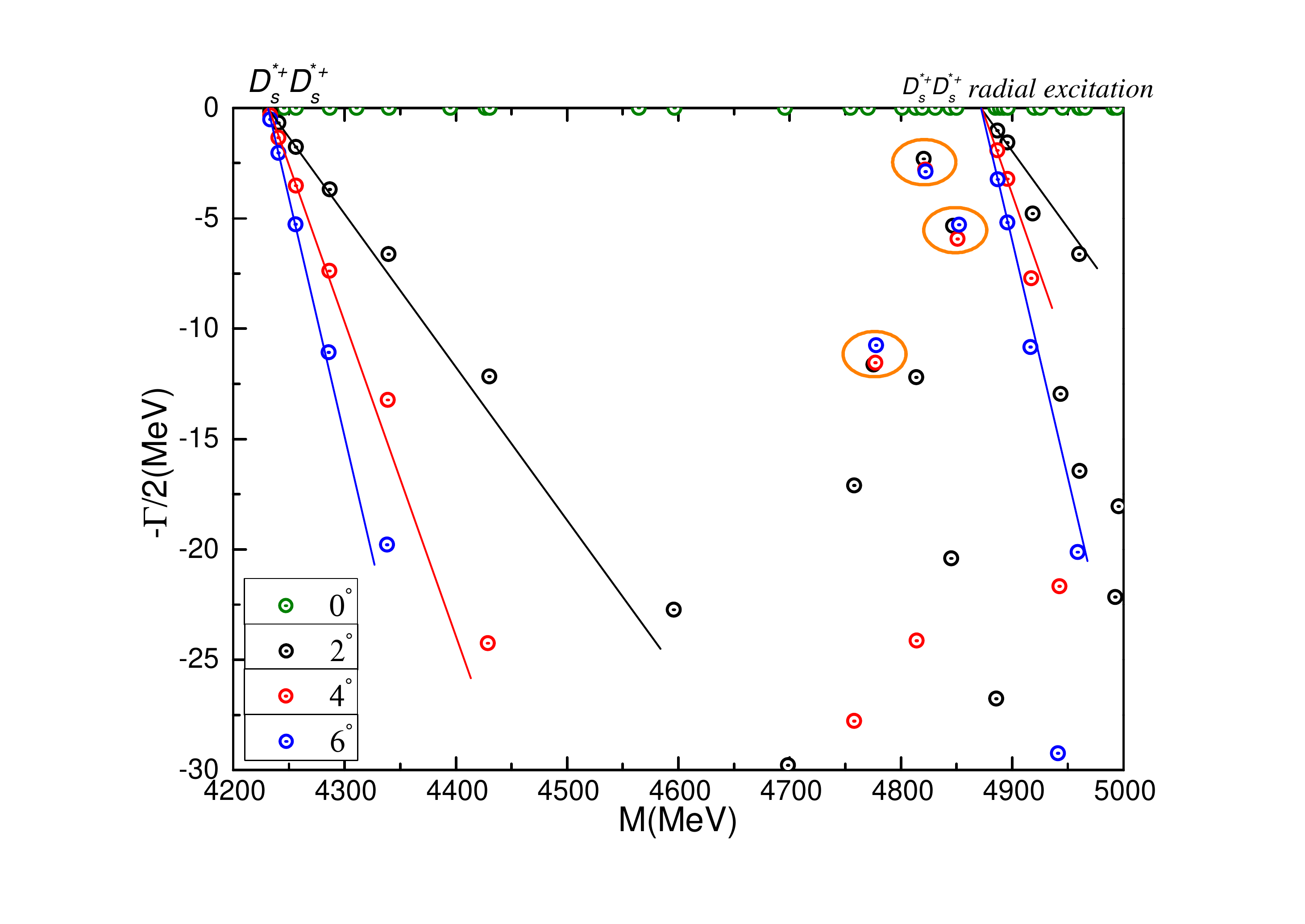}
\caption{Complex energies of $cc\bar{s}\bar{s}$ tetraquarks with $I(J^P)=0(2^+)$ in the coupled channels calculation, $\theta$ varying from $0^\circ$ to $6^\circ$ .} \label{PPCC3}
\end{figure}

\end{enumerate}

\item	$bb\bar{s}\bar{s}$ tetraquarks

In this part, we study the $I(J^P)=0(0^+)$, $0(1^+)$ and $0(2^+)$ states for $bb\bar{s}\bar{s}$ tetraquarks. It is similar to the $cc\bar{s}\bar{s}$ systems that only narrow resonances are found in $0(0^+)$ and $0(2^+)$ states.
Let us proceed to discuss them in detail.

\begin{enumerate}[leftmargin=*,labelsep=4.9mm, listparindent=2em]
\item	$I(J^P)=0(0^+)$ state

In this quantum state as shown in Table~\ref{GresultCCSS4}, there are two meson-meson, diquark-antidiquark and 12 K-type channels under investigated. Theoretical masses of each single channels locate in the region from 10.71 GeV to 11.45 GeV, besides, the coupled-mass $\sim$10.9 GeV in $\bar{B}^0_s \bar{B}^0_s$ and $\bar{B}^*_s \bar{B}^*_s$ channels is comparable with those of $K_1$, $K_2$, $K_3$ and $K_4$ channels. Although the color-singlet channels mass  is the lowest, on bound state is found in the coupled-channels calculation which includes the fully channels case.

Additionally, Fig.~\ref{PPCC4} shows the results of complete coupled-channels by complex scaling method. The states of ground and first radial excitation for $\bar{B}^0_s \bar{B}^0_s$ and $\bar{B}^*_s \bar{B}^*_s$ channels are generally presented from 10.7 GeV to 11.5 GeV. However, there are three narrow resonance poles obtained near the $2S$ state of $\bar{B}^0_s \bar{B}^0_s$. Particularly, in the two big orange circles, the calculated dots which the value of rotated angle $\theta$ are $2^\circ$, $4^\circ$ and $6^\circ$, respectively are almost overlapped. Their masses and widths are (11.31 GeV, 1.86 MeV), (11.33 GeV, 1.84 MeV) and (11.41 GeV, 1.54 MeV), respectively. Furthermore, with much heavier flavored quark included, more narrow molecular resonance state will be obtained when we compare these results with those of $cc\bar{s}\bar{s}$ tetraquarks in the $00^+$ state. Herein, we can identify the first two resonances which are about 0.5 GeV above the $\bar{B}^*_s \bar{B}^*_s$ threshold as two $\bar{B}^*_s$ mesons molecule state, and the third one as $\bar{B}^0_s \bar{B}^0_s$ resonance state due to it is $\sim$50 MeV above the $(1S)\bar{B}^0_s (2S)\bar{B}^0_s$ non-interacting threshold.

\begin{table}[H]
\caption{\label{GresultCCSS4} The lowest-lying eigen-energies of $bb\bar{s}\bar{s}$ tetraquarks with $I(J^P)=0(0^+)$ in the real range calculation. (unit: MeV)}
\centering
\begin{tabular}{ccccc}
\toprule
~~\textbf{Channel}~~   & ~~\textbf{Index}~~ & ~~$\chi_J^{\sigma_i}$;~$\chi_I^{f_j}$;~$\chi_k^c$~~ & ~~$\textbf{Mass}$~~ & ~~\textbf{Mixed}~~ \\
         &     &$[i; ~j; ~k]$&  \\
\midrule
$(\bar{B}^0_s \bar{B}^0_s)^1 (10734)$          & 1(S)  & $[1; ~2; ~1]$  & $10710$ & \\
$(\bar{B}^*_s \bar{B}^*_s)^1 (10830)$  & 2(S)  & $[2; ~2; ~1]$  & $10800$ & $10710$ \\[2ex]
$(\bar{B}^0_s \bar{B}^0_s)^8$          & 3(H)  & $[1; ~2; ~2]$  & $11184$ & \\
$(\bar{B}^*_s \bar{B}^*_s)^8$  & 4(H) & $[2; ~2; ~2]$   & $11205$ & $10943$ \\[2ex]
$(bb)(\bar{s}\bar{s})$      & 5  & $[3; ~2; ~4]$  & $10967$ & \\
$(bb)^*(\bar{s}\bar{s})^*$  & 6  & $[4; ~2; ~3]$  & $10901$ & $10896$ \\[2ex]
$K_1$  & 7  & $[5; ~2; ~5]$  & $11445$ & \\
$K_1$  & 8 & $[5; ~2; ~6]$  & $10928$ & \\
$K_1$  & 9 & $[6; ~2; ~5]$   & $11259$ & \\
$K_1$  & 10  & $[6; ~2; ~6]$  & $10863$ & $10843$ \\[2ex]
$K_2$  & 11  & $[7; ~2; ~7]$  & $10877$ & \\
$K_2$  & 12  & $[7; ~2; ~8]$  & $11445$ & \\
$K_2$  & 13 & $[8; ~2; ~7]$   & $10815$ & \\
$K_2$  & 14 & $[8; ~2; ~8]$   & $11441$ & $10802$ \\[2ex]
$K_3$  & 15  & $[9; ~2; ~10]$  & $10902$ & \\
$K_3$  & 16  & $[10; ~2; ~9]$  & $10960$ & $10895$ \\[2ex]
$K_4$  & 17  & $[11; ~2; ~12]$  & $10901$ & \\
$K_4$  & 18  & $[12; ~2; ~11]$  & $10980$ & $10897$ \\[2ex]
\multicolumn{3}{c}{All of the above channels:} & $10710$ \\
\bottomrule
\end{tabular}
\end{table}

\begin{figure}[H]
\centering
\includegraphics[width=14 cm]{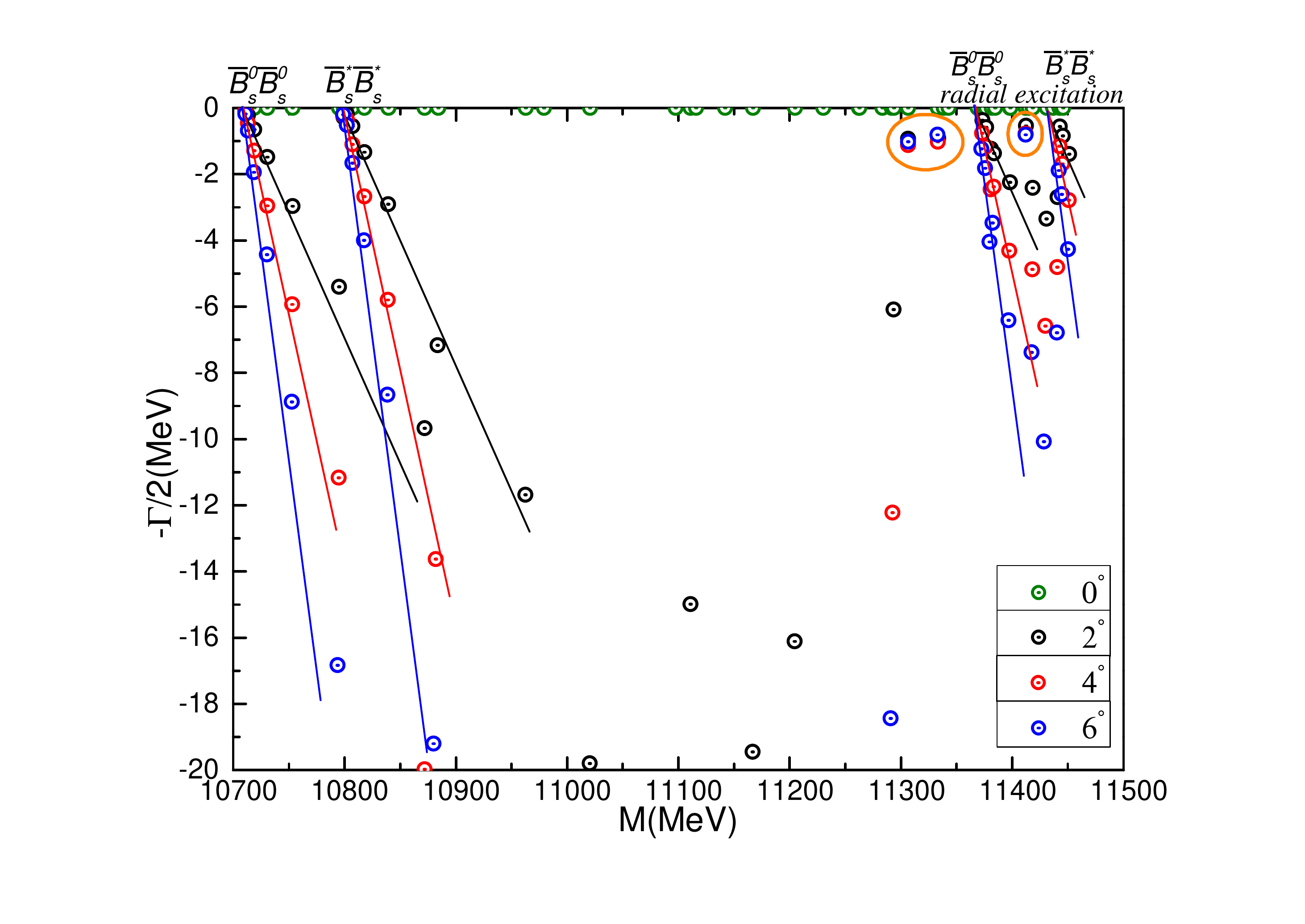}
\caption{Complex energies of $bb\bar{s}\bar{s}$ tetraquarks with $I(J^P)=0(0^+)$ in the coupled channels calculation, $\theta$ varying from $0^\circ$ to $6^\circ$ .} \label{PPCC4}
\end{figure}

\item	$I(J^P)=0(1^+)$ state

In contrast the obtained deeply bound and narrow resonance states of $bb\bar{q}\bar{q}$ tetraquarks in $01^+$ state,  bound state is forbidden as the $cc\bar{s}\bar{s}$ system. Firstly, in Table~\ref{GresultCCSS5}, masses of the meson-meson channels of $\bar{B}^0_s \bar{B}^*_s$ and $\bar{B}^*_s \bar{B}^*_s$ in color-singlet state are 10755 MeV and 10800 MeV, respectively. These results are not changed when their hidden-color channels included. As for the other exotic structures, $i. e.$, diquark-antidiquark $bb\bar{s}\bar{s}$, and K-type channels, the theoretical masses are all $\sim$10.9 GeV. Besides, these excited states do not help in forming a bound state in the complete coupled-channels calculation which the lowest mass is still 10755 MeV.

The above conclusion is clear shown in Fig.~\ref{PPCC5} which is the results in complex-range study. In the mass region from 10.7 GeV to 11.3 GeV, only two scattering state of $\bar{B}^0_s \bar{B}^*_s$ and $\bar{B}^*_s \bar{B}^*_s$ channels are obtained, the other bound or resonance state is unavailable in the present theoretical framework. Specifically, one may notice a gradually varied dots at 11.15 GeV, however, these unstable poles still can not be identified as a regular resonance state.

\begin{table}[H]
\caption{\label{GresultCCSS5} The lowest-lying eigen-energies of $bb\bar{s}\bar{s}$ tetraquarks with $I(J^P)=0(1^+)$ in the real range calculation. (unit: MeV)}
\centering
\begin{tabular}{ccccc}
\toprule
~~\textbf{Channel}~~   & ~~\textbf{Index}~~ & ~~$\chi_J^{\sigma_i}$;~$\chi_I^{f_j}$;~$\chi_k^c$~~ & ~~$\textbf{Mass}$~~ & ~~\textbf{Mixed}~~ \\
       &       &$[i; ~j; ~k]$&  \\
\midrule
$(\bar{B}^0_s \bar{B}^*_s)^1 (10782)$      & 1(S)  & $[1; ~2; ~1]$  & $10755$ & \\
$(\bar{B}^*_s \bar{B}^*_s)^1 (10830)$  & 2(S)  & $[3; ~2; ~1]$  & $10800$ & $10755$ \\[2ex]
$(\bar{B}^0_s \bar{B}^*_s)^8$          & 3(H) & $[1; ~2; ~2]$   & $10949$ & \\
$(\bar{B}^*_s \bar{B}^*_s)^8$     & 4(H)  & $[3; ~2; ~2]$   & $11185$ & $10949$ \\[2ex]
$(bb)^*(\bar{s}\bar{s})^*$   & 5 & $[6; ~2; ~3]$  & $10906$ & $10906$ \\[2ex]
$K_1$  & 6 & $[7; ~2; ~5]$  & $11041$ & \\
$K_1$  & 7 & $[8; ~2; ~5]$  & $11048$ & \\
$K_1$  & 8 & $[9; ~2; ~5]$  & $11038$ & \\
$K_1$  & 9 & $[7; ~2; ~6]$  & $10936$ & \\
$K_1$  & 10 & $[8; ~2; ~6]$  & $10949$ & \\
$K_1$  & 11  & $[9; ~2; ~6]$  & $10917$ & $10870$ \\[2ex]
$K_2$  & 12 & $[10; ~2; ~7]$  & $10911$ & \\
$K_2$  & 13 & $[11; ~2; ~7]$   & $10914$ & \\
$K_2$  & 14  & $[12; ~2; ~7]$  & $10879$ & \\
$K_2$  & 15  & $[10; ~2; ~8]$   & $11216$ & \\
$K_2$  & 16 & $[11; ~2; ~8]$   & $11483$ & \\
$K_2$  & 17 & $[12; ~2; ~8]$  & $11373$ & $10840$ \\[2ex]
$K_3$  & 18  & $[13; ~2; ~10]$  & $10928$ & \\
$K_3$  & 19  & $[14; ~2; ~10]$  & $10929$ & \\
$K_3$  & 20  & $[15; ~2; ~9]$  & $11557$ & $10907$ \\[2ex]
$K_4$  & 21  & $[16; ~2; ~12]$  & $10911$ & \\
$K_4$  & 22  & $[17; ~2; ~12]$  & $10908$ & \\
$K_4$  & 23  & $[18; ~2; ~11]$  & $11458$ & $10906$ \\[2ex]
\multicolumn{3}{c}{All of the above channels:} & $10755$ \\
\bottomrule
\end{tabular}
\end{table}

\begin{figure}[H]
\centering
\includegraphics[width=14 cm]{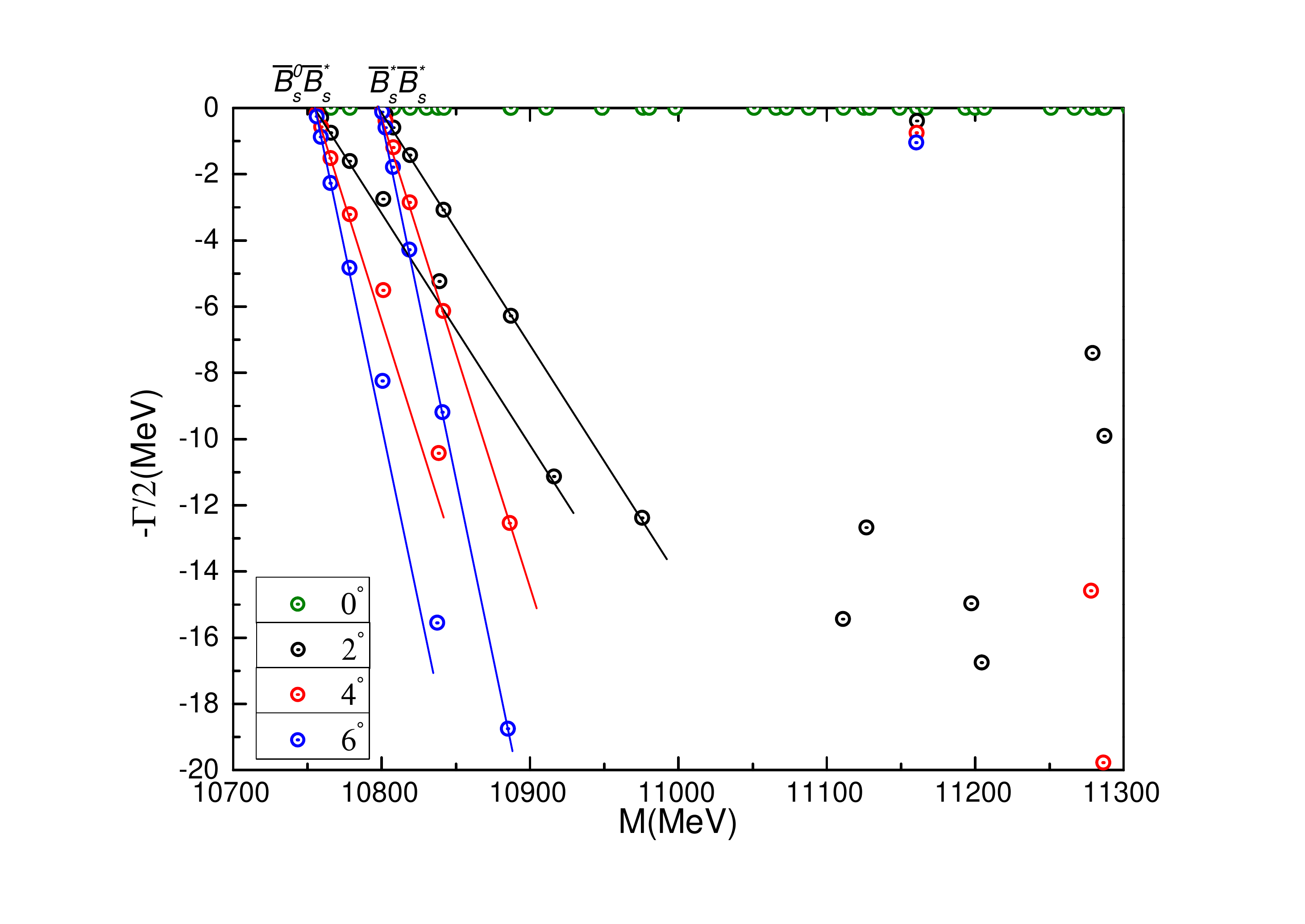}
\caption{Complex energies of $bb\bar{s}\bar{s}$ tetraquarks with $I(J^P)=0(1^+)$ in the coupled channels calculation, $\theta$ varying from $0^\circ$ to $6^\circ$ .} \label{PPCC5}
\end{figure}

\item	$I(J^P)=0(2^+)$ state

One $\bar{B}^*_s \bar{B}^*_s$ meson-meson channel is studied in Table~\ref{GresultCCSS6}, but the calculated mass 10.8 GeV in color-singlet channel is just the theoretical threshold value. The other hidden-color, diqaurk-antidiquark and K-type structures are all above 10.9 GeV in the coupled-channels calculation of each configurations except the 10.87 MeV for $K_1$ and $K_2$ channels. Meanwhile, it is the same as the several cases before that the lowest energy level is still unbound in the fully coupled-channels computation and the coupled mass is 10.8 GeV.

Nevertheless, when the investigation extended to the complex-range, new exotic states are obtained. In Fig.~\ref{PPCC6}, three narrow resonance states at $\sim$11.35 GeV are marked with orange circles which are near the real-axis. With a angle $\theta$ varied less than $6^\circ$, masses and widths of these three fixed poles are (11.33 GeV, 1.48 MeV), (11.36 GeV, 4.18 MeV) and (11.41 GeV, 2.52 MeV), respectively. Obviously, they are also around 0.6 GeV above the $\bar{B}^*_s \bar{B}^*_s$ threshold, therefore the nature of two $\bar{B}^*_s$ mesons molecule states can be drawn herein.

\begin{table}[H]
\caption{\label{GresultCCSS6} The lowest-lying eigen-energies of $bb\bar{s}\bar{s}$ tetraquarks with $I(J^P)=0(2^+)$ in the real range calculation. (unit: MeV)}
\centering
\begin{tabular}{ccccc}
\toprule
~~\textbf{Channel}~~   & ~~\textbf{Index}~~  & ~~$\chi_J^{\sigma_i}$;~$\chi_I^{f_j}$;~$\chi_k^c$~~ & ~~$\textbf{Mass}$~~ & ~~\textbf{Mixed}~~ \\
          &    &$[i; ~j; ~k]$&  \\
\midrule
$(\bar{B}^*_s \bar{B}^*_s)^1 (10830)$  & 1(S)  & $[1; ~2; ~1]$  & $10800$ & $10800$ \\[2ex]
$(\bar{B}^*_s \bar{B}^*_s)^8$  & 2(H)  & $[1; ~2; ~2]$  & $10959$ & $10959$ \\[2ex]
$(bb)^*(\bar{s}\bar{s})^*$  & 3 & $[1; ~2; ~3]$  & $10915$ & $10915$ \\[2ex]
$K_1$  & 4  & $[1; ~2; ~5]$  & $11023$ & \\
$K_1$  & 5  & $[1; ~2; ~6]$  & $10894$ & $10879$ \\[2ex]
$K_2$  & 6  & $[1; ~2; ~7]$  & $10870$ & \\
$K_2$  & 7  & $[1; ~2; ~8]$  & $11186$ & $10869$ \\[2ex]
$K_3$  & 8  & $[1; ~2; ~10]$  & $10918$ & $10918$ \\[2ex]
$K_4$  & 9  & $[1; ~2; ~12]$  & $10916$ & $10916$ \\[2ex]
\multicolumn{3}{c}{All of the above channels:} & $10800$ \\
\bottomrule
\end{tabular}
\end{table}

\begin{figure}[H]
\centering
\includegraphics[width=14 cm]{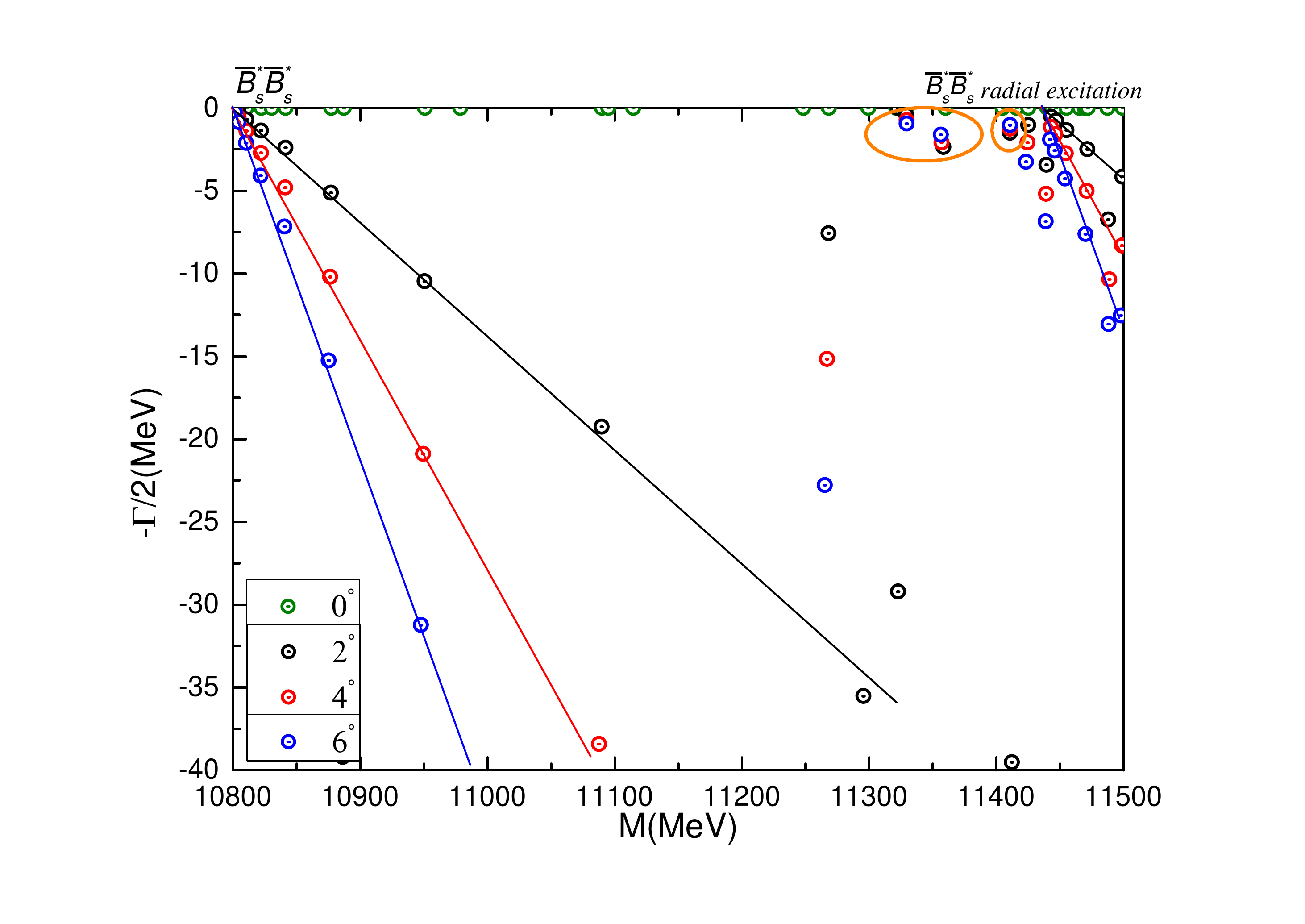}
\caption{Complex energies of $bb\bar{s}\bar{s}$ tetraquarks with $I(J^P)=0(2^+)$ in the coupled channels calculation, $\theta$ varying from $0^\circ$ to $6^\circ$ .} \label{PPCC6}
\end{figure}

\end{enumerate}

\item	$cb\bar{s}\bar{s}$ tetraquarks

Bound state is also not found in this sector, however, some narrow resonance states in $I(J^P)=0(0^+)$, $0(1^+)$ and $0(2^+)$ are obtained. The following parts are devoted to the discussions on them.

\begin{enumerate}[leftmargin=*,labelsep=4.9mm, listparindent=2em]
\item	$I(J^P)=0(0^+)$ state

As shown in Table~\ref{GresultCCSS7}, two meson-meson syructures, $D^+_s \bar{B}^0_s$ and $D^{*+}_s \bar{B}^*_s$, two diquark-antidiquark channels, $(cb)(\bar{s}\bar{s})$ and $(cb)^*(\bar{s}\bar{s})^*$, 14 K-type ones are considered in this quantum state. Firstly, the calculated masses of these channels are located in the energy region from 7.34 GeV to 8.67 GeV and no bound state is found. Then this result remains in the coupled-channels computations for the dimeson, diquark-antidiquark and K-type configurations. In particular, the coupling in $D^+_s \bar{B}^0_s$ and $D^{*+}_s \bar{B}^*_s$ channels is extremely weak. However, dozens to hundreds MeV decreased in the other configurations couplings, and they are all above 7.6 GeV. Finally, in the real-range fully coupled-channels calculation, the lowest energy level is still at 7344 MeV.

In a further investigation which the CSM is employed, one can find that two resonance states are marked in Fig.~\ref{PPCC7}. Around 0.5 GeV higher than the $D^{*+}_s \bar{B}^*_s$ threshold value, two narrow resonance poles with masses and widths equal to (7.92 GeV, 1.02 MeV) and (7.99 GeV, 3.22 MeV), respectively are stable against the variation of rotated angle $\theta$. Herein, we can identify them as the molecule states of $D^{*+}_s \bar{B}^*_s$.

\begin{table}[H]
\caption{\label{GresultCCSS7} The lowest-lying eigen-energies of $cb\bar{s}\bar{s}$ tetraquarks with $I(J^P)=0(0^+)$ in the real range calculation. (unit: MeV)}
\centering
\begin{tabular}{ccccc}
\toprule
~~\textbf{Channel}~~   & ~~\textbf{Index}~~ & ~~$\chi_J^{\sigma_i}$;~$\chi_I^{f_j}$;~$\chi_k^c$~~ & ~~$\textbf{Mass}$~~ & ~~\textbf{Mixed}~~ \\
           &   &$[i; ~j; ~k]$&  \\
\midrule
$(D^+_s \bar{B}^0_s)^1 (7336)$          & 1(S) & $[1; ~3; ~1]$   & $7344$ & \\
$(D^{*+}_s \bar{B}^*_s)^1 (7527)$  & 2(S) & $[2; ~3; ~1]$   & $7516$ & $7344$ \\[2ex]
$(D^+_s \bar{B}^0_s)^8$          & 3(H)  & $[1; ~3; ~2]$  & $7910$ & \\
$(D^{*+}_s \bar{B}^*_s)^8$  & 4(H) & $[2; ~3; ~2]$  & $7927$ & $7678$ \\[2ex]
$(cb)(\bar{s}\bar{s})$      & 5 & $[3; ~3; ~4]$  & $7726$ & \\
$(cb)^*(\bar{s}\bar{s})^*$  & 6  & $[4; ~3; ~3]$ & $7675$ & $7662$ \\[2ex]
$K_1$  & 7  & $[5; ~3; ~5]$ & $8171$ & \\
$K_1$  & 8 & $[5; ~3; ~6]$  & $8274$ & \\
$K_1$  & 9 & $[6; ~3; ~5]$  & $8369$ & \\
$K_1$  & 10  & $[6; ~3; ~6]$  & $8145$ & $7613$ \\[2ex]
$K_2$  & 11  & $[7; ~3; ~7]$  & $7896$ & \\
$K_2$  & 12  & $[7; ~3; ~8]$  & $8266$ & \\
$K_2$  & 13  & $[8; ~3; ~7]$  & $7758$ & \\
$K_2$  & 14  & $[8; ~3; ~8]$  & $8282$ & $7629$ \\[2ex]
$K_3$  & 15  & $[9; ~3; ~9]$  & $8647$ & \\
$K_3$  & 16  & $[9; ~3; ~10]$  & $8181$ & \\
$K_3$  & 17  & $[10; ~3; ~9]$  & $8321$ & \\
$K_3$  & 18  & $[10; ~3; ~10]$  & $8675$ & $8010$ \\[2ex]
$K_4$  & 19  & $[11; ~3; ~12]$  & $8199$ & \\
$K_4$  & 20  & $[12; ~3; ~11]$  & $8359$ & $8063$ \\[2ex]
\multicolumn{3}{c}{All of the above channels:} & $7344$ \\
\bottomrule
\end{tabular}
\end{table}

\begin{figure}[H]
\centering
\includegraphics[width=14 cm]{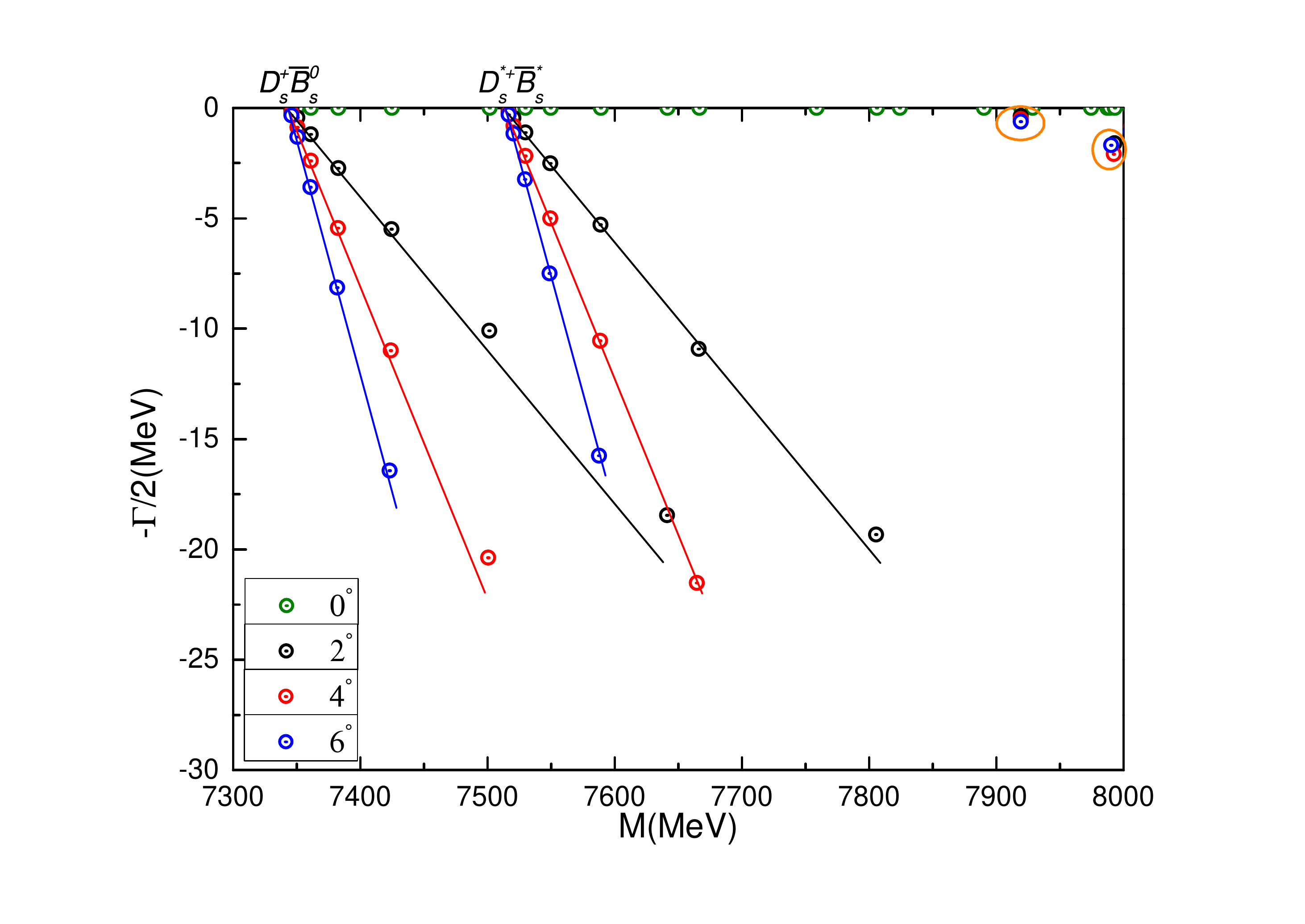}
\caption{Complex energies of $cb\bar{s}\bar{s}$ tetraquarks with $I(J^P)=0(0^+)$ in the coupled channels calculation, $\theta$ varying from $0^\circ$ to $6^\circ$ .} \label{PPCC7}
\end{figure}

\item	$I(J^P)=0(1^+)$ state

Table~\ref{GresultCCSS8} presents 30 channels which include the $D^+_s \bar{B}^*_s$, $D^{*+}_s \bar{B}^0_s$ and $D^{*+}_s \bar{B}^*_s$ dimeson channels, $(cb)(\bar{s}\bar{s})^*$, $(cb)^*(\bar{s}\bar{s})$ and $(cb)^*(\bar{s}\bar{s})^*$ diquark-antidiquark structures and 21 K-type configurations. In the first kind of study, masses of each channels are between 7.39 GeV and 8.23 GeV. Particularly, the lowest level is the scattering state of $D^+_s \bar{B}^*_s$ with $M=7389$ MeV. Furthermore, the coupled-masses are slightly modified in each configurations: mass of the meson-meson structure in color-singlet channel is still 7389 MeV and the other configurations' masses are $\sim$7.67 GeV except for the $K_2$-type channels with $M=7.51$ GeV. Bound state is not found in the fully coupled-channels study, however, two narrow resonances are obtained in the complex-range.

Fig.~\ref{PPCC8} shows the calculated complex energies in the region from 7.3 GeV to 8.0 GeV. Two points can be concluded accoedingly, (i) the majority poles belong to the scattering states of $D^+_s \bar{B}^*_s$, $D^{*+}_s \bar{B}^0_s$ and $D^{*+}_s \bar{B}^*_s$, (ii) two fixed resonance poles at 7.92 GeV and 7.99 GeV with widths equal to 1.20 MeV and 4.96 MeV, respectively are obtained. Because they are above the $D^{*+}_s \bar{B}^*_s$ threshold lines and farer from the other two scattering states, a nature of the $D^{*+}_s \bar{B}^*_s$ molecule state for the two resonances is reasonable.

\begin{table}[H]
\caption{\label{GresultCCSS8} The lowest-lying eigen-energies of $cb\bar{s}\bar{s}$ tetraquarks with $I(J^P)=0(1^+)$ in the real range calculation. (unit: MeV)}
\centering
\begin{tabular}{ccccc}
\toprule
~~\textbf{Channel}~~   & ~~\textbf{Index}~~ & ~~$\chi_J^{\sigma_i}$;~$\chi_I^{f_j}$;~$\chi_k^c$~~ & ~~$\textbf{Mass}$~~ & ~~\textbf{Mixed}~~ \\
          &    &$[i; ~j; ~k]$&  \\
\midrule
$(D^+_s \bar{B}^*_s)^1 (7384)$          & 1(S)  & $[1; ~3; ~1]$  & $7389$ & \\
$(D^{*+}_s \bar{B}^0_s)^1 (7479)$     & 2(S)  & $[2; ~3; ~1]$  & $7471$ & \\
$(D^{*+}_s \bar{B}^*_s)^1 (7527)$  & 3(S) & $[3; ~3; ~1]$  & $7516$ & $7389$ \\[2ex]
$(D^+_s \bar{B}^*_s)^8$          & 4(H)  & $[1; ~3; ~2]$  & $7900$ & \\
$(D^{*+}_s \bar{B}^0_s)^8$     & 5(H) & $[2; ~3; ~2]$   & $7891$ & \\
$(D^{*+}_s \bar{B}^*_s)^8$      & 6(H)  & $[3; ~3; ~2]$  & $7920$ & $7684$ \\[2ex]
$(cb)(\bar{s}\bar{s})^*$      & 7  & $[6; ~3; ~3]$  & $7683$ & \\
$(cb)^*(\bar{s}\bar{s})$      & 8 & $[5; ~3; ~3]$   & $7680$ & \\
$(cb)^*(\bar{s}\bar{s})^*$  & 9  & $[4; ~3; ~4]$  & $7725$ & $7671$ \\[2ex]
$K_1$  & 10  & $[7; ~3; ~5]$  & $7796$ & \\
$K_1$  & 11  & $[8; ~3; ~5]$ & $8172$ & \\
$K_1$  & 12 & $[9; ~3; ~5]$   & $8009$ & \\
$K_1$  & 13 & $[7; ~3; ~6]$    & $7695$ & \\
$K_1$  & 14  & $[8; ~3; ~6]$   & $7760$ & \\
$K_1$  & 15  & $[9; ~3; ~6]$   & $7634$ & $7620$ \\[2ex]
$K_2$  & 16 & $[10; ~3; ~7]$  & $7607$ & \\
$K_2$  & 17 & $[11; ~3; ~7]$  & $7621$ & \\
$K_2$  & 18  & $[12; ~3; ~7]$   & $7510$ & \\
$K_2$  & 19  & $[10; ~3; ~8]$  & $8137$ & \\
$K_2$  & 20 & $[11; ~3; ~8]$  & $8211$ & \\
$K_2$  & 21 & $[12; ~3; ~8]$   & $8209$ & $7505$ \\[2ex]
$K_3$  & 22 & $[13; ~3; ~10]$  & $7705$ & \\
$K_3$  & 23  & $[14; ~3; ~10]$  & $7706$ & \\
$K_3$  & 24  & $[15; ~3; ~10]$  & $7682$ & \\
$K_3$  & 25  & $[13; ~3; ~9]$  & $7734$ & \\
$K_3$  & 26  & $[14; ~3; ~9]$  & $7733$ & \\
$K_3$  & 27  & $[15; ~3; ~9]$  & $8298$ & $7666$ \\[2ex]
$K_4$  & 28  & $[16; ~3; ~12]$  & $7687$ & \\
$K_4$  & 29  & $[17; ~3; ~12]$  & $7677$ & \\
$K_4$  & 30  & $[18; ~3; ~11]$  & $7771$ & $7670$ \\[2ex]
\multicolumn{3}{c}{All of the above channels:} & $7389$ \\
\bottomrule
\end{tabular}
\end{table}

\begin{figure}[H]
\centering
\includegraphics[width=14 cm]{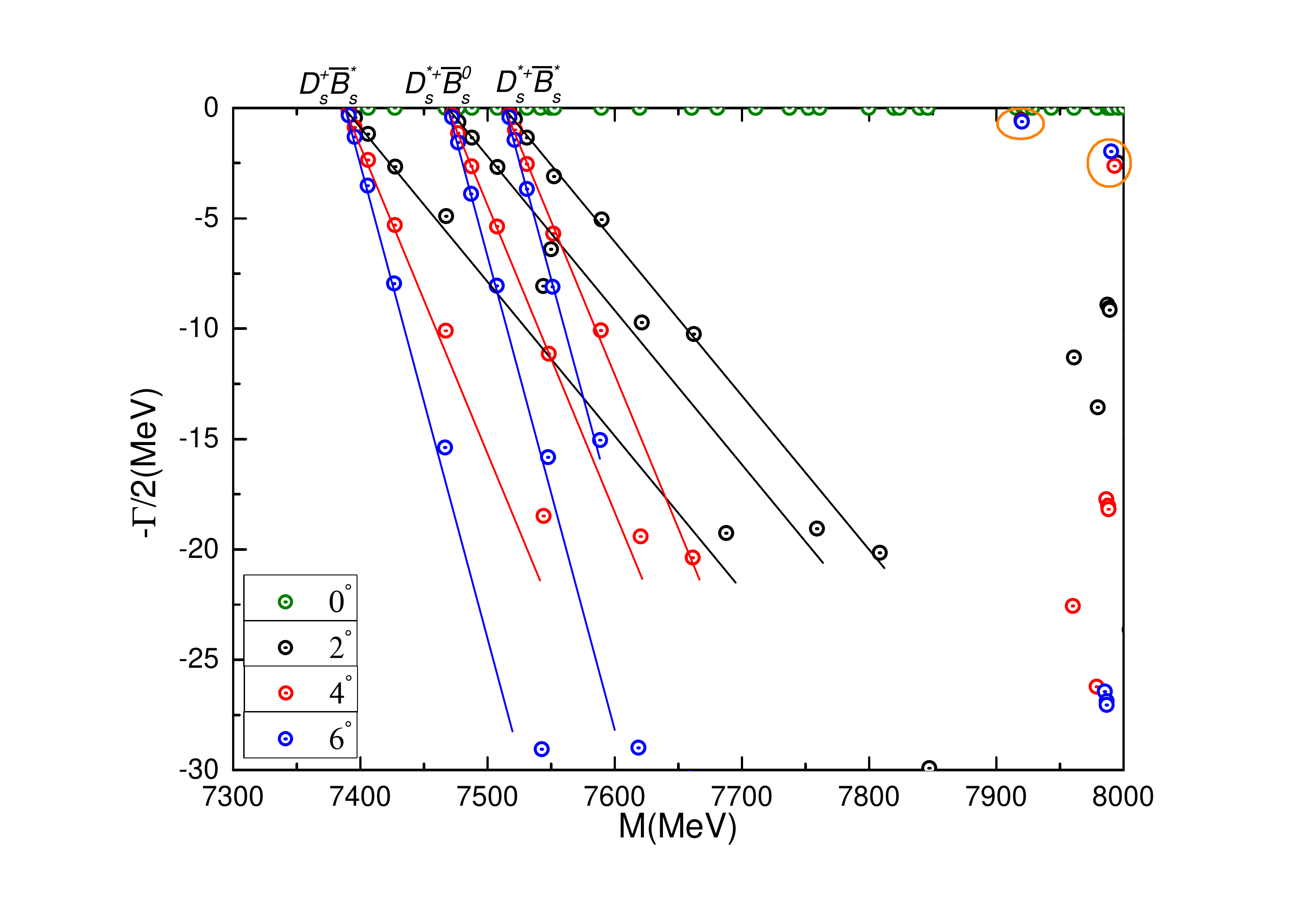}
\caption{Complex energies of $cb\bar{s}\bar{s}$ tetraquarks with $I(J^P)=0(1^+)$ in the coupled channels calculation, $\theta$ varying from $0^\circ$ to $6^\circ$ .} \label{PPCC8}
\end{figure}

\item	$I(J^P)=0(2^+)$ state

There is both one channel for the dimeson $D^{*+}_s \bar{B}^*_s$ and diquark-antidiquark $(cb)^*(\bar{s}\bar{s})^*$ configuration. Besides, 7 K-type channels are listed in Table~\ref{GresultCCSS9}. First of all, it is similar to the other $QQ\bar{s}\bar{s}$ state, bound state is impossible neither in the single channel calculation nor the coupled-channels case. Theoretical threshold of the lowest state $D^{*+}_s \bar{B}^*_s$ is 7516 MeV and the other excited states in coupled-channels are all above 7.72 GeV, in particular, there is a degeneration between $K_3$ and $K_4$ channels which coupled masses are 7697 MeV.

Unlike the results of $cb\bar{q}\bar{q}$ tetraquarks in $02^+$ state, two resonance states are obtained in the complex analysis of the complete coupled-channels of $cb\bar{s}\bar{s}$ tetraquarks. In Fig.~\ref{PPCC9}, two stable resonance poles circled with orange are $\sim$0.6 GeV above the $D^{*+}_s \bar{B}^*_s$ threshold value and  $\sim$0.1 GeV below its first radial excitation state. Hence, they can be identified as the $D^{*+}_s \bar{B}^*_s$ resonances with masses and widths are (8.05 GeV, 1.42 MeV) and (8.10 GeV, 2.90 MeV), respectively. The other energies points are the scattering state of $D^{*+}_s \bar{B}^*_s$.

\begin{table}[H]
\caption{\label{GresultCCSS9} The lowest-lying eigen-energies of $cb\bar{s}\bar{s}$ tetraquarks with $I(J^P)=0(2^+)$ in the real range calculation. (unit: MeV)}
\centering
\begin{tabular}{ccccc}
\toprule
~~\textbf{Channel}~~   & ~~\textbf{Index}~~ & ~~$\chi_J^{\sigma_i}$;~$\chi_I^{f_j}$;~$\chi_k^c$~~  & ~~$\textbf{Mass}$~~ & ~~\textbf{Mixed}~~ \\
        &      &$[i; ~j; ~k]$&  \\
\midrule
$(D^{*+}_s \bar{B}^*_s)^1 (7527)$  & 1(S)  & $[1; ~3; ~1]$  & $7516$ & $7516$ \\[2ex]
$(D^{*+}_s \bar{B}^*_s)^8$  & 2(H)  & $[1; ~3; ~2]$  & $7712$ & $7712$ \\[2ex]
$(cb)^*(\bar{s}\bar{s})^*$  & 3  & $[1; ~3; ~3]$  & $7698$ & $7698$ \\[2ex]
$K_1$  & 4 & $[1; ~3; ~5]$   & $7804$ & \\
$K_1$  & 5  & $[1; ~3; ~6]$  & $7705$ & $7704$ \\[2ex]
$K_2$  & 6 & $[1; ~3; ~7]$   & $7624$ & \\
$K_2$  & 7 & $[1; ~3; ~8]$   & $8205$ & $7622$ \\[2ex]
$K_3$  & 8 & $[1; ~3; ~9]$   & $8311$ & \\
$K_3$  & 9 & $[1; ~3; ~10]$   & $7701$ & $7696$ \\[2ex]
$K_4$  & 10 & $[1; ~3; ~12]$  & $7697$ & $7697$ \\[2ex]
\multicolumn{3}{c}{All of the above channels:} & $7516$ \\
\bottomrule
\end{tabular}
\end{table}

\begin{figure}[H]
\centering
\includegraphics[width=14 cm]{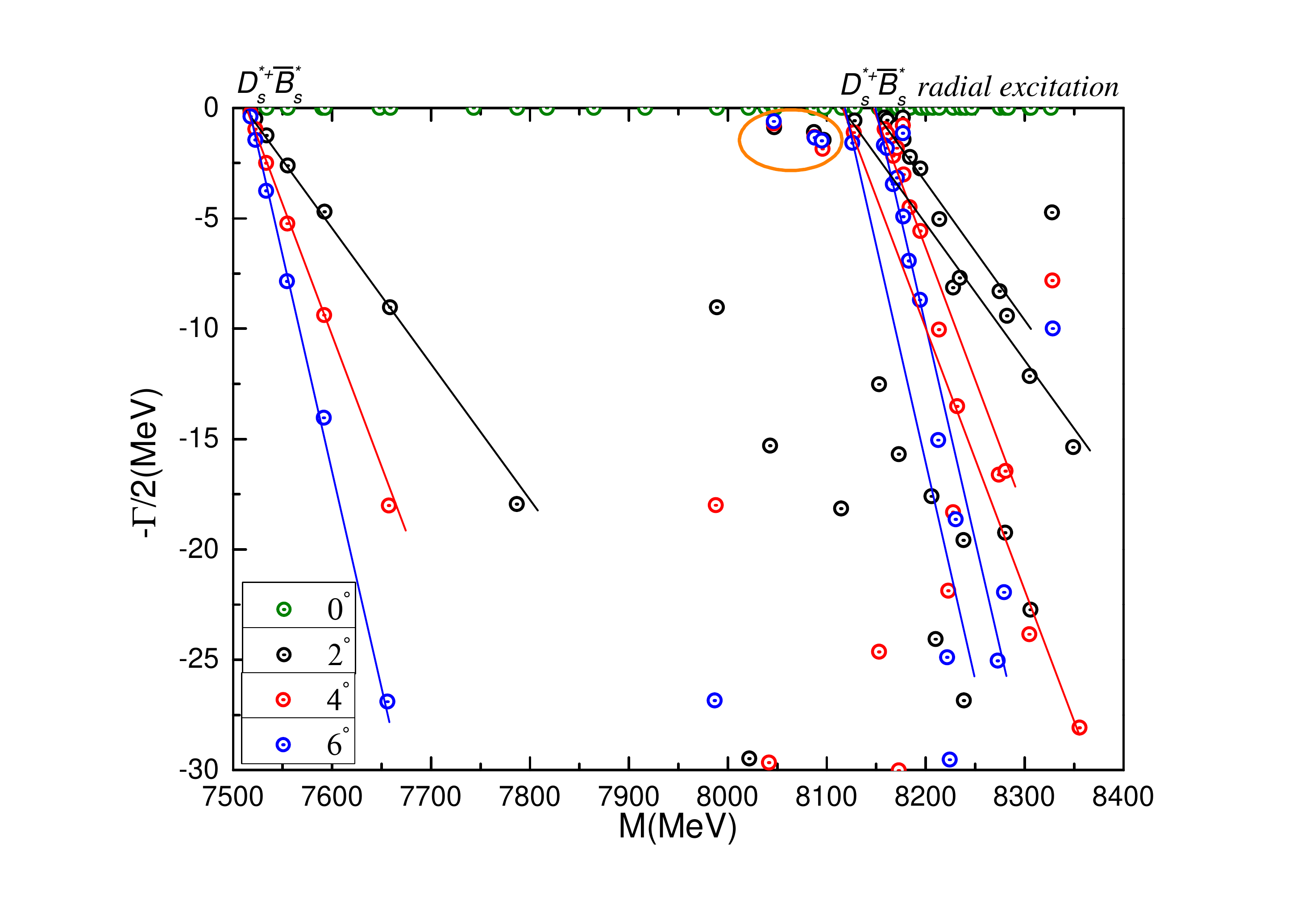}
\caption{Complex energies of $cb\bar{s}\bar{s}$ tetraquarks with $I(J^P)=0(2^+)$ in the coupled channels calculation, $\theta$ varying from $0^\circ$ to $6^\circ$ .} \label{PPCC9}
\end{figure}

\end{enumerate}

\end{itemize}

\subsubsection{$QQ\bar{Q}\bar{Q}$ Tetraquarks}

As the previous discussion on the fully heavy tetraquark states, recent experimental progress on the $cc\bar{c}\bar{c}$ system by the LHCb collaboration~\cite{LA2020CERNINDICO} triggers a revived interest in the $QQ\bar{Q}\bar{Q}$ system. Herein, a potential model which is based on the well investigated phenomenon of heavy quark pair by Lattice QCD are employed and two sectors, fully charm and fully bottom tetraquarks, will be discussed in the following parts.

\begin{itemize}[leftmargin=*,labelsep=5.8mm, listparindent=2em]
\item	Fully-charm tetraquarks

All of the three spin-parity channels, $J^P=0^+$, $1^+$ and $2^+$ are studied in the $cc\bar{c}\bar{c}$ tetraquark sector and no bound state is obtained. However, some resonance states are possible and we will introduce them individually.
 Particularly, in Tables~\ref{GresultCCQ1},~\ref{GresultCCQ2} and \ref{GresultCCQ3}, the first column presents the allowed channels, and the possible noninteracting meson-meson experimental threshold values are labeled in the parenthesis. Their indices are in the second column. The necessary bases which fulfill the Pauli principle with a combination in spin, flavor and color degrees of freedom are summarized in the third column. Theoretical masses of the obtained resonance states are in the last one.

\begin{enumerate}[leftmargin=*,labelsep=4.9mm, listparindent=2em]
\item	$I(J^P)=0(0^+)$ state

Two meson-meson channels, $\eta_c \eta_c$, $J/\psi J/\psi$ and two diquark-antidiquark structures, $(cc)(\bar{c}\bar{c})$, $(cc)^*(\bar{c}\bar{c})^*$ are investigated herein. Table~\ref{GresultCCQ1} lists the calculated masses for them along with the coupling.
  Clearly, the lowest energy level is 6469 MeV of $(cc)^*(\bar{c}\bar{c})^*$ channel and the other diquark-antidiquark structure of $(cc)(\bar{c}\bar{c})$ is at 6683 MeV. Besides, the calculated masses of color-singlet and hidden-color channels in the dimeson configurations, $\eta_c \eta_c$ and $J/\psi J/\psi$, are degenerated since there is no color-dependent interaction in the Cornell-like model. It is 6536 MeV in the $\eta_c \eta_c$ channel and 6657 MeV in the di-$J/\psi$ one. Herein, the fully coupled-channels calculation is important for different color configurations are not orthogonal. The bottom of Table~\ref{GresultCCQ1} presents two resonances' masses which are obtained in such computation. In particular, they are 6423 and 6650 MeV, respectively, and the first resonance state is around 50 MeV lower than the mass of $(cc)^*(\bar{c}\bar{c})^*$ channel.

In order to have better identifications of the nature of these two resonance states in the fully coupled-channels calculation. The components and their internal structures are analyzed in Table~\ref{GresultComp1} and \ref{tab:dis1}, respectively. 
Firstly, the couplings are strong for both of these two resonance states. Besides, the meson-meson configuration dominates them. From Table~\ref{GresultComp1} one can find that the components of $\eta_c \eta_c$ and $J/\psi J/\psi$, which are the sum of their corresponding singlet- and hidden-color channels, are (49\%, 45\%) for the first resonance state and (31\%, 48\%) for another one. Furthermore, they are both of a compact tetraquark configurations with sizes $\sim$0.34 fm.

We also extend the investigation on the complete coupled-channels from a real-range to a complex one, the results are shown in Fig.~\ref{PPQ1}. In the $6-10$ GeV energy region, there are two fixed poles in the real-axis when the rotated angle $\theta$ varied from $0^\circ$ to $6^\circ$. Actually, masses of these two stable poles are just 6423 MeV and 6650 MeV. This fact confirms the two previously obtained resonance states in the fully coupled-channels calculation of real-range. Moreover, the other energy points are unstable and always descend more or less along with the change of $\theta$.

\begin{table}[H]
\caption{\label{GresultCCQ1} Possible resonance states of fully-charm tetraquarks with quantum numbers $I(J^P)=0(0^+)$, unit in MeV.}
\centering
\begin{tabular}{cccc}
\toprule
~~\textbf{Channel}~~   & ~~\textbf{Index}~~ & ~~$\chi_J^{\sigma_i}$;~$\chi_I^{f_j}$;~$\chi_k^c$~~ & ~~$\textbf{Mass}$~~\\
 & &[i; ~j; ~k] &  \\
\midrule
~~$[\eta_c \eta_c]^1(5936)$      & 1  & $[1; ~1; ~1]$ & $6536$~~ \\
~~$[\eta_c \eta_c]^8$      & 2  & $[1; ~1; ~2]$ & $6536$~~ \\[2ex]
~~$[J/\psi J/\psi]^1 (6204)$       & 3  & $[2; ~1; ~1]$ & $6657$~~  \\
~~$[J/\psi J/\psi]^8$       & 4  & $[2; ~1; ~2]$ & $6657$~~  \\[2ex]
~~$(cc)(\bar{c}\bar{c})$   & 5 & $[1; ~1; ~4]$  & $6683$~~ \\
~~$(cc)^*(\bar{c}\bar{c})^*$   & 6 & $[2; ~1; ~3]$  & $6469$~~ \\[2ex]
~~Mixed  & & & $6423^{1st}$~~ \\
                & & & $6650^{2nd}$~~ \\
\bottomrule
\end{tabular}
\end{table}

\begin{table}[H]
\caption{{Component of each channel in the coupled-channels calculation of fully-charm resonance states with $I(J^P)=0(0^+)$.}  \label{GresultComp1}}
\centering
\begin{tabular}{ccccccc} 
\toprule
   &  ~~$[\eta_c \eta_c]^1$~~  & ~~$[\eta_c \eta_c]^8$~~ & ~~$[J/\psi J/\psi]^1$~~  &
   ~~$[J/\psi J/\psi]^8$~~ \\
 ~~$1{st}$~~  & ~~31.1\%~~  & ~~17.8\%~~  & 23.7\%~~  & 21.0\% \\
 ~~$2{nd}$~~ & ~~14.0\%~~  & ~~17.0\%~~  & 21.7\%~~  & 26.1\% \\[2ex]
        & ~~$(cc)(\bar{c}\bar{c})$~~ & ~~$(cc)^*(\bar{c}\bar{c})^*$~~  &    & \\
 ~~$1{st}$~~  & ~~3.0\%~~  & ~~3.4\%~~  &   &  \\
 ~~$2{nd}$~~ & ~~21.1\%~~  & ~~0.1\%~~  &   & \\
 \bottomrule
\end{tabular}
\end{table}

\begin{table}[H]
\caption{\label{tab:dis1} The distances, between any two quarks of the found fully-charm resonance states with $I(J^P)=0(0^+)$ in coupled-channels calculation, unit in fm.}
\centering
\begin{tabular}{cccc}
\toprule
 &  ~~~~$\textit{\textbf{r}}_{cc}$~~~~ & ~~~~$\textit{\textbf{r}}_{c\bar{c}}$~~~~ & ~~~~$\textit{\textbf{r}}_{\bar{c}\bar{c}}$~~~~  \\
\midrule
 ~~$1st$ & 0.325 & 0.342 & 0.342~~ \\
 ~~$2nd$ & 0.344 & 0.353 & 0.353~~ \\
\bottomrule
\end{tabular}
\end{table}

\begin{figure}[H]
\centering
\includegraphics[width=14 cm]{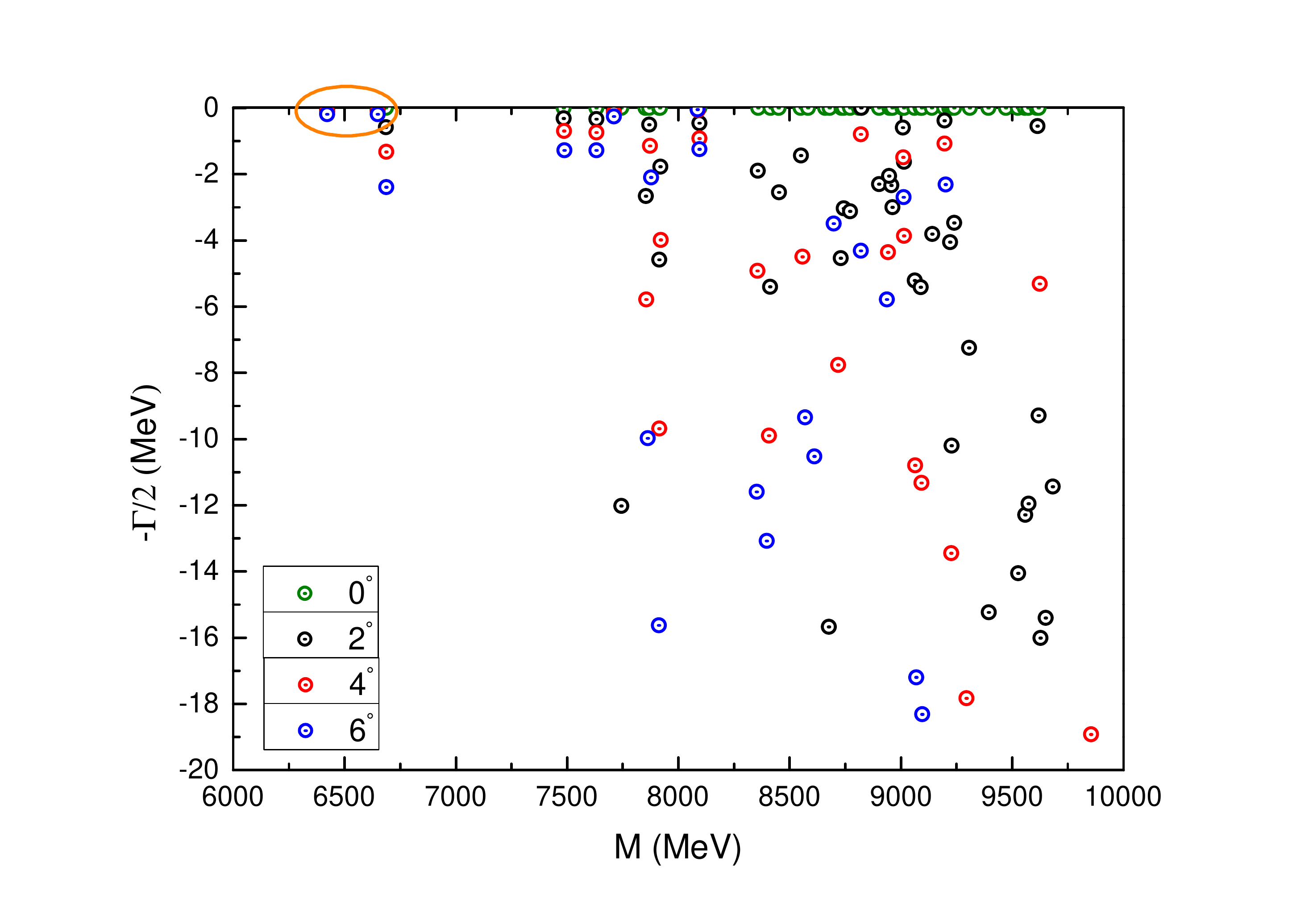}
\caption{Complex energies of fully-charm tetraquarks with $I(J^P)=0(0^+)$ in the coupled channels calculation, $\theta$ varying from $0^\circ$ to $6^\circ$ .} \label{PPQ1}
\end{figure}

\item	$I(J^P)=0(1^+)$ state

In this sector we still do not find any bound state of $cc\bar{c}\bar{c}$ system, however, resonance states with masses around 6.6 GeV are available. In particular, three almost degenerate states with masses $\sim$6.67 GeV are presented in Table~\ref{GresultCCQ2}, they are the color-singlet and hidden-color channel of $\eta_c J/\psi$ and $(cc)^*(\bar{c}\bar{c})^*$ diquark-antidiquark channel, respectively. However, no stable resonance state can be found in the di-$J/\psi$ channel. Then in a fully coupled-channels case, which both the meson-meson and diquark-antidiquark structures are considered, lower mass at 6627 MeV is obtained for the $I(J^P)=0(1^+)$ state. Herein, the coupling between color-singlet and hidden-color channels of $\eta_c J/\psi$ is also strong that the contributions are about 56\% and 35\% for them, respectively. Only less than 10\% is from the $(cc)^*(\bar{c}\bar{c})^*$ channel. From Table~\ref{tab:dis2} we can conclude that it is a compact tetraquark state, which size is $\sim$0.35 fm, in the $I(J^P)=0(1^+)$ state.

Additionally, the complete coupled-channels investigation is also performed in the complex-range framework where a rotated angle is varied from $0^\circ$ to $6^\circ$. The calculated results of complex energies from 6.5 GeV to 10.0 GeV are presented in Fig.~\ref{PPQ2}. Obviously, there is a fixed pole in the real-axis and circled with orange. It can be identified as a $\eta_c J/\psi$ resonance with mass at 6627 MeV. However, the other poles should be corresponded to the scattering states of $\eta_c J/\psi$ in higher energy region.

\begin{table}[H]
\caption{\label{GresultCCQ2} Possible resonance states of fully-charm tetraquarks with quantum numbers $I(J^P)=0(1^+)$, unit in MeV.}
\centering
\begin{tabular}{cccc}
\toprule
~~\textbf{Channel}~~   & ~~\textbf{Index}~~ & ~~$\chi_J^{\sigma_i}$;~$\chi_I^{f_j}$;~$\chi_k^c$~~ & ~~$\textbf{Mass}$~~\\
\midrule
~~$[\eta_c J/\psi]^1 (6070)$      & 1  & $[1; ~1; ~1]$  & $6671$~~ \\
~~$[\eta_c J/\psi]^8$      & 2  & $[1; ~1; ~2]$  & $6671$~~ \\
~~$(cc)^*(\bar{c}\bar{c})^*$   & 3 & $[3; ~1; ~3]$  & $6674$~~ \\[2ex]
~~Mixed  & & & $6627$~~ \\
\multicolumn{4}{l}{~~Component (1; 2; 3):~~~ 55.5\%; 34.7\%; 9.8\%}  \\
\bottomrule
\end{tabular}
\end{table}

\begin{table}[H]
\caption{\label{tab:dis2} The distances, between any two quarks of the found fully-charm resonance states with $I(J^P)=0(1^+)$ in coupled-channels calculation, unit in fm.}
\centering
\begin{tabular}{ccc}
\toprule
~~~~$\textit{\textbf{r}}_{cc}$~~~~ & ~~~~$\textit{\textbf{r}}_{c\bar{c}}$~~~~ & ~~~~$\textit{\textbf{r}}_{\bar{c}\bar{c}}$~~~~  \\
\midrule
0.342 & 0.357 & 0.357 \\
\bottomrule
\end{tabular}
\end{table}

\begin{figure}[H]
\centering
\includegraphics[width=14 cm]{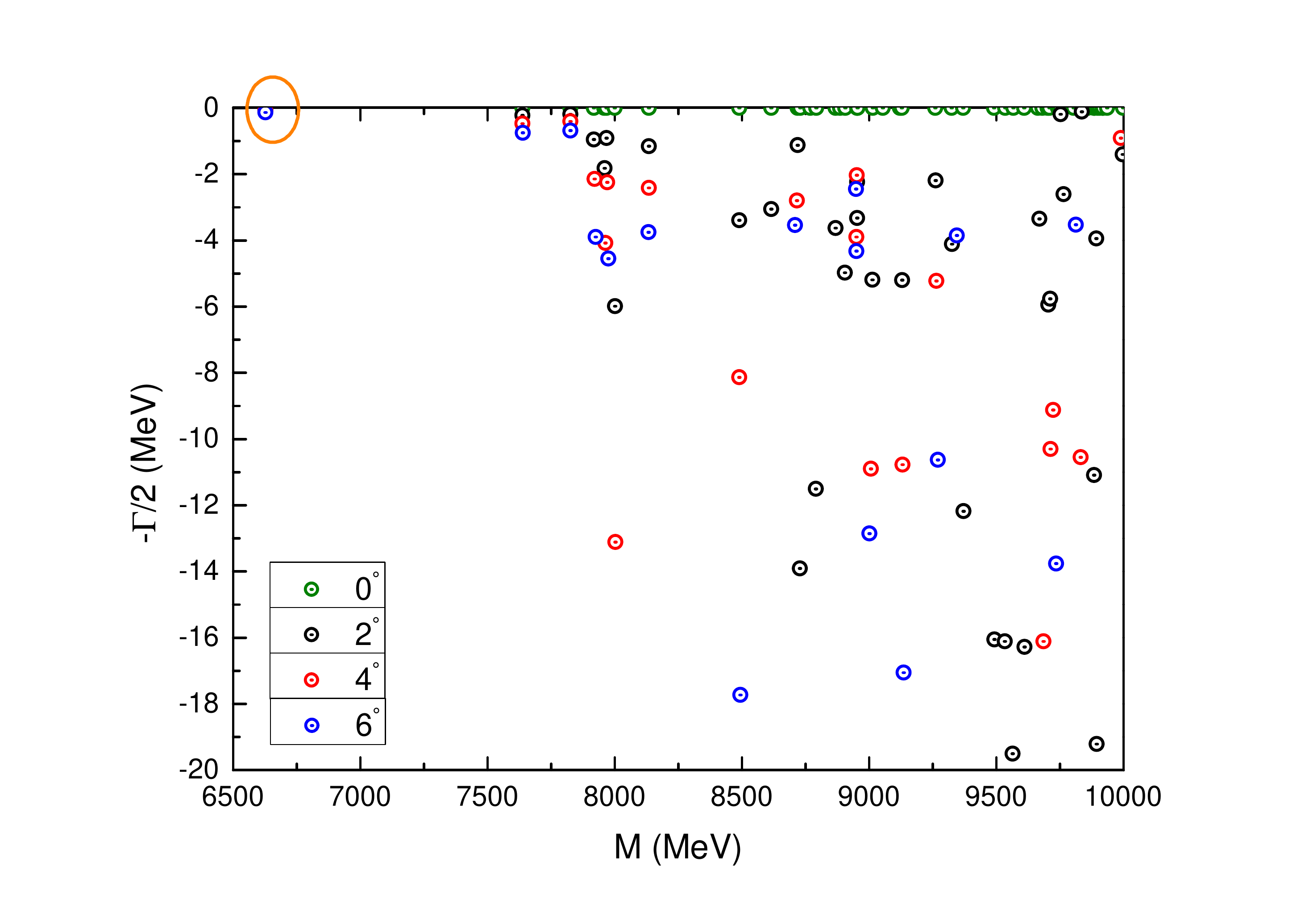}
\caption{Complex energies of fully-charm tetraquarks with $I(J^P)=0(1^+)$ in the coupled channels calculation, $\theta$ varying from $0^\circ$ to $6^\circ$ .} \label{PPQ2}
\end{figure}

\item	$I(J^P)=0(2^+)$ state

Similar to the $0(1^+)$ quantum state, two meson-meson $J/\psi  J/\psi$ channels and one diquark-antidiquark $(cc)^*(\bar{c}\bar{c})^*$ channel are considered in the highest spin state of tetraquark. In particular, masses of them are all around 7.03 GeV and the $(cc)^*(\bar{c}\bar{c})^*$ channel is the lowest one with mass at 7026 MeV. Then in their coupling calculation, the mixed mass is 7014 MeV shown in Table~\ref{GresultCCQ3}. Meanwhile, the percentages of $[J/\psi  J/\psi]^1$, $[J/\psi  J/\psi]^8$ and $(cc)^*(\bar{c}\bar{c})^*$ are around 53\%, 33\% and 14\%, respectively. This strong coupling fact also leads to a compact tetraquark configuration with size around 0.38 fm listed in Table~\ref{tab:dis3}.

By employing the complex scaling method in our model investigation of the complete coupled-channels, the above conclusions are confirmed too. Particularly, one stable resonance pole against the two-body strong decay is obtained at 7014 MeV of Fig.~\ref{PPQ3}. Obviously, this resonance mass is quite close to the reported new structure at 6.9 GeV by the LHCb collaboration~\cite{LA2020CERNINDICO}. Hence it can be identified as a compact fully charmed tetraquark in $0(2^+)$ state. However, the other complex energy points always move along with the variation of angle $\theta$.

\begin{table}[H]
\caption{\label{GresultCCQ3} Possible resonance states of fully-charm tetraquarks with quantum numbers $I(J^P)=0(2^+)$, unit in MeV.}
\centering
\begin{tabular}{cccc}
\toprule
~~\textbf{Channel}~~   & ~~\textbf{Index}~~ & ~~$\chi_J^{\sigma_i}$;~$\chi_I^{f_j}$;~$\chi_k^c$~~ & ~~$\textbf{Mass}$~~\\
\midrule
~~$[J/\psi  J/\psi]^1 (6204)$      & 1  & $[1; ~1; ~1]$ & $7030$~~ \\
~~$[J/\psi  J/\psi]^8$      & 2  & $[1; ~1; ~2]$ & $7030$~~ \\
~~$(cc)^*(\bar{c}\bar{c})^*$   & 3 & $[1; ~1; ~3]$  & $7026$~~ \\[2ex]
~~Mixed  & & & $7014$~~ \\
\multicolumn{4}{l}{~~Component (1; 2: 3):~~~ 52.8\%; 33.0\%; 14.2\%}  \\
\bottomrule
\end{tabular}
\end{table}

\begin{table}[H]
\caption{\label{tab:dis3} The distances, between any two quarks of the found fully-charm resonance states with $I(J^P)=0(2^+)$ in coupled-channels calculation, unit in fm.}
\centering
\begin{tabular}{ccc}
\toprule
~~~~$\textit{\textbf{r}}_{cc}$~~~~ & ~~~~$\textit{\textbf{r}}_{c\bar{c}}$~~~~ & ~~~~$\textit{\textbf{r}}_{\bar{c}\bar{c}}$~~~~  \\
\midrule
~~0.375 & 0.389 & 0.389~~ \\
\bottomrule
\end{tabular}
\end{table}

\begin{figure}[H]
\centering
\includegraphics[width=14 cm]{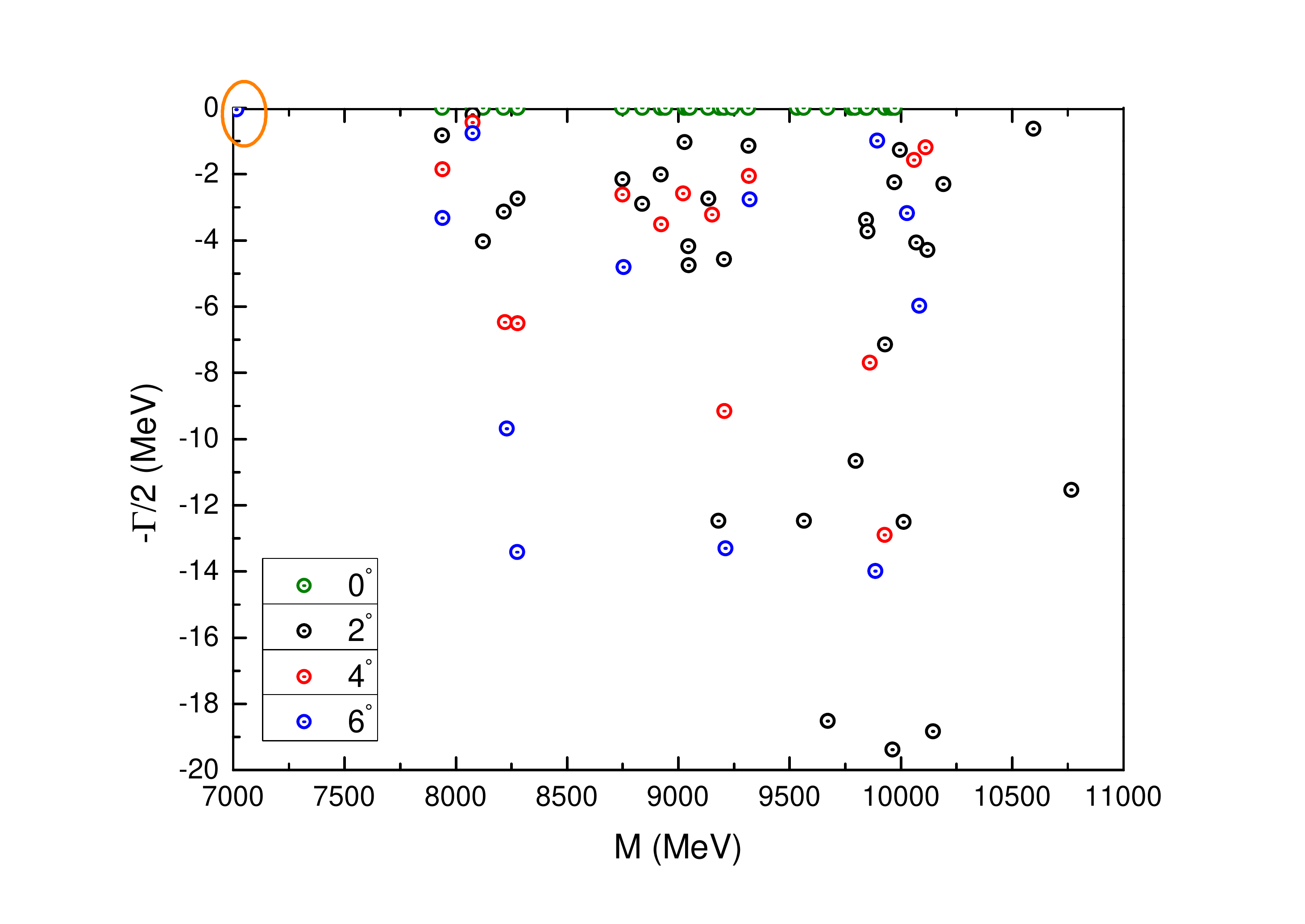}
\caption{Complex energies of fully-charm tetraquarks with $I(J^P)=0(2^+)$ in the coupled channels calculation, $\theta$ varying from $0^\circ$ to $6^\circ$ .} \label{PPQ3}
\end{figure}

\end{enumerate}

\item	Fully-bottom tetraquarks

Bound states and resonances in $J^P$=$0^+$, $1^+$ and $2^+$, $I=0$ are found in the $bb\bar{b}\bar{b}$ tetraquark systems. Furthermore, they are more compact than the $cc\bar{c}\bar{c}$ resonances. The calculated masses are listed in Tables~\ref{GresultCCQ4},~\ref{GresultCCQ5} and \ref{GresultCCQ6}. Particularly, the first column shows the allowed channels and, in the parenthesis, the noninteracting meson-meson experimental threshold values. Meson-meson and diquark-antidiquark channels are indexed in the second column, respectively. The necessary bases in spin, flavor and color degrees of freedom are listed in the third column. The fourth column refers to the theoretical mass of each channels and their couplings, binding energies of the dimeson channels are listed in the last column. The details are as follows.

\begin{enumerate}[leftmargin=*,labelsep=4.9mm, listparindent=2em]
\item	$I(J^P)=0(0^+)$ state

As shown in Table~\ref{GresultCCQ4}, four meson-meson configurations, which include the singlet- and hidden-color channels of $\eta_b \eta_b$ and $\Upsilon \Upsilon$, along with two diquark-antidiquark ones, $(bb)(\bar{b}\bar{b})$ and $(bb)^*(\bar{b}\bar{b})^*$, are calculated in the $0(0^+)$ state of fully bottom tetraquark. Firstly, two low-lying stable states are found in each single channel computations. Masses of the first energy level of them are $\sim$18.0 GeV, and the second one is around 19.0 GeV. Obviously, the lowest-lying state of each channel is deeply bound with $E_b$ more than $-800$ MeV and the higeher one is a resonance state which is $\sim$150 MeV above the threshold. Additionally, masses of three stable states in the complete coupled-channels case are also listed in Table~\ref{GresultCCQ4}, in particular, they are 17.92 GeV, 18.01 GeV and 19.28 GeV, respectively. Generally, they are still located at 18.0 GeV and 19.0 GeV.

Natures of these three exotic states can be indicated in Table~\ref{GresultComp4} and \ref{tab:dis4}. In particular, the percentages of each meson-meson and diquark-antidiquark channels of $bb\bar{b}\bar{b}$ tetraquark states are listed in Table~\ref{GresultComp4}. Therein, the diquark-antidiquark channels are less than 11\% for all of these three states. Nevertheless, the couplings between color-singlet and hidden-color channels of $\eta_b \eta_b$ and $\Upsilon \Upsilon$ are very strong. Besides, a compact fully-bottom tetraquark configuration is shown in Table~\ref{tab:dis4}, where the sizes of two bound states are around 0.16 fm and 0.29 fm for the resonance one. Apparently, the conclusions are consistent with the deeply binding energies obtained before.

Additionally, we also investigate the $bb\bar{b}\bar{b}$ system in a complex-range. With a complex scaling method employed in the fully coupled-channels calculation, the three fully-bottom tetraquark states, which are obtained in the real-range, are all well presented in Fig.~\ref{PPQ4} again. Therein, apart from the scattering points, which always descend with the variation of angle $\theta$, the three poles in the real-axis and circled orange are stable. Hence, the bound and resonance states in $bb\bar{b}\bar{b}$ sector are possible. However, as the statement on the fully-heavy tetraquarks in Sec. I, since no color-dependent interaction is considered in our present model, the obtained exotic states, especially the bound states of $bb\bar{b}\bar{b}$ are quite negotiable. Much more efforts both theoretical and experimental are deserved~\cite{KTCSL2020}.

\begin{table}[H]
\caption{\label{GresultCCQ4} Possible bound and resonance states of fully-bottom tetraquarks with quantum numbers $I(J^P)=0(0^+)$, unit in MeV.}
\centering
\begin{tabular}{ccccc}
\toprule
~~\textbf{Channel}~~   & ~~\textbf{Index}~~ & ~~$\chi_J^{\sigma_i}$;~$\chi_I^{f_j}$;~$\chi_k^c$~~ & ~~$\textbf{Mass}$~~ & ~~$\textbf{E}_\textbf{b}$~~\\
\midrule
~~$[\eta_b \eta_b]^1 (18802)$      & 1  & $[1; ~2; ~1]$  & $17999^{1st}$  & $-803$~~ \\
                                                  &   &  & $19036^{2nd}$  & $+234$~~ \\
~~$[\Upsilon \Upsilon]^1 (18926)$       & 2 & $[2; ~2; ~1]$   & $18038^{1st}$  & $-888$~~  \\
       &   &  & $19069^{2nd}$  & $+143$~~  \\
~~$[\eta_b \eta_b]^8$      & 3  & $[1; ~2; ~2]$  & $17999^{1st}$  & $-803$~~ \\
                                                  &   &  & $19036^{2nd}$  & $+234$~~ \\
~~$[\Upsilon \Upsilon]^8$       & 4 & $[2; ~2; ~2]$   & $18038^{1st}$  & $-888$~~  \\
       &   &  & $19069^{2nd}$  & $+143$~~  \\
~~$(bb)(\bar{b}\bar{b})$   & 5  & $[1; ~2; ~4]$  & $18068^{1st}$  &  \\
   &  &  & $19097^{2nd}$  &  \\
~~$(bb)^*(\bar{b}\bar{b})^*$   & 6 & $[2; ~2; ~3]$  & $17975^{1st}$  &  \\
   &   &  & $19033^{2nd}$  &  \\[2ex]
~~Mixed  & & & $17917^{1st}$ & \\
                & & & $18010^{2nd}$ & \\
                & & & $19280^{3rd}$ & \\
\bottomrule
\end{tabular}
\end{table}

\begin{table}[H]
\caption{{Component of each channel in the coupled-channels calculation of fully-bottom bound and resonance states with $I(J^P)=0(0^+)$.}  \label{GresultComp4}}
\centering
\begin{tabular}{ccccc}
\toprule
   &  ~~$[\eta_b \eta_b]^1$~~  &  ~~$[\eta_b \eta_b]^8$~~    & ~~$[\Upsilon \Upsilon]^1$~~  & ~~$[\Upsilon \Upsilon]^8$~~\\
 ~~$1{st}$~~  & ~~27.3\%~~  & ~~19.4\%~~  & 21.6\%~~  & 20.5\% \\
 ~~$2{nd}$~~  & ~~13.5\%~~  & ~~32.8\%~~  & 14.0\%~~  & 29.8\% \\
  ~~$3{rd}$~~  & ~~20.6\%~~  & ~~15.8\%~~  & 30.0\%~~  & 30.1\% \\[2ex]
    & ~~$(bb)(\bar{b}\bar{b})$~~ & ~~$(bb)^*(\bar{b}\bar{b})^*$~~  &  & \\
 ~~$1{st}$~~  & ~~0.9\%~~  & ~~10.3\%~~  &   & \\
 ~~$2{nd}$~~  & ~~9.7\%~~  & ~~0.2\%~~  &   &  \\
  ~~$3{rd}$~~  & ~~3.5\%~~  & ~~0.0\%~~  &   &  \\
\bottomrule
\end{tabular}
\end{table}

\begin{table}[H]
\caption{\label{tab:dis4} The distances, between any two quarks of the found fully-bottom bound and resonance states with $I(J^P)=0(0^+)$ in coupled-channels calculation, unit in fm.}
\centering
\begin{tabular}{cccc}
\toprule
 &  ~~~~$\textit{\textbf{r}}_{bb}$~~~~ & ~~~~$\textit{\textbf{r}}_{b\bar{b}}$~~~~ & ~~~~$\textit{\textbf{r}}_{\bar{b}\bar{b}}$~~~~  \\
\midrule
 ~~$1st$ & 0.160 & 0.166 & 0.166~~ \\
 ~~$2nd$ & 0.163 & 0.168 & 0.168~~ \\
  ~~$3rd$ & 0.246 & 0.292 & 0.292~~ \\
\bottomrule
\end{tabular}
\end{table}

\begin{figure}[H]
\centering
\includegraphics[width=14 cm]{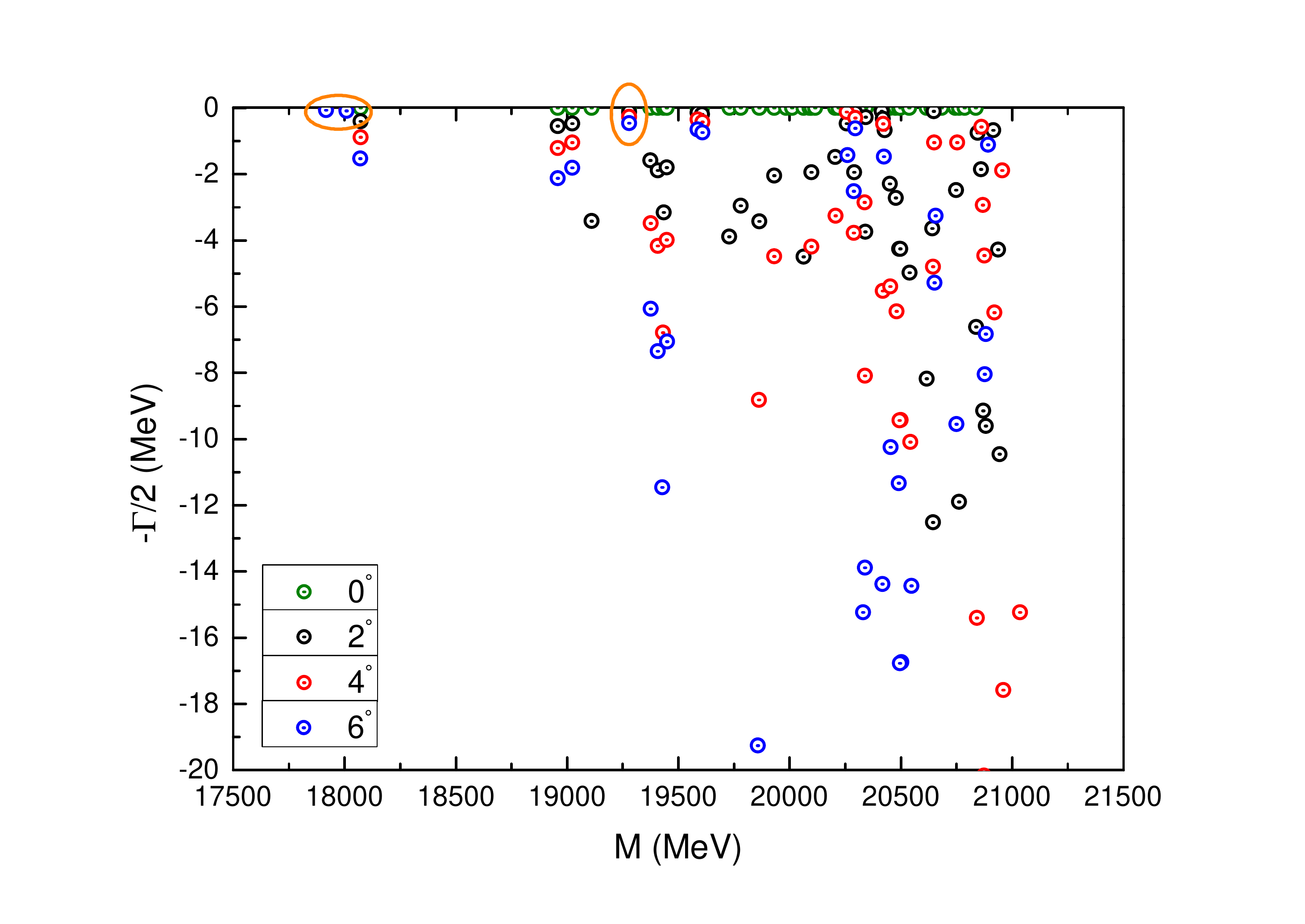}
\caption{Complex energies of fully-bottom tetraquarks with $I(J^P)=0(0^+)$ in the coupled channels calculation, $\theta$ varying from $0^\circ$ to $6^\circ$ .} \label{PPQ4}
\end{figure}

\item	$I(J^P)=0(1^+)$ state

It is similar to the $cc\bar{c}\bar{c}$ system in $0(1^+)$ state, di-$\Upsilon$ is a scattering state under investigation. Hence Table~\ref{GresultCCQ5} just lists the bound and resonance states obtained in $[\eta_b \Upsilon]^1$, $[\eta_b \Upsilon]^8$ and $(bb)^*(\bar{b}\bar{b})^*$ channels. Particularly, the three bound states with masses both around 18.06 GeV are found in these three configurations, and there is $-802$ MeV binding energy for the $\eta_b \Upsilon$ dimeson channel. The resonance states' masses are $\sim$19.09 GeV. Then in a complete coupled-channels calculation, the lowest bound state is found at 18.01 GeV, the resonances are at 19.34 GeV and 19.63 GeV, respectively. Their components are also listed in the bottom of Table~\ref{GresultCCQ5} where the $\eta_b \Upsilon$ meson-meson channel dominates the first two exotic states and a strong coupling between dimeson and diquark-antidiquark configurations is obtained for the third resonance states.

In Table~\ref{tab:dis5} we can notice that sizes of the three $bb\bar{b}\bar{b}$ tetraquark states are quite comparable with those in $0(0^+)$ case. It is $\sim$0.16 fm and 0.27 fm for the bound and resonance states, respectively. Meanwhile, Fig.~\ref{PPQ5} shows the distribution of complex energies in the complete coupled-channels calculation of $bb\bar{b}\bar{b}$ system. The bound state, which mass is 18.01 GeV, and the two resonances, which masses at 19.34 GeV and 19.63 GeV, respectively, are stable in the real-axis and independent of the variation of rotated angle $\theta$.

\begin{table}[H]
\caption{\label{GresultCCQ5} Possible bound and resonance states of fully-bottom tetraquarks with quantum numbers $I(J^P)=0(1^+)$, unit in MeV.}
\centering
\begin{tabular}{ccccc}
\toprule
~~\textbf{Channel}~~   & ~~\textbf{Index}~~ & ~~$\chi_J^{\sigma_i}$;~$\chi_I^{f_j}$;~$\chi_k^c$~~ & ~~$\textbf{Mass}$~~ & ~~$\textbf{E}_\textbf{b}$~~\\
\midrule
~~$[\eta_b \Upsilon]^1 (18864)$      & 1  & $[1; ~2; ~1]$  & $18062^{1st}$  &$-802$ ~~ \\
      &  &  & $19087^{2nd}$  &$+223$ ~~ \\
~~$[\eta_b \Upsilon]^8$      & 2  & $[1; ~2; ~2]$  & $18062^{1st}$  &$-802$ ~~ \\
      &  &  & $19087^{2nd}$  &$+223$ ~~ \\
~~$(bb)^*(\bar{b}\bar{b})^*$   & 3 & $[3; ~2; ~3]$   & $18065^{1st}$ & \\
   &  &  & $19093^{2nd}$ & \\[2ex]
~~Mixed  & & & $18009^{1st}$ & \\
  & & & $19338^{2nd}$ & \\
  & & & $19627^{3rd}$ & \\[2ex]
\multicolumn{4}{l}{~~Component $(1; 2; 3)^{1st}$:~~~ 50.9\%; 31.9\%; 17.2\%}  \\
\multicolumn{4}{l}{~~Component $(1; 2; 3)^{2nd}$:~~~ 48.6\%; 51.4\%; 0.0\%}  \\
\multicolumn{4}{l}{~~Component $(1; 2; 3)^{3rd}$:~~~ 28.1\%; 17.5\%; 54.4\%}  \\
\bottomrule
\end{tabular}
\end{table}

\begin{table}[H]
\caption{\label{tab:dis5} The distances, between any two quarks of the found fully-bottom bound and resonance states with $I(J^P)=0(1^+)$ in coupled-channels calculation, unit in fm.}
\centering
\begin{tabular}{cccc}
\toprule
 & ~~~~$\textit{\textbf{r}}_{bb}$~~~~ & ~~~~$\textit{\textbf{r}}_{b\bar{b}}$~~~~ & ~~~~$\textit{\textbf{r}}_{\bar{b}\bar{b}}$~~~~  \\
\midrule
~~$1st$ & ~~0.163 & 0.169 & 0.169~~ \\
~~$2nd$ & ~~0.248 & 0.295 & 0.295~~ \\
~~$3rd$ & ~~0.279 & 0.265 & 0.265~~ \\
\bottomrule
\end{tabular}
\end{table}

\begin{figure}[H]
\centering
\includegraphics[width=14 cm]{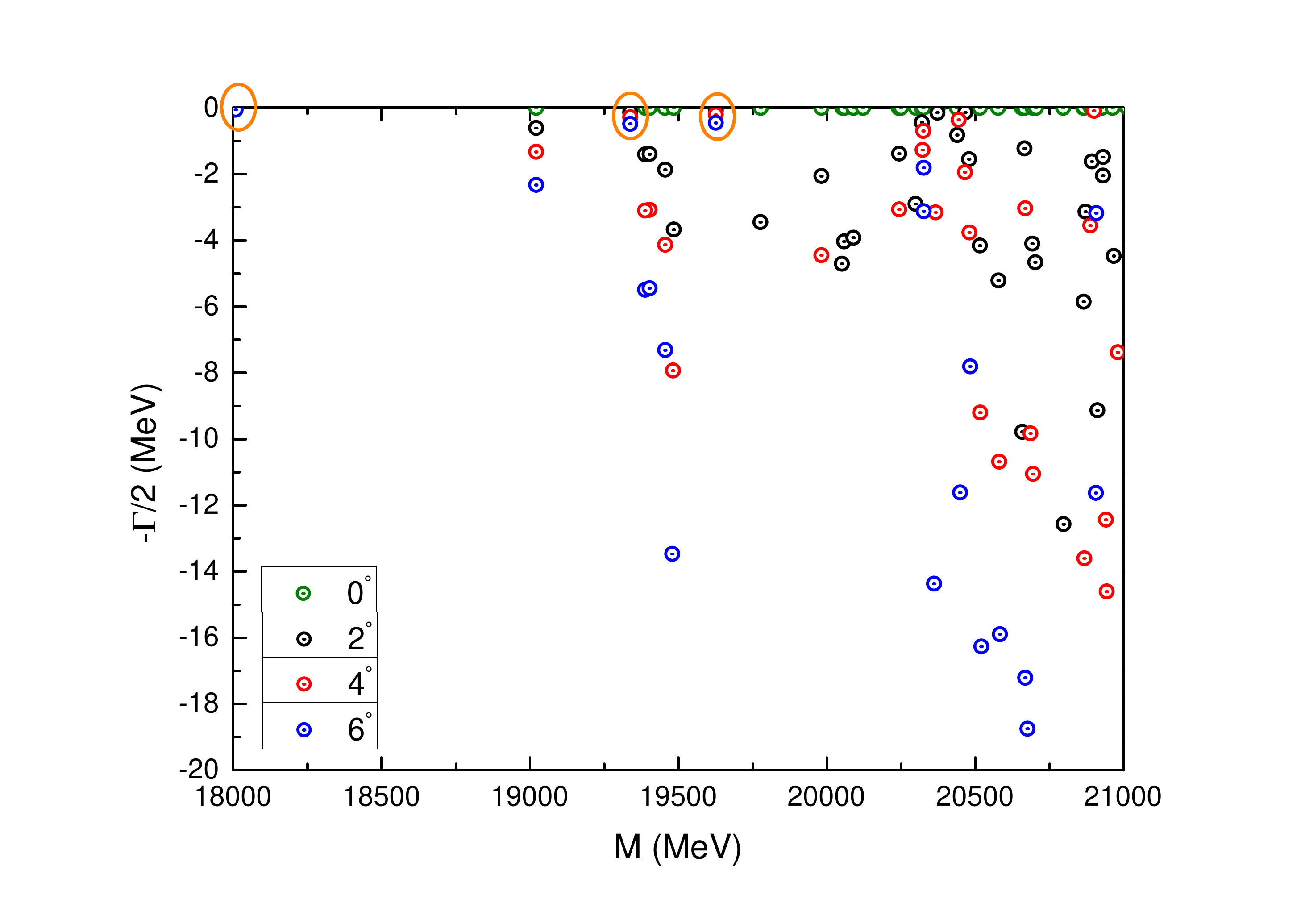}
\caption{Complex energies of fully-bottom tetraquarks with $I(J^P)=0(1^+)$ in the coupled channels calculation, $\theta$ varying from $0^\circ$ to $6^\circ$ .} \label{PPQ5}
\end{figure}

\item	$I(J^P)=0(2^+)$ state

In this highest spin state, one meson-meson configuration $\Upsilon(1S)  \Upsilon(1S)$, which include a color-singlet and a hidden-color channel, and one diquark-antidiquark configuration $(bb)^*(\bar{b}\bar{b})^*$ are considered. Firstly, bound states at $\sim$18.24 GeV are obtained in each single channels studies, and the binding energy $\sim$$-690$ MeV is a littler shallower than the other two quantum states. Furthermore, there are also three resonances found at $\sim$19.21 GeV. Then in a complete coupled-channels investigation, masses of the bound and resonance states are 18.19 GeV, 19.45 GeV and 19.71 GeV, respectively.
Then by comparing the components of these exotic states in $1^+$ and $2^+$ states, similar features can also be drawn herein. Furthermore, these compact configurations are also confirmed in analyzing the internal structure of $bb\bar{b}\bar{b}$ system in Table~\ref{tab:dis6}, in particular, the mean square radii are around 0.17, 0.30 and 0.27 fm for the one bound and two resonance states, respectively.

Additionally, the results obtained in a real-range investigation is supported by a complex-range one of Fig.~\ref{PPQ6}. Therein, three outstanding poles in the real-axis are just the bound and resonance states found in our study. In the varied region from $0^\circ$ to $6^\circ$ of angle $\theta$, these three poles are fixed. However, the other energy points are obviously unstable.

\begin{table}[H]
\caption{\label{GresultCCQ6} Possible bound and resonance states of fully-bottom tetraquarks with quantum numbers $I(J^P)=0(2^+)$, unit in MeV.}
\centering
\begin{tabular}{ccccc}
\toprule
~~\textbf{Channel}~~   & ~~\textbf{Index}~~ & ~~$\chi_J^{\sigma_i}$;~$\chi_I^{f_j}$;~$\chi_k^c$~~ & ~~$\textbf{Mass}$~~ & ~~$\textbf{E}_\textbf{b}$~~\\
\midrule
~~$[\Upsilon  \Upsilon]^1 (18926)$      & 1  & $[1; ~2; ~1]$  & $18238^{1st}$  & $-688$~~ \\
      &  &  & $19207^{2nd}$  & $+281$~~ \\
~~$[\Upsilon  \Upsilon]^8$      & 2  & $[1; ~2; ~2]$  & $18238^{1st}$  & $-688$~~ \\
      &  &  & $19207^{2nd}$  & $+281$~~ \\
~~$(bb)^*(\bar{b}\bar{b})^*$   & 3  & $[1; ~2; ~3]$ & $18241^{1st}$ & \\
   &  &  & $19211^{2nd}$ & \\[2ex]
~~Mixed  & & & $18189^{1st}$ & \\
  & & & $19451^{2nd}$ & \\
  & & & $19708^{3rd}$ & \\[2ex]
\multicolumn{4}{l}{~~Component $(1; 2; 3)^{1st}$:~~~ 53.2\%; 33.2\%; 13.6\%}  \\
\multicolumn{4}{l}{~~Component $(1; 2; 3)^{2nd}$:~~~ 48.8\%; 51.2\%; 0.0\%}  \\
\multicolumn{4}{l}{~~Component $(1; 2; 3)^{3rd}$:~~~ 28.3\%; 17.7\%; 54.0\%}  \\
\bottomrule
\end{tabular}
\end{table}

\begin{table}[H]
\caption{\label{tab:dis6} The distances, between any two quarks of the found fully-bottom bound and resonance states with $I(J^P)=0(2^+)$ in coupled-channels calculation, unit in fm.}
\centering
\begin{tabular}{cccc}
\toprule
 & ~~~~$\textit{\textbf{r}}_{bb}$~~~~ & ~~~~$\textit{\textbf{r}}_{b\bar{b}}$~~~~ & ~~~~$\textit{\textbf{r}}_{\bar{b}\bar{b}}$~~~~  \\
\midrule
~~$1st$ & 0.168 & 0.174 & 0.174~~ \\
~~$2nd$ & 0.254 & 0.302 & 0.302~~ \\
~~$3rd$ & 0.284 & 0.268 & 0.268~~ \\
\bottomrule
\end{tabular}
\end{table}

\begin{figure}[H]
\centering
\includegraphics[width=14 cm]{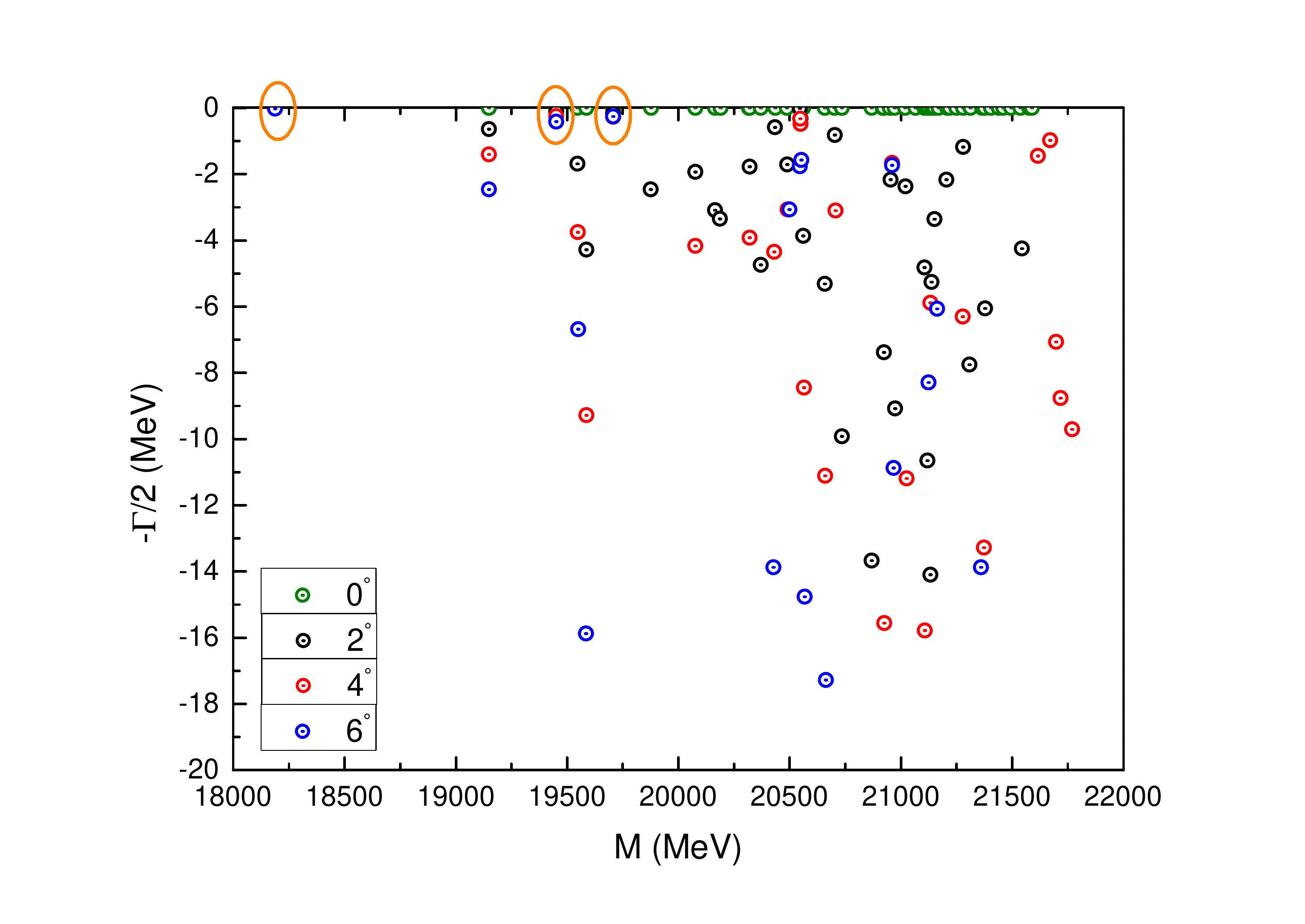}
\caption{Complex energies of fully-bottom tetraquarks with $I(J^P)=0(2^+)$ in the coupled channels calculation, $\theta$ varying from $0^\circ$ to $6^\circ$ .} \label{PPQ6}
\end{figure}

\end{enumerate}

\end{itemize}

\subsection{Hidden and Open Heavy Flavor Pentaquarks}

In this part three types of 5-quarks systems will be introduced according to the investigations by chiral quark model. Particularly, hidden charm $qqqc\bar{c}$, hidden bottom $qqqb\bar{b}$ and doubly charmed $ccq\bar{q}$ pentaquarks in spin-parity $J^P=\frac{1}{2}^-$, $\frac{3}{2}^-$ and $\frac{5}{2}^-$, $I=\frac{1}{2}$ or $\frac{3}{2}$ are calculated. The details are presented in each subsections.

\subsubsection{Hidden Charm Pentaquarks}

In the $qqqc\bar{c}$ ($q=u, d$) sector, several pentaquark states reported by the LHCb collaboration can be well explained in our work. In particular, the $P^+_c(4380)$ can be identified as the $\Sigma_c^* \bar{D}$ molecule state with $I(J^P)=\frac{1}{2}({\frac{3}{2}}^-)$, $P^+_c(4312)$, $P^+_c(4440)$ and $P^+_c(4457)$ can be explained as the molecule states of $\frac12 \frac12^-$ $\Sigma_c\bar{D}$, $\frac12 \frac12^-$ $\Sigma_c\bar{D}^*$ and $\frac12 \frac32^-$ $\Sigma_c\bar{D}^*$, respectively. Let us discuss them in the following parts.

\begin{itemize}[leftmargin=*,labelsep=5.8mm, listparindent=2em]
\item	$I(J^P)=\frac{1}{2}(\frac{1}{2}^-)$ state

Seven baryon-meson channels, $N\eta_c$, $NJ/\psi$, $\Lambda_c \bar{D}$, $\Lambda_c \bar{D}^*$, $\Sigma_c \bar{D}$, $\Sigma_c \bar{D}^*$ and $\Sigma_c^* \bar{D}^*$ are studied in the lowest spin state. Table~\ref{Gresult1} shows the calculated results. Particularly, the first column lists the necessary bases in spin, flavor and color degrees of freedom. Mass of 5-quark system is in the second one. The theoretical and experimental thresholds of each channels are in the third and fifth column, respectively. In the fourth column, the binding energy of state is presented. A modified mass $E'$ which is obtained by the summation of binding energy $E_B$ and experimental threshold value $E^{Exp}_{th}$ is in the last column.

Obviously, it is a scattering nature of the $N\eta_c$, $NJ/\psi$, $\Lambda_c \bar{D}$ and $\Lambda_c \bar{D}^*$ channels which $E_B$ is 0. However, resonance states are available in the $\Sigma_c \bar{D}$, $\Sigma_c \bar{D}^*$ and $\Sigma_c^* \bar{D}^*$ channels. Particularly, binding energies $E_B=-4$ MeV, $-2$ MeV and $-3$ MeV is found for the color-singlet channels of these three states, respectively. In additional, deeper binding energies are obtained through the couplings with hidden-color channels for the $\Sigma_c \bar{D}$, $\Sigma_c \bar{D}^*$ and $\Sigma_c^* \bar{D}^*$, the new $E_B$ is $-8$ MeV, $-41$ MeV and $-105$ MeV, respectively. The main contributions are color-singlet channels for the first two states (91\% in $\Sigma_c \bar{D}$ and 67\% in $\Sigma_c \bar{D}^*$), but $\sim$80\% component is the hidden-color channel for $\Sigma_c^* \bar{D}^*$ state. After a mass shift by considering the systematical error, the rescaled masses $E'$ for the three resonance states are 4312 MeV, 4421 MeV and 4422 MeV, respectively. Accordingly, the first one is nicely consistent with $P^+_c(4312)$ and can be identified as the $\Sigma_c \bar{D}$ state. 
The nature of $\Sigma_c$ baryon and $\bar{D}$ meson molecular state is confirmed in Table~\ref{PcDistance}, in which the distances between any two quarks are shown and the obtained 2.1 fm for the $c\bar{c}$ pair in hidden charm pentaquark state is quite comparable with the size of deuteron.
 Besides, there is a degeneration between the coupled masses of $\Sigma_c \bar{D}^*$ and $\Sigma_c^* \bar{D}^*$ states, however, the $P^+_c(4440)$ is preferred to be explained as a $\Sigma_c \bar{D}^*$ state since its threshold value and the mass of color-singlet channel are more reasonable.
  The molecular nature for this pentaquark state can also be guessed from Table~\ref{PcDistance}. Meanwhile, a complete coupled-channels mass which includes all baryon-meson structures is listed in the bottom of Table~\ref{Gresult1}, clearly, 3745 MeV is just the theoretical threshold value of $N\eta_c$. Therefore, no bound state is found in the fully coupled-channels case.

\begin{table}[H]
\caption{The lowest eigen-energies of the $udc{\bar{c}}u$ system with $J^P=\frac12^-$ (unit: MeV).
 The percentages of color-singlet (S) and hidden-color (H) channels are also given. \label{Gresult1}}
\centering
\begin{tabular}{cccccc}
\toprule
   ~~~~~~~~\textbf{Channel}~~   & ~~~~\textit{\textbf{E}}~~~~  & ~~$\textit{\textbf{E}}_{\textbf{th}}^{\textbf{Theo}}$~~   & ~~$\textit{\textbf{E}}_\textbf{B}$~~ &
    $\textit{\textbf{E}}_{\textbf{th}}^{\textbf{Exp}}$   &  \textit{\textbf{E'}} \\
\midrule
    $\chi^{\sigma i}_{1/2}\chi^f_j\chi^c_k~~i=4,5,~~j=3,4,~~k=1$  & 3745 & 3745 & 0    & 3919($N\eta_c$)       & 3919 \\
    $\chi^{\sigma i}_{1/2}\chi^f_j\chi^c_k~~i=4,5,~~j=3,4,~~k=2,3$  & 4714 &  &     &        &  \\
    color-singlet+hidden color & 3745 &  \multicolumn{4}{l}{ }  \\ [2ex]
    $\chi^{\sigma i}_{1/2}\chi^f_j\chi^c_k~~i=2,3,~~j=3,4,~~k=1$  & 3841 & 3841 & 0    & 4036($NJ/\psi$)       & 4036 \\
    $\chi^{\sigma i}_{1/2}\chi^f_j\chi^c_k~~i=2,3,~~j=3,4,~~k=2,3$  & 4964 &  &     &        &  \\
    color-singlet+hidden color  & 3841 & \multicolumn{4}{l}{ } \\ [2ex]
    $\chi^{\sigma i}_{1/2}\chi^f_j\chi^c_k~~i=4,5,~~j=2,~~k=1$  & 3996 & 3996 & 0    & 4151($\Lambda_c\bar{D}$)    & 4151 \\
    $\chi^{\sigma i}_{1/2}\chi^f_j\chi^c_k~~i=4,5,~~j=2,~~k=2,3$  & 4663 &  &     &     &  \\
    color-singlet+hidden color  & 3996 & \multicolumn{4}{l}{ }  \\ [2ex]
    $\chi^{\sigma i}_{1/2}\chi^f_j\chi^c_k~~i=2,3,~~j=2,~~k=1$  & 4115 & 4115 & 0    & 4293($\Lambda_c\bar{D}^*$)  & 4293 \\
    $\chi^{\sigma i}_{1/2}\chi^f_j\chi^c_k~~i=2,3,~~j=2,~~k=2,3$  & 4599 &  &     &   &  \\
    color-singlet+hidden color  & 4115 & \multicolumn{4}{l}{ }  \\ [2ex]
    $\chi^{\sigma i}_{1/2}\chi^f_j\chi^c_k~~i=4,5,~~j=1,~~k=1$  & 4398 & 4402 & $-4$ & 4320($\Sigma_c\bar{D}$)     & 4316 \\
    $\chi^{\sigma i}_{1/2}\chi^f_j\chi^c_k~~i=4,5,~~j=1,~~k=2,3$  & 4835 &  &  &      &  \\
    color-singlet+hidden color  & 4394 & 4402 & $-8$ & 4320 & 4312 \\
        &   & \multicolumn{4}{l}{percentage(S;H): 91.0\%; 7.0\%}  \\ [2ex]
    $\chi^{\sigma i}_{1/2}\chi^f_j\chi^c_k~~i=2,3,~~j=1,~~k=1$  & 4518 & 4520 & $-2$ & 4462($\Sigma_c\bar{D}^*$)   & 4460 \\
    $\chi^{\sigma i}_{1/2}\chi^f_j\chi^c_k~~i=2,3,~~j=1,~~k=2,3$  & 4728 &  &  &    &  \\
    color-singlet+hidden color  & 4479 & 4520  & $-41$ & 4462  & 4421 \\
       &   &  \multicolumn{4}{l}{percentage(S;H): 67.4\%; 32.6\%}  \\ [2ex]
    $\chi^{\sigma i}_{1/2}\chi^f_j\chi^c_k~~i=1,~~j=1,~~k=1$  & 4563 & 4566 & $-3$ & 4527($\Sigma^*_c\bar{D}^*$) & 4524 \\
    $\chi^{\sigma i}_{1/2}\chi^f_j\chi^c_k~~i=1,~~j=1,~~k=2,3$  & 4476 & &     &        &   \\
    color-singlet+hidden color  & 4461 & 4566  &  $-105$ & 4527  & 4422 \\
       &   &  \multicolumn{4}{l}{percentage(S;H): 23.0\%; 77.0\%}   \\ [2ex]
     mixed (only color singlet)  & 3745 & &  \\ [2ex]
     mixed (color singlet+hidden color)  & 3745 & & \\
\bottomrule
\end{tabular}
\end{table}

\item	$I(J^P)=\frac{1}{2}(\frac{3}{2}^-)$ state

Among the five baryon-meson channels listed in Table~\ref{Gresult2} for $\frac{1}{2}(\frac{3}{2}^-)$ state, $NJ/\psi$ and $\Lambda_c \bar{D}^*$ are still unbound neither in the single channel calculation nor the hidden-color channel included and this is similar to the $\frac{1}{2}(\frac{1}{2}^-)$ case. However, several resonance states with shallow binding energies of $\Sigma_c^{(*)} \bar{D}^{(*)}$ configurations are obtained. Specifically, binding energies of the three states of $\Sigma_c \bar{D}^*$, $\Sigma_c^* \bar{D}$ and $\Sigma^*_c \bar{D}^*$ in color-singlet channels are all $\sim$2 MeV. Besides, except $E_B=-3$ MeV for the $\Sigma_c \bar{D}^*$ state in a coupled-channels calculation, there are dozens MeV decreased when hidden-color channels of the other two states included. Furthermore, the most contributions are still from their color-singlet channels and the percentages of singlet are 96\%, 83\% and 61\% for $\Sigma_c \bar{D}^*$, $\Sigma_c^* \bar{D}$ and $\Sigma^*_c \bar{D}^*$, respectively.

 Although the lowest mass remains at 3841 MeV in a fully coupled channels computation, the obtained $\Sigma_c \bar{D}^*$ resonance with mass equals to 4459 MeV is a robust support for the $P^+_c(4457)$ which is also a new hidden charm pentaquark state reported by the LHCb collaboration. This exotic state can be identified as a $\Sigma_c$ baryon and $\bar{D}^*$ meson molecular state whose theoretical size is around 2.3 fm according to the results in Table~\ref{PcDistance}.
 
 \begin{table}[H]
\caption{Distances between any two quarks (unit: fm).  \label{PcDistance}}
\centering
\begin{tabular}{cccccc}
\toprule
  ~~$\textbf{J}^{\textbf{P}}$~~  & ~~~~~~~~\textbf{Channel}~~   & ~~$\textbf{r}_\textbf{{12}}$~~ & ~~$\textbf{r}_\textbf{{13}}$~~ & ~~$\textbf{r}_\textbf{{14}}$~~ & ~~$\textbf{r}_\textbf{{34}}$~~  \\
\midrule
 ${\frac{1}{2}}^{-}$
  & $\chi^{\sigma i}_{1/2}\chi^f_j~i=4,5,j=1,k=1$ ($\Sigma_c \bar{D}$)     & 0.8  & 0.7  & 2.1 & 2.1 \\
  & $\chi^{\sigma i}_{1/2}\chi^f_j~i=4,5,j=1,k=2,3$                  & 1.0  & 0.8  & 0.8 & 0.4 \\
  & $\chi^{\sigma i}_{1/2}\chi^f_j~i=2,3,j=1,k=1$ ($\Sigma_c \bar{D}^*$)   & 0.8  & 0.7  & 2.2 & 2.1 \\
  & $\chi^{\sigma i}_{1/2}\chi^f_j~i=2,3,j=1,k=2,3$                  & 0.9  & 0.8  & 0.8 & 0.4 \\
  & $\chi^{\sigma i}_{1/2}\chi^f_j~i=1,j=1,k=1$ ($\Sigma_c^* \bar{D}^*$)   & 0.9  & 0.8  & 2.1 & 2.0 \\
  & $\chi^{\sigma i}_{1/2}\chi^f_j~i=1,j=1,k=2,3$                    & 0.9  & 0.8  & 0.8 & 0.4 \\ [2ex]
 ${\frac{3}{2}}^{-}$
  & $\chi^{\sigma i}_{3/2}\chi^f_j~i=3,4,j=1,k=1$ ($\Sigma_c \bar{D}^*$)    & 0.8  & 0.7  & 2.4 & 2.3 \\
  & $\chi^{\sigma i}_{3/2}\chi^f_j~i=3,4,j=1,k=2,3$                   & 1.1  & 0.9  & 0.9 & 0.5 \\
  & $\chi^{\sigma i}_{3/2}\chi^f_j~i=2,j=1,k=1$ ($\Sigma_c^* \bar{D}$)      & 0.9  & 0.8  & 2.2 & 2.2 \\
  & $\chi^{\sigma i}_{3/2}\chi^f_j~i=2,j=1,k=2,3$                     & 1.0  & 0.9  & 0.9 & 0.5 \\
  & $\chi^{\sigma i}_{3/2}\chi^f_j~i=1,j=1,k=1$ ($\Sigma_c^* \bar{D}^*$)    & 0.9  & 0.8  & 2.6 & 2.4 \\
  & $\chi^{\sigma i}_{3/2}\chi^f_j~i=1,j=1,k=2,3$                     & 0.9  & 0.9  & 0.8 & 0.4 \\  [2ex]
 ${\frac{5}{2}}^{-}$
  & $\chi^{\sigma i}_{5/2}\chi^f_j~i=1,j=1,k=1$ ($\Sigma_c^* \bar{D}^*$)    & 0.9  & 0.8 & 2.4 & 2.3 \\
  & $\chi^{\sigma i}_{5/2}\chi^f_j~i=1,j=1,k=2,3$                     & 1.3  & 1.4 & 1.3 & 0.8 \\
\bottomrule
\end{tabular}
\end{table}

\item	$I(J^P)=\frac{1}{2}(\frac{5}{2}^-)$ state

There is only one baryon-meson channel $\Sigma^*_c \bar{D}^*$ under consideration in the highest spin state. Firstly, there is -3 MeV binding energy in the color-singlet channel calculation and after a mass shift with $E'=E^{Exp}_{th}+E_B$, 4524 MeV resonance mass is obtained. However, the hidden-color channel mass is higher and at 5002 MeV. A strong coupling between this two channels leads to a deeper binding energy which is -89 MeV, and the modified mass in coupled-channels is 4438 MeV. The components of this resonance state is comparable with 66\% for the color-singlet channel of $\Sigma^*_c \bar{D}^*$ and 34\% of its hidden-color one. Hence it is also a good candidate for the hidden charm pentaquark in high spin state.

\begin{table}[H]
\caption{The lowest eigen-energies of the $udc{\bar{c}}u$ system with $J^P=\frac32^-$ and $\frac52^-$(unit: MeV).
  \label{Gresult2}}
\centering
\begin{tabular}{cccccc}
\toprule
   ~~~~~~~~\textbf{Channel}~~   & ~~~~\textit{\textbf{E}}~~~~  & ~~$\textit{\textbf{E}}_{\textbf{th}}^{\textbf{Theo}}$~~   & ~~$\textit{\textbf{E}}_\textbf{B}$~~ &
    $\textit{\textbf{E}}_{\textbf{th}}^{\textbf{Exp}}$   &  \textit{\textbf{E'}} \\
\midrule
  \multicolumn{6}{c}{$J^P=3/2^{-}$} \\
    $\chi^{\sigma i}_{3/2}\chi^f_j\chi^c_k~~i=3,4,~~j=3,4,~~k=1$  & 3841 & 3841 & 0    & 4036($NJ/\psi$)       & 4036 \\
    $\chi^{\sigma i}_{3/2}\chi^f_j\chi^c_k~~i=3,4,~~j=3,4,~~k=2,3$  & 4722 &  &     &        &  \\
    color-singlet+hidden color  & 3841 & \multicolumn{4}{l}{ }  \\ [2ex]
    $\chi^{\sigma i}_{3/2}\chi^f_j\chi^c_k~~i=3,4,~~j=2,~~k=1$  & 4115 & 4115 & 0    & 4293($\Lambda_c\bar{D}^*$)   & 4293 \\
    $\chi^{\sigma i}_{3/2}\chi^f_j\chi^c_k~~i=3,4,~~j=2,~~k=2,3$  & 4680 &  &     &    &  \\
    color-singlet+hidden color  & 4115 &  \multicolumn{4}{l}{ } \\ [2ex]
    $\chi^{\sigma i}_{3/2}\chi^f_j\chi^c_k~~i=3,4,~~j=1,~~k=1$  & 4518 & 4520 & $-2$ & 4462($\Sigma_c\bar{D}^*$)    & 4460 \\
    $\chi^{\sigma i}_{3/2}\chi^f_j\chi^c_k~~i=3,4,~~j=1,~~k=2,3$  & 4961 &  &  &     &  \\
    color-singlet+hidden color  & 4517 & 4520 &  $-3$ & 4462  & 4459 \\
      &  &   \multicolumn{4}{l}{percentage(S;H): 96.3\%; 3.7\%}  \\ [2ex]
    $\chi^{\sigma i}_{3/2}\chi^f_j\chi^c_k~~i=2,~~j=1,~~k=1$  & 4444 & 4447 & $-3$ & 4385($\Sigma^*_c\bar{D}$)   & 4382 \\
    $\chi^{\sigma i}_{3/2}\chi^f_j\chi^c_k~~i=2,~~j=1,~~k=2,3$  & 4754 &  &  &    &  \\
    color-singlet+hidden color  & 4432 & 4447  &  $-15$ & 4385 & 4370 \\
      &  &  \multicolumn{4}{l}{percentage(S;H): 82.6\%; 17.4\%} \\ [2ex]
    $\chi^{\sigma i}_{3/2}\chi^f_j\chi^c_k~~i=1,~~j=1,~~k=1$  & 4564 & 4566 & $-2$ & 4527($\Sigma^*_c\bar{D}^*$)  & 4525 \\
    $\chi^{\sigma i}_{3/2}\chi^f_j\chi^c_k~~i=1,~~j=1,~~k=2,3$  & 4623 &  &  &   &  \\
    color-singlet+hidden color  & 4549 & 4566  & $-17$  & 4527  & 4510 \\
      &  & \multicolumn{4}{l}{percentage(S;H): 61.1\%; 38.9\%}  \\ [2ex]
    mixed (only color-singlet) & 3841 & & & \\  [2ex]
    mixed (color-singlet+hidden color) & 3841 & & & \\
\midrule
  \multicolumn{6}{c}{$J^P=5/2^{-}$} \\
    $\chi^{\sigma i}_{5/2}\chi^f_j\chi^c_k~~i=1,~~j=1,~~k=1$  & 4563 & 4566 & $-3$ & 4527($\Sigma^*_c\bar{D}^*$)  & 4524 \\
    $\chi^{\sigma i}_{5/2}\chi^f_j\chi^c_k~~~i=1,~~j=1,~~k=2,3$  & 5002 &   &  &   & \\
    color-singlet+hidden color  & 4477 &  4566  &  $-89$  & 4527  & 4438 \\
      &   & \multicolumn{4}{l}{percentage(S;H): 66.2\%; 33.8\%}  \\
\bottomrule
\end{tabular}
\end{table}

\end{itemize}

\subsubsection{Hidden Bottom Pentaquarks}

In this part, we extend the investigation on hidden charm pentaquark state to the hidden bottom sector. Spin-parity $J^P=\frac{1}{2}^+$, $\frac{3}{2}^-$ and $\frac{5}{2}^-$, and both in the isospin $I=\frac{1}{2}$ and $\frac{3}{2}$ of $qqqb\bar{b}$ systems are considered. The allowed baryon-meson channels in each $I(J^P)$ quantum state are summarized in Table~\ref{GCC}. Several resonance states are found in these quantum states except the $\frac{3}{2}(\frac{1}{2}^-)$ case. 

Particularly, from Table~\ref{GresultPb1} to \ref{GresultPb5}, the first column lists the channels on our obtained bound states, their experimental values of the noninteracting baryon-meson threshold ($E^{ex}_{th}$) are also shown in parentheses. The second column labels the color-singlet (S), hidden-color (H) and coupled-channels (S+H) computation. Then, the calculated theoretical mass ($M$) and binding energy ($E_B$) of pentaquark state is presented in the third and fourth columns, respectively. A re-scaled mass ($M'$), which is obtained by attending to the experimental baryon-meson threshold, $M'=E^{ex}_{th}+E_B$, and with a purpose of removing the theoretical uncertainty, is listed in the last column.
 The percentages of color-singlet (S) and hidden-color (H) channels are also given when the coupled-channels calculation is performed. Hence, we will discuss them according to the $I(J^P)$ state, respectively.

\begin{table}[H]
\caption{\label{GCC} All allowed channels for hidden-bottom pentaquark systems with negative parity.}
\centering
\begin{tabular}{cccccc}
\toprule
& & \multicolumn{2}{c}{$\textit{\textbf{I}=}\frac{\textbf{1}}{\textbf{2}}$} & \multicolumn{2}{c}{$\textit{\textbf{I}=}\frac{\textbf{3}}{\textbf{2}}$} \\[2ex]
~~$\textit{\textbf{J}}^\textbf{P}$~~&~~\textbf{Index}~~ & ~~$\chi_J^{\sigma_i}$;~$\chi_I^{f_j}$;~$\chi_k^c$~~ & ~~\textbf{Channel}~~ & ~~$\chi_J^{\sigma_i}$;~$\chi_I^{f_j}$;~$\chi_k^c$~~ & ~~\textbf{Channel}~~ \\
&&$[i; ~j; ~k]$& &$[i; ~j; ~k]$&  \\
\midrule
$\frac{1}{2}^-$ & 1  & $[4,5; ~3,4; ~1]$   & $(N \eta_b)^1$ & $[1; ~1; ~1]$   & $(\Delta \Upsilon)^1$ \\
&  2 & $[4,5; ~3,4; ~2,3]$ & $(N \eta_b)^8$ & $[1; ~1; ~3]$  & $(\Delta \Upsilon)^8$ \\
&  3 & $[2,3; ~3,4; ~1]$   & $(N \Upsilon)^1$  & $[4; ~2; ~1]$   & $(\Sigma_b \bar{B})^1$ \\
&  4 & $[2,3; ~3,4; ~2,3]$ & $(N \Upsilon)^8$ & $[4,5; ~2; ~2,3]$   & $(\Sigma_b \bar{B})^8$   \\
&  5 & $[5; ~2; ~1]$     & $(\Lambda_b \bar{B})^1$  & $[2; ~2; ~1]$   & $(\Sigma_b \bar{B}^*)^1$ \\
&  6 & $[4,5; ~2; ~2,3]$ & $(\Lambda_b \bar{B})^8$  & $[2,3; ~2; ~2,3]$   & $(\Sigma_b \bar{B}^*)^8$ \\
&  7 & $[3; ~2; ~1]$     & $(\Lambda_b \bar{B}^*)^1$  & $[1; ~2; ~1]$   & $(\Sigma^*_b \bar{B}^*)^1$ \\
&  8 & $[2,3; ~2; ~2,3]$ & $(\Lambda_b \bar{B}^*)^8$  & $[1; ~2; ~3]$   & $(\Sigma^*_b \bar{B}^*)^8$ \\
&  9 & $[4; ~1; ~1]$     & $(\Sigma_b \bar{B})^1$     & & \\
& 10 & $[4,5; ~1; ~2,3]$ & $(\Sigma_b \bar{B})^8$     & & \\
& 11 & $[2; ~1; ~1]$     & $(\Sigma_b \bar{B}^*)^1$   & & \\
& 12 & $[2,3; ~1; ~2,3]$ & $(\Sigma_b \bar{B}^*)^8$   & & \\
& 13 & $[1; ~1; ~1]$     & $(\Sigma^*_b \bar{B}^*)^1$ & & \\
& 14 & $[1; ~1; ~3]$     & $(\Sigma^*_b \bar{B}^*)^8$ & & \\[2ex]
$\frac{3}{2}^-$ & 1  & $[3,4; ~3,4; ~1]$   & $(N \Upsilon)^1$ & $[2; ~1; ~1]$   & $(\Delta \eta_b)^1$\\
& 2  & $[3,4; ~3,4; ~2,3]$  & $(N \Upsilon)^8$ & $[2; ~1; ~3]$  & $(\Delta \eta_b)^8$ \\
& 3  & $[4; ~2; ~1]$     & $(\Lambda_b \bar{B}^*)^1$  & $[1; ~1; ~1]$   & $(\Delta \Upsilon)^1$ \\
& 4  & $[3,4; ~2; ~2,3]$  & $(\Lambda_b \bar{B}^*)^8$ & $[1; ~1; ~3]$ & $(\Delta \Upsilon)^8$ \\
& 5  & $[3; ~1; ~1]$     & $(\Sigma_b \bar{B}^*)^1$  & $[3; ~2; ~1]$   & $(\Sigma_b \bar{B}^*)^1$ \\
& 6  & $[3,4; ~1; ~2,3]$  & $(\Sigma_b \bar{B}^*)^8$  & $[3,4; ~2; ~2,3]$   & $(\Sigma_b \bar{B}^*)^8$  \\
& 7  & $[2; ~1; ~1]$     & $(\Sigma^*_b \bar{B})^1$  & $[2; ~2; ~1]$   & $(\Sigma^*_b \bar{B})^1$ \\
& 8  & $[2; ~1; ~3]$  & $(\Sigma^*_b \bar{B})^8$  & $[2; ~2; ~3]$   & $(\Sigma^*_b \bar{B})^8$  \\
& 9  & $[1; ~1; ~1]$   & $(\Sigma^*_b \bar{B}^*)^1$ & $[1; ~2; ~1]$  & $(\Sigma^*_b \bar{B}^*)^1$ \\
& 10 & $[1; ~1; ~3]$  & $(\Sigma^*_b \bar{B}^*)^8$  & $[1; ~2; ~3]$   &  $(\Sigma^*_b \bar{B}^*)^8$  \\[2ex]
$\frac{5}{2}^-$ & 1  & $[1; ~1; ~1]$   & $(\Sigma^*_b \bar{B}^*)^1$ & $[1; ~1; ~1]$   & $(\Delta \Upsilon)^1$ \\
& 2  & $[1; ~1; ~3]$  & $(\Sigma^*_b \bar{B}^*)^8$ & $[1; ~1; ~3]$  & $(\Delta \Upsilon)^8$ \\
& 3  &   &   & $[1; ~2; ~1]$   & $(\Sigma^*_b \bar{B}^*)^1$ \\
& 4  &   & & $[1; ~2; ~3]$ & $(\Sigma^*_b \bar{B}^*)^8$ \\
\bottomrule
\end{tabular}
\end{table}

\begin{itemize}[leftmargin=*,labelsep=5.8mm, listparindent=2em]
\item	$I(J^P)=\frac{1}{2}(\frac{1}{2}^-)$ state

There are 14 possibilities which include $N\eta_b$, $N\Upsilon$, $\Lambda_b \bar{B}$, $\Lambda_b \bar{B}^*$, $\Sigma_b \bar{B}$, $\Sigma_b \bar{B}^*$ and $\Sigma_b^* \bar{B}^*$ both in singlet- and hidden-color channel. However, resonance states are only obtained in the $\Sigma_b^{(*)} \bar{B}^{(*)}$ configurations and they are listed in Table~\ref{GresultPb1}. Particularly, there are around $-20$ MeV binding energy for the color-singlet channels of $\Sigma_b \bar{B}$, $\Sigma_b \bar{B}^*$ and $\Sigma_b^* \bar{B}^*$. Their modified masses $M'$ are 11.07 GeV, 11.11 GeV and 11.13 GeV, respectively. As for their hidden-color channels, only the $\Sigma_b^* \bar{B}^*$ channel with $E_B=-232$ MeV is found, the other two are at least 120 MeV above their corresponding theoretical threshold values.

In additional, within a calculation which the coupling on color-singlet and hidden-color channels are considered, the $\Sigma_b \bar{B}^*$ and $\Sigma_b^* \bar{B}^*$ resonances with much deeper binding energies (more than $-90$ MeV) are obtained, their coupled masses are 11.04 GeV and 10.86 GeV, respectively. Therefore, compact configuration of hidden bottom pentaquark state is favored herein. However, the coupling in $\Sigma_b \bar{B}$ resonance is quite weak with only 2 MeV binding energy decreased and the rescaled mass is still $\sim$11.07 GeV, this can be identified as a molecule state. Moreover, these results are confirmed by two facts: (i) the color-singlet channel is almost 99\% in $\Sigma_b \bar{B}$, 58\% in $\Sigma_b \bar{B}^*$ and only 16\% in $\Sigma_b^* \bar{B}^*$, (ii) in Table~\ref{tab:disqqq}, the calculated size for $\Sigma_b \bar{B}$ resonance state is around 1 fm. However, $\sim$ 0.7 fm for the other two resonance states and especially, the distance between a $b\bar{b}$ pair in $\Sigma_b \bar{B}^*$ and $\Sigma_b^* \bar{B}^*$ states is only $\sim$0.3 fm.

\begin{table}[H]
\caption{\label{GresultPb1} Lowest-lying states of hidden-bottom pentaquarks with quantum numbers $I(J^P)=\frac12(\frac12^-)$, unit in MeV. The baryon-meson channels that do not appear here have been also considered in the computation but no bound states were found.}
\centering
\begin{tabular}{ccccc}
\toprule
~~\textbf{Channel}~~   & ~~\textbf{Color}~~ & ~~\textit{\textbf{M}}~~ & ~~$\textit{\textbf{E}}_\textbf{B}$~~ & ~~\textit{\textbf{M'}}~~ \\
\midrule
$\Sigma_b\bar{B}$ & S   & $11080$ & $-15$  & $11074$ \\
$(11089)$         & H   & $11364$ & $+269$ & $11358$ \\
                  & S+H & $11078$ & $-17$  & $11072$ \\
                  & \multicolumn{4}{c}{Percentage (S;H): 98.5\%; 1.5\%}  \\[2ex]
$\Sigma_b\bar{B}^*$ & S   & $11115$ & $-21$  & $11113$ \\
$(11134)$           & H   & $11257$ & $+121$ & $11255$ \\
                    & S+H & $11043$ & $-93$  & $11041$ \\
                    & \multicolumn{4}{c}{Percentage (S;H): 57.9\%; 42.1\%} \\[2ex]
$\Sigma^*_b\bar{B}^*$ & S   & $11127$ & $-26$  & $11128$ \\
$(11154)$             & H   & $10921$ & $-232$ & $10922$ \\
                      & S+H & $10861$ & $-292$ & $10862$ \\
                      & \multicolumn{4}{c}{Percentage (S;H): 15.8\%; 84.2\%} \\ 
\bottomrule
\end{tabular}
\end{table}

\item	$I(J^P)=\frac{1}{2}(\frac{3}{2}^-)$ state

We only find resonances in the $\Sigma_b \bar{B}^*$, $\Sigma_b^* \bar{B}$ and $\Sigma_b^* \bar{B}^*$ states. Table~\ref{GresultPb2} presents the predicted mass, binding energy and components of these hidden bottom pentaquarks. Firstly, in the color-singlet channels calculation,  the binding energies are $-12\,\text{MeV}$, $-15\,\text{MeV}$ and $-15\,\text{MeV}$ for the $\Sigma_b \bar{B}^*$, $\Sigma_b^* \bar{B}$ and $\Sigma_b^* \bar{B}^*$ channels, respectively. 

However, when the hidden-color channels are incorporated into computation, a strong coupling effect is obtained in the later two states and the binding energies are $-67$ MeV and $-195$ MeV, respectively. In contrast, the coupling is very weak and only 2 MeV decreased when hidden-color channel included in $\Sigma_b \bar{B}^*$ state. The modified masses are 11.12 GeV, 11.04 GeV and 10.96 GeV for the $\Sigma_b \bar{B}^*$, $\Sigma_b^* \bar{B}$ and $\Sigma_b^* \bar{B}^*$ states in the coupled-channels study, respectively. Furthermore, Table~\ref{GresultPb2} also presents the percentages of singlet- and hidden-channels of the three resonance states. In particular, it is almost the complete color-singlet component in the $\Sigma_b \bar{B}^*$ state. But the main part (78\%) is the hidden-color channel in $\Sigma_b^* \bar{B}^*$ state, and also a considerable percentage (45\%) in the $\Sigma_b^* \bar{B}$.

These results are consistent with the analysis on the internal structures of $\Sigma_b^{(*)} \bar{B}^{(*)}$ resonance states. In Table~\ref{tab:disqqq} one can notice that a compact configuration within 0.7 fm for both the $\Sigma_b^* \bar{B}$ and $\Sigma_b^* \bar{B}^*$ states confirms the previous facts of deep binding energy and strong coupling. However, it is a $\Sigma_b \bar{B}^*$ molecule state which size is around 1.1 fm in the coupled-channels calculation.

\begin{table}[H]
\caption{\label{GresultPb2} Lowest-lying states of hidden-bottom pentaquarks with quantum numbers $I(J^P)=\frac12(\frac32^-)$, unit in MeV. The baryon-meson channels that do not appear here have been also considered in the computation but no bound states were found.}
\centering
\begin{tabular}{ccccc}
\toprule
~~\textbf{Channel}~~   & ~~\textbf{Color}~~ & ~~\textit{\textbf{M}}~~ & ~~$\textit{\textbf{E}}_\textbf{B}$~~ & ~~\textit{\textbf{M'}}~~ \\
\midrule
$\Sigma_b\bar{B}^*$ & S   & $11124$ & $-12$  & $11122$ \\
$(11134)$           & H   & $11476$ & $+340$ & $11475$ \\
                    & S+H & $11122$ & $-14$  & $11120$ \\
                    & \multicolumn{4}{c}{Percentage (S;H): 99.6\%; 0.4\%} \\[2ex]
$\Sigma^*_b\bar{B}$ & S   & $11097$ & $-15$ & $11094$ \\
$(11109)$           & H   & $11175$ & $+63$ & $11172$ \\
                    & S+H & $11045$ & $-67$ & $11042$ \\
                    & \multicolumn{4}{c}{Percentage (S;H): 55.5\%; 44.5\%} \\[2ex]
$\Sigma^*_b\bar{B}^*$ & S   & $11138$ & $-15$  & $11139$ \\
$(11154)$             & H   & $11051$ & $-102$ & $11052$ \\
                      & S+H & $10958$ & $-195$ & $10959$ \\
                      & \multicolumn{4}{c}{Percentage (S;H): 22.2\%; 77.8\%} \\ 
\bottomrule
\end{tabular}
\end{table}

\item	$I(J^P)=\frac{1}{2}(\frac{5}{2}^-)$ state

Only one $\Sigma_b^* \bar{B}^*$ configuration need to be considered in the highest spin state and Table~\ref{GresultPb3} shows the calculated mass of this resonance state. Clearly, there are $-12$ MeV binding energy in the color-singlet channel. Besides, this result is comparable with the colorless channel of $\Sigma_b^* \bar{B}^*$ in the other two spin-parity states.

In a further step with hidden-color channel included, only a weak coupling with $E_B=-13$ MeV is obtained and the rescaled mass which is obtained as the previous procedure is 11.14 GeV. Accordingly, this resonance can be identified as the $\Sigma_b^* \bar{B}^*$ molecule state with 99.6\% color-singlet component and $\sim$1.1 fm size shown in Table~\ref{tab:disqqq}.

\begin{table}[H]
\caption{\label{GresultPb3} Lowest-lying states of hidden-bottom pentaquarks with quantum numbers $I(J^P)=\frac12(\frac52^-)$, unit in MeV. The baryon-meson channels that do not appear here have been also considered in the computation but no bound states were found.}
\centering
\begin{tabular}{ccccc}
\toprule
~~\textbf{Channel}~~   & ~~\textbf{Color}~~ & ~~\textit{\textbf{M}}~~ & ~~$\textit{\textbf{E}}_\textbf{B}$~~ & ~~\textit{\textbf{M'}}~~ \\
\midrule
$\Sigma^*_b\bar{B}^*$ & S   & $11141$ & $-12$  & $11151$ \\
$(11154)$             & H   & $11547$ & $+394$ & $11548$ \\
                      & S+H & $11140$ & $-13$  & $11141$ \\
                      & \multicolumn{4}{c}{Percentage (S;H): 99.6\%; 0.4\%} \\
\bottomrule
\end{tabular}
\end{table}

\item	$I(J^P)=\frac{3}{2}(\frac{3}{2}^-)$ state

In this spin-parity channel, neither bound nor resonance state is found in the $\Delta \eta_b$ and $\Delta \Upsilon$ configurations, however, resonance states are possible in the $\Sigma_b^{(*)}\bar{B}^{(*)}$ configurations. The details are listed in Table~\ref{GresultPb4}.

Firstly, only scattering states are found in the $\Sigma_b\bar{B}^\ast$, $\Sigma_b^\ast\bar{B}$ and $\Sigma_b^\ast\bar{B}^\ast$ channels when only color-singlet configurations are included. However, hidden-color structures help in obtaining resonance states with deep binding energies. Namely, coupled-masses of these three resonance states are all below the corresponding theoretical thresholds about 110 MeV. The modified masses are 11.02 GeV, 10.99 GeV and 11.05 GeV for the $\Sigma_b\bar{B}^\ast$, $\Sigma_b^\ast\bar{B}$ and $\Sigma_b^\ast\bar{B}^\ast$ states, respectively. Furthermore, their calculated sizes listed in Table~\ref{tab:disqqq} are $\sim$0.8 fm which is a compact 5-quarks configuration. This feature is supported by the fact that the coupling between singlet- and hidden-channels is strong and the percentage of the later configuration is quite considerable, 35\% in $\Sigma_b\bar{B}^\ast$ and both more than 81\% in $\Sigma_b^\ast\bar{B}$ and $\Sigma_b^\ast\bar{B}^\ast$.

\begin{table}[H]
\caption{\label{GresultPb4} Lowest-lying states of hidden-bottom pentaquarks with quantum numbers $I(J^P)=\frac32(\frac32^-)$, unit in MeV. The baryon-meson channels that do not appear here have been also considered in the computation but no bound states were found.}
\centering
\begin{tabular}{ccccc}
\toprule
~~\textbf{Channel}~~   & ~~\textbf{Color}~~ & ~~\textit{\textbf{M}}~~ & ~~$\textit{\textbf{E}}_\textbf{B}$~~ & ~~\textit{\textbf{M'}}~~ \\
\midrule
$\Sigma_b\bar{B}^*$ & S   & $11136$ & $0$    & $11134$ \\
$(11134)$           & H   & $11310$ & $+174$ & $11308$  \\
                    & S+H & $11021$ & $-115$ & $11019$ \\
                    & \multicolumn{4}{c}{Percentage (S;H): 64.7\%; 35.3\%} \\[2ex]
$\Sigma^*_b\bar{B}$ & S   & $11112$ & $0$    & $11109$ \\
$(11109)$           & H   & $11041$ & $-71$  & $11038$ \\
                    & S+H & $10999$ & $-113$ & $10996$ \\
                    & \multicolumn{4}{l}{Percentage (S;H): 18.4\%; 81.6\%} \\[2ex]
$\Sigma^*_b\bar{B}^*$ & S   & $11153$ & $0$    & $11154$ \\
$(11154)$             & H   & $11102$ & $-51$  & $11103$ \\
                      & S+H & $11048$ & $-105$ & $11049$ \\
                      & \multicolumn{4}{c}{Percentage (S;H): 15.7\%; 84.3\%} \\ 
\bottomrule
\end{tabular}
\end{table}

\item	$I(J^P)=\frac{3}{2}(\frac{5}{2}^-)$ state

There are two baryon-meson channels devote to the highest spin and isospin state, $\Delta\Upsilon$ and $\Sigma_b^\ast \bar{B}^\ast$. Table~\ref{GresultPb5} just lists the results of $\Sigma_b^\ast \bar{B}^\ast$ resonance state, however, it is a scattering state of $\Delta\Upsilon$ channel. Firstly, in the single channel calculation of color-singlet and hidden-color configuration, there are more than $-100$ MeV binding energy obtained and especially $E_B= -179$ MeV for the later structure. Additionally, a deeper binding energy which is $-222$ MeV is obtained in their coupled-channels computation and the modified mass is 10.93 GeV. Apparently, this can be identified a color resonance which about 80\% component is the hidden-color channel of $\Sigma_b^\ast \bar{B}^\ast$. Moreover, the size $\sim$0.8 fm in Table~\ref{tab:disqqq} confirms a compact structure too.

\begin{table}[H]
\caption{\label{GresultPb5} Lowest-lying states of hidden-bottom pentaquarks with quantum numbers $I(J^P)=\frac32(\frac52^-)$, unit in MeV. The baryon-meson channels that do not appear here have been also considered in the computation but no bound states were found.}
\centering
\begin{tabular}{ccccc}
\toprule
~~\textbf{Channel}~~   & ~~\textbf{Color}~~ & ~~\textit{\textbf{M}}~~ & ~~$\textit{\textbf{E}}_\textbf{B}$~~ & ~~\textit{\textbf{M'}}~~ \\
\midrule
$\Sigma^*_b\bar{B}^*$ & S   & $11052$ & $-101$ & $11053$ \\
$(11154)$             & H   & $10974$ & $-179$ & $10975$ \\
                      & S+H & $10931$ & $-222$ & $10932$ \\
                      & \multicolumn{4}{c}{Percentage (S;H): 19.9\%; 80.1\%} \\ 
\bottomrule
\end{tabular}
\end{table}

\begin{table}[H]
\caption{\label{tab:disqqq} The distance, in fm, between any two quarks of the found pentaquark bound-states.}
\centering
\begin{tabular}{ccccccc}
\toprule
~~$\textbf{I}(\textbf{J}^{\textbf{P}})$~~ & ~~\textbf{Channel}~~ & ~~\textbf{Mixing}~~ & ~~$\textit{\textbf{r}}_{\textbf{qq}}$~~ & ~~$\textit{\textbf{r}}_{\textbf{qQ}}$~~ & ~~$\textit{\textbf{r}}_{\textbf{q}\bar{\textbf{Q}}}$~~ & ~~$\textit{\textbf{r}}_{\textbf{Q}\bar{\textbf{Q}}}$~~  \\
\midrule
$\frac12({\frac{1}{2}}^{-})$
& $\Sigma_b \bar{B}$     & S   & 1.17 & 0.87 & 1.02 & 1.00 \\
&                        & S+H & 1.13 & 0.84 & 0.98 & 0.94 \\
& $\Sigma_b \bar{B}^*$   & S   & 1.09 & 0.81 & 0.92 & 0.82 \\
&                        & S+H & 0.94 & 0.70 & 0.71 & 0.34 \\
& $\Sigma_b^* \bar{B}^*$ & S   & 1.06 & 0.79 & 0.88 & 0.75 \\
&                        & S+H & 0.91 & 0.71 & 0.70 & 0.24 \\[2ex]
$\frac12({\frac{3}{2}}^{-})$
& $\Sigma_b \bar{B}^*$   & S   & 1.23 & 0.90 & 1.09 & 1.09 \\
&                        & S+H & 1.21 & 0.90 & 1.07 & 1.07 \\
& $\Sigma_b^* \bar{B}$   & S   & 1.18 & 0.88 & 1.04 & 1.01 \\
&                        & S+H & 0.98 & 0.74 & 0.74 & 0.34 \\
& $\Sigma_b^* \bar{B}^*$ & S   & 1.17 & 0.87 & 1.02 & 0.97 \\
&                        & S+H & 0.95 & 0.72 & 0.72 & 0.25 \\[2ex]
$\frac12({\frac{5}{2}}^{-})$
& $\Sigma_b^* \bar{B}^*$ & S   & 1.25  & 0.92 & 1.11 & 1.13 \\
&                        & S+H &1.25  & 0.92 & 1.11 & 1.11 \\[2ex]
$\frac32({\frac{3}{2}}^{-})$ 
& $\Sigma_b \bar{B}^*$   & S+H & 1.02 & 0.78  & 0.77 & 0.27 \\
& $\Sigma_b^* \bar{B}$   & S+H & 1.02 & 0.84  & 0.82 & 0.26 \\
& $\Sigma_b^* \bar{B}^*$ & S+H & 1.05 & 0.83  & 0.81 & 0.26 \\[2ex]
$\frac32({\frac{5}{2}}^{-})$ 
& $\Sigma_b^* \bar{B}^*$ & S   & 1.03 & 0.86 & 0.86 & 0.29 \\                             
&                        & S+H & 1.00 & 0.86 & 0.84 & 0.26 \\
\bottomrule
\end{tabular}
\end{table}

\end{itemize}

\subsubsection{Doubly Charmed Pentaquarks}

Along with the hidden-charm pentaquark states announced experimentally, an open-flavor 5-quark system is also quite charming. In this part, the $ccqq\bar{q}$ ($q=u, d$) pentaquarks with spin-parity $J^P=\frac{1}{2}^-$, $\frac{3}{2}^-$ and $\frac{5}{2}^-$, isospin $I=\frac{1}{2}$ and $\frac{3}{2}$ are all investigated. Table~\ref{GCC1} and \ref{GCC2} listed the allowed channels in each $I(J^P)$ states. However, due to the fact that a large amount of computational effort is needed for exactly solving the 5-body Schr\"{o}dinger equation, we still perform the study in the baryon-meson sector in which both color-singlet and hidden-color channels are considered. Nevertheless, as shown in Fig.~\ref{5QCOF2}, there are four configurations in the 3+2 clusters. In particular, each five quarks are fixed and we only need to consider the different coupling sequences.

Moreover, the complex scaling method is also applied in this sector in order to have a better classification of bound and resonance states in the multiquark systems. Table~\ref{Rsum} summarized our results on the obtained $ccqq\bar{q}$ pentaquark states in advance, particularly, the second column shows the channels of bound or resonance states with their theoretical masses marked in the bracket, and the last column lists the binding energy or resonance width for these exotic states. The details are going to be discussed according to the $I(J^P)$ quantum states.

\begin{table}[H]
\caption{\label{GCC1} All possible channels for open-charm pentaquark systems with $J^P=1/2^-$.}
\centering
\begin{tabular}{cccccc}
\toprule
& & \multicolumn{2}{c}{$I=\frac{1}{2}$} & \multicolumn{2}{c}{$I=\frac{3}{2}$} \\
~~$\textbf{J}^\textbf{P}$~~&~~\textbf{Index}~~ & ~~$\chi_J^{n \sigma_i}$;~$\chi_I^{n f_j}$;~$\chi_k^{n c}$;~~ & ~~\textbf{Channel}~~ & ~~$\chi_J^{n \sigma_i}$;~$\chi_I^{n f_j}$;~$\chi_k^{n c}$;~~ & ~~\textbf{Channel}~~ \\
&&$[i; ~j; ~k;~n]$& &$[i; ~j; ~k;~n]$&  \\
\midrule
$\frac{1}{2}^-$ & 1  & $[4; ~3; ~1; ~1,2]$   & $(\Xi_{cc} \eta)^1$ & $[4; ~2; ~1; ~1,2]$   & $(\Xi_{cc} \pi)^1$ \\
&  2 & $[4,5; ~3; ~2,3; ~1,2]$ & $(\Xi_{cc} \eta)^8$ & $[4,5; ~2; ~2,3; ~1,2]$  & $(\Xi_{cc} \pi)^8$ \\
&  3 & $[2; ~3; ~1; ~1,2]$   & $(\Xi_{cc} \omega)^1$  & $[2; ~2; ~1; ~1,2]$   & $(\Xi_{cc} \rho)^1$ \\
&  4 & $[2,3; ~3; ~2,3; ~1,2]$ & $(\Xi_{cc} \omega)^8$ & $[2,3; ~2; ~2,3; ~1,2]$   & $(\Xi_{cc} \rho)^8$   \\
&  5 & $[4; ~4; ~1; ~1,2]$     & $(\Xi_{cc} \pi)^1$  & $[1; ~2; ~1; ~1,2]$   & $(\Xi^*_{cc} \rho)^1$ \\
&  6 & $[4,5; ~4; ~2,3; ~1,2]$ & $(\Xi_{cc} \pi)^8$  & $[1; ~2; ~3; ~1,2]$   & $(\Xi^*_{cc} \rho)^8$ \\
&  7 & $[2; ~4; ~1; ~1,2]$     & $(\Xi_{cc} \rho)^1$  & $[4; ~3; ~1; ~3,4]$   & $(\Sigma_c D)^1$ \\
&  8 & $[2,3; ~4; ~2,3; ~1,2]$ & $(\Xi_{cc} \rho)^8$  & $[4,5; ~3; ~2,3; ~3,4]$   & $(\Sigma_c D)^8$ \\
&  9 & $[1; ~3; ~1, ~1,2]$     & $(\Xi^*_{cc} \omega)^1$  & $[2; ~3; ~1; ~3,4]$   & $(\Sigma_c D^*)^1$ \\
& 10 & $[1; ~3; ~3, ~1,2]$ & $(\Xi^*_{cc} \omega)^8$  & $[2,3; ~3; ~2,3; ~3,4]$   & $(\Sigma_c D^*)^8$ \\
& 11 & $[1; ~4; ~1; ~1,2]$     & $(\Xi^*_{cc} \rho)^1$  & $[1; ~3; ~1; ~3,4]$   & $(\Sigma^*_c D^*)^1$ \\
& 12 & $[1; ~4; ~3; ~1,2]$ & $(\Xi^*_{cc} \rho)^8$  & $[1; ~3; ~3; ~3,4]$   & $(\Sigma^*_c D^*)^8$ \\
& 13 & $[5; ~2; ~1; ~3,4]$     & $(\Lambda_c D)^1$ & & \\
& 14 & $[4,5; ~2; ~2,3; ~3,4]$     & $(\Lambda_c D)^8$ & & \\
& 15 & $[3; ~2; ~1; ~3,4]$     & $(\Lambda_c D^*)^1$ & & \\
& 16 & $[2,3; ~2; ~2,3; ~3,4]$     & $(\Lambda_c D^*)^8$ & & \\
& 17 & $[4; ~1; ~1; ~3,4]$     & $(\Sigma_c D)^1$ & & \\
& 18 & $[4,5; ~1; ~2,3; ~3,4]$     & $(\Sigma_c D)^8$ & & \\
& 19 & $[2; ~1; ~1; ~3,4]$     & $(\Sigma_c D^*)^1$ & & \\
& 20 & $[2,3; ~1; ~2,3; ~3,4]$     & $(\Sigma_c D^*)^8$ & & \\
& 21 & $[1; ~1; ~1; ~3,4]$     & $(\Sigma^*_c D^*)^1$ & & \\
& 22 & $[1; ~1; ~3; ~3,4]$     & $(\Sigma^*_c D^*)^8$ & & \\
\bottomrule
\end{tabular}
\end{table}

\begin{table}[H]
\caption{\label{GCC2} All possible channels for open-charm pentaquark systems with $J^P=3/2^-$ and $5/2^-$.}
\centering
\begin{tabular}{cccccc}
\toprule
& & \multicolumn{2}{c}{$I=\frac{1}{2}$} & \multicolumn{2}{c}{$I=\frac{3}{2}$} \\
~~$\textbf{J}^\textbf{P}$~~&~~\textbf{Index}~~ & ~~$\chi_J^{n \sigma_i}$;~$\chi_I^{n f_j}$;~$\chi_k^{n c}$;~~ & ~~\textbf{Channel}~~ & ~~$\chi_J^{n \sigma_i}$;~$\chi_I^{n f_j}$;~$\chi_k^{n c}$;~~ & ~~\textbf{Channel}~~ \\
&&$[i; ~j; ~k;~n]$& &$[i; ~j; ~k;~n]$&  \\
\midrule
$\frac{3}{2}^-$ & 1  & $[3; ~3; ~1; ~1,2]$   & $(\Xi_{cc} \omega)^1$ & $[3; ~2; ~1; ~1,2]$ & $(\Xi_{cc} \rho)^1$\\
& 2  & $[3,4; ~3; ~2,3; ~1,2]$  & $(\Xi_{cc} \omega)^8$ & $[3,4; ~2; ~2,3; ~1,2]$  & $(\Xi_{cc} \rho)^8$ \\
& 3  & $[3; ~4; ~1; ~1,2]$     & $(\Xi_{cc} \rho)^1$  & $[2; ~2; ~1; ~1,2]$   & $(\Xi^*_{cc} \pi)^1$ \\
& 4  & $[3,4; ~4; ~2,3; ~1,2]$  & $(\Xi_{cc} \rho)^8$ & $[2; ~2; ~3; ~1,2]$ & $(\Xi^*_{cc} \pi)^8$ \\
& 5  & $[2; ~4; ~1; ~1,2]$     & $(\Xi^*_{cc} \pi)^1$  & $[1; ~2; ~1; ~1,2]$   & $(\Xi^*_{cc} \rho)^1$ \\
& 6  & $[2; ~4; ~3; ~1,2]$  & $(\Xi^*_{cc} \pi)^8$  & $[1; ~2; ~3; ~1,2]$   & $(\Xi^*_{cc} \rho)^8$  \\
& 7  & $[1; ~3; ~1; ~1,2]$     & $(\Xi^*_{cc} \omega)^1$  & $[3; ~3; ~1; ~3,4]$   & $(\Sigma_c D^*)^1$ \\
& 8  & $[1; ~3; ~3; ~1,2]$  & $(\Xi^*_{cc} \omega)^8$  & $[3,4; ~3; ~2,3; ~3,4]$   & $(\Sigma_c D^*)^8$  \\
& 9  & $[1; ~4; ~1; ~1,2]$   & $(\Xi^*_{cc} \rho)^1$ & $[2; ~3; ~1; ~3,4]$  & $(\Sigma^*_c D)^1$ \\
& 10 & $[1; ~4; ~3; ~1,2]$  & $(\Xi^*_{cc} \rho)^8$  & $[3; ~3; ~3; ~3,4]$   &  $(\Sigma^*_c D)^8$  \\
& 11  & $[4; ~2; ~1; ~3,4]$   & $(\Lambda_c D^*)^1$ & $[1; ~3; ~1; ~3,4]$  & $(\Sigma^*_c D^*)^1$ \\
& 12 & $[3,4; ~2; ~2,3; ~3,4]$  & $(\Lambda_c D^*)^8$  & $[1; ~3; ~3; ~3,4]$   &  $(\Sigma^*_c D^*)^8$  \\
& 13  & $[3; ~1; ~1; ~3,4]$   & $(\Sigma_c D^*)^1$ &   &  \\
& 14 & $[3,4; ~1; ~1; ~3,4]$  & $(\Sigma_c D^*)^8$  &   &  \\
& 15  & $[2; ~1; ~1; ~3,4]$   & $(\Sigma^*_c D)^1$ &   &  \\
& 16 & $[2; ~1; ~3; ~3,4]$  & $(\Sigma^*_c D)^8$  &   &  \\
& 17  & $[1; ~1; ~1; ~3,4]$   & $(\Sigma^*_c D^*)^1$ &   &  \\
& 18 & $[1; ~1; ~3; ~3,4]$  & $(\Sigma^*_c D^*)^8$  &   &  \\ [2ex]
$\frac{5}{2}^-$ & 1 & $[1; ~3; ~1; ~1,2]$  & $(\Xi^*_{cc} \omega)^1$ & $[1; ~2; ~1; ~1,2]$ & $(\Xi^*_{cc} \rho)^1$ \\
& 2  & $[1; ~3; ~3; ~1,2]$  & $(\Xi^*_{cc} \omega)^8$ & $[1; ~2; ~3; ~1,2]$  & $(\Xi^*_{cc} \rho)^8$ \\
& 3  & $[1; ~4; ~1; ~1,2]$   & $(\Xi^*_{cc} \rho)^1$   & $[1; ~3; ~1; ~3,4]$   & $(\Sigma^*_c D^*)^1$ \\
& 4  & $[1; ~4; ~3; ~1,2]$   & $(\Xi^*_{cc} \rho)^8$  & $[1; ~3; ~3; ~3,4]$ & $(\Sigma^*_c D^*)^8$ \\
& 5  & $[1; ~1; ~1; ~3,4]$   & $(\Sigma^*_c D^*)^1$  &  &  \\
& 6  & $[1; ~1; ~3; ~3,4]$   & $(\Sigma^*_c D^*)^8$  &  &  \\
\bottomrule
\end{tabular}
\end{table}

\begin{table}[H]
\caption{\label{Rsum} Possible bound and resonance states of doubly charm pentaquarks. The last column listed the binding energy or resonance width of each states.}
\centering
\begin{tabular}{cccc}
\toprule
                  & \textbf{Quantum state} &  $\textbf{E}_\textbf{B}$/\textbf{$\Gamma$} \textbf{(in MeV)} \\
\midrule
Bound state     & $\frac{1}{2}\frac{1}{2}^-$ $\Lambda_c D^*(4291)$ &  -2 \\
                 & $\frac{1}{2}\frac{3}{2}^-$ $\Sigma_c D^*(4461)$       & -1 \\
                 & $\frac{3}{2}\frac{1}{2}^-$ $\Sigma^*_c D^*(4523)$   & -4 \\
                 & $\frac{3}{2}\frac{3}{2}^-$ $\Sigma^*_c D^*(4524)$   & -3 \\
                 & $\frac{1}{2}\frac{3}{2}^-$ $\Xi^*_{cc} \pi(3757)$       &  -3 \\[2ex]
Resonance state  & $\frac{1}{2}\frac{1}{2}^-$ $\Sigma_c D(4356)$   & 4.8 \\
                 & $\frac{1}{2}\frac{3}{2}^-$ $\Sigma^*_c D(4449)$         & 8.0\\
                 & $\frac{3}{2}\frac{1}{2}^-$ $\Sigma_c D(4431)$             & 2.6\\
                 & $\frac{3}{2}\frac{1}{2}^-$ $\Sigma_c D(4446)$            & 2.2 \\
                 & $\frac{3}{2}\frac{3}{2}^-$ $\Sigma_c D^*(4514)$        & 4.0 \\
                 & $\frac{3}{2}\frac{5}{2}^-$ $\Xi^*_{cc} \rho(4461)$     & 3.0 \\
\bottomrule
\end{tabular}
\end{table}

\begin{itemize}[leftmargin=*,labelsep=5.8mm, listparindent=2em]
\item	$I(J^P)=\frac{1}{2}(\frac{1}{2}^-)$ state

There are eleven baryon-meson channels under investigation, $\Xi_{cc} \eta$, $\Xi_{cc} \omega$, $\Xi_{cc} \pi$, $\Xi_{cc} \rho$, $\Xi^*_{cc} \omega$, $\Xi^*_{cc} \rho$, $\Lambda_c D$, $\Lambda_c D^*$, $\Sigma_c D$, $\Sigma_c D^*$ and $\Sigma^*_c D^*$. The lowest channel is $\Xi_{cc} \pi$ which experimental threshold value is 3657 MeV, but it is unbound in our study. The scattering nature is also found in the other channels no matter in the color-singlet, hidden-color structures calculation or the coupling of them. However, a possible resonance state is obtained in the $\Lambda_c D^*$ channel which binding energy $E_B=-2\,\text{MeV}$ in the color-singlet computation. The coupling with hidden-color channel is extremely weak and the coupled-mass shown in Table~\ref{GresultCCP2} is 4291 MeV.

In the complete coupled-channels calculation which the CSM is employed and the rotated angle $\theta$ is varied from $0^\circ$ to $6^\circ$, we find that the previous obtained resonance state of $\Lambda_c D^*$ at $4291\,\text{MeV}$ in the single channel calculation is pushed above its threshold. The scattering nature is clear shown in Fig.~\ref{PPCCP1} that the calculated poles are alway moving along with the $\Lambda_c D^*$ threshold lines. Besides, this feature is also reflected by the other baryon-meson channels $e. g.$, $\Xi_{cc} \pi$, $\Lambda_c D$, $\Lambda_c D^*$, $etc$.

Since there is a dense distribution in the mass region from 4.35 to 4.46 GeV, an enlarged part is present in the bottom panel of Fig.~\ref{PPCCP1}. Therein, the small separations among $\Xi_{cc} \eta$, $\Xi_{cc} \omega$, $\Sigma_c D$, $\Xi^*_{cc} \omega$ and $\Xi_{cc} \rho$ channels can be distinguished clearly. Apart from the most continuum states, one resonance, whose mass and width are $\sim$$4416\,\text{MeV}$ and $\sim$$4.8\,\text{MeV}$, is obtained and has been circled with green in the bottom panel. We can identify it as the $\Sigma_c D$ molecular resonance state for this pole is above the threshold lines, a re-scaled mass, which is according to the systematic error between theoretical and experimental threshold values, equals to 4356 MeV. Meanwhile, several properties of the  $\Sigma_c D(4356)$ remind the nature of $P^+_c(4312)$, $i. e.$, the $I(J^P)$ quantum numbers are $\frac12 \frac12^-$ for both the $P^+_c(4312)$ and the $\Sigma_c D(4356)$ resonances in our calculations, their mass and width are also quite comparable.

\begin{table}[H]
\caption{\label{GresultCCP1} The lowest eigen-energies of doubly-charm pentaquarks with $I(J^P)=\frac12(\frac12^-)$, and the rotated angle $\theta=0^\circ$. (unit: MeV) }
\centering
\begin{tabular}{ccccc}
\toprule
~~\textbf{Channel}~~   & ~~\textbf{Color}~~ & ~~\textit{\textbf{M}}~~ & ~~\textbf{Channel}~~ & ~~\textit{\textbf{M}}~~ \\
\midrule
$\Xi_{cc}\eta$ & S   & $4351$ & $\Xi_{cc}\omega$  & $4358$ \\
$(4065)$         & H   & $4787$ & $(4300)$ & $4608$ \\
                  & S+H & $4351$ & & $4358$ \\[2ex]
$\Xi_{cc}\pi$ & S   & $3812$ & $\Xi_{cc}\rho$  & $4434$ \\
$(3657)$         & H   & $4620$ & $(4293)$ & $4613$ \\
                  & S+H & $3812$ & & $4434$ \\[2ex]
$\Xi^*_{cc}\omega$ & S   & $4412$ & $\Xi^*_{cc}\rho$  & $4488$ \\
$(4403)$         & H   & $4568$ & $(4396)$ & $4576$ \\
                  & S+H & $4412$ & & $4488$ \\[2ex]
$\Lambda_c D$ & S   & $3981$ & $\Sigma^*_c D^*$  & $4551$ \\
$(4155)$         & H   & $4299$ & $(4527)$ & $4779$ \\
                  & S+H & $3981$ & & $4551$ \\[2ex]
$\Sigma_c D$ & S   & $4384$ & $\Sigma_c D^*$  & $4503$ \\
$(4324)$         & H   & $4701$ & $(4462)$ & $4691$ \\
                  & S+H & $4384$ & & $4503$ \\
\bottomrule
\end{tabular}
\end{table}

\begin{table}[H]
\caption{\label{GresultCCP2} The lowest eigen-energies of $\Lambda_c D^*$ with $I(J^P)=\frac12(\frac12^-)$, and the rotated angle $\theta=0^\circ$. (unit: MeV) }
\centering
\begin{tabular}{ccccc}
\toprule
~~\textbf{Channel}~~   & ~~\textbf{Color}~~ & ~~\textit{\textbf{M}}~~ & $\textbf{E}_\textbf{B}$ & ~~\textit{\textbf{M'}}~~ \\
\midrule
$\Lambda_c D^*$ & S   & $4098$ & $-2$  & $4291$ \\
$(4293)$         & H   & $4312$ & $+212$ & $4505$ \\
                  & S+H & $4098$ & $-2$  & $4291$ \\
\bottomrule
\end{tabular}
\end{table}

\begin{figure}[H]
\centering
\includegraphics[clip, trim={3.0cm 2.0cm 3.0cm 1.0cm}, width=0.75\textwidth]{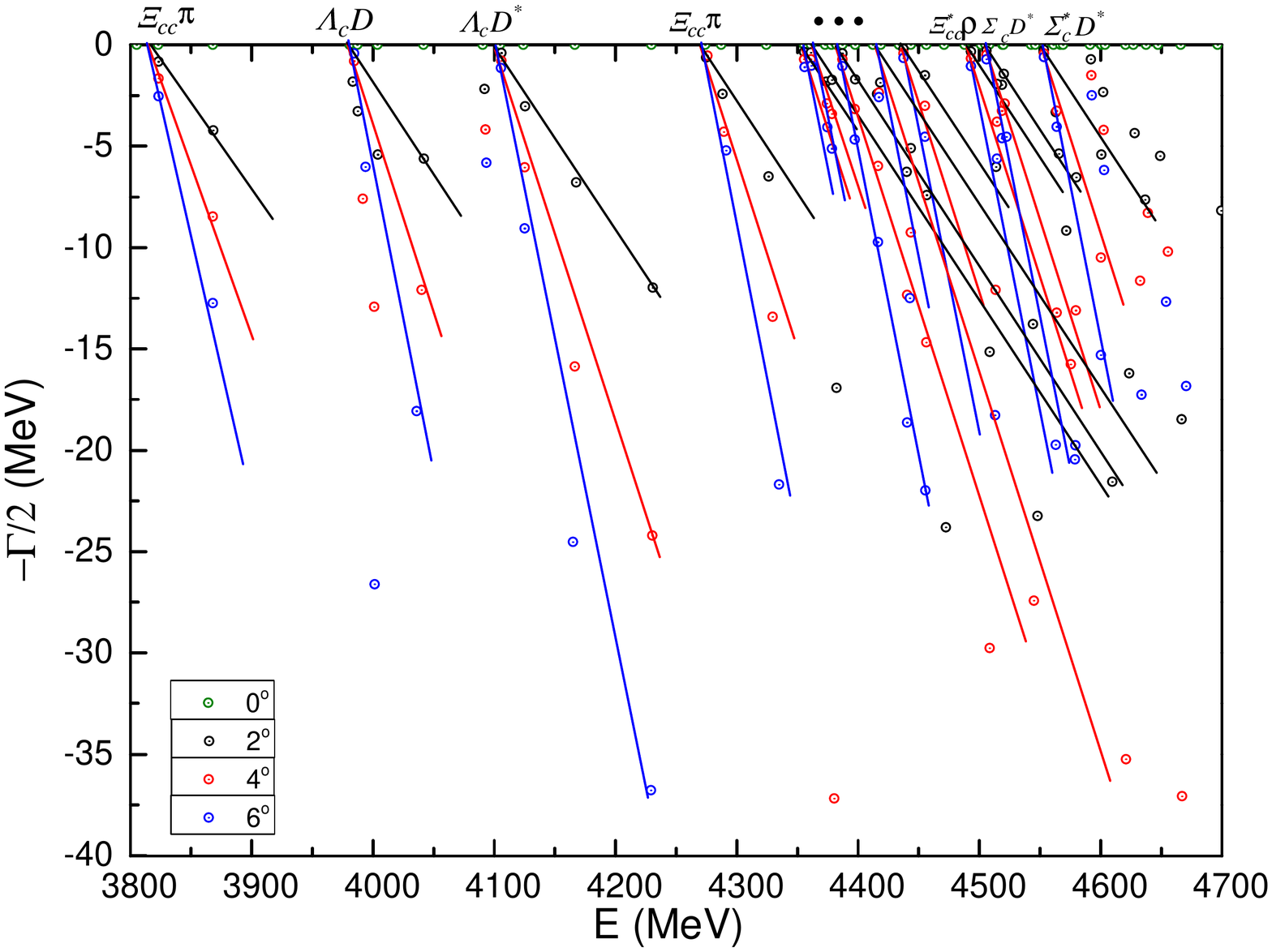} \\
\includegraphics[clip, trim={3.0cm 2.0cm 3.0cm 1.0cm}, width=0.75\textwidth]{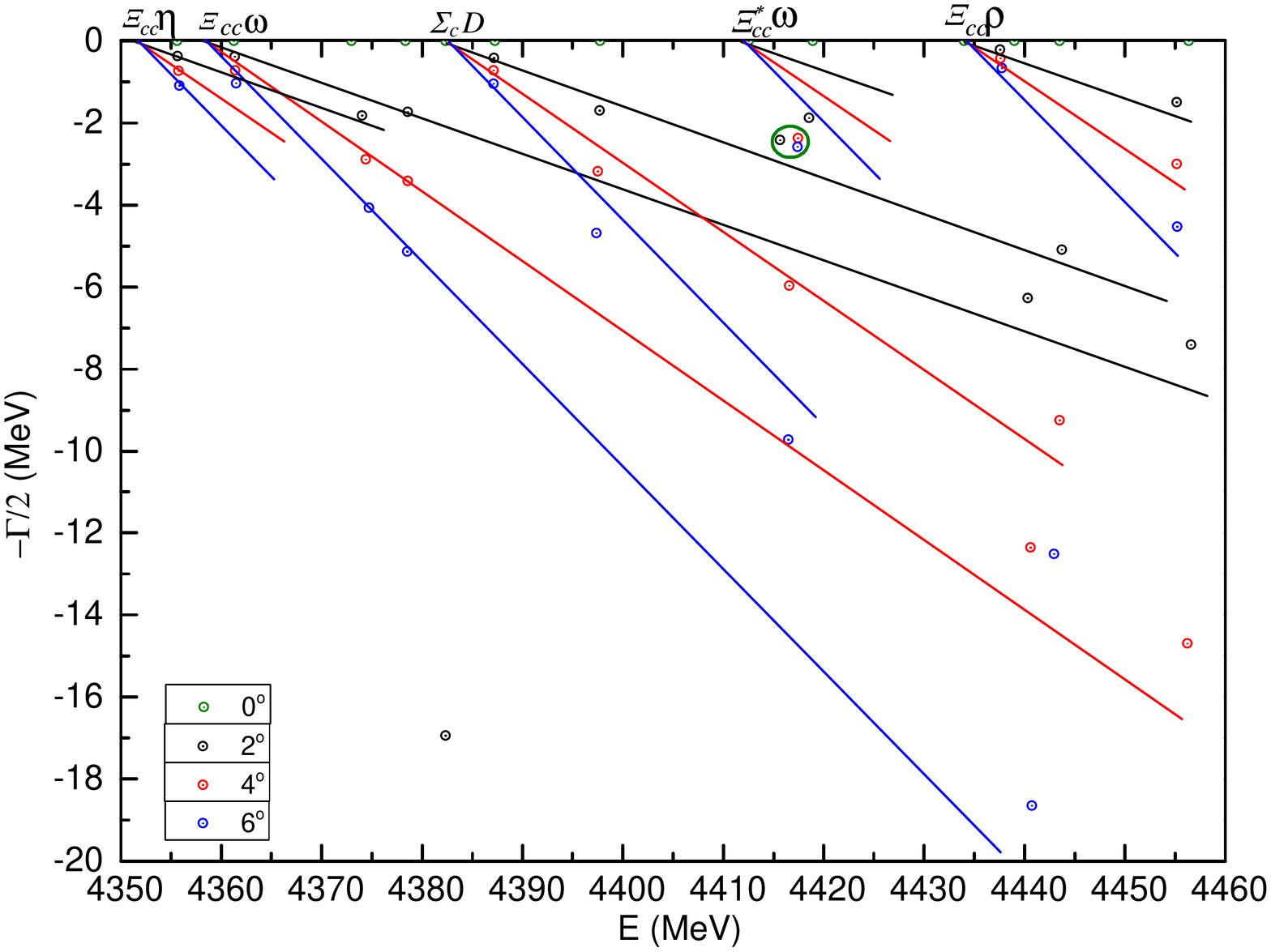}
\caption{\label{PPCCP1} {\it Top panel:} Pentaquark's complex energies of coupled-channels calculation with quantum numbers $I(J^P)=\frac12(\frac12^-)$ and for $\theta(^\circ)=0$ (green), $2$ (black), $4$ (red) and $6$ (blue). {\it Bottom panel:} Enlarged top panel, with real values of energy ranging from $4.35\,\text{GeV}$ to $4.46\,\text{GeV}$.} 
\end{figure}

\item	$I(J^P)=\frac{1}{2}(\frac{3}{2}^-)$ state

Tables~\ref{GresultCCP3} and \ref{GresultCCP4} present the calculated mass of each baryon-meson channels in the color-singlet, hidden-color and their coupling cases. In particular, they are all scattering states listed in Table~\ref{GresultCCP3} and one resonance state of $\Sigma_c D^*$ in Table~\ref{GresultCCP4}. Obviously, all the color-singlet channels of these $\Xi^{(*)}_{cc} \omega$, $\Xi^{(*)}_{cc} \rho$, $\Xi^*_{cc} \pi$, $\Lambda_c D^*$ and $\Sigma^{(*)}_c D^{(*)}$ states are unbound, their hidden-color channels are all above the corresponding threshold values. Only when the hidden-color channel of $\Sigma_c D^*$ included, there is a weak binding energy $E_B=-1$ MeV. The modified mass is 4461 MeV.

However, it is similar to the $\Lambda_c D^*(4291)$ in $I(J^P)=\frac12(\frac12^-)$ state, $\Sigma_c D^*(4461)$ turns to be a scattering state in the complete coupled-channels calculation and the evidence can be found in Fig.~\ref{PPCCP2} which the distributions of complex energies of each channels are well presented, no fixed resonance pole is near the $\Sigma_c D^*$ threshold lines. Nevertheless, one bound state with $E_B=-3$ MeV is obtained in the top panel of Fig.~\ref{PPCCP2}. Within a $0^\circ-6^\circ$ variation region of the angle $\theta$, a stable pole circled with purple is at 3863 MeV of the real-axis. The main component should be $\Xi^*_{cc} \pi$ and the rescaled mass is 3757 MeV. 

Additionally, a clearer assignments of the channels in a dense energy region $4.4-4.6$ GeV is presented in the bottom panel. Therein, one resonance pole is obtained and circled with green. In that small region, the calculated three dots with a rotated angle $\theta$ is $2^\circ$, $4^\circ$ and $6^\circ$, respectively, is almost overlapped. Hence the estimated mass and width of the resonance is $4492\,\text{MeV}$ and $8.0\,\text{MeV}$, respectively. It is more reasonable to identify this exotic state as a $\Sigma^*_c D$ molecule state: (i) this is above the $\Sigma^*_c D$ threshold lines, (ii) larger mass shift for an explanation in $\Xi_{cc} \rho$ channel, (iii) the $P^+_c(4457)$ which can be explained as the $\Sigma_c \bar{D}^*$ molecular state also has the quantum numbers $I(J^P)=\frac12(\frac32^-)$ and the width is $6.4\pm 2.0^{+5.7}_{-1.9}$ MeV. Accordingly, a rescaled mass which is respect to the $\Sigma_c \bar{D}^*$ experimental threshold value is 4449 MeV.

\begin{table}[H]
\caption{\label{GresultCCP3} The lowest eigen-energies of doubly-charm pentaquarks with $I(J^P)=\frac12(\frac32^-)$, and the rotated angle $\theta=0^\circ$. (unit: MeV)}
\centering
\begin{tabular}{ccccc}
\toprule
~~\textbf{Channel}~~   & ~~\textbf{Color}~~ & ~~\textit{\textbf{M}}~~ & ~~\textbf{Channel}~~ & ~~\textit{\textbf{M}}~~ \\
\midrule
 $\Xi_{cc}\omega$ & S & $4358$ & $\Xi_{cc}\rho$  & $4434$ \\
 $(4300)$  & H       & $4619$ & $(4293)$ & $4648$ \\
  & S+H   & $4358$ & & $4434$ \\[2ex]
$\Xi^*_{cc}\pi$ & S   & $3866$ & $\Xi^*_{cc}\omega$ & $4412$ \\
$(3760)$        & H   & $4671$ & $(4403)$           & $4614$ \\
                & S+H & $3866$ &                    & $4412$ \\[2ex]
$\Xi^*_{cc}\rho$ & S   & $4488$ & $\Lambda_c D^*$ & $4100$ \\
$(4396)$         & H   & $4641$ & $(4293)$        & $4284$ \\
                 & S+H & $4488$ &                 & $4100$ \\[2ex]
$\Sigma_c D^*$    & S   & $4503$ & $\Sigma^*_c D$ & $4432$ \\
$(4462)$          & H   & $4689$ & $(4389)$       & $4702$ \\
                  & S+H & $4503$ &                & $4432$ \\[2ex]
$\Sigma^*_c D^*$  & S   & $4551$ & & \\
$(4527)$          & H   & $4729$ & & \\
                  & S+H & $4551$ & & \\
\bottomrule
\end{tabular}
\end{table}

\begin{table}[H]
\caption{\label{GresultCCP4} The lowest eigen-energies of $\Sigma_c D^*$ with $I(J^P)=\frac12(\frac32^-)$, and the rotated angle $\theta=0^\circ$. (unit: MeV) }
\centering
\begin{tabular}{ccccc}
\toprule
~~\textbf{Channel}~~   & ~~\textbf{Color}~~ & ~~\textit{\textbf{M}}~~ & $\textbf{E}_\textbf{B}$ & ~~\textit{\textbf{M'}}~~ \\
\midrule
$\Sigma_c D^*$ & S     & $4503$ & $0$    & $4462$ \\
$(4462)$       & H     & $4689$ & $+186$ & $4648$ \\
               & S+H   & $4502$ & $-1$   & $4461$ \\
\bottomrule
\end{tabular}
\end{table}

\begin{figure}[H]
\centering
\includegraphics[clip, trim={3.0cm 2.0cm 3.0cm 1.0cm}, width=0.75\textwidth]{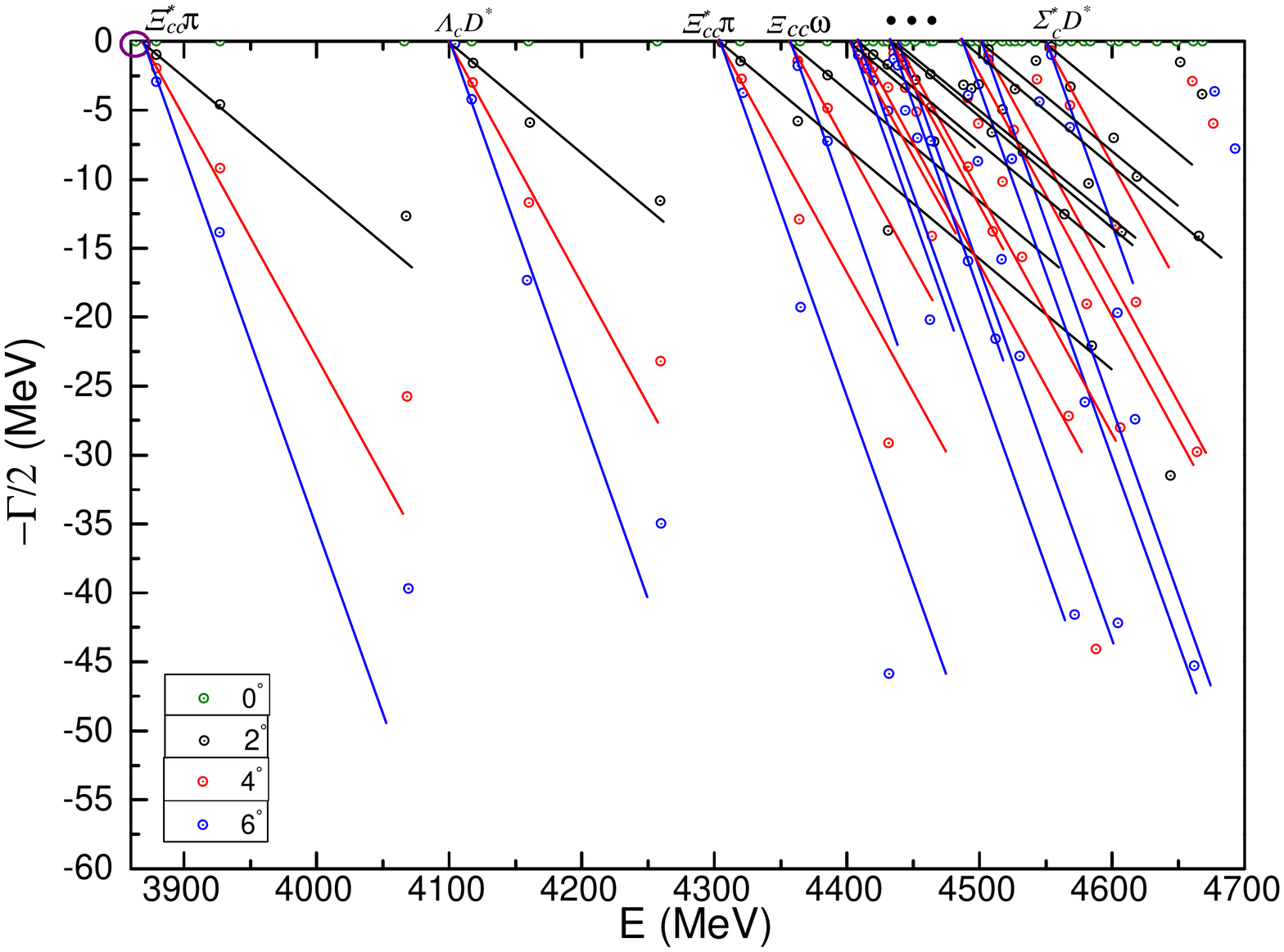} \\
\includegraphics[clip, trim={3.0cm 2.0cm 3.0cm 1.0cm}, width=0.75\textwidth]{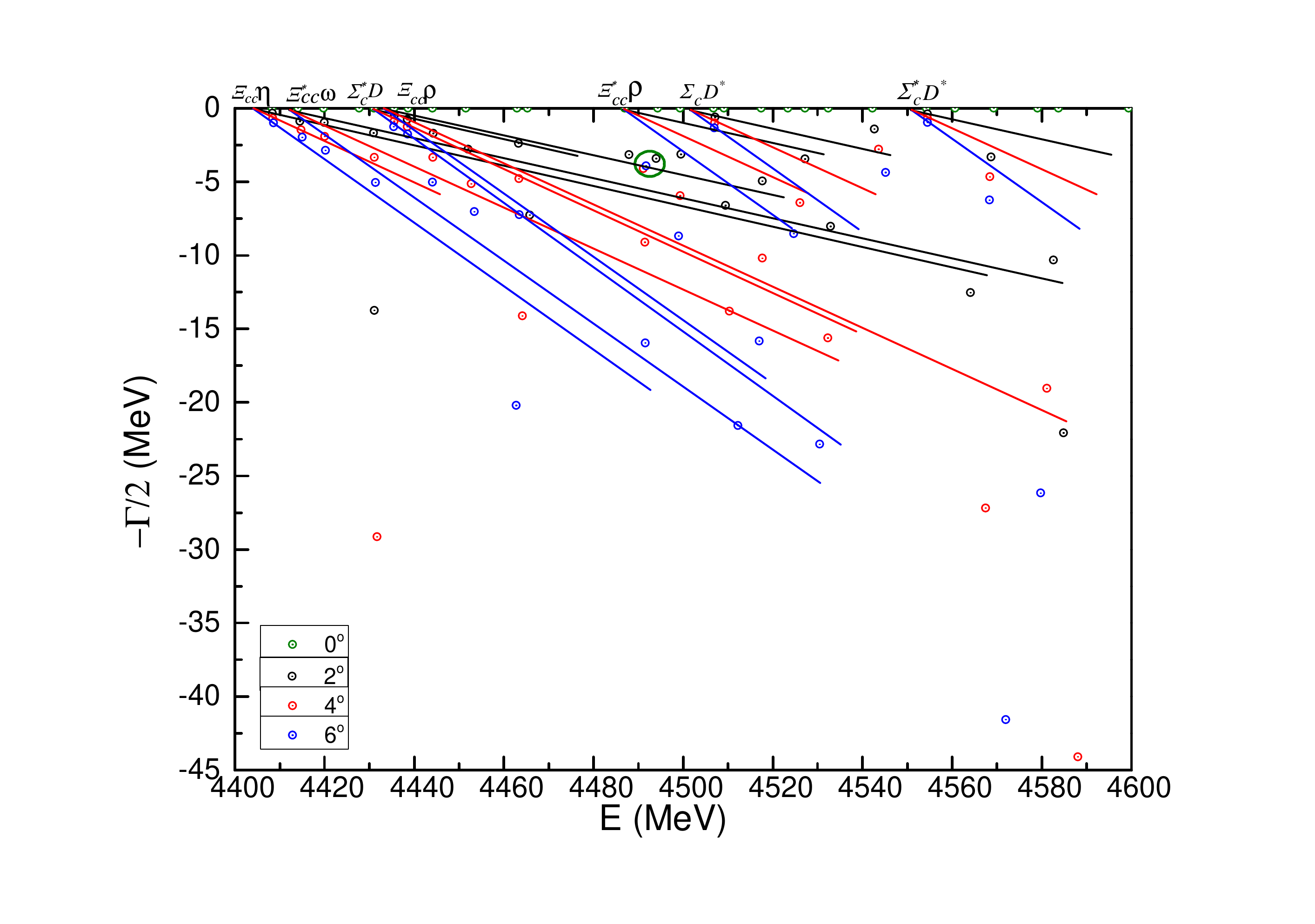}
\caption{\label{PPCCP2} {\it Top panel:} Pentaquark's complex energies of coupled-channels calculation with quantum numbers $I(J^P)=\frac12(\frac32^-)$ and for $\theta(^\circ)=0$ (green), $2$ (black), $4$ (red) and $6$ (blue). {\it Bottom panel:} Enlarged top panel, with real values of energy ranging from $4.40\,\text{GeV}$ to $4.60\,\text{GeV}$.} 
\end{figure}

\item	$I(J^P)=\frac{1}{2}(\frac{5}{2}^-)$ state

Three baryon-meson channels $\Xi^*_{cc} \omega$, $\Xi^*_{cc} \rho$ and $\Sigma^*_c D^*$ contribute to the highest spin state and the calculated results on their masses in the color-singlet channel, hidden-color one and their couplings in real-range are listed in Table~\ref{GresultCCP5}. Meanwhile, Fig.~\ref{PPCCP5} shows the distributions of the complex energies in the fully coupled-channels calculation of complex-range. Obviously, no bound state is obtained neither in the single channel nor the coupled cases and resonances are also not found. Hence, it is the scattering nature of $\Xi^*_{cc} \omega$, $\Xi^*_{cc} \rho$ and $\Sigma^*_c D^*$ in the $\frac{1}{2}(\frac{5}{2}^-)$ state.

\begin{table}[H]
\caption{\label{GresultCCP5} The lowest eigen-energies of doubly-charm pentaquarks with $I(J^P)=\frac12(\frac52^-)$, and the rotated angle $\theta=0^\circ$. (unit: MeV) }
\centering
\begin{tabular}{ccccc}
\toprule
~~\textbf{Channel}~~   & ~~\textbf{Color}~~ & ~~\textit{\textbf{M}}~~ & ~~\textbf{Channel}~~ & ~~\textit{\textbf{M}}~~ \\
\midrule
$\Xi^*_{cc}\omega$ & S   & $4412$ & $\Xi^*_{cc}\rho$ & $4488$ \\
$(4403)$           & H   & $4683$ & $(4396)$         & $4741$ \\
                   & S+H & $4412$ &                  & $4488$ \\[2ex]
$\Sigma^*_c D^*$ & S   & $4551$ & & \\
$(4527)$         & H   & $4655$ & & \\
                 & S+H & $4551$ & & \\
\bottomrule
\end{tabular}
\end{table}

\begin{figure}[H]
\centering
\includegraphics[clip, trim={3.0cm 2.0cm 3.0cm 1.0cm}, width=0.75\textwidth]{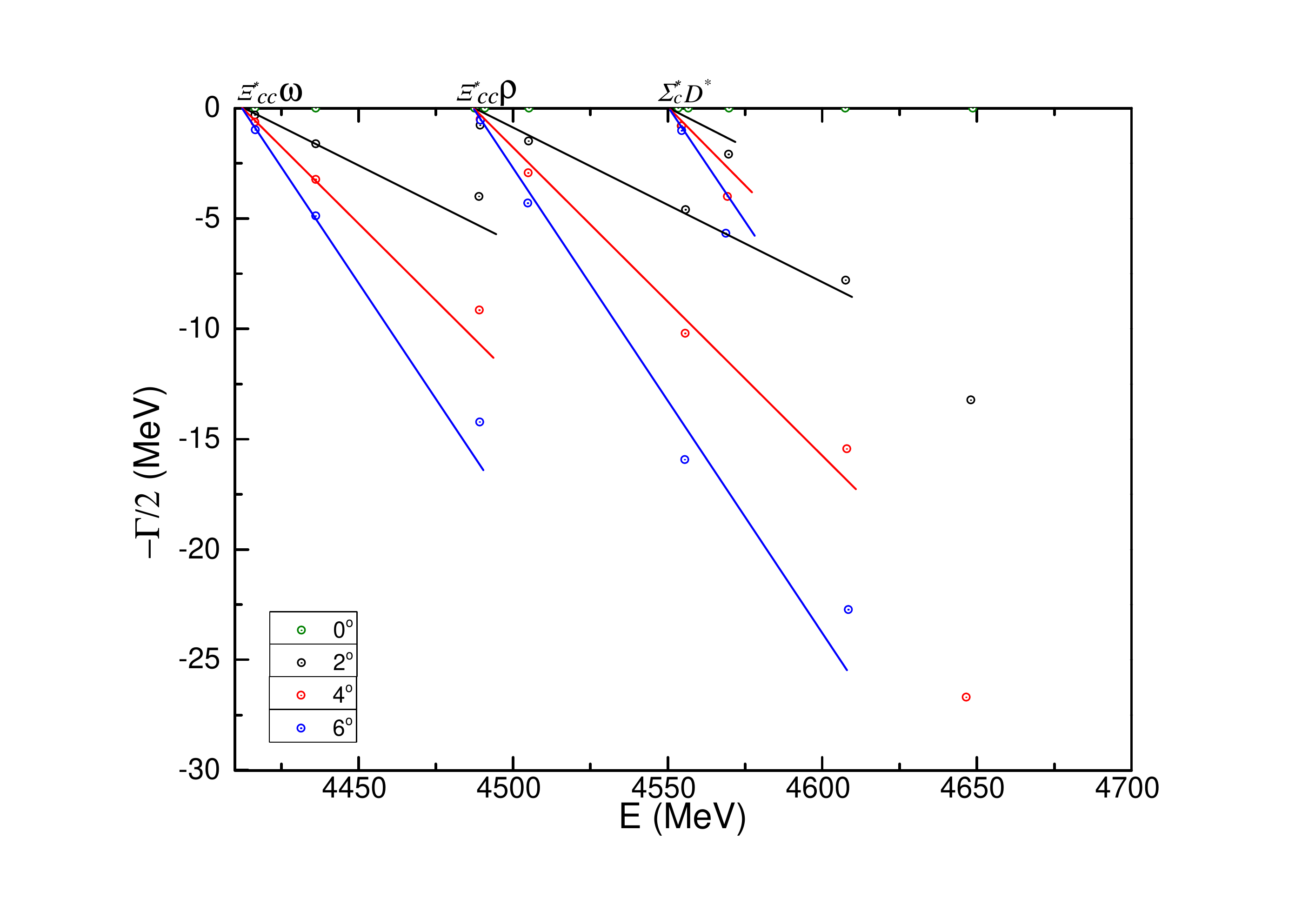}
\caption{\label{PPCCP5} Pentaquark's complex energies of coupled-channels calculation with quantum numbers $I(J^P)=\frac12(\frac52^-)$ and for $\theta(^\circ)=0$ (green), $2$ (black), $4$ (red) and $6$ (blue).} 
\end{figure}

\item	$I(J^P)=\frac{3}{2}(\frac{1}{2}^-)$ state

In this quantum state, the six allowed baryon-meson channels $\Xi_{cc}\pi$, $\Xi^{(*)}_{cc}\rho$ and $\Sigma^{(*)}_c D^{(*)}$ are investigated in the singlet- and hidden-color configurations, firstly. From Tables~\ref{GresultCCP6} and~\ref{GresultCCP7} one can find that resonance only obtained in the $\Sigma_c^* D^*$ channel with color-singlet state and the binding energy is $-3$ MeV. In a subsequent calculation which its hidden-color channel included, only 1 MeV increased for the $E_B$. The modified mass $M'$ of $\Sigma_c^* D^*$ resonance is 4523 MeV.

Apart from the real scaling calculation, Fig.~\ref{PPCCP3} presents the results of fully coupled-channels calculation in the complex-range. Particularly, the calculated energy points of the six channels $\Xi_{cc}\pi$, $\Xi^{(*)}_{cc}\rho$ and $\Sigma^{(*)}_c D^{(*)}$ distribute in the mass region $3.8-4.8$ GeV of the top panel. When the $\theta$ varies from $0^\circ$ to $6^\circ$, they generally present the scattering nature which the poles always moving along their threshold lines. However, much plenty of energy poles located in the $4.45-4.62$ GeV region, so the middle panel of Fig.~\ref{PPCCP3} shows an enlarged result on this part. Clearly, one resonance pole circled green is obtained herein and the theoretical mass and width is 4491 MeV and 2.6 MeV, respectively. It is both above the $\Sigma_c D$ and $\Xi_{cc} \rho$ threshold lines, however, there is much more systematic error of the later channel than $\Sigma_c D$. Hence this resonance is preferred to be identified as the $\Sigma_c$ baryon and $D$ meson molecular state. After a mass shift according to the experimental threshold value, the resonance mass is 4431 MeV.

Furthermore, there seems to be more structures between the threshold of $\Xi_{cc} \rho$ and $\Sigma_c D^*$, so a further enlarged part from 4.50 GeV to 4.53 GeV is shown in the bottom panel. Apparently, another $\Sigma_c D$ resonance state is obtained, the mass and width is $4506\,\text{MeV}$ and $2.2\,\text{MeV}$, respectively. Then through a mass shift with respect to its threshold value, the rescaled mass is 4446 MeV.

\begin{table}[H]
\caption{\label{GresultCCP6} The lowest eigen-energies of doubly-charm pentaquarks with $I(J^P)=\frac32(\frac12^-)$, and the rotated angle $\theta=0^\circ$. (unit: MeV) }
\centering
\begin{tabular}{ccccc}
\toprule
~~\textbf{Channel}~~   & ~~\textbf{Color}~~ & ~~\textit{\textbf{M}}~~ & ~~\textbf{Channel}~~ & ~~\textit{\textbf{M}}~~ \\
\midrule
$\Xi_{cc}\pi$ & S   & $3812$ & $\Xi_{cc}\rho$ & $4434$ \\
$(3657)$      & H   & $4682$ & $(4293)$       & $4685$ \\
              & S+H & $3812$ &                & $4434$ \\[2ex]
$\Xi^*_{cc}\rho$ & S   & $4488$ & $\Sigma_c D$ & $4384$ \\
$(4396)$         & H   & $4647$ & $(4324)$     & $4714$ \\
                 & S+H & $4488$ &              & $4384$ \\[2ex]
$\Sigma_c D^*$ & S   & $4503$ & & \\
$(4462)$       & H   & $4627$ & & \\
               & S+H & $4503$ & & \\
\bottomrule
\end{tabular}
\end{table}

\begin{table}[H]
\caption{\label{GresultCCP7} The lowest eigen-energies of $\Sigma^*_c D^*$ with $I(J^P)=\frac32(\frac12^-)$, and the rotated angle $\theta=0^\circ$. (unit: MeV) }
\centering
\begin{tabular}{ccccc}
\toprule
~~\textbf{Channel}~~   & ~~\textbf{Color}~~ & ~~\textit{\textbf{M}}~~ & $\textbf{E}_\textbf{B}$ & ~~\textit{\textbf{M'}}~~ \\
\midrule
$\Sigma^*_c D^*$ & S   & $4548$ & $-3$   & $4524$ \\
$(4527)$         & H   & $4693$ & $+142$ & $4669$ \\
                 & S+H & $4547$ & $-4$   & $4523$ \\
\bottomrule
\end{tabular}
\end{table}

\begin{figure}[H]
\centering
\includegraphics[clip, trim={2.0cm 2.0cm 2.0cm 1.0cm}, width=0.65\textwidth]{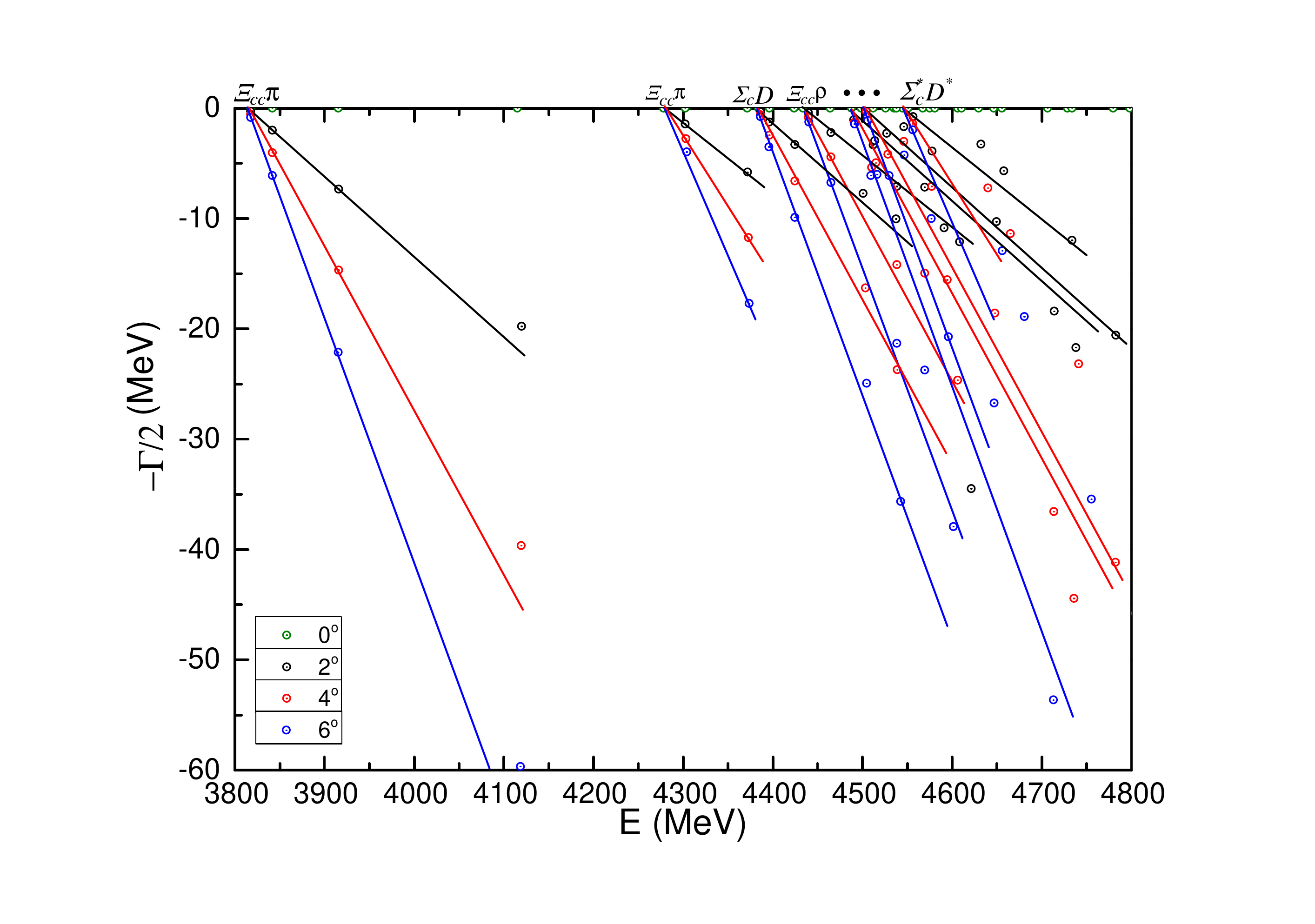} \\
\includegraphics[clip, trim={2.0cm 2.0cm 2.0cm 1.0cm}, width=0.65\textwidth]{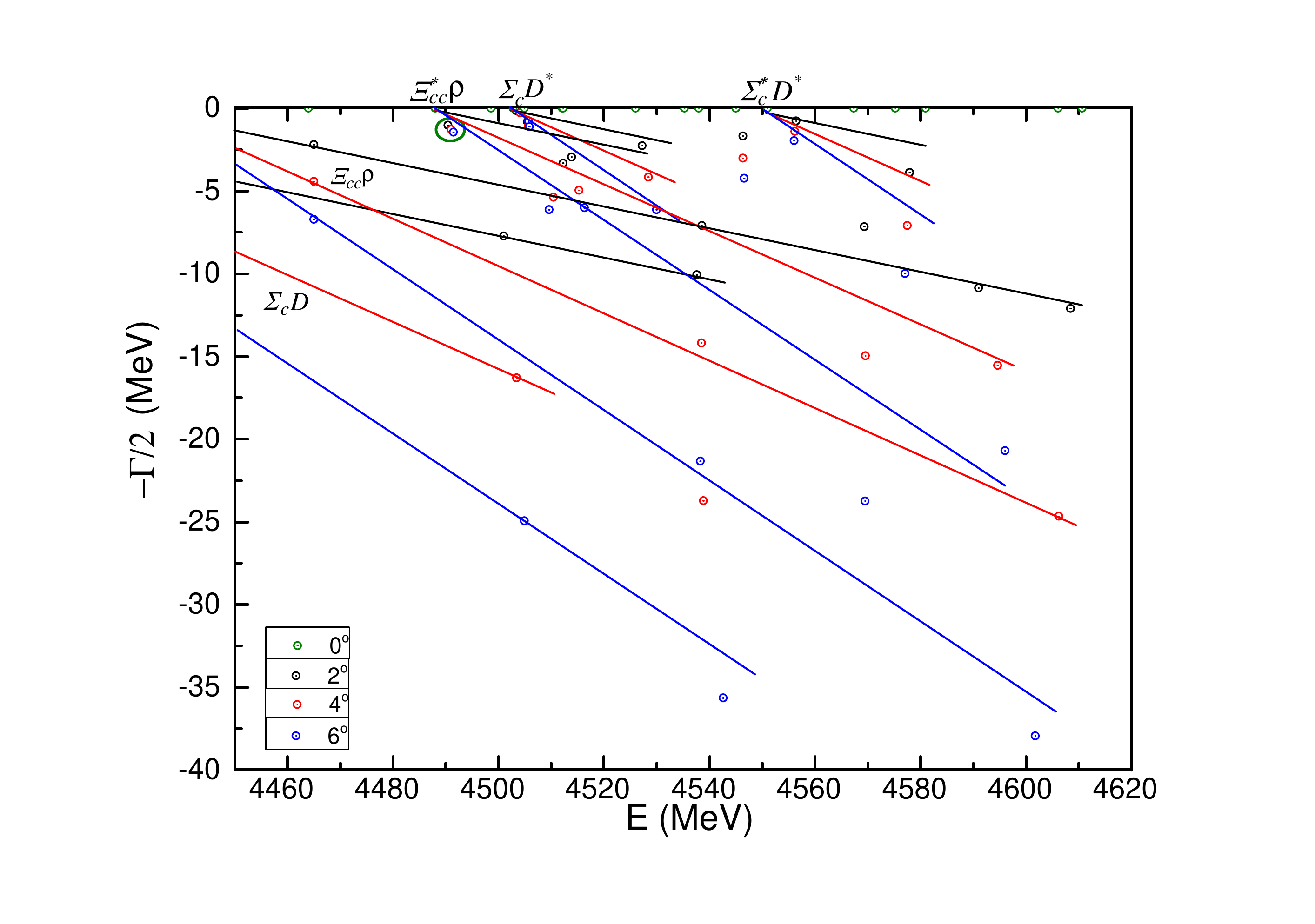}\\
\includegraphics[clip, trim={2.0cm 2.0cm 2.0cm 1.0cm}, width=0.65\textwidth]{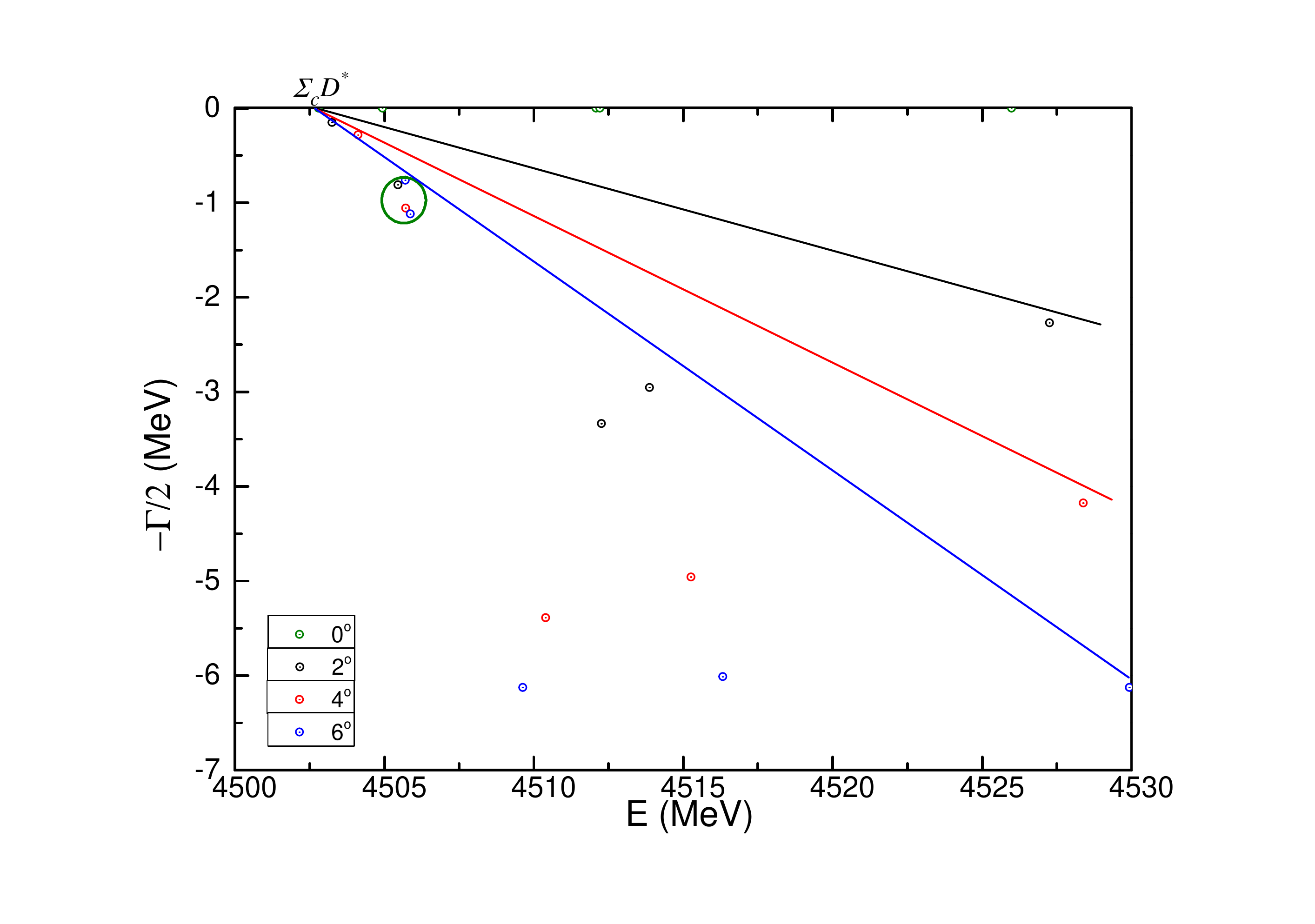}
\caption{\label{PPCCP3} {\it Top panel:} Pentaquark's complex energies of coupled-channels calculation with quantum numbers $I(J^P)=\frac32(\frac12^-)$ and for $\theta(^\circ)=0$ (green), $2$ (black), $4$ (red) and $6$ (blue). {\it Middle panel:} Enlarged top panel, with real values of energy ranging from $4.45\,\text{GeV}$ to $4.62\,\text{GeV}$. {\it Bottom panel:} Enlarged top panel, with real values of energy ranging from $4.50\,\text{GeV}$ to $4.53\,\text{GeV}$.} 
\end{figure}

\item	$I(J^P)=\frac{3}{2}(\frac{3}{2}^-)$ state

In the $J=I=\frac{3}{2}$ state, there are six channels listed in Tables~\ref{GresultCCP8} and \ref{GresultCCP9} under consideration, namely, $\Xi^{(*)}_{cc} \rho$, $\Xi^*_{cc} \pi$ and $\Sigma^{(*)}_c D^{(*)}$. Firstly, all of the channels in the color-singlet structure are unbound. Besides, it is the same conclusion in the hidden-color channel calculations. However, $\Sigma^*_c D^*$ resonance state is obtained in the coupled-channels study which both the singlet- and hidden-color channels are employed. The binding energy is $-3$ MeV and a modified mass is 4524 MeV. Clearly, there is a degeneration between this state and $\Sigma^*_c D^*$ resonance in the $\frac32(\frac12^-)$ state.

Within a complex-range computation in the complete coupled-channels, the scattering states of $\Xi^{(*)}_{cc} \rho$, $\Xi^*_{cc} \pi$ and $\Sigma^{(*)}_c D^{(*)}$ are generally presented in Fig.~\ref{PPCCP10}. Particularly, the obtained  $\Sigma^*_c D^*$ resonance at 4524 MeV in the real-range calculation turns to a scattering state. Nevertheless, a possible $\Sigma_c D^*$ resonance state is found in the bottom panel of Fig.~\ref{PPCCP10}. In the big green circle, three calculated poles are almost overlapped and the predicted mass and width is $4555\,\text{MeV}$ and $4.0\,\text{MeV}$, respectively. After a mass shift, the resonance can be identified as $\Sigma_c D^*(4514)$ molecular state.

\begin{table}[H]
\caption{\label{GresultCCP8} The lowest eigen-energies of doubly-charm pentaquarks with $I(J^P)=\frac32(\frac32^-)$, and the rotated angle $\theta=0^\circ$. (unit: MeV) }
\centering
\begin{tabular}{ccccc}
\toprule
~~\textbf{Channel}~~   & ~~\textbf{Color}~~ & ~~\textit{\textbf{M}}~~ & ~~\textbf{Channel}~~ & ~~\textit{\textbf{M}}~~ \\
\midrule
$\Xi_{cc}\rho$ & S   & $4434$ & $\Xi^*_{cc}\pi$ & $3866$ \\
$(4293)$       & H   & $4708$ & $(3760)$        & $4692$ \\
               & S+H & $4434$ &                 & $3866$ \\[2ex]
$\Xi^*_{cc}\rho$ & S   & $4488$ & $\Sigma_c D^*$ & $4503$ \\
$(4396)$         & H   & $4678$ & $(4462)$       & $4719$ \\
                 & S+H & $4488$ &                & $4503$ \\[2ex]
$\Sigma^*_c D$ & S   & $4432$ & & \\
$(4389)$       & H   & $4695$ & & \\
               & S+H & $4432$ & & \\
\bottomrule
\end{tabular}
\end{table}

\begin{table}[H]
\caption{\label{GresultCCP9} The lowest eigen-energies of $\Sigma^*_c D^*$ with $I(J^P)=\frac32(\frac32^-)$, and the rotated angle $\theta=0^\circ$. (unit: MeV) }
\centering
\begin{tabular}{ccccc}
\toprule
~~\textbf{Channel}~~   & ~~\textbf{Color}~~ & ~~\textit{\textbf{M}}~~ & $\textbf{E}_\textbf{B}$ & ~~\textit{\textbf{M'}}~~ \\
\midrule
$\Sigma^*_c D^*$ & S   & $4551$ & $0$  & $4527$ \\
$(4527)$         & H   & $4667$ & $+116$ & $4643$ \\
                  & S+H & $4548$ & $-3$  & $4524$ \\
\bottomrule
\end{tabular}
\end{table}

\begin{figure}[H]
\centering
\includegraphics[clip, trim={3.0cm 2.0cm 3.0cm 1.0cm}, width=0.75\textwidth]{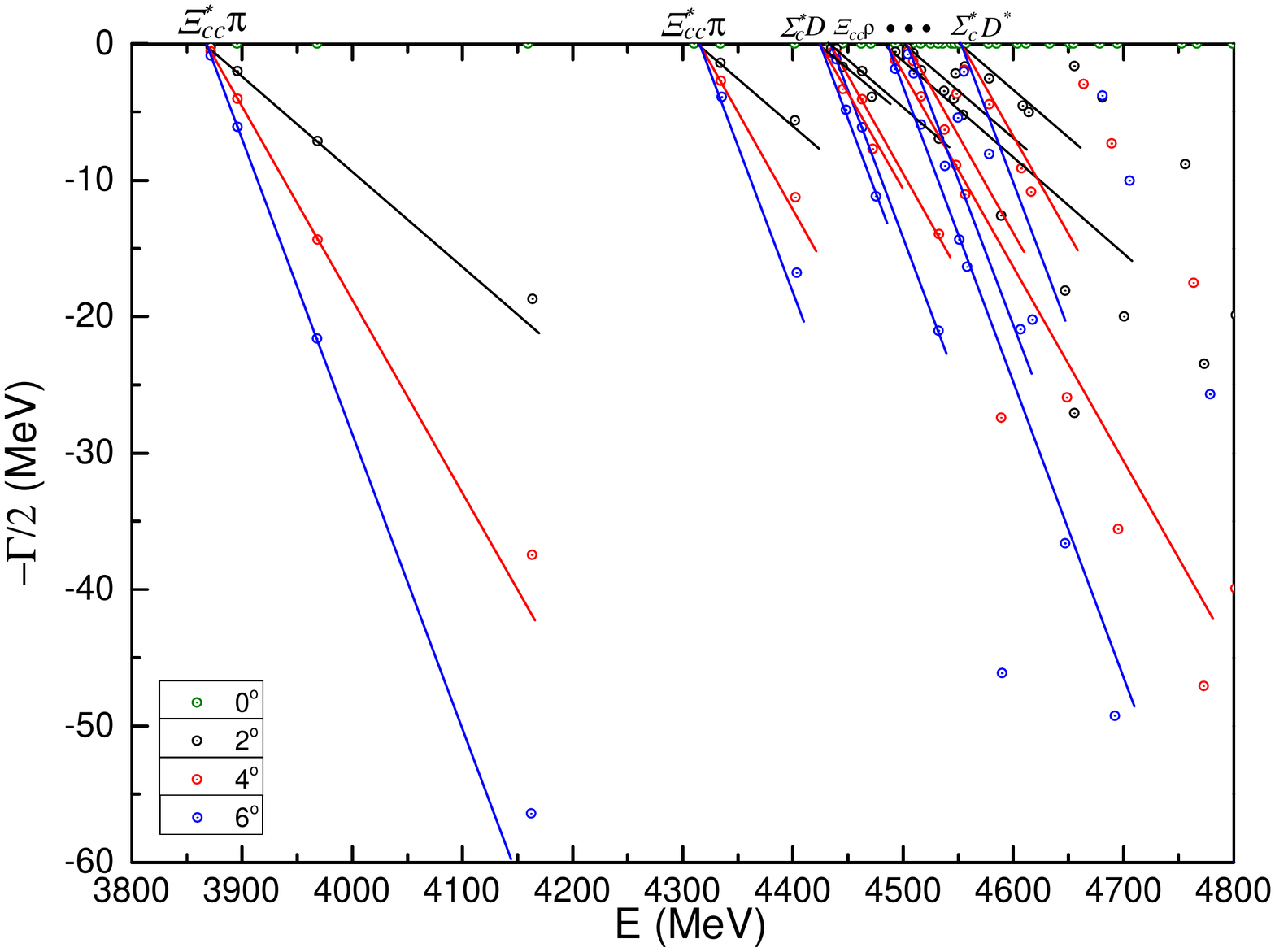} \\
\includegraphics[clip, trim={3.0cm 2.0cm 3.0cm 1.0cm}, width=0.75\textwidth]{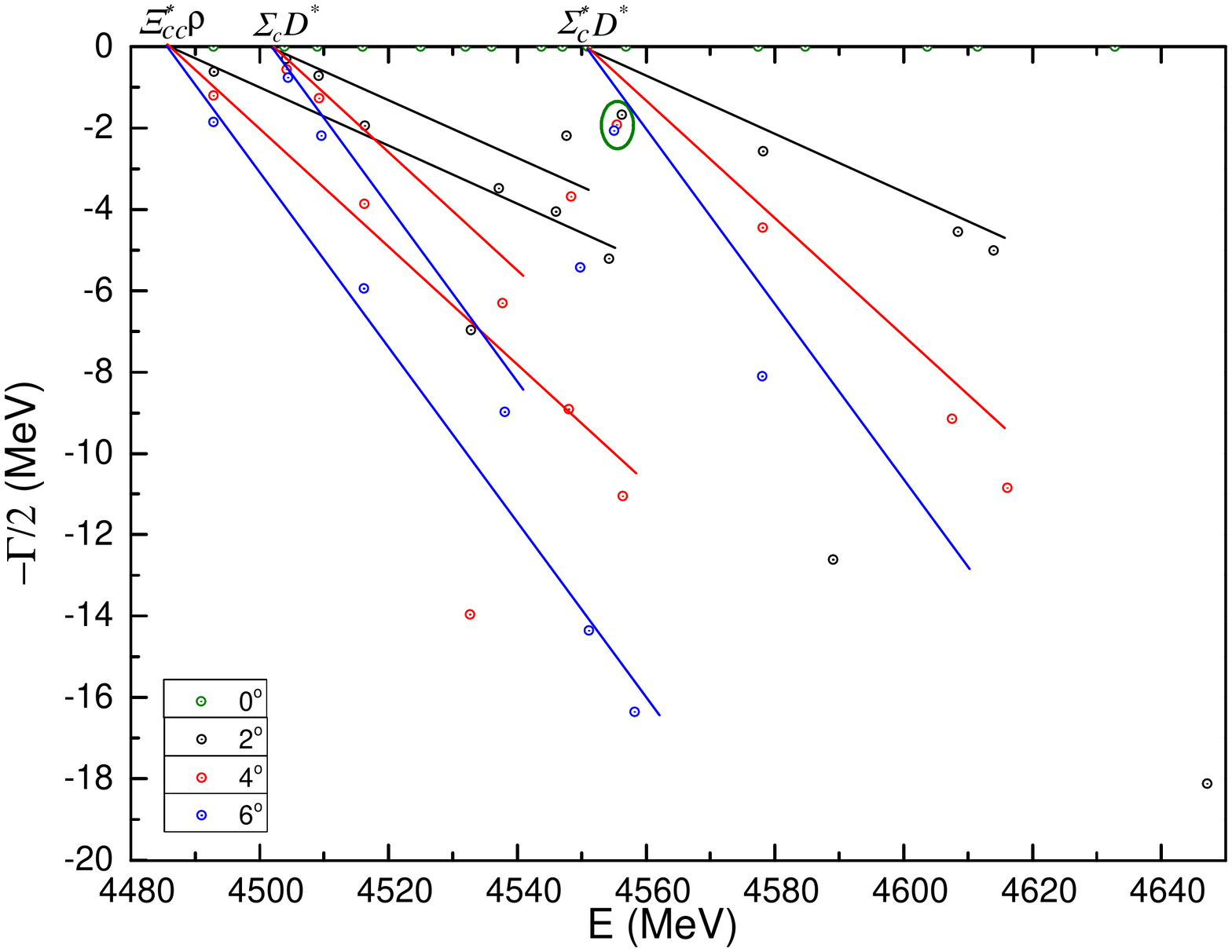}
\caption{\label{PPCCP10} {\it Top panel:} Pentaquark's complex energies of coupled-channels calculation with quantum numbers $I(J^P)=\frac32(\frac32^-)$ and for $\theta(^\circ)=0$ (green), $2$ (black), $4$ (red) and $6$ (blue). {\it Bottom panel:} Enlarged top panel, with real values of energy ranging from $4.48\,\text{GeV}$ to $4.65\,\text{GeV}$.} 
\end{figure}

\item	$I(J^P)=\frac{3}{2}(\frac{5}{2}^-)$ state

There are only two channels under investigation in the highest spin and isospin pentaquark state, $\Xi^*_{cc} \rho$ and $\Sigma^*_c D^*$. In Table~\ref{GresultCCP10} one can see that their lowest theoretical mass in the real-range is 4488 MeV and 4551 MeV, respectively. Hence, no bound state is obtained in this case.

Fig.~\ref{PPCCP11} shows the distributions of complex energies of these two channels. Apart from the scattering states, one stable resonance pole is found in the green circle. It can be identified as the $\Xi^*_{cc} \rho$ resonance with the modified mass and width is $4461\,\text{MeV}$ and $3.0\,\text{MeV}$, respectively. 

\begin{table}[H]
\caption{\label{GresultCCP10} The lowest eigen-energies of doubly-charm pentaquarks with $I(J^P)=\frac32(\frac52^-)$, and the rotated angle $\theta=0^\circ$. (unit: MeV) }
\centering
\begin{tabular}{ccccc}
\toprule
~~\textbf{Channel}~~   & ~~\textbf{Color}~~ & ~~\textit{\textbf{M}}~~ & ~~\textbf{Channel}~~ & ~~\textit{\textbf{M}}~~ \\
\midrule
$\Xi^*_{cc}\rho$ & S   & $4488$ & $\Sigma^*_c D^*$  & $4551$ \\
$(4396)$         & H   & $4727$ & $(4527)$ & $4706$ \\
                  & S+H & $4488$ & & $4551$ \\
\bottomrule
\end{tabular}
\end{table}

\begin{figure}[H]
\centering
\includegraphics[clip, trim={3.0cm 2.0cm 3.0cm 1.0cm}, width=0.75\textwidth]{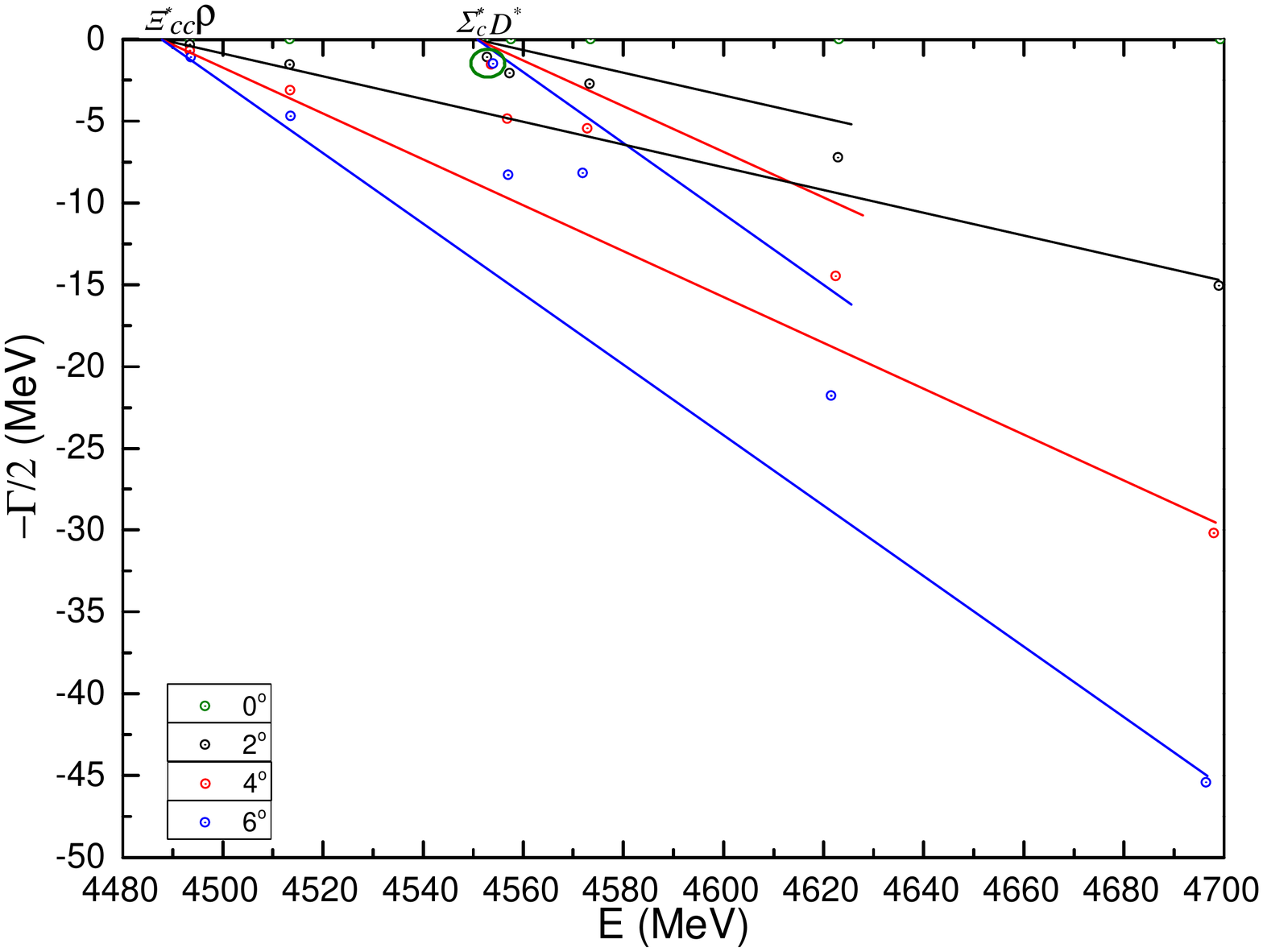}
\caption{\label{PPCCP11} Pentaquark's complex energies of coupled-channels calculation with quantum numbers $I(J^P)=\frac32(\frac52^-)$ and for $\theta(^\circ)=0$ (green), $2$ (black), $4$ (red) and $6$ (blue).} 
\end{figure}

\end{itemize}



\section{Summary}\label{sec:summary}

Motivated by the dozens of exotic hadron states recently reported in experiments worldwide, $e.g.$, the charmonium-like $X(3872)$ observed by the Belle collaboration in 2003 and the pentaquark $P^+_c(4380)$ announced by the LHCb collaboration in 2015, a constituent quark model formalism along with an exact and high efficiency computation method, Gaussian expansion method, are employed in systematically studying the double-, full-heavy tetraquark and hidden-charm, -bottom and doubly charmed pentaquark states. In particular, the spin-parity $J^P=0^+$, $1^+$ and $2^+$ tetraquark of $QQ\bar{q}\bar{q}$ and $QQ\bar{Q}\bar{Q}$ ($q=u, d, s$ and $Q=c, b$) states in siospin $I=0$ or $1$, besides, $J^P=\frac{1}{2}^-$, $\frac{3}{2}^-$ and $\frac{5}{2}^-$ $qqq\bar{Q}\bar{Q}$ and $ccqq\bar{q}$ pentaquarks in isospin $I=\frac{1}{2}$ or $\frac{3}{2}$ are systematically investigated in both the real- and complex-range.

Table~\ref{TPMSum} summarizes our theoretical findings for these tetra- and pent-aquark states. In particular, the $P^+_c(4380)$ can be well explained as a $\Sigma^*_c \bar{D}$ molecule state with $I(J^P)=\frac{1}{2}(\frac{3}{2}^-)$, besides, the $P^+_c(4312)$, $P^+_c(4440)$ and $P^+_c(4457)$ can also be identified as the $\frac{1}{2}\frac{1}{2}^-$ $\Sigma_c\bar{D}$, $\frac{1}{2}\frac{1}{2}^-$ $\Sigma_c\bar{D}^*$ and $\frac{1}{2}\frac{3}{2}^-$ $\Sigma_c\bar{D}^*$ molecule states, respectively. Meanwhile, the recently reported new structure at 6.9 GeV in the di-$J/\psi$ invariant mass spectrum by the LHCb collaboration can be regarded as a compact fully charmed tetraquark in $02^+$ state, and another broad structure around 6.2$\sim$6.8 GeV is also supported by us with several compact $cc\bar{c}\bar{c}$ resonances in $00^+$ and $01^+$ states. Furthermore, many other exotic states predicted in Table~\ref{TPMSum} should be expected to be confirmed in the future experiments, $e. g.$, the LHCb, ATLAS, CMS, BESIII, Belle II, JLAB, PANDA, EIC, $etc$. 

\begin{table}[H]
\caption{\label{TPMSum} Bound and resonance states obtained in tetraquark and pentaquark systems. Mass $M$, resonance width $\Gamma$ and binding energy $E_B$ are listed in the third, fourth and fifth column, respectively, unit in MeV.}
\centering
\begin{tabular}{ccccccccc}
\toprule
  ~~~~$\textit{\textbf{I}}\textit{\textbf{J}}^\textit{\textbf{P}}$~~~~ & ~~~~\textit{\textbf{Channel}}~~~~    &  ~~~~\textit{\textbf{M}}~~~~    &  ~~~~\textbf{$\Gamma$}~~~~    &  ~~~~$\textit{\textbf{E}}_\textit{\textbf{B}}$~~~~  & ~~~~\textit{\textbf{Channel}}~~~~    &  ~~~~\textit{\textbf{M}}~~~~    &  ~~~~\textbf{$\Gamma$}~~~~    &  ~~~~$\textit{\textbf{E}}_\textit{\textbf{B}}$~~~~ \\
\midrule
  $00^+$  & $cb\bar{q}\bar{q}$      & 6980  & $-$    & -196  & $D^{*0}\bar{B}^{*0}$  & 7726  & 12.00  & 550 \\
                & $bb\bar{b}\bar{b}$  & 17955  & $-$  & -847     & $D^+_s D^+_s$  & 4902  & 3.54  & 924 \\
                & $bb\bar{b}\bar{b}$  & 18030  & $-$  & -772     & $D^{*+}_s \bar{B}^*_s$  & 7919  & 1.02  & 575 \\
                &     &    &    &    & $D^{*+}_s \bar{B}^*_s$  & 7993  & 3.22  & 649 \\             
                &     &    &    &   & $\bar{B}^*_s \bar{B}^*_s$  & 11306  & 1.86  & 596 \\
                &     &    &    &   & $\bar{B}^*_s \bar{B}^*_s$  & 11333  & 1.84  & 623 \\
                &     &    &    &   & $\bar{B}^0_s \bar{B}^0_s$  & 11412  & 1.54  & 702 \\
                &     &    &    &   & $cc\bar{c}\bar{c}$  & 6449  & $-$  & 513 \\
                &     &    &    &   & $cc\bar{c}\bar{c}$  & 6659  & $-$  & 723 \\
                &     &    &    &   & $bb\bar{b}\bar{b}$  & 19005  & $-$  & 203 \\
                &     &    &    &   & $bb\bar{b}\bar{b}$  & 19049  & $-$  & 247 \\[2ex]
  $01^+$  & $cc\bar{q}\bar{q}$      & 3726  & $-$    & -189  & $D^{*+}D^{*0}$   & 4312  & 16.00   & 397 \\
                & $cb\bar{q}\bar{q}$     & 6997   & $-$   & -220  & $D^{*0}\bar{B}^0$  & 7327   & 2.40   & 110 \\
                & $bb\bar{q}\bar{q}$     & 10238  & $-$   & -359 & $B^{*-}\bar{B}^{*0}$  & 10814  & 2.00  & 217 \\ 
                & $bb\bar{q}\bar{q}$     & 10524  & $-$   & -73  & $D^{*+}_s\bar{B}^*_s$  & 7920  & 1.20  & 531 \\
                & $bb\bar{b}\bar{b}$     & 18046  & $-$   & -818  & $D^{*+}_s\bar{B}^*_s$   & 7995  & 4.96  & 606 \\
                &     &    &    &   & $cc\bar{c}\bar{c}$  & 6657  & $-$  & 587 \\
                &     &    &    &   & $bb\bar{b}\bar{b}$  & 19067  & $-$  & 203 \\[2ex]
 $02^+$   & $cb\bar{q}\bar{q}$      & 7333  & $-$    & -3  & $D^{*+}_s D^{*+}_s$  & 4821  & 5.58  & 589 \\
                & $bb\bar{b}\bar{b}$      & 18223  & $-$  & -703  & $D^{*+}_s D^{*+}_s$  & 4846  & 10.68  & 614 \\
                &     &    &    &    & $D^{*+}_s D^{*+}_s$  & 4775  & 23.26  & 543 \\
                &     &    &    &    & $D^{*+}_s \bar{B}^*_s$  & 8046  & 1.42  & 530 \\
                &     &    &    &    & $D^{*+}_s \bar{B}^*_s$  & 8096  & 2.90  & 580 \\
                &     &    &    &    & $\bar{B}^*_s \bar{B}^*_s$  & 11329  & 1.48  & 529 \\
                &     &    &    &    & $\bar{B}^*_s \bar{B}^*_s$  & 11356  & 4.18  & 556 \\
                &     &    &    &    & $\bar{B}^*_s \bar{B}^*_s$  & 11410  & 2.52  & 610 \\
                &     &    &    &   & $cc\bar{c}\bar{c}$  & 7022  & $-$  & 818 \\
                &     &    &    &   & $bb\bar{b}\bar{b}$ & 19189  & $-$  & 263 \\[2ex]
 $\frac{1}{2}\frac{1}{2}^-$ & $\Sigma_c \bar{D}$ & 4312 & $-$ & -8 & $\Sigma_c D$ & 4356 & 4.8 & 699 \\
                      & $\Sigma_c \bar{D}^*$ & 4421 & $-$ & -41 & $\Lambda_c \bar{D}^*$ & 4291 & $-$ & -2 \\
                      & $\Sigma^*_c \bar{D}^*$ & 4422 & $-$ & -105 & $\Sigma_b \bar{B}^*$ & 11072 & $-$ & -17 \\
                      & $\Sigma_b \bar{B}^*$ & 11041 & $-$ & -93 & $\Sigma^*_b \bar{B}^*$ & 10862 & $-$ & -292 \\[2ex]
 $\frac{1}{2}\frac{3}{2}^-$ & $\Sigma_c \bar{D}^*$ & 4459 & $-$ & -3 & $\Sigma^*_c D$ & 4449 & 8.0 & 689 \\
                             & $\Sigma^*_c \bar{D}$ & 4370 & $-$ & -15 & $\Sigma_c D^*$ & 4461 & $-$ & -1 \\
                             & $\Sigma^*_c \bar{D}^*$ & 4510 & $-$ & -17 & $\Xi^*_{cc} \pi$ & 3757 & $-$ & -3 \\
                             & $\Sigma_b \bar{B}^*$ & 11120 & $-$ & -14 & $\Sigma^*_b \bar{B}$ & 11042 & $-$ & -67 \\
                             & $\Sigma^*_b \bar{B}^*$ & 10959 & $-$ & -195 & & &  & \\[2ex]
 $\frac{1}{2}\frac{5}{2}^-$ & $\Sigma^*_c \bar{D}^*$ & 4438 & $-$ & -89 & $\Sigma^*_b \bar{B}^*$ & 11141 & $-$ & -13 \\[2ex]
 $\frac{3}{2}\frac{1}{2}^-$ & $\Sigma^*_c \bar{D}^*$ & 4523 & $-$ & -4 & $\Sigma_c D$ & 4431 & 2.6 & 774 \\ 
                                                             &     &    &    &   & $\Sigma_c D$ & 4446 & 2.2 & 789 \\ [2ex]
 $\frac{3}{2}\frac{3}{2}^-$ & $\Sigma_b \bar{B}^*$ & 11019 & $-$ & -115 & $\Sigma_c D^*$ & 4514 & 4.0 & 754 \\
                              & $\Sigma^*_b \bar{B}$ & 10996 & $-$ & -113 & $\Sigma^*_c D^*$ & 4524 & $-$ & -3 \\
                              & $\Sigma^*_b \bar{B}^*$ & 11049 & $-$ & -105 &  &  &  &  \\[2ex]
 $\frac{3}{2}\frac{5}{2}^-$ & $\Sigma^*_b \bar{B}^*$ & 10932 & $-$ & -222 & $\Xi^*_{cc}\rho$ & 4461 & 3.0 & 65 \\
\bottomrule
\end{tabular}
\end{table}

\vspace{6pt} 

\funding{Work partially financed by: the National Natural Science Foundation of China under Grant No. 11535005 and No. 11775118; the Ministerio Espa\~nol de Ciencia e Innovaci\'on under grant No. PID2019-107844GB-C22; and the Junta de Andaluc\'ia under contract No. Operativo FEDER Andaluc\'ia 2014-2020 UHU-1264517.}

\acknowledgments{We are particularly grateful to F. Fern\'andez, D. R. Entem, P. G. Ortega for fruitful discussions on this topic.}

\conflictsofinterest{The authors declare no conflict of interest. Moreover, the funders had no role in the design of the study; in the collection, analyses, or interpretation of data; in the writing of the manuscript, or in the decision to publish the results.} 





\reftitle{References}


\begin{thebibliography}{999}
\bibitem{Vijande:2004he} J. Vijande, F. Fernandez and A. Valcarce. Constituent quark model study of the meson spectra. J. Phys. G {\bf 31}, 481 (2005).
\bibitem{YYC2008BS} Y. C. Yang, C. Deng, H. Huang and J. Ping, Dynamical study of heavy-baryon spectroscopy. Moden Phys. Lett. A, Vol. 23, No. 22 (2008) 1819-1828.
\bibitem{Yang:2017xpp} G. Yang, J. Ping and J. Segovia. The S- and P-Wave Low-Lying Baryons in the Chiral Quark Model. Few-Body Syst. {\bf 59}, 113 (2018).
\bibitem{MGMPL1964} M. Gell-Mann. A schematic model of baryons and mesons, Phys. Lett. {\bf 8}, 214 (1964).
\bibitem{skc:2003prl} S. K. Choi {\em et al.} (LEPS Collaboration). Observation of a Narrow Charmoniumlike State in Exclusive $B^{\pm}\rightarrow K^{\pm}\pi^+\pi^-J/\psi$ Decays. Phys. Rev. Lett. {\bf 91}, 262001 (2003).
\bibitem{CDFX3872} D. Acosta {\em et al.} (CDF II Collaboration). Observation of the Narrow State $X(3872)\rightarrow J/\psi\pi^+\pi^-$ in $\bar{p}p$ Collisions at $\sqrt{s}=1.96$ TeV. Phys. Rev. Lett.  {\bf 93}, 072001 (2004).
\bibitem{D0X3872} V. M. Abazov {\em et al.} (D0 Collaboration). Observation and Properties of the $X(3872)$ Decaying to $J/\psi\pi^+\pi^-$ in $p\bar{p}$ Collisions at $\sqrt{s}=1.96$ TeV. Phys. Rev. Lett. {\bf 93}, 162002 (2004).
\bibitem{BABARX3872} B. Aubert {\em et al.} (BABAR Collaboration). Study of the $B^-\rightarrow J/\psi K^- \pi^+\pi^-$ decay and measurement of the $B^-\rightarrow X(3872)K^-$ branching fraction. Phys. Rev. D {\bf 71}, 071103 (2005).
\bibitem{BelleY3940} K. Abe {\em et al.} (Belle Collaboration). Observation of a near-threshold $\omega J/\psi$ mass enhancement in exclusive $B\rightarrow K\omega J/\psi$ decays. Phys. Rev. Lett. {\bf 94}, 182002 (2005).
\bibitem{BellZ4430} K. Abe {\em et al.} (Belle Collaboration). Observation of a resonance-like structure in the $\pi^{\pm}\psi'$ mass distribution in exclusive $B\rightarrow K\pi^{\pm}\psi'$ decays, Phys. Rev. Lett. {\bf 100}, 142001 (2008).
\bibitem{BABARY4260} B. Aubert {\em et al.} (BaBar Collaboration). Observation of a broad structure in the $\pi^+ \pi^- J/\psi$ mass spectrum around 4.26 $GeV/c^2$, Phys. Rev. Lett. {\bf 95}, 142001 (2005).
\bibitem{CLEOY4260} Q. He {\em et al.} (CLEO Collaboration). Confirmation of the Y(4260) resonance production in ISR. Phys. Rev. D {\bf 74}, 091104 (2006).
\bibitem{BelleY4260} C. Z. Yuan {\em et al.} (Belle Collaboration), Measurement of $e^+ e^-\rightarrow \pi^+ \pi^- J/\psi$ cross-section via initial state radiation at Belle. Phys. Rev. Lett. {\bf 99}, 182004 (2007).
\bibitem{PDG2018} M. Tanabashi {\em et al.} (Particle Data Group). Review of Particle Physics. Phys. Rev. D {\bf 98}, 030001 (2018).
\bibitem{vkams:2017jhep} V. Khachatryan {\em et al.} (CMS Collaboration). Observation of $\Upsilon(1S)$ pair production in proton-proton collisions at $\sqrt{s}=8$ TeV. J. High Energ. Phys. {\bf 05}, 013 (2017).
\bibitem{SD2018PHD} S. Durgut, Ph. D. thesis at University of Iowa. Evidence of a narrow structure in $\Upsilon(1S)1^+1^-$ mass spectrum and CMS Phase I and II silicon detector upgrade studies. https://ir.uiowa.edu/etd/6411/.
\bibitem{SD2018II} S. Durgut, APS April Meeting 2018. Search for Exotic Mesons at CMS. http://meetings.aps.org/Meeting/APR18/Session/U09.6.
\bibitem{KY2018} K. Yi. Things that go bump in the night: From $J/\psi\phi$ to other mass spectrum. Intl. J. Mod. Phys. A Vol. 33, No. 36, 1850224 (2018).
\bibitem{LCBland2019} L. C. Bland {\em et al.} ($A_N$DY Collaboration). Observation of Feynman scaling violations and evidence for a new resonance at RHIC. arXiv: 1909.03124 [nucl-ex].
\bibitem{raba:2018jhep} R. Aaij {\em et al.} (LHCb Collaboration). Search for beautiful tetraquarks in the $\Upsilon(1S)\mu^+\mu^-$ invariant-mass spectrum. J. High Energ. Phys. {\bf 10}, 086 (2018).
\bibitem{LA2020CERNINDICO}  R. Aaij {\em et al.} (LHCb Collaboration). Observation of structure in the $J/\psi$-pair mass spectrum. arXiv: 2006.16957 [hep-ex].
\bibitem{Aaij:2015tga} R. Aaij {\em et al.} (LHCb Collaboration). Observation of $J/\psi p$ Resonances Consistent with Pentaquark States in $\Lambda_b^0 \to J/\psi K^- p$ Decays. Phys. Rev. Lett. {\bf 115}, 072001 (2015).
\bibitem{A2016MI} R. Aaij {\em et al.} (LHCb Collaboration). Model-independent evidence for $J/\psi p$ contributions to $\Lambda_b^0 \to J/\psi K^- p$ decays, Phys. Rev. Lett. {\bf 117}, 082002 (2016).
\bibitem{lhcb:2019pc} R. Aaij {\em et al.} (LHCb Collaboration). Observation of a narrow pentaquark state, $P_c(4312)^+$, and of two-peak structure of the $P_c(4450)^+$. Phys. Rev. Lett. {\bf 122}, 222001 (2019).
\bibitem{Aaij:2017jgf} R. Aaij {\em et al.} (LHCb Collaboration). Search for weakly decaying $b$-flavored pentaquarks. Phys. Rev. D {\bf 97}, 032010 (2018).
\bibitem{jclh:1988prd} J. Carlson, L. Heller and J. A. Tjon. Stability of dimesons. Phys. Rev. D {\bf 37}, 744 (1988).
\bibitem{ejecq:2017prl} E. J. Eichten and C. Quigg. Heavy-Quark Symmetry Implies Stable Heavy Tetraquark Mesons $Q_iQ_j\bar{q}_k\bar{q}_l$. Phys. Rev. Lett. {\bf 119}, 202002 (2017).
\bibitem{mkjlr:2017prl} M. Karliner and J. L. Rosner. Discovery of the Doubly Charmed $\Xi_{cc}$ Baryon Implies a Stable $bb\bar{u}\bar{d}$ Tetraquark. Phys. Rev. Lett. {\bf 119}, 202001 (2017).
\bibitem{EHJVAVJMR2019} E. Hern\'andez, J. Vijande, A. Valcarce and J.-M. Richard. Spectroscopy, lifetime and decay modes of the $T^-_{bb}$ tetraquark. Phys. Lett. B {\bf 800}, 135073 (2020) .
\bibitem{cefgk:2019prd} C. E. Fontoura, G. Krein, A. Valcarce and J. Vijande. Production of exotic tetraquarks $QQ\bar{q}\bar{q}$ in heavy-ion collisions at the LHC. Phys. Rev. D {\bf 99}, 094037 (2019).
\bibitem{dernfvogwl2007} D. Ebert, R. N. Faustov, V. O. Galkin and W. Lucha. Masses of tetraquarks with two heavy quarks in the relativistic quark model. Phys. Rev. D {\bf 76}, 114015 (2007).
\bibitem{llsm:2019prd} L. Leskovec, S. Meinel, M. Pflaumer and M. Wagner. Lattice QCD investigation of a doubly-bottom $\bar{b}\bar{b}ud$ tetraquark with quantum numbers $I(J^P)=0(1^+)$. Phys. Rev. D {\bf 100}, 014503 (2019).
\bibitem{afrjhrlkm2017} A. Francis, R. J. Hudspith, R. Lewis and K. Maltman. Lattice Prediction for Deeply Bound Doubly Heavy Tetraquarks. Phys. Rev. Lett. {\bf 118}, 142001 (2017).
\bibitem{pjnmmp2019} P. Junnarkar, N. Mathur and M. Padmanath. Study of doubly heavy tetraquarks in lattice QCD. Phys. Rev. D {\bf 99}, 034507 (2019).
\bibitem{ssaka:2019arx} S. S. Agaev, K. Azizi and H. Sundu. Double-heavy axial-vector tetraquark $T^0_{bc;\bar{u}\bar{d}}$. Nucl. Phys. B {\bf 951}, 114890 (2020).
\bibitem{afrjhrlkm2019} A. Francis, R. J. Hudspith, R. Lewis and K. Maltman. Evidence for charm-bottom tetraquarks and the mass dependence of heavy-light tetraquark states from lattice QCD. Phys. Rev. D {\bf 99}, 054505 (2019).
\bibitem{CDHCJP2019} C. Deng, H. Chen and J. Ping. Towards the understanding of fully-heavy tetraquark states from various models. arXiv: 2003.05154 [hep-ph].
\bibitem{BWLMSZ2020} B. Wang, L. Meng and S. Zhu. Deciphering the charged heavy quarkoniumlike states in chiral effective field theory. arXiv: 2009.01980 [hep-ph].
\bibitem{ssakahs:2019prd} S. S. Agaev, K. Azizi and H. Sundu. Strong decays of double-charmed pseudoscalar and scalar $cc\bar{u}\bar{d}$ tetraquarks. Phys. Rev. D {\bf 99}, 114016 (2019).
\bibitem{yyjp:2019prd} Y. Yang and J. Ping. Investigation of $cs\bar{c}\bar{s}$ tetraquark in the chiral quark model. Phys. Rev. D {\bf 99}, 094032 (2019).
\bibitem{zgw:2019arx} Z. -G. Wang. Analysis of the axial vector $B_c$-like tetraquark states with the QCD sum rules. EuroPhys. Lett. {\bf 128}, 11001 (2019), arXiv: 1907.10921 [hep-ph].
\bibitem{YXRZ2018} Y. Xing and R. Zhu. Weak decays of stable doubly heavy tetraquark states. Phys. Rev. D {\bf 98}, 053005 (2018).
\bibitem{YXFSYRZ2019} Y. Xing, F. Yu and R. Zhu. Weak decays of stable open-bottom tetraquark by SU(3) symmetry analysis. Eur. Phys. J. C {\bf 79}, 373 (2019).
\bibitem{aaaypqqww2018} A. Ali, A. Ya. Parkhomenko, Q. Qin and W. Wang. Prospects of discovering stable double-heavy tetraquarks at a Tera-Z factory. Phys. Lett. B {\bf 782}, 412 (2018).
\bibitem{aaqqww2018} A. Ali, Q. Qin and W. Wei. Discovery potential of stable and near-threshold doubly heavy tetraquarks at the LHC. Phys. Lett. B {\bf 785}, 605 (2018).
\bibitem{GYangQQSS} G. Yang, J. Ping and J. Segovia. $QQ\bar{s}\bar{s}$ tetraquarks in the chiral quark model. arXiv: 2007.05190 [hep-ph].
\bibitem{YTWLJP2020} Y. Tan, W. Lu, J. Ping. $QQ\bar{q}\bar{q}$ in a chiral constituent quark model. arXiv: 2004.02106 [hep-ph].
\bibitem{mna:2018epjc} M. N. Anwar, J. Ferretti, F. -K. Guo, E. Santopinto and B. -S. Zou. Spectroscopy and decays of the fully-heavy tetraquarks. Eur. Phys. J. C {\bf 78}, 647 (2018).
\bibitem{zgwqqqq:2017epjc} Z. -G. Wang. Analysis the $QQ\bar{Q}\bar{Q}$ tetraquark states with QCD sum rules. Eur. Phys. J. C {\bf 77}, 432 (2017).
\bibitem{yb:2019plb} Y. Bai, S. Lu and J. Osborne. Beauty-full tetraquarks. Phys. Lett. B {\bf 798}, 134930 (2019).
\bibitem{mksnjl:2017prd} M. Karliner, S. Nussinov and J. L. Rosner. $QQ\bar{Q}\bar{Q}$ states: Masses, production, and decays. Phys. Rev. D {\bf 95}, 034011 (2017).
\bibitem{mabjfcdres2019} M. A. Bedolla, J. Ferretti, C. D. Roberts and E. Santopinto. Spectrum of fully-heavy tetraquarks from a diquark+antidiquark perspective. arXiv: 1911.00960 [hep-ph].
\bibitem{avb:2012prd} A. V. Berezhnoy, A. V. Luchinsky and A. A. Novoselov. Heavy tetraquarks production at the LHC. Phys. Rev. D {\bf 86}, 034004 (2012).
\bibitem{wchxc:2017plb} W. Chen, H. -X. Chen, X. Liu, T. G. Steele and S. -L. Zhu. Hunting for exotic doubly hidden-charm/bottom tetraquark states. Phys. Lett. B {\bf 773}, 247 (2017).
\bibitem{aeap:2018epjc} A. Esposito and A. D. Polosa. A $bb\bar{b}\bar{b}$ di-bottomonium ar the LHC? Eur. Phys. J. C {\bf 78}, 782 (2018).
\bibitem{TKSSPRD2012} T. Kawanai and S. Sasaki. Charmonium potential from full lattice QCD. Phys. Rev. D {\bf 85}, 091503 (2012).
\bibitem{GYangFHT2020} G. Yang, J. Ping, L. He and Q. Wang. A potential model prediction of fully-heavy tetraquarks $QQ\bar{Q}\bar{Q}$ ($Q=c, b$). arXiv: 2006.13756 [hep-ph].
\bibitem{avbakl:2011prd} A. V. Berezhnoy, A. K. Likhoded, A. V. Luchinsky and A. A. Novoselov. Production of $J/\psi$-meson pairs and 4c tetraquark at the LHC. Phys. Rev. D {\bf 84}, 094023 (2011).
\bibitem{whge:2012plb} W. Heupel, G. Eichmann and C. S. Fischer. Tetraquark bound states in a Bethe-Salpeter approach. Phys. Lett. B {\bf 718}, 545 (2012).
\bibitem{vrdfsn:2019cpc} V. R. Debastiani and F. S. Navarra. A non-relativistic model for the [cc][$\bar{c}\bar{c}$] tetraquark. Chin. Phys. C {\bf 43}, 013105 (2019).
\bibitem{PLTO2006} P. Lundhammar and T. Ohlsson. A Non-Relativistic Model of Tetraquarks and Predictions for Their Masses from Fits to Charmed and Bottom Meson Data. arXiv: 2006.09393 [hep-ph].
\bibitem{MSLFXLXHZQZ2020} M. S. Liu, F. X. Liu, X. H. Zhong and Q. Zhao. Full-heavy tetraquark states and their evidences in the LHCb di-$J/\psi$ spectrum. arXiv: 2006.11952 [hep-ph].
\bibitem{JFGRFL2008} J. F. Giron and R. F. Lebed. The Simple Spectrum of $cc\bar{c}\bar{c}$ States in the Dynamical Diquark Model. arXiv: 2008.01631 [hep-ph].
\bibitem{ZGW2020FHT} Z. G. Wang. Tetraquark candidates in the LHCb's di-$J/\psi$ mass spectrum. arXiv: 2006.13028 [hep-ph].
\bibitem{QDYFHT2020} Q. L$\ddot{u}$, D. Chen and Y. Bing. Masses of fully heavy tetraquarks $QQ\bar{Q}\bar{Q}$ in an extended relativized quark model. arXiv: 2006.14445 [hep-ph].
\bibitem{MKJLR2020FC} M. Karliner and J. L. Rosner. Interpretation of structure in the di-$J/\psi$ spectrum. arXiv: 2009.04429 [hep-ph].
\bibitem{XJTX2020} X. Jin, Y. Xue, H. Huang and J. Ping. Full-heavy tetraquarks in constituent quark models. arXiv: 2006.13745 [hep-ph].
\bibitem{RMASNARDRGR2008} R. M. Albuquerque, S. Narison, A. Rabemananjara, D. Rabetiarivony and G. Randriamanatrika. Doubly-hidden scalar heavy molecules and tetraquarks states from QCD at NLO. arXiv: 2008.01569 [hep-ph].
\bibitem{JSDW2008} J. Sonnenschein and D. Weissman. Deciphering the recently discovered tetraquark candidates around 6.9 GeV. arXiv: 2008.01095 [hep-ph].
\bibitem{JR2008FHT} J. Richard. About the $J/\psi$ $J/\psi$ peak of LHCb: fully-charmed tetraquark? arXiv: 2008.01962 [hep-ph].
\bibitem{KCSZ2008} K. Chao and S. Zhu. The possible tetraquark states $cc\bar{c}\bar{c}$ observed by the LHCb experiment. arXiv: 2008.07670 [hep-ph].
\bibitem{HWXSFHT2020} H. Chen, W. Chen, X. Liu and S. Zhu. Strong decays of fully-charm tetraquarks into di-charmonia. arXiv: 2006.16027 [hep-ph].
\bibitem{XWQLHXYXYHXQ2007} X. Wang, Q. Lin, H. Xu, Y. Xie, Y. Huang and X. Rong. Discovery potential for the LHCb fully-charm tetraquark X(6900) state via   annihilation reaction. arXiv: 2007.09697 [hep-ph].
\bibitem{JZDCXLTM2008} J. Zhang, D. Chen, X. Liu and T. Matsuki. Producing fully-charm structures in the $J/\psi$-pair invariant mass spectrum. arXiv: 2008.07430 [hep-ph].
\bibitem{KY2013} K. Yi. EXPERIMENTAL REVIEW OF STRUCTURES IN THE $J/\psi\phi$ MASS SPECTRUM. Intl. J. Mod. Phys. A Vol. 28, No. 18 (2013) 1330020.
\bibitem{jmrav:2017prd} J. -M. Richard, A. Valcarce and J. Vijande. String dynamics and metastability of all-heavy tetraquarks. Phys. Rev. D {\bf 95}, 054019 (2017).
\bibitem{jwyrl:2018prd} J. Wu, Y. -R. Liu, K. Chen, X. Liu and S. -L. Zhu. Heavy-flavored tetraquark states with the $QQ\bar{Q}\bar{Q}$ configuration. Phys. Rev. D {\bf 97}, 094015 (2018).
\bibitem{xc:2019epja} X. Chen. Analysis of hidden-bottom $bb\bar{b}\bar{b}$ states. Eur. Phys. J. A {\bf 55}, 106 (2019).
\bibitem{mslqfl:2019prd} M. -S. Liu, Qi -F. L{\"u} and X. -H. Zhong and Q. Zhao. All-heavy tetraquarks. Phys. Rev. D {\bf 100}, 016006 (2019).
\bibitem{gjw:2019arx} G. -J. Wang, L. Meng and S. -L. Zhu. Spectrum of the fully-heavy tetraquark state $QQ\bar{Q}'\bar{Q}'$. Phys. Rev. D {\bf 100}, 096013 (2019).
\bibitem{jmravjv2018} J. -M. Richard, A. Valcarce and J. Vijande. Few-body quark dynamics for doubly heavy baryons and tetraquarks. Phys. Rev. C {\bf 97}, 035211 (2018).
\bibitem{cheec:2018prd} C. Hughes, E. Eichten and C. T. H. Davies. Searching for beauty-fully bound tetraquarks using lattice nonrelativistic QCD. Phys. Rev. D {\bf 97}, 054505 (2018).
\bibitem{SALHCbM20201} R. Aaij {\em et al.} (LHCb Collaboration). Amplitude analysis of the $B^+\rightarrow D^+D^-K^+$ decay. arXiv: 2009.00026 [hep-ex].
\bibitem{SALHCbM20202} R. Aaij {\em et al.} (LHCb Collaboration). A model-independent study of resonant structure in $B^+\rightarrow D^+D^-K^+$ decays. arXiv: 2009.00025 [hep-ex].
\bibitem{JCSLYLZSTY2020} J. Cheng, S. Li, Y. Liu, Y. Liu, Z. Si and T. Yao. Spectrum and rearrangement decays of tetraquark states with four different flavors. Phys. Rev. D {\bf 101}, 114017 (2020).
\bibitem{QMTBEO2010} R. Molina, T. Branz and E. Oset. New interpretation for the $D^*_{s2}(2573)$ and the prediction of novel exotic charmed mesons. Phys. Rev. D {\bf 82}, 014010 (2010).
\bibitem{XHWWRZ2020} X. He, W. Wang and R. Zhu. Open-charm tetraquark $X_c$ and open-bottom tetraquark $X_b$. arXiv: 2008.07145 [hep-ph].
\bibitem{HCWCRDNS2020} H. Chen, W. Chen, R. Dong and N. Su. $X_0(2900)$ and $X_1(2900)$: hadronic molecules or compact tetraquarks. arXiv: 2008.07516 [hep-ph].
\bibitem{SSAKAHS2020} S. S. Agaev, K. Azizi and H. Sundu. New scalar resonance $X_0(2900)$ as a $\bar{D}^*K^*$ molecule: Mass and width. arXiv: 2008.13027 [hep-ph].
\bibitem{MLJXLG2020} M. Liu, J. Xie and L. Geng. $X_0(2866)$ as a $D^*\bar{K}^*$ molecular state. arXiv: 2008.07389 [hep-ph].
\bibitem{JHDC2020} J. He and D. Chen. Molecular picture for $X_0(2900)$ and $X_1(2900)$. arXiv: 2008.07782 [hep-ph].
\bibitem{YXXJHHJP2020} Y. Xue, X. Jin, H. Huang and J. Ping. Tetraquark with open charm favor. arXiv: 2008.09516 [hep-ph].
\bibitem{RMEO2020} R. Molina and E. Oset. Molecular picture for the $X_0(2866)$ as a $D^*\bar{K}^*$ $J^P=0^+$ state and related $1^+$, $2^+$ states. arXiv: 2008.11171 [hep-ph].
\bibitem{YHJLJXLG2020} Y. Huang, J. Lu, J. Xie and L. Geng. Strong decays of $\bar{D}^*K^*$ molecules and the newly observed $X_{0,1}$ states. arXiv: 2008.07959 [hep-ph].
\bibitem{JZ2020OC} J. Zhang. An open charm tetraquark candidate: note on $X_0(2900)$. arXiv: 2008.07295 [hep-ph].
\bibitem{ZW2020OC} Z. Wang. Analysis of the $X_0(2900)$ as the scalar tetraquark state via the QCD sum rules. arXiv: 2008.07833 [hep-ph].
\bibitem{MKJLR2020OC} M. Karliner and J. L. Rosner. First exotic hadron with open heavy flavor: $cs\bar{u}\bar{d}$ tetraquark. arXiv: 2008.05993 [hep-ph].
\bibitem{MHXLPLQW2020OC} M. Hu, X. Lao, P. Ling and Q. Wang. The molecular nature of the $X_0(2900)$. arXiv: 2008.06894 [hep-ph].
\bibitem{XLMYHKGLJX2020OC} X. Liu, M. Yan, H. Ke, G. Li and J. Xie. Triangle singularity as the origin of $X_0(2900)$ and $X_1(2900)$ observed in $B\rightarrow D^+D^-K^+$. arXiv: 2008.07190 [hep-ph].
\bibitem{TJBESS2020} T. J. Burns and E. S. Swanson. Triangle Cusp and Resonance Interpretations of the X(2900). arXiv: 2008.12838 [hep-ph].
\bibitem{LDCYD2020OC} Q. L$\ddot{u}$, D. Chen and Y. Dong. Open charm and bottom tetraquarks in an extended relativized quark model. arXiv: 2008.07340 [hep-ph].

\bibitem{PRL1051} J. Wu, R. Molina, E. Oset and B. S. Zou. Prediction of Narrow $N^*$ and $\Lambda^*$ Resonances with Hidden Charm above 4 GeV. Phys. Rev. Lett. {\bf 105}, 232001 (2010).
\bibitem{PRL1052} J. Wu, R. Molina, E. Oset and B. S. Zou. Dynamically generated $N^*$ and $\Lambda^*$ resonances in the hidden charm sector around 4.3 GeV. Phys. Rev. C {\bf 84}, 015202 (2011).
\bibitem{Oset1} J. Wu, T. S. H. Lee and B. S. Zou. Nucleon resonances with hidden charm in coupled-channels models. Phys. Rev. C {\bf 85}, 044002 (2012).
\bibitem{Oset2}C. Garcia-Recio, J. Nieves, O. Romanets, L. L. Salcedo and L. Tolos. Hidden charm $N$ and $\Delta$ resonances with heavy-quark symmetry. Phys. Rev. D {\bf 87}, 074034 (2013).
\bibitem{Oset3}C. W. Xiao, J. Nieves and E. Oset. Combining heavy quark spin and local hidden gauge symmetries in the dynamical generation of hidden charm baryons. Phys. Rev. D {\bf 88}, 056012 (2013).
\bibitem{CPC36} Z. C. Yang, Z. F. Sun, J. He, X. Liu and S. L. Zhu. Possible hidden-charm molecular baryons composed of an anti-charmed meson and a charmed baryon. Chin. Phys. C {\bf 36(1)}, 6 (2012).
\bibitem{RChen} R. Chen, X. Liu, X. Q. Li and S. L. Zhu. Identifying Exotic Hidden-Charm Pentaquarks. Phys. Rev. Lett. {\bf 115}, 132002 (2015).
\bibitem{Yang:2015bmv} G. Yang, J. Ping and F. Wang. The structure of pentaquarks $P_c^+$ in the chiral quark model. Phys. Rev. D {\bf 95}, 014010 (2017).
\bibitem{HHXPC2016} H. Huang, C. Deng, J. Ping, and F. Wang. Possible pentaquarks with heavy quarks. Eur. Phys. J. C {\bf 76}, 624 (2016).
\bibitem{JHe} J. He. $\bar{D}\Sigma^*_c$ and $\bar{D}^*\Sigma_c$ interactions and the LHCb hidden-charmed pentaquarks. Phys. Lett. B {\bf 753}, 547 (2016).
\bibitem{HXChen} H. X. Chen, W. Chen, X. Liu, T. G. Steel and S. L. Zhu. Towards Exotic Hidden-Charm Pentaquarks in QCD. Phys. Rev. Lett. {\bf 115}, 172001 (2015).
\bibitem{ZGWang} Z. G. Wang. Analysis of $P_c(4380)$ and $P_c(4450)$ as pentaquark states in the diquark model with QCD sum rules. Eur. Phys. J. C {\bf 76}, 70 (2016).
\bibitem{KAYSHS2017} K. Azizi, Y. Sarac, and H. Sundu. Analysis of $P^+_C(4380)$ and $P^+_c(4450)$ as pentaquark states in the molecular picture with QCD sum rules. Phys. Rev. D {\bf 95}, 094016 (2017).
\bibitem{RZCQ2016} R. Zhu and C. Qiao. Pentaquark states in a diquark-triquark model. Phys. Lett. B {\bf 756}, 259 (2016).
\bibitem{MBPCTA2016} M. Bayar, F. Aceti, F. Guo, and E. Oset. Discussion on triangle singularities in the $\Lambda_b \rightarrow J/\psi K^-p$ reaction. Phys. Rev. D {\bf 94}, 074039 (2016).
\bibitem{XHLPCTS20162} X. Liu, Q. Wang, and Q. Zhao. Understanding the newly observed heavy pentaquark candidates. Phys. Lett. B {\bf 757}, 231 (2016).
\bibitem{XHLPCTS2016} X. Liu and M. Oka. Understanding the nature of heavy pentaquarks and searching for them in pion-induced recations. Nucl. Phys. A {\bf 954}, 352 (2016).
\bibitem{KAYSHS2018SD} K. Azizi, Y. Sarac and H. Sundu. Strong decay of $P_c(4380)$ pentaquark in a molecular picture. Phys. Lett. B {\bf 782}, 694 (2018).
\bibitem{MZL190311560} M. Liu, Y. Pan, F. Peng, M. S$\acute{a}$nchez, L. Geng, A. Hosaka and M. P. Valderrama. Emergence of a complete heavy-quark spin symmetry multiplet: seven molecular pentaquarks in light of the latest LHCb analysis. Phys. Rev. Lett. {\bf 122}, 242001 (2019).
\bibitem{JH190311872} J. He. Study of $P_c(4457)$, $P_c(4440)$, and $P_c(4312)$ in a quasipotential Bethe-Salpeter equation approach. Eur. Phys. J. C {\bf 79}, 393 (2019).
\bibitem{CWX190401296} C. W. Xiao, J. Nieves and E. Oset. Heavy quark spin symmetric molecular states from $\bar{D}^{(*)}\Sigma^{(*)}_c$ and other coupled channels in the light of the recent LHCb pentaquarks. Phys. Rev. D {\bf 100}, 014021 (2019).
\bibitem{CJX190400872} C. Xiao, Y. Huang, Y. Dong, L. Geng and D. Chen. Exploring the molecular scenario of Pc(4312), Pc(4440), and Pc(4457). Phys. Rev. D {\bf 100}, 014022 (2019).
\bibitem{YS190400587} Y. Shimizu, Y. Yamaguchi and M. Harada. Heavy quark spin multiplet structure of $P_c(4457)$, $P_c(4440)$, and $P_c(4312)$. arXiv: 1904.00587 [hep-ph].
\bibitem{ZHG190400851} Z. Guo and J. A. Oller. Anatomy of the newly observed hidden-charm pentaquark states: $P_c(4457)$, $P_c(4440)$, and $P_c(4312)$. Phys. Lett. B {\bf 793}, 144 (2019).
\bibitem{HH190400221} H. Huang, J. He and J. Ping. Looking for the hidden-charm pentaquark resonances in $J/\psi p$ scattering. arXiv: 1904.00221 [hep-ph].
\bibitem{HM1904.09756} H. Mutuk. Neural Network Study of Hidden-Charm Pentaquark Resonances. Chin. Phys. C {\bf 43}, 093103 (2019).
\bibitem{RZ190410285} R. Zhu, X. Liu, H. Huang and C. Qiao. Analyzing doubly heavy tetra- and penta-quark states by variational method. Phys. Lett. B {\bf 797}, 134869 (2019).
\bibitem{MIE190411616} M. I. Eides, V. Yu. Petrov and M. V. Polyakov. New LHCb pentaquarks as Hadrocharmonium States. Mod. Phys. Lett. A  Vol. 35, No. 18, 2050151 (2020).
\bibitem{XZW190409891} X. Weng, X. Chen, W. Deng and S. Zhu. Hidden-charm pentaquarks and $P_c$ states. Phys. Rev. D {\bf 100}, 016014 (2019).
\bibitem{FLW190503636} F. Wang, R. Chen, Z. Liu and X. Liu. Probing new types of $P_c$ states inspired by the interaction between an S-wave charmed baryon and an anticharmed meson in a $\bar{T}$ doublet state. Phys. Rev. C {\bf 101}, 025201 (2020).
\bibitem{LM190504113} L. Meng, B. Wang, G. Wang and S. Zhu. The hidden charm pentaquark states and $\Sigma_c\bar{D}^*$ interaction in chiral perturbation theory. Phys. Rev. D {\bf 100}, 014031 (2019).
\bibitem{ZGW190502892} Z. Wang. Analysis of the $P_c(4312)$, $P_c(4440)$, $P_c(4457)$ and related hidden-charm pentaquark states with QCD sum rules. Int. J. Mod. Phys. A {\bf35} 2050003 (2020).
\bibitem{XC190406015} X. Cao and J.-P. Dai. Confronting pentaquark photoproduction with new LHCb observations. Phys. Rev. D {\bf 100}, 054033 (2019).
\bibitem{XYW190411706} X.-Y. Wang, X.-R. Chen and J. He. Possibility to study the pentaquark states $P_c(4312)$, $P_c(4440)$, and $P_c(4457)$ in the reaction $\gamma p\rightarrow J/\psi p$. Phys. Rev. D {\bf 99}, 114007 (2019).
\bibitem{JMR190103578} J.-M. Richard, A. Valcarce and J. Vijande. Pentaquarks with anticharm or beauty revisited. Phys. Lett. B {\bf 790}, 248 (2019).
\bibitem{QSZ180104557} Q.-S. Zhou, K. Chen, X. Liu, Y.-R. Liu and S.-L. Zhu. Surveying exotic pentaquarks with the typical $QQqq\bar{q}$ configuration.Phys. Rev. C {\bf 98}, 045204 (2018).
\bibitem{FG190304430} F. Giannuzzi. Heavy pentaquark spectroscopy in the diquark model. Phys. Rev. D {\bf 99}, 094006 (2019).
\bibitem{gy:2020dcp} G. Yang, J. Ping and J. Segovia. Doubly charmed pentaquarks. Phys. Rev. D {\bf 101}, 074030 (2020).
\bibitem{FLW190101542} F. Wang, R. Chen, Z. Liu and X. Liu. Possible triple-charm molecular pentaquarks from $\Xi_{cc}D_1/\Xi_{cc}D^*_2$ interactions. Phys. Rev. D {\bf 99}, 054021 (2019).
\bibitem{KAYSHS2018STP} K. Azizi, Y. Sarac, and H. Sundu. Possible molecular pentaquark states with different spin and quark configurations. Phys. Rev. D {\bf 98}, 054002 (2018).
\bibitem{JV190209799} J. Vijande, J. Richard and A. Valcarce. Few-body insights of multiquark exotic hadrons. arXiv:1902.09799 [hep-ph].
\bibitem{YRL190311976} Y. Liu, H. Chen, W. Chen, X. Liu and S. Zhu. Pentaquark and Tetraquark states. Prog. Part. Nucl. Phys. {\bf 107}, 237 (2019).
\bibitem{Valcarce:1995dm} A. Valcarce, F. Fernandez, P. Gonzalez and V. Vento. Chiral quark cluster model study of the low-energy baryon spectrum. Phys. Lett. B {\bf 367}, 35 (1996).
\bibitem{Segovia:2008zza} J. Segovia, D. R. Entem and F. Fernandez. Is chiral symmetry restored in the excited meson spectrum? Phys. Lett. B {\bf 662}, 33 (2008).
\bibitem{Segovia:2008zz} J. Segovia, A. M. Yasser, D. R. Entem and F. Fern\'{a}ndez. $J^{PC}=1^{--}$ hidden charm resonances. Phys. Rev. D {\bf 78}, 114033 (2008).
\bibitem{Ortega:2016hde} P. G. Ortega, J. Segovia, D. R. Entem and F. Fern\'{a}ndez. Canonical description of the new LHCb resonances. Phys. Rev. D {\bf 94}, 114018 (2016).
\bibitem{Fernandez:1993hx} F. Fern\'{a}ndez, A. Valcarce, U. Straub and A. Faessler. The Nucleon-nucleon interaction in terms of quark degrees of freedom. J. Phys. G {\bf 19}, 2013 (1993).
\bibitem{Valcarce:1994nr} A. Valcarce, F. Fern\'{a}ndez, A. Buchmann and A. Faessler. Can one simultaneously describe the deuteron properties and the nucleon-nucleon phase shifts in the quark cluster model? Phys. Rev. C {\bf 50}, 2246 (1994).
\bibitem{Ortega:2009hj} P. G. Ortega, J. Segovia, D. R. Entem and F. Fern\'{a}ndez. Coupled channel approach to the structure of the X(3872). Phys. Rev. D {\bf 81}, 054023 (2010).
\bibitem{Ortega:2016mms} P. G. Ortega, J. Segovia, D. R. Entem and F. Fern\'{a}ndez. Molecular components in P-wave charmed-strange mesons. Phys. Rev. D {\bf 94}, 074037 (2016).
\bibitem{Ortega:2016pgg} P. G. Ortega, J. Segovia, D. R. Entem and F. Fern\'{a}ndez. Threshold effects in P-wave bottom-strange mesons. Phys. Rev. D {\bf 95}, 034010 (2017).
\bibitem{Vijande:2006jf} J. Vijande, A. Valcarce and K. Tsushima. Dynamical study of $QQ\bar{u}\bar{d}$ mesons. Phys. Rev. D {\bf 74}, 054018 (2006).
\bibitem{Yang:2017rpg} G. Yang and J. Ping. Dynamical study of $\Omega_c^0$ in the chiral quark model. Phys. Rev. D {\bf 97}, 034023 (2018).
\bibitem{gy:2020dht} G. Yang, J. Ping and J. Segovia. Double-heavy tetraquarks. Phys. Rev. D {\bf 101}, 014001 (2020).
\bibitem{Yang:2018oqd} G. Yang, J. Ping and J. Segovia. Hidden-bottom pentaquarks. Phys. Rev. D {\bf 99}, 014035 (2019).
\bibitem{Alexandrou:2001ip} C. Alexandrou, P. De Forcrand and A. Tsapalis. The Static three quark SU(3) and four quark SU(4) potentials. Phys. Rev. D {\bf 65}, 054503 (2002).
\bibitem{Okiharu:2004wy} F. Okiharu, H. Suganuma and T. T. Takahashi. First study for the pentaquark potential in SU(3) lattice QCD. Phys. Rev. Lett. {\bf 94}, 192001 (2005).
\bibitem{Prelovsek:2014swa} S. Prelovsek, C. B. Lang, L. Leskovec and D. Mohler. Study of the $Z_c^+$ channel using lattice QCD. Phys. Rev. D {\bf 91}, 014504 (2015).
\bibitem{Lang:2014yfa} C. B. Lang, L. Leskovec, D. Mohler, S. Prelovsek and R. M. Woloshyn. Ds mesons with DK and D*K scattering near threshold. Phys. Rev. D {\bf 90}, 034510 (2014).
\bibitem{Briceno:2017max} R. A. Briceno, J. J. Dudek and R. D. Young. Scattering processes and resonances from lattice QCD. Rev. Mod. Phys. {\bf 90}, 025001 (2018).
\bibitem{JA22269} J. Aguilar and J. M. Combes. A class of analytic perturbations for one-body Schr\"{o}dinger Hamiltonians. Commun. Math. Phys. {\bf 22}, 269 (1971).
\bibitem{EB22280} E. Balslev and J. M. Combes. Spectral properties of many-body Schr\"{o}dinger operators with dilatation-analytic interactions. Commun. Math. Phys. {\bf 22}, 280 (1971).
\bibitem{TMPPNP7912014} T. Myo, Y. Kikuchi, H. Masui and K. Kato. Recent development of complex scaling method for many-body resonances and continua in light nuclei. Prog. Part. Nucl. Phys. {\bf 79}, 1 (2014).
\bibitem{Bali:2005fu} G. S. Bali, H. Neff, T. Duessel, T. Lippert and K. Schilling. Observation of string breaking in QCD. Phys. Rev. D {\bf 71}, 114513 (2005).
\bibitem{Segovia:2013wma} J. Segovia, D. R. Entem, F. Fernandez and E. Hernandez. Constituent quark model description of charmonium phenomenology. Int. J. Mod. Phys. {\bf E22}, 1330026 (2013).
\bibitem{Scadron:1982eg} M. D. Scadron. SPONTANEOUS BREAKDOWN AND THE SCALAR NONET. Phys. Rev. D {\bf 26}, 239 (1982).
\bibitem{HFUECHENC2020} J. Zhao, K. Zhou, S. Chen and P. Zhuang. Heavy flavors under extreme conditions in high energy nuclear collisions. arXiv: 2005.08277 [nucl-th].
\bibitem{CAPFOJ2002} C. Alexandrou, Ph. de Forcrand and O. Jahn. The ground state of three quarks. Nucl. Phys. B (Proc. Suppl.) {\bf 119}, 667 (2003), arXiv: hep-lat/0209062v1 [hep-lat].
\bibitem{Harvey:1980rva} M. Harvey. Effective nuclear forces in the quark model with Delta and hidden color channel coupling. Nucl. Phys. A  {\bf 352}, 326 (1981).
\bibitem{Vijande:2009kj} J. Vijande, A. Valcarce and N. Barnea. Exotic meson-meson molecules and compact four-quark states. Phys. Rev. D {\bf 79}, 074010 (2009).
\bibitem{Hiyama:2003cu} E. Hiyama, Y. Kino and M. Kamimura. Gaussian expansion method for few-body systems. Prog. Part. Nucl. Phys. {\bf 51}, 223 (2003).
\bibitem{tfcjvav2019} T. F. Caram\'es, J. Vijande and A. Valcarce. Exotic $bc\bar{q}\bar{q}$ four-quark states. Phys. Rev. D {\bf 99}, 014006 (2019).
\bibitem{KTCSL2020} K. Chao and S. Zhu. The possible tetraquark states $cc\bar{c}\bar{c}$ observed by the LHCb experiment. arXiv: 2008.07670 [hep-ph].
\end{thebibliography}
\end{document}